\documentclass[final,sort&compress]{elsarticle}

\usepackage{graphicx}% Include figure files
\usepackage{dcolumn}% Align table columns on decimal point
\usepackage{bm}% bold math
\usepackage{verbatim}
\usepackage{amssymb}
\usepackage{amsmath}
\usepackage[top=2.0cm,bottom=2.0cm,left=2.0cm,right=2.0cm]{geometry}

\journal{Annals of Physics}

\allowdisplaybreaks

\begin{document}

\begin{frontmatter}

\title{Two-particle bound states on a lattice}

\author{Pavel E. Kornilovitch}
 \ead{pavel.kornilovich@gmail.com}
 \affiliation{Department of Physics, Oregon State University, Corvallis, Oregon 97331, USA} 
 %Lines break automatically or can be forced with \\

\begin{abstract}

Two-particle lattice states are important for physics of magnetism, superconducting oxides, and cold quantum gases. The quantum-mechanical lattice problem is exactly solvable for finite-range interaction potentials. A two-body Schr\"odinder equation can be reduced to a system of linear equations whose numbers scale with the number of interacting sites. For the simplest cases such as on-site or nearest-neighbor attractions, many pair properties can be derived analytically, although final expressions can be quite complicated. In this work, we systematically investigate bound pairs in one-, two-, and three-dimensional lattices. We derive pairing conditions, plot phase diagrams, and compute energies, effective masses and radii. Along the way, we analyze nontrivial physical effects such as light pairs and the dependence of binding thresholds on pair momenta. At the end, we discuss the preformed-pair mechanism of superconductivity and stability of many-pair systems against phase separation. The paper is a combination of original work and pedagogical tutorial.    

\end{abstract}

%\keywords{Two-particle states, bound states, lattice, preformed pairs, superconductivity}       % Use showkeys class option if keyword display desired

%\maketitle


\end{frontmatter}

%\begin{onecolumn}

\tableofcontents

%\end{onecolumn}

%\begin{twocolumn}

\section{\label{twopart:sec:one}
Introduction
}

In 1986, Daniel Mattis published a review on few-body lattice problems~\cite{Mattis1986}. The motivation for that work was deeper understanding of magnetism. Many successful models of magnetism are formulated in terms of localized spins arranged in regular lattices~\cite{Mattis1981}. Exact results obtained for two and three interacting magnons provided valuable physical insights. Mattis also mentioned other areas that could benefit from similar analysis, specifically excitons and superconductivity (`$\ldots$ do Cooper pairs have bound states of ``Cooper molecules''?', \cite{Mattis1986}, p. 362). At about the same time, high-temperature superconductivity (HTSC) was discovered by Bednorz and M\"uller \cite{Bednorz1986}, which generated enormous interest in unconventional theories of superconductivity. The unusual HTSC properties such as a low carrier density, a short coherence length~\cite{Worthington1987}, and a Bose-gas-like scaling of the magnetic penetration depth with $T_c$~\cite{Uemura1989,Uemura1991} revived the {\em pre-BCS} proposal~\cite{Ogg1946,Schafroth1954a,Schafroth1954b,Schafroth1957} that superconductivity could be a Bose-Einstein condensation (BEC) of charged bosons formed by binding of electrons or holes into real-space pairs. Many properties of superconducting oxides have been shown to be well described by charged Bose-gas phenomenology~\cite{Micnas1990,Alexandrov1993b,Alexandrov1994,Geshkenbein1997,Alexandrov1999b,Chen2005}. More recently, normal-state pairs were observed in iron-based superconductors~\cite{Seo2019,Kang2020} and in the shot noise in copper oxide junctions~\cite{Zhou2019}. The debates about the {\em preformed pairs} mechanism of superconductivity and its relevance to HTSC and pseudogap physics continue today~\cite{Chen2022b}.

Physics of real-space pairs is tightly linked with the pairing mechanism. In superconducting oxides, main candidates are: strong electron-phonon interaction~\cite{Alexandrov2013}, the Jahn--Teller effect~\cite{Bednorz1986,Mihailovich2001,Kabanov2002,Mihailovic2017}, and spin fluctuations~\cite{Scalapino2012}, although other mechanisms and combinations have been proposed~\cite{Micnas1990,Callaway1989,Bussmann1989,Berciu2009,Yacoby2021}. However, the complexity of the problem suggests splitting one big puzzle into smaller ones. By postulating a simple phenomenological attraction of {\em some} kind, one can investigate the BEC-BCS crossover~\cite{Chen2005,Nozieres1985,Paiva2004}, the physics of pseudogap, the role of anisotropy~\cite{Kornilovitch2015}, phase separation~\cite{Emery1990,Dagotto1993,Kornilovitch2022}, electrodynamics, and other nontrivial topics, all without arguing about the specific nature of pairing interaction. 

An added benefit of this approach is that the two-particle lattice problem is exactly solvable for a wide class of non-retarded interaction potentials, which provides a welcome rigor and often an analytical formula. The binding threshold, pair energy, dispersion, effective mass $m^{\ast}_{p}$, effective radius $r^{\ast}_{p}$, and wave function: all can be determined without approximations. These properties provide important physical insights. For example, mass anisotropy of {\em pairs} differs from the bare anisotropy of the member particles, see, e.g., Sections~\ref{twopart:sec:threesix} and \ref{twopart:sec:eleven}. This relationship may be helpful in understanding the anisotropy of transport and electromagnetic properties of HTSC. Additionally, pair wave function is directly proportional to a macroscopic superconducting order parameter~\cite{Bogoliubov1970}, In particular, both share the same orbital symmetry. Knowledge of the exact pair wave function helps elucidate the relationship between the order parameter and other elements of the system. For example, it was proposed~\cite{Bak1999} that correlation-induced diagonal hopping may produce a $d$-symmetric ground-state pair.

\begin{figure*}[t]
\begin{center}
\includegraphics[width=0.95\textwidth]{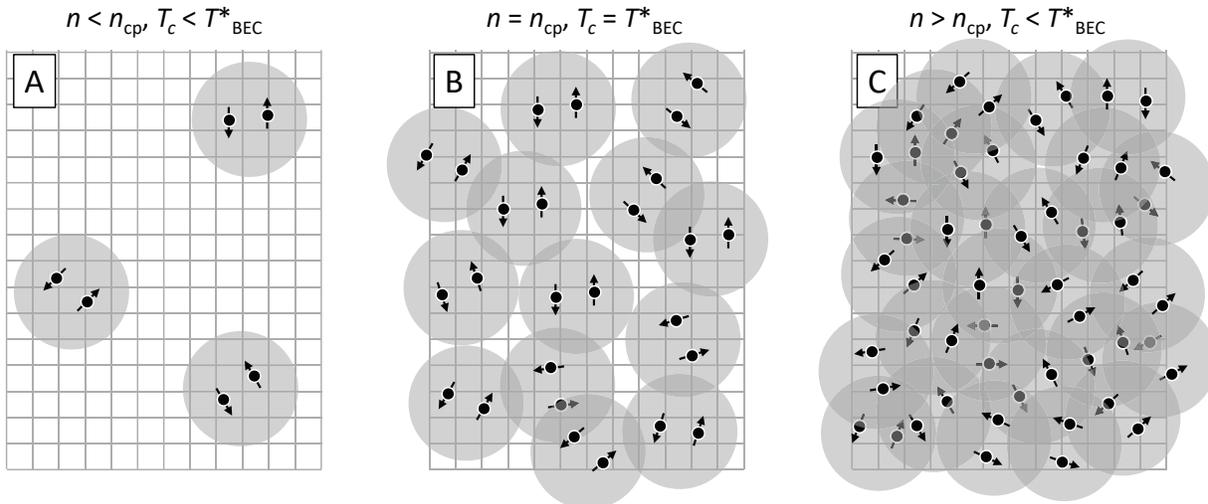}
\caption{Evolution of BEC of real-space pairs with density. $n_{\rm cp} = \Omega^{-1}_{p}$ is the close-packing density, $T_{c}$ is a ``critical temperature of phase transition''. $T^{\ast}_{\rm BEC}$ is determined by the mass and size of a single pair, see Eq.~(\ref{twopart:eq:oneone}). The symmetry of a macroscopic order parameter is defined by the pair wave function~\cite{Bogoliubov1970}. } 
\end{center}
\label{twopart:fig:thirtyone}
\end{figure*}

A striking application of two-particle properties to unconventional superconductivity comes from analyzing the BEC temperature of pairs:           
\begin{equation}
T_{\rm BEC} = C \, \frac{\hbar^2}{m^{\ast}_{p}} \, n^{2/3}_p \: .  
\label{twopart:eq:one}
\end{equation}
Here, $C$ is a numerical coefficient and $n_p$ is the pair number density assumed to be known from chemical composition. $m^{\ast}_{p}$ is the pair mass provided by an exact solution. Consider system's evolution with $n_{p}$ shown in Fig.~\ref{twopart:fig:thirtyone}. (A) At low $n_{p}$, the average distance between pairs is larger than their size, and Eq.~(\ref{twopart:eq:one}) applies. (B) With $n_{p}$ increasing, the system reaches ``close-packing'' when the pairs begin to overlap. The corresponding density is approximately given by an inverse pair volume, $n_{\rm cp} = \Omega^{-1}_{p} = (r^{\ast}_{p})^{-3}$. The pairs interact strongly but the phase transition is still of BEC type with a transition temperature approximately given by 
\begin{equation}
T_{\rm BEC}(n_{\rm cp}) = T^{\ast}_{\rm BEC} = 
C \, \frac{\hbar^2}{m^{\ast}_{p}} \frac{1}{r^{\ast 2}_{p}} \propto 
\frac{1}{ m^{\ast}_{p} \, r^{\ast 2}_{p} } \: .  
\label{twopart:eq:oneone}
\end{equation}
(C) Upon further increase of density, the pairs overlap much more, with the average distance smaller than the pair size. The constituent fermions begin to form a Fermi sea~\cite{Chen2019}, and the phase transition gradually shifts to the BCS type. Equation~(\ref{twopart:eq:one}) no longer applies and $T_{c}$ begins to fall. Thus, the ``maximal attainable'' transition temperature is given by Eq.~(\ref{twopart:eq:oneone})~\cite{Kornilovitch2015,Ivanov1994,Zhang2022}. Remarkably, the latter contains only the properties of a single pair, with both $m^{\ast}_{p}$ and $r^{\ast}_{p}$ supplied by the exact solution. Then, various effects on $T^{\ast}_{\rm BEC}$ may be studied rigorously. This methodology was applied, for example, to understanding the effects of interlayer hopping in Ref.~\cite{Kornilovitch2015}. This topic is the subject of Section~\ref{twopart:sec:eleven}.  

Early developments of short-range attractive models of superconductivity~\cite{Micnas1988a,Micnas1988b,Micnas1989} were summarized by Micnas, Ranninger, and Robaszkiewicz~\cite{Micnas1990}. Those authors considered two groups of models. The first group included {\em on-site} attraction, and was essentially derived from the attractive Hubbard model. Those models were useful to follow the BCS-BEC crossover, to understand system's thermodynamics, electrodynamics and other properties. At the same time, those models were too simplistic to describe real superconducting materials. It is hard to come up with a physical mechanism strong enough to overcome Coulomb repulsion between charge carriers and still keep the pairs mobile. The second group of models included {\em intersite} attraction and were more realistic. Moving attraction to finite distances allowed keeping a strong repulsive core (Hubbard repulsion) that could model a screened Coulomb repulsion. In addition, intersite models add the possibility of antiferromagnetic ordering, of $p$- and $d$-pairing, and make a better connection to lattice models of magnetism. Such models with on-site repulsion and intersite attraction will be the main focus of the present work. They will be referred hereafter as ``$UV$ models''. These models continue to be actively investigated today~\cite{Nayak2018,Boudjada2020,Kheirkhah2020,Jiang2021,Zhang2021,Qu2021,Singh2021,Huang2021,Chen2022,Peng2023,Linner2023,Anjos2023,Sun2023}.

Even nearest-neighbor attraction is an approximation to inter-electron or inter-hole attractions of real materials. The latter results from mediation by an intermediary subsystem such as phonons, and typically extend beyond nearest neighbors~\cite{Alexandrov2002A,Alexandrov2002B}. Such two-particle models are still exactly solvable but the complexity of solution increases sharply with the range. One example is analyzed in Section~\ref{twopart:sec:fivethree}. What is important, however, is that $UV$ models comprise the two most essential elements: {\em some} repulsion $U$ representing strong correlations and {\em some} attraction $V$ representing a mediating subsystem. As such, the $UV$ potential is the simplest representative of an entire class of potentials that possess common properties. For example, all such models have a threshold function $V_{\rm cr}(U)$ that separates bound and unbound states and saturates in the $U \rightarrow \infty$ limit. Effects of other factors, such as anisotropy or pair motion, on the delicate balance between the two forces can be studied within the $UV$ model, at least qualitatively.     

Intersite attraction also appear within another popular theory of unconventional superconductivity based on the {\it t-J} model~\cite{Chao1978,Emery1990,Lin1991,Petukhov1992,Dagotto1993,Kagan1994,Chernyshev1999}. In the dilute limit (many holes and few electrons), the exchange interaction is equivalent to a nearest-neighbor attraction and the system reduces to a gas of bound pairs. As the {\it t-J} model has mostly been studied on square lattices, this case will be covered in Section~\ref{twopart:sec:five}. 
        
In the last two decades, optical lattices and cold atoms emerged as another physical realization of models with short-range attraction~\cite{Bloch2008,Blume2012,Sowinski2019,Mistakidis2022}. Whereas in the cuprates the $UV$ model is an approximation to real inter-particle potentials, in optical lattices it can be precisely engineered and studied in pure form. The onsite interaction is controlled via Feshbach resonances~\cite{Chin2010} and can be made either repulsive~\cite{Joerdens2008} or attractive~\cite{Strohmaier2007}. The intersite interaction can be controlled either by exciting dressed Rydberg atoms to large quantum numbers~\cite{Hague2012} or via proper alignment of dipolar quantum gases~\cite{Lahaye2009}. Precise manipulation of few particles in optical traps~\cite{Serwane2011,Zurn2012,Zurn2013} and BEC of {\em molecules} in an attractive Fermi gas~\cite{Greiner2003} have been demonstrated. Local pairing can now be measured directly using gas microscopy~\cite{Mitra2018,Hartke2022}.

Bound states of two spin waves also appear in models of quantum magnetism~\cite{Mattis1981,Wortis1963,Hanus1963,Akhiezer1968,Kuzian2007,Mook2023} including frustrated magnetics~\cite{Ueda2009,Zhitomirsky2010,Jiang2022b}.

Basic properties of two-body states in $UV$ models were discussed by Micnas, Ranninger, and Robaszkiewicz~\cite{Micnas1990}. In particular, conditions for pair formation and binding energies were found for 1D, 2D square, and 3D simple cubic $UV$ models. Since then, the field has seen several developments. For example, it was realized that threshold of pair formation depends on the pair momentum~\cite{Kornilovitch2004}. $UV$ models have been solved for triangular~\cite{Bak2007}, tetragonal~\cite{Kornilovitch2015}, BCC~\cite{Adebanjo2021}, and FCC~\cite{Adebanjo2022} lattices. New analytical results obtained for cubic Watson integrals by Joyce and coworkers~\cite{Joyce1994,Joyce1998,Glasser2000,Joyce2001a,Joyce2001b,Joyce2002,Joyce2003,Zucker2011} suggest revisiting the cubic models for deeper analysis. Given also that the HTSC puzzle remains largely unsolved and the recent observation of real-space pairs in the normal state~\cite{Seo2019,Kang2020,Zhou2019}, a fresh review of two-body states on a lattice seems to be worthwhile.        
  
The purpose of this work is twofold. First, we collect results obtained for different lattices~\cite{Kornilovitch2004,Kornilovitch2015,Kornilovitch1995,Kornilovitch1997,Bak2007,Hague2010,Davenport2012,Adebanjo2021,Adebanjo2022} and develop them systematically in one place and with unified formalism. Second, we present a large body of new results that were developed over the last 30 years but remained unpublished until now. Most of the material in the main text is original, including the general theory of Section~\ref{twopart:sec:two}. If a particular result or a formula is not accompanied by a specific reference to prior work, one should assume the result is published for the first time. The paper has a pedagogical side as well as we included in Appendixes some textbook material to make exposition self-contained. In particular, we collected available information on lattice Green's functions in different geometries. We also included in \ref{twopart:sec:appbthree} an example of utilizing group theory to separate states of different orbital symmetries. In general, we find the two-particle problem to be an excellent primer on non-relativistic quantum mechanics that teaches wave-function symmetry, emergence of complex band dispersions, the Galilean invariance and lack thereof, multi-component wave functions, scattering states, and other topics.   

The paper is organized as follows. After formulating the general theory (Section~\ref{twopart:sec:two}), we analyze bound states in the attractive Hubbard model in different dimensions (Section~\ref{twopart:sec:three}), and then the $UV$ model in 1D (Section~\ref{twopart:sec:four}), 2D (Sections~\ref{twopart:sec:five}, \ref{twopart:sec:twenty}, and \ref{twopart:sec:six}), and 3D (Sections~\ref{twopart:sec:seven} and \ref{twopart:sec:eight}). The next two sections are devoted to two special topics: light pairs (Section~\ref{twopart:sec:appf}), and the stability of pairs against phase separation (Section~\ref{twopart:sec:ten}). Section~\ref{twopart:sec:twelveone} summarizes the most important common physical properties of lattice bound states. Sections~\ref{twopart:sec:eleven} and \ref{twopart:sec:twelvetwo} discuss relevance to HTSC. Extensive Appendixes contain additional information intended for subject matter expects. \ref{twopart:sec:nine} briefly summarizes two-particle problems not covered in this work in detail. \ref{twopart:sec:twothree} explains how the present formalism can be extended to multi-orbital models~\cite{Ivanov1994,Alexandrov1992,Alexandrov1993}. The latter are excluded from this review because they are not yet ready for systematic exposition. The other Appendixes provide explicit expressions for analytically known lattice Green's functions, recipes for efficient numerical integration when Green's functions are not known analytically, a group theory example, and details of several most cumbersome derivations.

\begin{figure}[t]
\begin{center}
\includegraphics[width=0.48\textwidth]{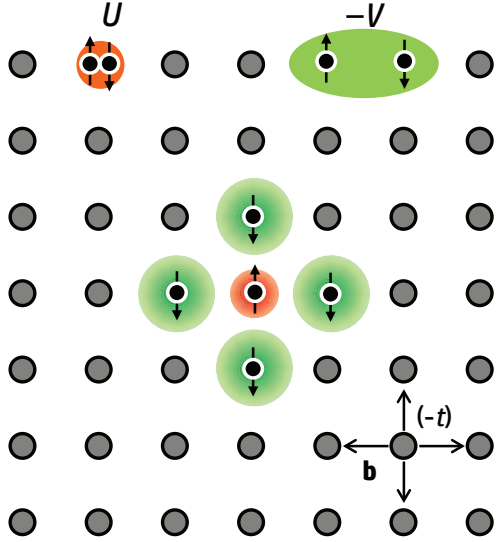}
\end{center}
\caption{Illustration of a $UV$ model on the square lattice with nearest-neighbor attraction and hopping. The cartoon in the center illustrates an $s$-symmetric wave function in which the on-site amplitude is suppressed relative to the inter-site ones.}
\label{twopart:fig:one}
\end{figure}

\section{\label{twopart:sec:two}
General theory 
}

\subsection{\label{twopart:sec:twoone}
Model 
}

We begin with formulating an underlying model. We restrict consideration to Bravais lattices only. (Two-body problems in more complex lattices can also be solved~\cite{Alexandrov1992,Alexandrov1993,Ivanov1994} but the resulting algebraic expressions are much more cumbersome, see \ref{twopart:sec:twothree}.) The inversion symmetry of Bravais lattices allows for easier separation of singlet and triplet pair states, which greatly reduces the complexity of equations. Second, we will consider only finite-range potentials. The two-particle Schr\"odinger equation reduces to an algebraic matrix equation with a size equal to the number of nonzero elements in the inter-particle potential. An infinite-range potential will lead to an infinitely large matrix and the advantages of an exact solution will be lost. Even for finite ranges, the complexity of the matrix solution grows rapidly with the interaction range. In practice, only solutions with short-range potentials are simple enough to produce analytical results. Most space in this paper will be devoted to models with on-site and nearest-neighbor interactions, although one case of a longer range interaction will be discussed in Section~\ref{twopart:sec:fivethree}. A typical $UV$ model is illustrated in Fig.~\ref{twopart:fig:one}. Third, we will consider only spin-$\frac{1}{2}$ fermions. Their coordinate wave function can be either symmetric or antisymmetric; so the solutions will also cover spin-0 bosons and spinless fermions. Finally, bound states will be the main focus, and scattering states will not be discussed.

A second-quantized model Hamiltonian reads 
\begin{equation}
\hat{H} = - \sum_{{\bf m},\bar{\bf b},\sigma} t_{\bar{\bf b}} \, 
 c^{\dagger}_{{\bf m}+\bar{\bf b},\sigma} c_{{\bf m} \sigma} 
+ \frac{U}{2} \sum_{\bf m} \hat{n}_{\bf m} \left( \hat{n}_{\bf m} - 1 \right)  
+ \frac{1}{2} \sum_{{\bf m},{\bf b}} V_{\bf b} \, 
  \hat{n}_{{\bf m} + {\bf b}} \, \hat{n}_{\bf m} \: .  
\label{twopart:eq:two}
\end{equation}
Here $\hat{n}_{\bf m} = \sum_{\sigma} \hat{n}_{{\bf m} \sigma} = c^{\dagger}_{{\bf m}\uparrow} c_{{\bf m} \uparrow} + c^{\dagger}_{{\bf m}\downarrow} c_{{\bf m} \downarrow}$ is the fer\-mi\-on number operator on site ${\bf m}$. $\bar{\bf b}$ and ${\bf b}$ are hopping and interaction neighbor vectors, respectively. The first term in Eq.~(\ref{twopart:eq:two}) is kinetic energy of free fermions defined by spin-independent hopping integrals $t_{\bar{\bf b}}$. We will invariably consider only negative hopping integrals, which is reflected by explicitly writing a negative sign in front of the sum. Therefore, $t_{\bar{\bf b}} > 0$ and $t_{-\bar{\bf b}} = t_{\bar{\bf b}}$ for all $\bar{\bf b}$. The energy of atomic orbitals is used as a zero energy, and the corresponding term is not written. The second term is the on-site (Hubbard) repulsion with amplitude $U$. Because of the property $\hat{n}^{2}_{\sigma} = \hat{n}_{\sigma}$, it is equivalent to the usual form $U \sum_{\bf m} \hat{n}_{{\bf m} \uparrow} \hat{n}_{{\bf m} \downarrow}$. The last term in Eq.~(\ref{twopart:eq:two}) represents nearest-neighbor interaction, where $V_{-{\bf b}} = V_{\bf b}$. The prefactor $\frac{1}{2}$ is included to compensate double-counting. Equation~(\ref{twopart:eq:two}) is written in the $UV$ form to emphasize the special role of on-site interaction $U$. For most of the paper, we will set the lattice spacing to one, $a = 1$, only restoring $a$ in places where it provides additional physical insights.

\subsection{\label{twopart:sec:twotwo}
Unsymmetrized solution
}

In first quantization, two-body wave function $\Psi({\bf m}_1,{\bf m}_2)$ must satisfy the Schr\"odinger equation:
\begin{equation}
- \sum_{\bar{\bf b}} t_{\bar{\bf b}} \left[ \Psi( {\bf m}_1 + \bar{\bf b} , {\bf m}_2 ) + 
                                \Psi( {\bf m}_1 , {\bf m}_2 + \bar{\bf b} ) \right] 
+ U \, \delta_{{\bf m}_1 , {\bf m}_2} \Psi({\bf m}_1,{\bf m}_2)        
+ \sum_{\bf b} V_{\bf b} \, \delta_{{\bf m}_1 - {\bf m}_2 , {\bf b}} \Psi({\bf m}_1,{\bf m}_2) 
= E \, \Psi({\bf m}_1,{\bf m}_2) \: ,  
\label{twopart:eq:four}
\end{equation}
where $E$ is the total energy. Equation~(\ref{twopart:eq:four}) is converted in momentum space using the transformation
\begin{align}
\Psi({\bf m}_1, {\bf m}_2) & = \frac{1}{N} \sum_{{\bf k}_1 {\bf k}_2} 
\psi_{{\bf k}_1 {\bf k}_2} \, e^{ i {\bf k}_1 {\bf m}_1 + i {\bf k}_2 {\bf m}_2 }   \, ,  
\label{twopart:eq:five} \\
\psi_{{\bf k}_1 {\bf k}_2} & = \frac{1}{N} \! \sum_{{\bf m}_1 {\bf m}_2} \!\! 
\Psi({\bf m}_1, {\bf m}_2) \, e^{ - i {\bf k}_1 {\bf m}_1 - i {\bf k}_2 {\bf m}_2 } \, ,
\label{twopart:eq:fiveone}
\end{align}
where $N$ is the total number of lattice unit cells. A transformed equation reads 
\begin{equation}
( E - \varepsilon_{{\bf k}_1} - \varepsilon_{{\bf k}_2} ) \, \psi_{{\bf k}_1 {\bf k}_2} =  
U \frac{1}{N} \sum_{\bf q} \psi_{{\bf q}, {\bf k}_1 + {\bf k}_2 - {\bf q} } 
+ \frac{1}{N} \sum_{{\bf b} {\bf q}} V_{\bf b} \, e^{ i ( {\bf q} - {{\bf k}_1} ) {\bf b} } \, 
\psi_{{\bf q}, {\bf k}_1 + {\bf k}_2 - {\bf q} } \: .  
\label{twopart:eq:six}
\end{equation}
Here,
\begin{equation}
\varepsilon_{\bf k} = - \sum_{\bar{\bf b}} t_{\bar{\bf b}} \, e^{ i {\bf k} \bar{\bf b} }   
\label{twopart:eq:seven}
\end{equation}
is the one-particle dispersion of the model. The right-hand-side of Eq.~(\ref{twopart:eq:six}) is a linear combination of quantities 
\begin{align}
\Phi_{\bf 0}({\bf k}_1 + {\bf k}_2) & = \Phi_{\bf 0}({\bf P}) \equiv 
\frac{1}{N} \sum_{\bf q} \psi_{ {\bf q} , {\bf k}_1 + {\bf k}_2 - {\bf q} } 
= \frac{1}{N} \sum_{\bf q} \psi_{ {\bf q} , {\bf P} - {\bf q} } \: ,    
\label{twopart:eq:eight} \\
\Phi_{\bf b}({\bf k}_1 + {\bf k}_2) & = \Phi_{\bf b}({\bf P}) \equiv 
\frac{1}{N} \sum_{\bf q} e^{ i {\bf q} {\bf b} } \, \psi_{ {\bf q} , {\bf k}_1 + {\bf k}_2 - {\bf q} } 
= \frac{1}{N} \sum_{\bf q} e^{ i {\bf q} {\bf b} } \, \psi_{ {\bf q} , {\bf P} - {\bf q} } \: ,    
\label{twopart:eq:eightone}
\end{align}
where 
\begin{equation}
{\bf P} = {\bf k}_1 + {\bf k}_2 \: ,    
\label{twopart:eq:nine}
\end{equation}
is the total lattice momentum of the two fermions. It is critically important for the existence of an exact solution that $\Phi_{\bf 0}$ and $\Phi_{\bf b}$ are functions of only {\em one} argument ${\bf P}$ rather than two separate arguments ${\bf k}_1$ and ${\bf k}_2$. Utilizing the definitions, Eqs.~(\ref{twopart:eq:eight}) and (\ref{twopart:eq:eightone}), the wave function is expressed from Eq.~(\ref{twopart:eq:six}) as follows
\begin{equation}
\psi_{{\bf k}_1 {\bf k}_2} =  
\frac{ U }{ E - \varepsilon_{{\bf k}_1} - \varepsilon_{{\bf k}_2} } \, \Phi_{\bf 0}( {\bf P} )
+ \sum_{\bf b} V_{\bf b} \, 
\frac{e^{ - i {\bf k}_1 {\bf b} }}{ E - \varepsilon_{{\bf k}_1} - \varepsilon_{{\bf k}_2} } \,   
\Phi_{\bf b}( {\bf P} ) \: .  
\label{twopart:eq:ten}
\end{equation}
Substitution of Eq.~(\ref{twopart:eq:ten}) back in Eq.~(\ref{twopart:eq:eight}) yields a system of linear {\em algebraic} equations for $\Phi$:
\begin{align}
\Phi_{\bf 0}({\bf P}) & = - U M_{\bf 00} \, \Phi_{\bf 0}({\bf P}) 
- \sum_{{\bf b}^{\prime}} V_{{\bf b}^{\prime}} \, 
M_{{\bf 0} {\bf b}^{\prime}}(E,{\bf P}) \, \Phi_{{\bf b}^{\prime}}( {\bf P} ) \: ,  
\label{twopart:eq:eleven} \\
\Phi_{\bf b}({\bf P}) & = - U M_{\bf b0} \, \Phi_{\bf 0}({\bf P})  
- \sum_{{\bf b}^{\prime}} V_{{\bf b}^{\prime}} \, 
M_{{\bf b} {\bf b}^{\prime}}(E,{\bf P}) \, \Phi_{{\bf b}^{\prime}}( {\bf P} ) \: ,  
\label{twopart:eq:elevenone}
\end{align}
where
\begin{align}
M_{\bf 00}( E , {\bf P} ) & = \frac{1}{N} \sum_{\bf q}  
\frac{ 1 }{ - E + \varepsilon_{\bf q} + \varepsilon_{ {\bf P} - {\bf q} } } \: ,   
\hspace{1.0cm} 
M_{{\bf 0} {\bf b}^{\prime}}( E , {\bf P} ) = \frac{1}{N} \sum_{\bf q}  
\frac{ e^{ - i {\bf q} {\bf b}^{\prime} } }
     { - E + \varepsilon_{\bf q} + \varepsilon_{ {\bf P} - {\bf q} } } \: ,   
\label{twopart:eq:twelve}      \\
M_{\bf b0}( E , {\bf P} ) & = \frac{1}{N} \sum_{\bf q}  
\frac{ e^{ i {\bf q} {\bf b} } }
     { - E + \varepsilon_{\bf q} + \varepsilon_{ {\bf P} - {\bf q} } } \: ,   
\hspace{1.0cm}
M_{{\bf b} {\bf b}^{\prime}}( E , {\bf P} ) = \frac{1}{N} \sum_{\bf q}  
\frac{ e^{ i {\bf q} ( {\bf b} - {\bf b}^{\prime} ) } }
     { - E + \varepsilon_{\bf q} + \varepsilon_{ {\bf P} - {\bf q} } } \: .   
\label{twopart:eq:twelveone}
\end{align}
Notice how in the process of substitution, ${\bf k}_1$ is replaced by ${\bf q}$, and ${\bf k}_2$ by ${\bf P} - {\bf q}$. But the argument of $\Phi$ remains unchanged: ${\bf P} = {\bf k}_1 + {\bf k}_2 \rightarrow {\bf q} + {\bf P} - {\bf q} = {\bf P}$. As a result, $\Phi({\bf P})$ can be moved outside the ${\bf q}$ sums, and the equations reduce from integral to algebraic. Thus the entire method is predicated on conservation of total momentum.  

Quantities $M_{{\bf b} {\bf b}^{\prime}}$ in Eqs.~(\ref{twopart:eq:twelve}) and (\ref{twopart:eq:twelveone}) are {\em two-body} Green's functions of underlying lattices. In some lattices, $M_{{\bf b} {\bf b}^{\prime}}$ can be reduced to one-body Green's functions by an appropriate transformation. In solving a typical two-body problem, most effort is spent on calculating and analyzing $M_{{\bf b} {\bf b}^{\prime}}$, and the existence of close-form final formulas depends on whether $M_{{\bf b} {\bf b}^{\prime}}$ can be evaluated analytically. In 1D, integration is elementary leading to algebraic expressions. In 2D, $M_{{\bf b} {\bf b}^{\prime}}$ can be usually reduced to the complete elliptic integrals of different kinds. And in 3D, $M_{{\bf b} {\bf b}^{\prime}}$ are generalized Watson integrals for which many new analytical results have been obtained in the last 30 years~\cite{Joyce1994,Joyce1998,Glasser2000,Joyce2001a,Joyce2001b,Joyce2002,Joyce2003,Zucker2011}. (As mentioned in the Introduction, the latest advances in mathematical physics is one of the reasons that prompted this paper.) Much of the Appendixes are devoted to deriving and listing available results for $M_{{\bf b} {\bf b}^{\prime}}$ on different lattices. Note that we have flipped the sign of the energy denominator in Eq.~(\ref{twopart:eq:twelve}) for future convenience. In this work, we will be interested only in bounds states with $E < 0$. The definitions, Eqs.~(\ref{twopart:eq:twelve}) and (\ref{twopart:eq:twelveone}), render most of $M_{{\bf b} {\bf b}^{\prime}}$ positive. In the following, we will often write $\vert E \vert$ instead of $- E$ to avoid any confusion.       

The consistency condition of Eqs.~(\ref{twopart:eq:eleven}) and (\ref{twopart:eq:elevenone}), 
\begin{equation}
{\rm det} \left\vert \!\! 
\begin{array}{cc}
  U M_{\bf 00}(E,{\bf P}) + 1 &   V_{{\bf b}^{\prime}} \, M_{{\bf 0} {\bf b}^{\prime}}(E,{\bf P}) \\
  U M_{\bf b0}(E,{\bf P})     &   V_{{\bf b}^{\prime}} \, 
M_{{\bf b} {\bf b}^{\prime}}( E , {\bf P} ) + \delta_{{\bf b} {\bf b}'} 
\end{array} \!\! \right\vert = 0 \: ,  
\label{twopart:eq:thirteen}
\end{equation}
determines system's energy $E$ as a function of total momentum ${\bf P}$ and, consequently, the pair's energy, dispersion, and effective mass. If $n_{\bf b}$ is the number of vectors ${\bf b}$ with nonzero $V_{\bf b}$, then in Eq.~(\ref{twopart:eq:thirteen}), $M_{\bf b0}$ is an $( n_{\bf b} \times 1 )$ column, $V_{{\bf b}^{\prime}} \, M_{{\bf 0} {\bf b}^{\prime}}$ is a $( 1 \times n_{\bf b} )$ row, and $V_{{\bf b}^{\prime}} \, M_{{\bf b} {\bf b}^{\prime}}$ is an $( n_{\bf b} \times n_{\bf b} )$ matrix where each column is multiplied by its respective $V_{{\bf b}^{\prime}}$. The eigenvector of Eqs.~(\ref{twopart:eq:eleven}) and (\ref{twopart:eq:elevenone}) determines a pair wave function via Eq.~(\ref{twopart:eq:ten}). Equations~(\ref{twopart:eq:ten})-(\ref{twopart:eq:thirteen}) constitute the general solution of the two-body lattice problem.

\subsection{\label{twopart:sec:twofour}
Symmetrized solution. Spin singlets 
}

The size and complexity of the main system, Eq.~(\ref{twopart:eq:thirteen}), rapidly grows with the radius of interaction. Consider for example the square lattice. For contact, zero-range interaction, Eq.~(\ref{twopart:eq:thirteen}) is a $( 1 \times 1 )$ matrix, for nearest-neighbor $UV$ model it is a $( 5 \times 5 )$ matrix (one zero-range potential plus four nearest-neighbor potentials), and for next-nearest-neighbor $UV$ model it is already a $( 9 \times 9 )$ matrix. The ability to perform analytical calculations diminishes rapidly, especially for ${\bf P}$'s away from special symmetry points. In this situation, {\em permutation} symmetry offers a way to simplify the solution. The two-fermion wave function must be either symmetric or antisymmetric with respect to argument exchange: $\Psi({\bf m}_2,{\bf m}_1) = \pm \Psi({\bf m}_1,{\bf m}_2)$, corresponding to spin-singlet ($+$) and spin-triplet ($-$) pair states. Including this symmetry from the beginning reduces the final system's size by about half. An additional benefit is the $(+)$ solutions also describe bound pairs of spin-$0$ bosons and the $(-)$ solutions describe spinless fermions. In the rest of this section we derive the $(+)$ solution. The $(-)$ solution is derived in Section~\ref{twopart:sec:twofive}. 

In order to restrict the wave function symmetry, we use the following transformation instead of Eq.~(\ref{twopart:eq:five}):
\begin{align}
\Psi({\bf m}_1, {\bf m}_2) & = \frac{1}{2N} \sum_{{\bf k}_1 {\bf k}_2 } 
\left( e^{ i {\bf k}_1 {\bf m}_1 + i {\bf k}_2 {\bf m}_2 } + 
       e^{ i {\bf k}_1 {\bf m}_2 + i {\bf k}_2 {\bf m}_1 } \right)   
\phi^{+}_{ {\bf k}_1 {\bf k}_2 } \: , 
\hspace{0.5cm}
\Psi( {\bf m}_2 , {\bf m}_1 ) = + \Psi( {\bf m}_1 , {\bf m}_2 ) \: ,
\label{twopart:eq:thirteenone} \\
\phi^{+}_{ {\bf k}_1 {\bf k}_2 } & = \frac{1}{2N} \sum_{{\bf m}_1 {\bf m}_2 } 
\left( e^{ - i {\bf k}_1 {\bf m}_1 - i {\bf k}_2 {\bf m}_2 } + 
       e^{ - i {\bf k}_1 {\bf m}_2 - i {\bf k}_2 {\bf m}_1 } \right)   
\Psi({\bf m}_1, {\bf m}_2) \: ,
\hspace{0.5cm}
\phi^{+}_{ {\bf k}_2 {\bf k}_1 } = + \phi^{+}_{ {\bf k}_1 {\bf k}_2 } \: .
\label{twopart:eq:thirteentwo}       
\end{align}
Next, we multiply the Schr\"odinger equation, Eq.~(\ref{twopart:eq:four}), by the expression in parentheses in Eq.~(\ref{twopart:eq:thirteentwo}) and apply the operation $(2N)^{-1} \sum_{{\bf m}_1 {\bf m}_2}$. The result is  
\begin{align}
( E - \varepsilon_{{\bf k}_1} - \varepsilon_{{\bf k}_2} ) \, \phi^{+}_{{\bf k}_1 {\bf k}_2} & =   
U \frac{1}{N} \sum_{\bf q} \phi^{+}_{{\bf q}, {\bf k}_1 + {\bf k}_2 - {\bf q} } 
\nonumber \\
& + \sum_{\bf b} V_{\bf b} \frac{1}{4N} \sum_{\bf q} \! \left[ 
\left( e^{   i {\bf k}_1 {\bf b} } + e^{   i {\bf k}_2 {\bf b} } \right) e^{ - i {\bf q} {\bf b} } \! + 
\left( e^{ - i {\bf k}_1 {\bf b} } + e^{ - i {\bf k}_2 {\bf b} } \right) e^{   i {\bf q} {\bf b} }
\right] \phi^{+}_{{\bf q}, {\bf k}_1 + {\bf k}_2 - {\bf q} } \: .  
\label{twopart:eq:thirteenfour}
\end{align}
The $V$ term contains two groups of ${\bf q}$ integrals: one that contains $\exp{(i{\bf qb})}$ and another $\exp{(-i{\bf qb})}$. One now observes that the two groups transform into each other when ${\bf b} \rightarrow -{\bf b}$. To make use of this symmetry, we arrange all vectors ${\bf b}$ into pairs $( {\bf b} , - {\bf b} )$, then select only one vector from each pair and collect them in new group ${\bf b}_{+}$. Thus, the full group splits in two subgroups: 
\begin{equation}
\{ {\bf b} \} \rightarrow \{ {\bf b}_{+} \} , - \{ {\bf b}_{+} \} \: .  
\label{twopart:eq:thirteenfive}
\end{equation}
For example, in the square lattice with nearest-neighbor interaction, $\{ {\bf b} \} = \{ + {\bf x} , + {\bf y} , - {\bf x} , - {\bf y} \}$ can be split in two pairs, $\{ ( + {\bf x} , - {\bf x} ) , ( + {\bf y} , - {\bf y} ) \}$. Then one can choose $\{ {\bf b}_{+} \}$ out of four equivalent possibilities: $\{ {\bf b}_{+} \} = \{ + {\bf x} , + {\bf y} \}$, $\{ {\bf b}_{+} \} = \{ + {\bf x} , - {\bf y} \}$, $\{ {\bf b}_{+} \} = \{ - {\bf x} , + {\bf y} \}$, and $\{ {\bf b}_{+} \} = \{ - {\bf x} , - {\bf y} \}$. The ${\bf b}$ sum in the $V$ term in Eq.~(\ref{twopart:eq:thirteenfour}) splits in two: 
\begin{align}
V \: {\rm term}  = & \; \sum_{{\bf b}_{+}} V_{{\bf b}_{+}} \frac{1}{4N} \sum_{\bf q} \left[ 
\left( e^{   i {\bf k}_1 {\bf b}_{+} } + e^{   i {\bf k}_2 {\bf b}_{+} } \right) e^{ - i {\bf q} {\bf b}_{+} } + 
\left( e^{ - i {\bf k}_1 {\bf b}_{+} } + e^{ - i {\bf k}_2 {\bf b}_{+} } \right) e^{   i {\bf q} {\bf b}_{+} }
\right] \phi^{+}_{{\bf q}, {\bf k}_1 + {\bf k}_2 - {\bf q} }   
\nonumber \\
& + \sum_{{\bf b}_{+}} V_{-{\bf b}_{+}} \frac{1}{4N} \sum_{\bf q} \left[ 
\left( e^{ - i {\bf k}_1 {\bf b}_{+} } + e^{ - i {\bf k}_2 {\bf b}_{+} } \right) e^{   i {\bf q} {\bf b}_{+} } + 
\left( e^{   i {\bf k}_1 {\bf b}_{+} } + e^{   i {\bf k}_2 {\bf b}_{+} } \right) e^{ - i {\bf q} {\bf b}_{+} }
\right] \phi^{+}_{{\bf q}, {\bf k}_1 + {\bf k}_2 - {\bf q} }  
\nonumber \\
 = & \; \sum_{{\bf b}_{+}} V_{{\bf b}_{+}} \frac{1}{2N} \sum_{\bf q} \left[ 
\left( e^{   i {\bf k}_1 {\bf b}_{+} } + e^{   i {\bf k}_2 {\bf b}_{+} } \right) e^{ - i {\bf q} {\bf b}_{+} } + 
\left( e^{ - i {\bf k}_1 {\bf b}_{+} } + e^{ - i {\bf k}_2 {\bf b}_{+} } \right) e^{   i {\bf q} {\bf b}_{+} }
\right] \phi^{+}_{{\bf q}, {\bf k}_1 + {\bf k}_2 - {\bf q} } \: .  
\label{twopart:eq:thirteensix}
\end{align}
The last equality is true because we consider only symmetric potentials, $V_{{\bf b}_{+}} = V_{-{\bf b}_{+}}$. In the next crucial step, we apply a variable change ${\bf q}^{\prime} = {\bf k}_1 + {\bf k}_2 - {\bf q}$ to the first half of Eq.~(\ref{twopart:eq:thirteensix}). That renders the exponential terms equal to the exponential terms of the second half but the wave function transforms to $\phi^{+}_{{\bf k}_1 + {\bf k}_2 - {\bf q} , {\bf q} }$. However due to the permutation symmetry it is equal to $\phi^{+}_{{\bf q}, {\bf k}_1 + {\bf k}_2 - {\bf q} }$. This proves that the two halves of Eq.~(\ref{twopart:eq:thirteensix}) are equal. The entire $V$ term can be written as twice the second half (for example). Returning to the full equation, Eq.~(\ref{twopart:eq:thirteenfour}), it reads
\begin{equation}
( E - \varepsilon_{{\bf k}_1} - \varepsilon_{{\bf k}_2} ) \, \phi^{+}_{{\bf k}_1 {\bf k}_2}  =    
U \frac{1}{N} \sum_{\bf q} \phi^{+}_{{\bf q}, {\bf k}_1 + {\bf k}_2 - {\bf q} } 
+ \sum_{{\bf b}_{+}} V_{{\bf b}_{+}} \left( e^{ - i {\bf k}_1 {\bf b}_{+} } + e^{ - i {\bf k}_2 {\bf b}_{+} } \right)
\frac{1}{N} \sum_{\bf q} e^{ i {\bf q} {\bf b}_{+} }
\phi^{+}_{{\bf q}, {\bf k}_1 + {\bf k}_2 - {\bf q} } \: .  
\label{twopart:eq:thirteenseven}
\end{equation}
Equation~(\ref{twopart:eq:thirteenseven}) has only half as many $V$ terms as its unsymmetrized counterpart, Eq.~(\ref{twopart:eq:six}). Accordingly, we introduce auxiliary functions
\begin{align}
\Phi^{+}_{\bf 0}({\bf k}_1 + {\bf k}_2) & = \Phi^{+}_{\bf 0}({\bf P}) \equiv 
\frac{1}{N} \sum_{\bf q} \phi^{+}_{ {\bf q} , {\bf k}_1 + {\bf k}_2 - {\bf q} } =    
\frac{1}{N} \sum_{\bf q} \phi^{+}_{ {\bf q} , {\bf P} - {\bf q} } \: ,    
\label{twopart:eq:thirteeneight} \\
\Phi^{+}_{\bf b_{+}}({\bf k}_1 + {\bf k}_2) & = \Phi^{+}_{\bf b_{+}}({\bf P}) \equiv 
\frac{1}{N} \sum_{\bf q} e^{ i {\bf q} {\bf b}_{+} } \, \phi^{+}_{ {\bf q} , {\bf k}_1 + {\bf k}_2 - {\bf q} } =    
\frac{1}{N} \sum_{\bf q} e^{ i {\bf q} {\bf b}_{+} } \, \phi^{+}_{ {\bf q} , {\bf P} - {\bf q} } \: .   
\label{twopart:eq:thirteennine}
\end{align}
The wave function is expressed from Eq.~(\ref{twopart:eq:thirteenseven}) 
\begin{equation}
\phi^{+}_{{\bf k}_1 {\bf k}_2} = 
\frac{ U }{ E - \varepsilon_{{\bf k}_1} - \varepsilon_{{\bf k}_2} } \, \Phi^{+}_{\bf 0}( {\bf P} )
+ \sum_{{\bf b}_{+}} V_{{\bf b}_{+}} \, 
\frac{ e^{ - i {\bf k}_1 {\bf b}_{+}} + e^{ - i {\bf k}_2 {\bf b}_{+}} }
{ E - \varepsilon_{{\bf k}_1} - \varepsilon_{{\bf k}_2} } \,   
\Phi^{+}_{{\bf b}_{+}}( {\bf P} ) \: .  
\label{twopart:eq:thirteenten}
\end{equation}
Substituting $\phi^{+}$ back in the definitions of $\Phi^{+}$, one obtains:
\begin{align}
\Phi^{+}_{\bf 0}({\bf P}) & =  - U M^{+}_{\bf 00} \, \Phi^{+}_{\bf 0}({\bf P}) 
- \sum_{{\bf b}^{\prime}_{+}} V_{{\bf b}^{\prime}_{+}} \, 
M^{+}_{{\bf 0} {\bf b}^{\prime}_{+}}(E,{\bf P}) \, \Phi^{+}_{{\bf b}^{\prime}_{+}}( {\bf P} ) \: ,  
\label{twopart:eq:thirteeneleven} \\
\Phi^{+}_{{\bf b}_{+}}({\bf P}) & =  - U M^{+}_{{\bf b}_{+} {\bf 0}} \, \Phi^{+}_{\bf 0}({\bf P})  
- \sum_{{\bf b}^{\prime}_{+}} V_{{\bf b}^{\prime}_{+}} \, 
M^{+}_{{\bf b}_{+} {\bf b}^{\prime}_{+}}(E,{\bf P}) \, \Phi^{+}_{{\bf b}^{\prime}_{+}}( {\bf P} ) \: ,  
\label{twopart:eq:thirteentwelve}
\end{align}
where
\begin{align}
M^{+}_{\bf 00}( E , {\bf P} ) & = \frac{1}{N} \sum_{\bf q}  
\frac{ 1 }{ - E + \varepsilon_{\bf q} + \varepsilon_{ {\bf P} - {\bf q} } } \: ,   
\hspace{0.5cm}
M^{+}_{{\bf 0} {\bf b}^{\prime}_{+}}( E , {\bf P} ) = \frac{1}{N} \sum_{\bf q}  
\frac{ e^{ - i {\bf q} {\bf b}^{\prime}_{+} } + e^{ - i ( {\bf P} - {\bf q} ) {\bf b}^{\prime}_{+} } }
     { - E + \varepsilon_{\bf q} + \varepsilon_{ {\bf P} - {\bf q} } } \: ,   
\label{twopart:eq:thirteenthirtheen} \\
M^{+}_{{\bf b}_{+} {\bf 0}}( E , {\bf P} ) & = \frac{1}{N} \sum_{\bf q}  
\frac{ e^{ i {\bf q} {\bf b}_{+} } }
     { - E + \varepsilon_{\bf q} + \varepsilon_{ {\bf P} - {\bf q} } } \: ,   
\hspace{0.5cm}
M^{+}_{{\bf b}_{+} {\bf b}^{\prime}_{+}}( E , {\bf P} ) = \frac{1}{N} \sum_{\bf q}  
\frac{ e^{ i {\bf q} ( {\bf b}_{+} - {\bf b}^{\prime}_{+} ) } + 
       e^{ i {\bf q} {\bf b}_{+} } e^{ - i ( {\bf P} - {\bf q} ) {\bf b}^{\prime}_{+} } } 
     { - E + \varepsilon_{\bf q} + \varepsilon_{ {\bf P} - {\bf q} } } \: .   
\label{twopart:eq:thirteenfourteen}
\end{align}
For nonzero ${\bf P}$, it may be more convenient to use an equivalent but a more symmetric formulation. Let us perform a variable change in Eqs.~(\ref{twopart:eq:thirteenthirtheen}) and (\ref{twopart:eq:thirteenfourteen}):  
\begin{equation}
{\bf q} \to {\bf q} + \frac{\bf P}{2} \: , \hspace{1.0cm} 
{\bf P} - {\bf q} \to \frac{\bf P}{2} - {\bf q} \: .
\label{twopart:eq:tenone}
\end{equation}
That leads to the appearance of factors like $\exp{( \pm i {\bf P} {\bf b}_{+}/2 )}$ in front of the ${\bf q}$ sums. They can be absorbed into a new definition of $\Phi^{+}$:
\begin{equation}
{\tilde \Phi}^{+}_{0}({\bf P}) \equiv \Phi^{+}_{0}({\bf P}) \: , \hspace{1.0cm} 
{\tilde \Phi}^{+}_{{\bf b}_{+}}({\bf P}) \equiv e^{ - i \frac{\bf P}{2} {\bf b}_{+} }
\Phi^{+}_{{\bf b}_{+}}({\bf P}) \: .
\label{twopart:eq:tentwo}
\end{equation}
In terms of the new amplitudes, Eqs.~(\ref{twopart:eq:thirteeneleven}) and (\ref{twopart:eq:thirteentwelve}) assume the form
\begin{align}
{\tilde \Phi}^{+}_{\bf 0}({\bf P}) & =  - U {\tilde M}^{+}_{\bf 00} \, {\tilde \Phi}^{+}_{\bf 0}({\bf P}) 
- \sum_{{\bf b}^{\prime}_{+}} V_{{\bf b}^{\prime}_{+}} \, 
{\tilde M}^{+}_{{\bf 0} {\bf b}^{\prime}_{+}}(E,{\bf P}) \, 
{\tilde \Phi}^{+}_{{\bf b}^{\prime}_{+}}( {\bf P} ) \: ,  
\label{twopart:eq:tenthree} \\
{\tilde \Phi}^{+}_{{\bf b}_{+}}({\bf P}) & =  
- U {\tilde M}^{+}_{{\bf b}_{+} {\bf 0}} \, {\tilde \Phi}^{+}_{\bf 0}({\bf P})  
- \sum_{{\bf b}^{\prime}_{+}} V_{{\bf b}^{\prime}_{+}} \, 
{\tilde M}^{+}_{{\bf b}_{+} {\bf b}^{\prime}_{+}}(E,{\bf P}) \, 
{\tilde \Phi}^{+}_{{\bf b}^{\prime}_{+}}( {\bf P} ) \: ,  
\label{twopart:eq:tenfour}
\end{align}
where
\begin{align}
{\tilde M}^{+}_{\bf 00}( E , {\bf P} ) & = \frac{1}{N} \sum_{\bf q}  
\frac{ 1 }{ - E + \varepsilon_{{\bf q} + \frac{\bf P}{2}} + \varepsilon_{ {\bf q} - \frac{\bf P}{2} } } \: ,   
\hspace{0.5cm}
{\tilde M}^{+}_{{\bf 0} {\bf b}^{\prime}_{+}}( E , {\bf P} ) = \frac{1}{N} \sum_{\bf q}  
\frac{ 2 \cos{ ( {\bf q} {\bf b}_{+} ) } }
     { - E + \varepsilon_{{\bf q} + \frac{\bf P}{2}} + \varepsilon_{ {\bf q} - \frac{\bf P}{2} } } \: ,   
\label{twopart:eq:tenfive} \\
{\tilde M}^{+}_{{\bf b}_{+} {\bf 0}}( E , {\bf P} ) & = \frac{1}{N} \sum_{\bf q}  
\frac{ \cos{ ( {\bf q} {\bf b}_{+} ) } }
     { - E + \varepsilon_{{\bf q} + \frac{\bf P}{2}} + \varepsilon_{ {\bf q} - \frac{\bf P}{2} } } \: ,   
\hspace{0.5cm}
{\tilde M}^{+}_{{\bf b}_{+} {\bf b}^{\prime}_{+}}( E , {\bf P} ) = \frac{1}{N} \sum_{\bf q}  
\frac{ 2 \cos{ ( {\bf q} {\bf b}_{+} ) } \cos{ ( {\bf q} {\bf b}^{\prime}_{+} ) } } 
     { - E + \varepsilon_{{\bf q} + \frac{\bf P}{2}} + \varepsilon_{ {\bf q} - \frac{\bf P}{2} } } \: ,   
\label{twopart:eq:tensix}
\end{align}
and the condition $\varepsilon_{- {\bf q}} = \varepsilon_{\bf q}$ has been used. There are several advantages of Eqs.~(\ref{twopart:eq:tenthree})-(\ref{twopart:eq:tensix}) over Eqs.~(\ref{twopart:eq:thirteeneleven})-(\ref{twopart:eq:thirteenfourteen}): (i) All ${\tilde M}$'s are manifestly real which simplifies analytics; (ii) Quite generally, ${\tilde M}^{+}_{{\bf 0} {\bf b}^{\prime}_{+}} = 2 {\tilde M}^{+}_{{\bf b}_{+} {\bf 0}}$; and (iii) Pair momentum ${\bf P}$ only enters the denominators of ${\tilde M}$'s, which simplifies analysis of pair effective mass and dispersion in some cases. Otherwise, the two formulations are equivalent.   

The consistency condition of the linear system, Eqs.~(\ref{twopart:eq:thirteeneleven})-(\ref{twopart:eq:thirteentwelve}) or Eqs.~(\ref{twopart:eq:tenthree})-(\ref{twopart:eq:tenfour}), defines the pair's energy, and its eigenvector defines the pair's wave function. The system comprises only $1 + \frac{1}{2} n_{\bf b}$ linear equations versus $1 + n_{\bf b}$ equations in the unsymmetrized method. This is significant simplification. For example, in the simple-cubic $UV$ model with nearest neighbor interaction, symmetrization reduces the consistency condition from a $(7 \times 7)$ matrix to a $(4 \times 4)$ matrix. The symmetrized solution describes only spin-singlet pairs. Spin triplets are discussed next.

\subsection{\label{twopart:sec:twofive}
Anti-symmetrized solution. Spin triplets 
}

In this section, we repeat the derivation of Section~\ref{twopart:sec:twofour} for anti-symmetric wave functions that describe spin-triplet pairs. The corresponding Fourier transformation reads
\begin{align}
\Psi({\bf m}_1, {\bf m}_2) & =  \frac{1}{2N} \sum_{{\bf k}_1 {\bf k}_2 } 
\left( e^{ i {\bf k}_1 {\bf m}_1 + i {\bf k}_2 {\bf m}_2 } - 
       e^{ i {\bf k}_1 {\bf m}_2 + i {\bf k}_2 {\bf m}_1 } \right)   
\phi^{-}_{ {\bf k}_1 {\bf k}_2 } \: , 
\hspace{0.5cm}
\Psi( {\bf m}_2 , {\bf m}_1 ) = - \Psi( {\bf m}_1 , {\bf m}_2 ) \: ,
\label{twopart:eq:thirteenfifteen} \\
\phi^{-}_{ {\bf k}_1 {\bf k}_2 } & =  \frac{1}{2N} \sum_{{\bf m}_1 {\bf m}_2 } 
\left( e^{ - i {\bf k}_1 {\bf m}_1 - i {\bf k}_2 {\bf m}_2 } - 
       e^{ - i {\bf k}_1 {\bf m}_2 - i {\bf k}_2 {\bf m}_1 } \right)   
\Psi({\bf m}_1, {\bf m}_2) \: ,
\hspace{0.5cm}
\phi^{-}_{ {\bf k}_2 {\bf k}_1 } = - \phi^{-}_{ {\bf k}_1 {\bf k}_2 } \: .
\label{twopart:eq:thirteensixteen}       
\end{align}
We multiply the Schr\"odinger equation, Eq.~(\ref{twopart:eq:four}), by the expression in parentheses in Eq.~(\ref{twopart:eq:thirteensixteen}) and apply operation $(2N)^{-1} \sum_{{\bf m}_1 {\bf m}_2}$. The result is  
\begin{equation}
( E - \varepsilon_{{\bf k}_1} - \varepsilon_{{\bf k}_2} ) \, \phi^{-}_{{\bf k}_1 {\bf k}_2}  =    
\sum_{\bf b} V_{\bf b} \frac{1}{4N} \sum_{\bf q} \! \left[ 
\left( e^{   i {\bf k}_1 {\bf b} } - e^{   i {\bf k}_2 {\bf b} } \right) e^{ - i {\bf q} {\bf b} } \! + 
\left( e^{ - i {\bf k}_1 {\bf b} } - e^{ - i {\bf k}_2 {\bf b} } \right) e^{   i {\bf q} {\bf b} }
\right] \phi^{-}_{{\bf q}, {\bf k}_1 + {\bf k}_2 - {\bf q} } \: .  
\label{twopart:eq:thirteenseventeen}
\end{equation}
Of note here is the absence of a $U$ term that cancels out due to antisymmetry. Next, we split the sum over ${\bf b}$ in two partial sums: over ${\bf b}_{+}$ and over $-{\bf b}_{+}$, and then change variables ${\bf q}^{\prime} = {\bf k}_1 + {\bf k}_2 - {\bf q}$ in the $\exp{(-i{\bf q}{\bf b}_{+})}$ terms. The result is
\begin{equation}
( E - \varepsilon_{{\bf k}_1} - \varepsilon_{{\bf k}_2} ) \, \phi^{-}_{{\bf k}_1 {\bf k}_2}  =    
\sum_{{\bf b}_{+}} V_{{\bf b}_{+}} \left( e^{ - i {\bf k}_1 {\bf b}_{+} } - e^{ - i {\bf k}_2 {\bf b}_{+} } \right)
\frac{1}{N} \sum_{\bf q} e^{ i {\bf q} {\bf b}_{+} }
\phi^{-}_{{\bf q}, {\bf k}_1 + {\bf k}_2 - {\bf q} } \: .  
\label{twopart:eq:thirteeneighteen}
\end{equation}
To convert this to linear equations, we introduce $\frac{1}{2} n_{\bf b}$ auxiliary functions
\begin{equation}
\Phi^{-}_{\bf b_{+}}({\bf k}_1 + {\bf k}_2) = \Phi^{-}_{\bf b_{+}}({\bf P}) \equiv 
\frac{1}{N} \sum_{\bf q} e^{ i {\bf q} {\bf b}_{+} } \, \phi^{-}_{ {\bf q} , {\bf k}_1 + {\bf k}_2 - {\bf q} } =    
\frac{1}{N} \sum_{\bf q} e^{ i {\bf q} {\bf b}_{+} } \, \phi^{-}_{ {\bf q} , {\bf P} - {\bf q} } \: .   
\label{twopart:eq:thirteennineteen}
\end{equation}
Using these definitions, the pair wave function follows from Eq.~(\ref{twopart:eq:thirteeneighteen}):
\begin{equation}
\phi^{-}_{{\bf k}_1 {\bf k}_2} = 
\sum_{{\bf b}_{+}} V_{{\bf b}_{+}} \, 
\frac{ e^{ - i {\bf k}_1 {\bf b}_{+}} - e^{ - i {\bf k}_2 {\bf b}_{+}} }
{ E - \varepsilon_{{\bf k}_1} - \varepsilon_{{\bf k}_2} } \,   
\Phi^{-}_{{\bf b}_{+}}( {\bf P} ) \: .  
\label{twopart:eq:thirteentwenty}
\end{equation}
Substituting $\phi^{-}_{{\bf k}_1 {\bf k}_2}$ back in Eq.~(\ref{twopart:eq:thirteennineteen}), one obtains the final system
\begin{equation}
\Phi^{-}_{{\bf b}_{+}}({\bf P}) =   
- \sum_{{\bf b}^{\prime}_{+}} V_{{\bf b}^{\prime}_{+}} \, 
M^{-}_{{\bf b}_{+} {\bf b}^{\prime}_{+}}(E,{\bf P}) \, \Phi^{-}_{{\bf b}^{\prime}_{+}}( {\bf P} ) \: ,  
\label{twopart:eq:thirteentwentyone}
\end{equation}
\begin{equation}
M^{-}_{{\bf b}_{+} {\bf b}^{\prime}_{+}}( E , {\bf P} ) = \frac{1}{N} \sum_{\bf q}  
\frac{ e^{ i {\bf q} ( {\bf b}_{+} - {\bf b}^{\prime}_{+} ) } - 
       e^{ i {\bf q} {\bf b}_{+} } e^{ - i ( {\bf P} - {\bf q} ) {\bf b}^{\prime}_{+} } } 
     { - E + \varepsilon_{\bf q} + \varepsilon_{ {\bf P} - {\bf q} } } \: .     
\label{twopart:eq:thirteentwentytwo}
\end{equation}
Applying the variable change defined in Eq.~(\ref{twopart:eq:tenone}) and introducing new amplitudes: 
\begin{equation}
{\tilde \Phi}^{-}_{{\bf b}_{+}}({\bf P}) \equiv e^{ - i \frac{\bf P}{2} {\bf b}_{+} }
\Phi^{-}_{{\bf b}_{+}}({\bf P}) \: ,
\label{twopart:eq:teneleven}
\end{equation}
the system is cast in an alternative form: 
\begin{equation}
{\tilde \Phi}^{-}_{{\bf b}_{+}}({\bf P}) =   
- \sum_{{\bf b}^{\prime}_{+}} V_{{\bf b}^{\prime}_{+}} \, 
{\tilde M}^{-}_{{\bf b}_{+} {\bf b}^{\prime}_{+}}(E,{\bf P}) \, 
{\tilde \Phi}^{-}_{{\bf b}^{\prime}_{+}}( {\bf P} ) \: ,  
\label{twopart:eq:tentwelve}
\end{equation}
\begin{equation}
{\tilde M}^{-}_{{\bf b}_{+} {\bf b}^{\prime}_{+}}( E , {\bf P} ) = \frac{1}{N} \sum_{\bf q}  
\frac{ 2 \sin{ ( {\bf q} {\bf b}_{+} ) } \sin{ ( {\bf q} {\bf b}^{\prime}_{+} ) } } 
     { - E + \varepsilon_{{\bf q} + \frac{\bf P}{2}} + \varepsilon_{ {\bf q} - \frac{\bf P}{2} } } \: .     
\label{twopart:eq:tenthirteen}
\end{equation}
The size of the triplet system is one less that of the singlet system because of the lack of the Hubbard term. Therefore, triplet pairs are usually easier to deal with than singlets.

\subsection{\label{twopart:sec:twosix}
Pair size  
}

In this section, we derive a general recipe of calculating pair's effective radius $r^{\ast}_{p}$. We consider only spin-singlet pairs with ${\bf P} = 0$. Other situations can be analyzed similarly. Radius components are defined as follows
\begin{equation}
\left( r^{\ast}_{pj} \right)^2 = \langle m^2_{j} \rangle = 
\frac{ \sum_{\bf m} m^2_{j} \: \Psi^{\ast}({\bf m},0) \Psi({\bf m},0) }
     { \sum_{\bf m}            \Psi^{\ast}({\bf m},0) \Psi({\bf m},0) } 
\equiv \frac{J_{j}}{J_{0}}     \: .   
\label{twopart:eq:twentytwothree}
\end{equation}
The ${\bf k}_1 + {\bf k}_2 = 0$ wave function follows from Eq.~(\ref{twopart:eq:thirteenone}):
\begin{equation}
\Psi( {\bf m} , 0 ) = \frac{1}{2N} \sum_{\bf k} 
\left( e^{ i {\bf k} {\bf m} } + e^{ - i {\bf k} {\bf m} } \right) \phi^{+}_{{\bf k} , - {\bf k} } \: .   
\label{twopart:eq:tentwenty}
\end{equation}
Substitution in the denominator of Eq.~(\ref{twopart:eq:twentytwothree}) yields
\begin{equation}
J_{0} = \sum_{\bf m} \Psi^{\ast}({\bf m},0) \Psi({\bf m},0) 
= \frac{1}{N} \sum_{\bf k} \phi^{+}_{{\bf k} , - {\bf k} } \phi^{+ \ast}_{{\bf k} , - {\bf k} } 
= \frac{1}{N} \sum_{\bf k} \left\vert  \phi^{+}_{{\bf k} , - {\bf k} }  \right\vert^2  .   
\label{twopart:eq:tentwentyone}
\end{equation}
Upon substitution in $J_{j}$, one notices that $m_{j}$ can be expresses as a $k_{j}$ derivative of $e^{ i {\bf k} {\bf m} }$. Taking into account the permutation symmetry of $\phi^{+}_{{\bf k} , - {\bf k} }$, one obtains 
\begin{equation}
J_{j} = \frac{1}{N^2} \sum_{\bf m} 
\left( \sum_{\bf k} \phi^{+}_{{\bf k} , - {\bf k} } \frac{\partial}{\partial k_{j}} e^{ i {\bf k} {\bf m} } \right) 
\left( \sum_{{\bf k}^{\prime}} 
\phi^{+\ast}_{{\bf k}^{\prime} , - {\bf k}^{\prime} } 
\frac{\partial}{\partial k^{\prime}_{j}} e^{ - i {\bf k}^{\prime} {\bf m} } \right) .   
\label{twopart:eq:tentwentytwo}
\end{equation}
Both expressions in parentheses can be transformed utilizing the Green's theorem for periodic functions, see for example Ref.~\cite{Ashcroft1976}. That transfers the derivatives to $\phi^{+}$. The final result for the pair radius reads
\begin{equation}
\left( r^{\ast}_{pj} \right)^2 = \langle m^2_{j} \rangle = \frac{J_{j}}{J_{0}} =
\frac{ \frac{1}{N} \sum_{\bf k} \frac{\partial}{\partial k_{j}} ( \phi^{+}_{ {\bf k} , - {\bf k} } ) 
                                \frac{\partial}{\partial k_{j}} ( \phi^{+\ast}_{ {\bf k} , - {\bf k} } )   }
     { \frac{1}{N} \sum_{\bf k} ( \phi^{+}_{ {\bf k} , - {\bf k} } ) ( \phi^{+\ast}_{ {\bf k} , - {\bf k} } ) }  \: .   
\label{twopart:eq:tentwentythree}
\end{equation}
The wave function is expressible via amplitudes $\Phi^{+}$ using Eq.~(\ref{twopart:eq:thirteenten})
\begin{equation}
\phi^{+}_{{\bf k} , - {\bf k}} = 
\frac{ U }{ E - 2 \varepsilon_{\bf k} } \, \Phi^{+}_{\bf 0}( 0 )
+ \sum_{{\bf b}_{+}} V_{{\bf b}_{+}} \, 
\frac{ 2 \cos{ ( {\bf k} {\bf b}_{+} ) } }{ E - 2 \varepsilon_{\bf k} } \,   
\Phi^{+}_{{\bf b}_{+}}( 0 ) \: .  
\label{twopart:eq:tentwentyfour}
\end{equation}
From here, one derives
\begin{equation}
\frac{\partial}{\partial k_{j}}( \phi^{+}_{{\bf k} , - {\bf k}} ) = 
\left[ \frac{ 2 U }{ ( E - 2 \varepsilon_{\bf k} )^2 } 
\frac{\partial \varepsilon_{\bf k}}{\partial k_{j}} \right] \Phi^{+}_{\bf 0}( 0 )
+ \sum_{{\bf b}_{+}} V_{{\bf b}_{+}} \left[ 
\frac{ - 2 ( b_{+j} ) \sin{ ( {\bf k} {\bf b}_{+} ) } }{ E - 2 \varepsilon_{\bf k} } 
+ \frac{ 4 \cos{ ( {\bf k} {\bf b}_{+} ) } }{ ( E - 2 \varepsilon_{\bf k} )^2 }
\frac{\partial \varepsilon_{\bf k}}{\partial k_{j}} \right] \Phi^{+}_{{\bf b}_{+}}( 0 ) \: .  
\label{twopart:eq:tentwentyfive}
\end{equation}
Thus, calculation of pair size comprises the following steps. (i) For given $U$, $V_{{\bf b}_{+}}$, and $\varepsilon_{\bf k}$, one solves the eigenvalue problem, Eqs.~(\ref{twopart:eq:thirteeneleven})-(\ref{twopart:eq:thirteentwelve}) or Eqs.~(\ref{twopart:eq:tenthree})-(\ref{twopart:eq:tenfour}), for ${\bf P} = 0$. That provides pair energy $E$ and eigenvector $\{ \Phi^{+}_{\bf 0} , \Phi^{+}_{{\bf b}_{+}} \}$. The eigenvector does not have to be normalized since the normalization constant cancels out in the ratio of two integrals. (ii) The wave function and its derivatives are calculated according to Eqs.~(\ref{twopart:eq:tentwentyfour}) and (\ref{twopart:eq:tentwentyfive}). (iii) Everything is substituted in Eq.~(\ref{twopart:eq:tentwentythree}) and the two integrals are computed numerically. This method is more efficient than direct calculation of $\Psi({\bf m},0)$ as a Fourier transform of $\phi^{+}$ followed by a summation over ${\bf m}$. In general, one expects $r^{\ast}_{p}$ to diverge near the binding threshold and to be of order one lattice constant in the strong coupling limit. In the simplest cases, $r^{\ast}_{p}$ can be calculated analytically as shown in the next section.

\section{\label{twopart:sec:three}
Negative-$U$ Hubbard model 
}

\subsection{\label{twopart:sec:threezero}
General expressions  
}

Before getting to more complex $UV$ models, it is instructive to consider the simpler case of zero-range interaction, that is the negative Hubbard model. Several characteristic features of lattice bound states show up already at this level. Additionally, due to the model's relative simplicity, analytical calculations can be carried out to the fullest extent. The model is defined by the potential  
\begin{equation}
U = - \vert U \vert \: , \hspace{1.0cm} V_{\bf b} = 0 \: .   
\label{twopart:eq:twentyone}
\end{equation}
One expects only one singlet bound state, so either unsymmetrized solution, Eq.~(\ref{twopart:eq:eleven}), or symmetrized one, Eq.~(\ref{twopart:eq:thirteeneleven}), can be applied. In both cases, the system reduces to a single equation for $\Phi_{\bf 0}$ with the consistency condition 
\begin{equation}
\vert U \vert \, M_{\bf 00}(E,{\bf P}) = 1 \: ,   
\label{twopart:eq:twentytwo}
\end{equation}
which defines pair energy $E({\bf P})$. The pair wave function is 
\begin{equation}
\psi_{{\bf k}_1 {\bf k}_2} = \frac{1}{ E - \varepsilon_{{\bf k}_1} - \varepsilon_{{\bf k}_2} } \: ,   
\label{twopart:eq:twentyoneone}
\end{equation}
up to a normalization constant. Since total momentum ${\bf P}$ is fixed, $\psi$ is a function of only one argument:
\begin{equation}
\psi_{{\bf P}}({\bf q}) = \frac{1}{ E - \varepsilon_{\bf q} - \varepsilon_{{\bf P} - {\bf q}} } \: .   
\label{twopart:eq:twentyonetwo}
\end{equation}
The real-space wave function follows from Eq.~(\ref{twopart:eq:five})
\begin{equation}
\Psi({\bf m}_1, {\bf m}_2) = \frac{1}{N} \sum_{\bf q}
\frac{ e^{ i {\bf q} {\bf m}_1 + i ( {\bf P} - {\bf q} ) {\bf m}_2 } }
     { E - \varepsilon_{\bf q} - \varepsilon_{{\bf P} - {\bf q}} } 
= e^{ i {\bf P} \frac{ ( {\bf m}_1 + {\bf m}_2 )}{2} } \frac{1}{N} \sum_{\bf q} 
\frac{ e^{ i {\bf q} ( {\bf m}_1 - {\bf m}_2 ) } }
     { E - \varepsilon_{ \frac{{\bf P}}{2} + {\bf q} } - \varepsilon_{ \frac{{\bf P}}{2} - {\bf q}} }  \: .      
\label{twopart:eq:twentyonethree}
\end{equation}
The first factor describes center-of-mass motion while the integral over ${\bf q}$ describes internal structure of the pair. Analysis will continue for different lattices separately.

\subsection{\label{twopart:sec:threetwo}
1D. One dimensional chain  
}

The 1D attractive Hubbard model provides the simplest example of a lattice bound state. Many pair properties can be derived analytically. The basic integral, $M_{\bf 00}$ in Eq.~(\ref{twopart:eq:twelve}), is 
\begin{equation}
M^{\rm 1D}_{\bf 00}  =  \int^{\pi}_{-\pi} \frac{{\rm d} q}{2\pi} 
\frac{1}{ \vert E \vert - 4 t \cos{\left( \frac{P}{2} \right)} \cos{q}} 
= \frac{1}{\sqrt{ E^2 - 16 \, t^2 \cos^2{\left( \frac{P}{2} \right)} }}\: .   
\label{twopart:eq:twentyfive}
\end{equation}
Substitution in Eq.~(\ref{twopart:eq:twentytwo}) yields pair energy
\begin{equation}
E(P) = - \sqrt{ \vert U \vert^2 + 16 \, t^2 \cos^2{ \left( \frac{P}{2} \right) } } \: .   
\label{twopart:eq:twentysix}
\end{equation}
This is a rare case when pair energy is known as an explicit formula. Based on it, a number of interes\-ting properties can be established. (i) The minimum energy of two {\em free} particles with total momentum $P$ is $E_{11} = - 4 t \cos{(P/2)}$. Comparing that with Eq.~(\ref{twopart:eq:twentysix}), one finds $E(P) < E_{11}(P)$ for any $\vert U \vert > 0$. In other words, $\vert U \vert = 0$ is the threshold of pair formation for any $P$. The same conclusion can be reached by noting that $M_{\bf 00}$ diverges at $E \rightarrow E_{11}$, see Eq.~(\ref{twopart:eq:twentyfive}). (ii) Energy $E(P)$ is periodic with period $(2\pi)$, despite $P$ being a sum of two single-particle momenta, each of which varying between $-\pi$ and $\pi$. Thus, pairing leads to Brillouin zone (BZ) folding and the pair behaves as one particle with $-\pi \leq P \leq \pi$. (iii) Pair energy in BZ corners is $E(\pm \pi) = - \vert U \vert$. (iv) The pair binding energy is quadratic in coupling near the threshold:
\begin{equation}
E(\vert U \vert \ll t) = 
- 4 t \cos{\left( \frac{P}{2} \right)} - \frac{\vert U \vert^2}{8t \cos{ \left( \frac{P}{2} \right) }} \: .   
\label{twopart:eq:twentyseven}
\end{equation}
The first term here is the minimum energy of two free particles with total momentum $P$. (v) Expansion of Eq.~(\ref{twopart:eq:twentysix}) for small $P$ yields the pair effective mass [in units of the bare one-particle mass $m_0 = \hbar^2/(2ta^2)$]:
\begin{equation}
\frac{m^{\ast}_p}{m_0} = \frac{ \sqrt{ \vert U \vert^2 + 16 \, t^2 }}{2t} \: .   
\label{twopart:eq:twentyeight}
\end{equation}
The pair mass is not constant but increases with the binding energy. This is a common property of bound states~\cite{Mattis1986} related to the lack of Galilean invariance on the lattice. Comparison between Eqs.~(\ref{twopart:eq:twentyeight}) and (\ref{twopart:eq:twentysix}) reveals a curious relationship between the pair mass and its ``rest energy'' $E(0)$. Restoring for a moment the intersite distance $a$, and using $m_0 = \hbar^2/(2 t a^2)$, one obtains   
\begin{equation}
\vert E(0) \vert = m^{\ast}_{p} \left( \frac{ 2 t a }{ \hbar } \right)^2 .   
\label{twopart:eq:twentyeightone}
\end{equation}
The expression in parentheses is recognized as the maximum group velocity on the lattice. Thus, Eq.~(\ref{twopart:eq:twentyeightone}) has the form of $E = mc^2$ of relativistic physics. 

\begin{figure*}[t]
\includegraphics[width=1.00\textwidth]{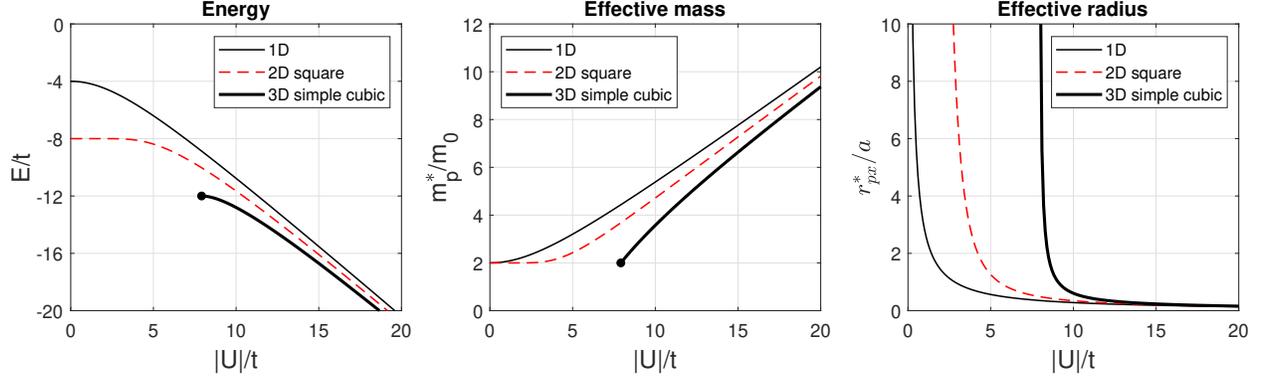}
\caption{Properties of ${\bf P} = 0$ bound pairs in the attractive Hubbard model on the 1D chain, 2D square lattice, and 3D simple cubic lattice. The black circle marks the pair formation threshold in 3D, Eq.~(\ref{twopart:eq:sixtytwo}).}
\label{twopart:fig:twozero}
\end{figure*}

Transitioning to the wave function, the integral in Eq.~(\ref{twopart:eq:twentyonethree}) can be calculated explicitly~\cite{Prudnikov1998}:
\begin{equation}
\int^{\pi}_{-\pi} \frac{{\rm d} q}{2\pi} \, \frac{ \cos{[ q ( m_1 - m_2 ) ]} }
{ \vert E \vert - 4 t \cos{ \frac{P}{2} } \, \cos{q} } 
= \frac{1}{ \sqrt{ \vert E \vert^2 - \alpha^2 } }   
\left[ \frac{\alpha}{ \vert E \vert + \sqrt{ \vert E \vert^2 - \alpha^2 } } 
\right]^{\vert m_1 - m_2 \vert} 
= \frac{1}{ \vert U \vert }
\left[ \frac{ \vert U \vert + \sqrt{ \vert U \vert^2 + \alpha^2 } } 
            {\alpha} \right]^{- \vert m_1 - m_2 \vert} .
\label{twopart:eq:twentyeighttwo}
\end{equation}
where $\alpha \equiv 4t \cos(P/2)$ has been set for brevity. Thus, the un-normalized wave function can be written as 
\begin{equation}
\Psi( m_1 , m_2 ) = e^{ i \frac{P}{2} ( m_1 + m_2 )} 
\cdot e^{ - \gamma \vert m_1 - m_2 \vert } \: ,   
\label{twopart:eq:twentynine}
\end{equation}
where 
\begin{equation}
\sinh{\gamma} = \frac{\vert U \vert}{\alpha} = \frac{\vert U \vert}{ 4t \cos{ \frac{P}{2} } } \: ,   
\label{twopart:eq:thirty}
\end{equation}
and
\begin{equation}
E = - 4t \cos{ \frac{P}{2} } \cosh{\gamma} \: .    
\label{twopart:eq:thirtyone}
\end{equation}
The same expressions can be derived directly from a real-space Schr\"odinger equation by means of two-particle Bethe ansatz. Using the explicit form of $\Psi$, it is straightforward to compute potential energy, kinetic energy, and effective radius of a moving pair:
\begin{equation}
E_{\rm pot} = \left\langle - \vert U \vert \, \delta_{m_1 , m_2} \right\rangle 
= - \frac{ \vert U \vert^2 }{\sqrt{ \vert U \vert^2 + 16 \, t^2 \cos^2{ \! \frac{P}{2} } }} \: ,    
\label{twopart:eq:thirtytwo}
\end{equation}
\begin{equation}
E_{\rm kin} = E - E_{\rm pot} = 
- \frac{ 16 \, t^2 \cos^2{ \! \frac{P}{2} } }
       {\sqrt{ \vert U \vert^2 + 16 \, t^2 \cos^2{ \! \frac{P}{2} } }} \: ,    
\label{twopart:eq:thirtythree}
\end{equation}
\begin{equation}
r^{\ast}_{p} = \left\langle \left( m_1 - m_2 \right)^2 \right\rangle^{1/2} = 
\frac{ 4 t \cos{ \! \frac{P}{2} } }{ \sqrt{2} \, \vert U \vert } \: .    
\label{twopart:eq:thirtyfour}
\end{equation}
Interestingly, $r^{\ast}_{p}( P \rightarrow \pi ) \rightarrow 0$, which means the pair shrinks to a point. The same conclusion can also be derived directly from Eq.~(\ref{twopart:eq:twentyonethree}). At $P = \pi$, the two kinetic terms in the denominator cancel out which renders the internal wave function to be $\propto \delta_{{\bf m}_1 {\bf m}_2}$. The pair energy, mass, and radius of the 1D Hubbard pair are plotted in Fig.~\ref{twopart:fig:twozero}.

\subsection{\label{twopart:sec:threethree}
2D. Square lattice  
}

For the square lattice, the basic integral $M_{\bf 00}$ in Eq.~(\ref{twopart:eq:twelve}) can be expressed via complete elliptic integral of the first kind ${\bf K}(z)$, see \ref{twopart:sec:appafive}, Eq.~(\ref{twopart:eq:appafiftytwo}), for details. Denoting $\alpha \equiv 4 t \cos{ \frac{P_x}{2} }$ and $\beta \equiv 4 t \cos{ \frac{P_y}{2} }$, one has
\begin{equation}
M^{\rm sq}_{\bf 00}  = 
\int\limits^{\pi}_{-\pi} \!\! \int\limits^{\pi}_{-\pi} \frac{ {\rm d} q_x \, {\rm d} q_y}{(2\pi)^2} 
\frac{1}{ \vert E \vert - \alpha \cos{q_x} - \beta \cos{q_y} } 
= \frac{2}{\pi  \sqrt{ \vert E \vert^2 - ( \alpha - \beta )^2 }} \,
{\bf K} \!\! \left[ \sqrt{\frac{4 \alpha \beta}{ \vert E \vert^2 - ( \alpha - \beta )^2 }} \right] \! .
\label{twopart:eq:thirtyfive}
\end{equation}
This result applies not only to the isotropic square model at arbitrary ${\bf P}$, but also to the rectangular model with $t_x \neq t_y$. Inserting Eq.~(\ref{twopart:eq:thirtyfive}) in Eq.~(\ref{twopart:eq:twentytwo}) defines pair energy in the most general case. The minimum energy of two free particles is $E_{11} = 2 \varepsilon_{{\bf P}/2} = - ( \alpha + \beta )$, at which the argument of ${\bf K}$ reaches 1 and $M^{\rm sq}_{\bf 00}$ diverges logarithmically. Similar to 1D, the divergence is interpreted as the existence of a bound state at any nonzero $\vert U \vert$. Thus, $\vert U \vert = 0$ is pair formation threshold at {\em any} ${\bf P}$. Let us determine pair energy near threshold. Setting $E = - \alpha - \beta - \Delta$, $\Delta \ll \alpha, \beta$, Eqs.~(\ref{twopart:eq:thirtyfive}) and (\ref{twopart:eq:twentytwo}) produce in the leading order        
\begin{equation}
\frac{\vert U \vert}{\pi \sqrt{ \alpha \beta }} \,  
{\bf K} \! \left( 1 - \frac{ \alpha + \beta }{ 4 \alpha \beta } \, \Delta \right) = 1 \: .    
\label{twopart:eq:thirtysix}
\end{equation}
Using the asymptote ${\bf K}( 1 - z ) \sim \frac{1}{2} \ln{ \frac{8}{z} }$, at $z \rightarrow +0$, one obtains
\begin{equation}
\Delta = \frac{32 \, \alpha \beta }{ \alpha + \beta } \: 
{\rm exp} \left( - \frac{2\pi \sqrt{ \alpha \beta }}{\vert U \vert} \right) .    
\label{twopart:eq:thirtyseven}
\end{equation}
In the ground state, $\alpha = \beta = 4t$, and the binding energy is
\begin{equation}
\Delta_0 = 64 \, t \: {\rm exp} \left( - \frac{8 \pi t}{\vert U \vert} \right) .    
\label{twopart:eq:thirtyeight}
\end{equation}
Near BZ corners, pair energy is $E( \pm \pi , \pm \pi ) = - \vert U \vert$, like in the 1D case.    
  
Pair effective mass is derived next. It is most convenient to consider BZ diagonal, $P_x = P_y = P$, where Eq.~(\ref{twopart:eq:thirtyfive}) simplifies considerably. Writing $\vert E \vert = \vert E_0 \vert - \frac{\hbar^2 P^2}{m^{\ast}_{p}}$, expanding Eq.~(\ref{twopart:eq:thirtyfive}) for $P \ll 1$, and applying the formula 
\begin{equation}
 \frac{d {\bf K}( z ) }{ d z } = \frac{ {\bf E}( z ) }{ z ( 1 - z^2 ) } 
- \frac{ {\bf K}( z ) }{ z } \: ,   
\label{twopart:eq:fortyfive}
\end{equation}
one obtains from Eq.~(\ref{twopart:eq:twentytwo}):
\begin{equation}
\frac{m_0}{m^{\ast}_p} = \frac{ \vert E_0 \vert }{ 16 t } 
\left\{ 1 - \left[ 1 - \frac{(8t)^2}{ \vert E_0 \vert^2 } \right] 
\frac{ {\bf K} \left( \frac{8t}{\vert E_0 \vert} \right)}
     { {\bf E} \left( \frac{8t}{\vert E_0 \vert} \right)} \right\} ,   
\label{twopart:eq:fortysix}
\end{equation}
where ${\bf E}(\kappa)$ is complete elliptic integral of the second kind. Equation~(\ref{twopart:eq:fortysix}) has correct limits: $m^{\ast}_{p}(\vert U \vert \rightarrow 0) = 2 m_0$ and $m^{\ast}_{p}(\vert U \vert \rightarrow \infty) = \frac{\vert U \vert}{2t} \, m_0$. The factor ${\bf K} \left( \frac{8t}{\vert E_0 \vert} \right)$ in Eq.~(\ref{twopart:eq:fortysix}) can be written as $\frac{\pi \vert E_0 \vert }{ 2 \vert U \vert }$, which follows from Eq.~(\ref{twopart:eq:twentytwo}).     

Pair effective radius is discussed next. Unlike 1D, there is no explicit formula for the pair wave function for {\em all} ${\bf m}$ in 2D. However, for each given ${\bf m}$, the wave function can be derived from a few basic integrals using recurrence relations, as explained in \ref{twopart:sec:appa}. On BZ diagonals, $\Psi({\bf m}_1 , {\bf m}_2)$ is always a linear combination of ${\bf K}$ and ${\bf E}$. At arbitrary ${\bf P}$, the wave function is a linear combination of all three complete elliptic integrals ${\bf K}$, ${\bf E}$, and ${\bf \Pi}$. To calculate effective radius, we substitute Eq.~(\ref{twopart:eq:twentyonethree}) in Eq.~(\ref{twopart:eq:twentytwothree}) and perform the following transformation 
\begin{equation}
r^{\ast 2}_{pj}  = \frac{ \sum_{\bf m} \sum_{{\bf q}_1} 
\frac{ \frac{\partial}{\partial q_{1j}} ( e^{-i {\bf q}_1{\bf m}} ) }{ E - \xi_{ {\bf P} , {\bf q}_1} } 
\sum_{{\bf q}_2} 
\frac{ \frac{\partial}{\partial q_{2\alpha}} ( e^{i {\bf q}_2{\bf m}} ) }{ E - \xi_{ {\bf P} , {\bf q}_2} } }
{ \sum_{\bf m} \sum_{{\bf q}_1} \frac{ e^{-i {\bf q}_1{\bf m}} }{ E - \xi_{ {\bf P} , {\bf q}_1} }  
               \sum_{{\bf q}_2} \frac{ e^{ i {\bf q}_2{\bf m}} }{ E - \xi_{ {\bf P} , {\bf q}_2} } }    
= \frac{ \sum_{\bf q} \frac{ \left( \frac{\partial \xi_{{\bf P},{\bf q}}}{\partial q_{j}} \right)^2 }
{ ( E - \xi_{{\bf P},{\bf q}} )^4 } }
{ \sum_{\bf q} \frac{1}{ ( E - \xi_{{\bf P},{\bf q}} )^2 } } =
\frac{ \left( - \frac{1}{6} \right) \frac{\partial^3}{\partial |E|^3}
\sum_{\bf q} \frac{ \left( \frac{\partial \xi_{{\bf P},{\bf q}}}{\partial q_{j}} \right)^2 }
{ |E| + \xi_{{\bf P},{\bf q}} } }
{(-1) \frac{\partial}{\partial |E|} \sum_{\bf q} \frac{1}{ |E| + \xi_{{\bf P},{\bf q}} } } \, ,
\label{twopart:eq:twentytwofour}
\end{equation}
where $\xi_{{\bf P},{\bf q}} \equiv \varepsilon_{ \frac{{\bf P}}{2} + {\bf q} } + \varepsilon_{ \frac{{\bf P}}{2} - {\bf q}}$. Both sums in Eq.~(\ref{twopart:eq:twentytwofour}) are recognized as integrals $M^{\rm sq}_{nm}$ of the square lattice evaluated in \ref{twopart:sec:appa}. Let us limit consideration to BZ diagonals, $P_x = P_x \equiv P$. With the notation of \ref{twopart:sec:appathree}, one writes
\begin{equation}
r^{\ast 2}_{px} = r^{\ast 2}_{py} = \frac{( 4t \cos{\frac{P}{2}} )^2 }{12} 
\frac{ \frac{\partial^3}{\partial |E|^3} \left( M^{\rm sq}_{00} - M^{\rm sq}_{20} \right) }
     { \frac{\partial}{\partial |E|}     \left( M^{\rm sq}_{00} \right) }  \: .
\label{twopart:eq:twentytwofive}
\end{equation}
Using explicit expressions, Eqs.~(\ref{twopart:eq:appathirtyone}) and (\ref{twopart:eq:appathirtyfour}), one derives, after transformations, a final formula 
\begin{equation}
r^{\ast 2}_{px} = r^{\ast 2}_{py} = \frac{1}{12} \left\{ 
\frac{1 + \left( \frac{ 8t \cos{\frac{P}{2}} }{\vert E \vert} \right)^2}
     {1 - \left( \frac{ 8t \cos{\frac{P}{2}} }{\vert E \vert} \right)^2 } - 
\frac{{\bf K} \left( \frac{ 8t \cos{\frac{P}{2}} }{\vert E \vert} \right)}
     {{\bf E} \left( \frac{ 8t \cos{\frac{P}{2}} }{\vert E \vert} \right)}   
\right\}  .    
\label{twopart:eq:fiftyoneone}
\end{equation}
The pair shrinks to a point in the strong coupling limit, $E \rightarrow -\infty$, and at $P = \pm \pi$ for any $\vert U \vert$. The radius diverges at the threshold, $E \rightarrow - 8t \cos{\frac{P}{2}}$, as expected on physical reasoning. In the ground state, ${\bf P} = 0$, the asymptotes are $r^{\ast}_{px} (|U| \rightarrow \infty) \approx \frac{4 t}{\sqrt{2} |U|}$ and $r^{\ast}_{px} (|U| \rightarrow 0) \approx 96^{-\frac{1}{2}} \, \exp{\frac{4\pi t}{|U|}}$. The energy, mass, and radius of the 2D Hubbard pair are plotted in Fig.~\ref{twopart:fig:twozero}.

\subsection{\label{twopart:sec:threefour}
2D. Triangular lattice  
}

In this section, we consider the two-dimensional triangular lattice with nearest-neighbor isotropic hopping, $t_{\bf b} = t$. Single particle dispersion, Eq.~(\ref{twopart:eq:seven}), is
\begin{equation}
\varepsilon_{\bf k} = - 2 t \cos{k_x} 
- 4 t \cos{ \left( k_x/2 \right) \cos{ ( \sqrt{3}k_y/2 ) }} .   
\label{twopart:eq:fiftytwo}
\end{equation}
The double integral $M^{\rm tr}_{\bf 00}$, Eq.~(\ref{twopart:eq:twelve}), corresponding to this $\varepsilon_{\bf k}$ is evaluated in \ref{twopart:sec:appbone}. For the ground state, $P_x = P_y = 0$, the result is  
\begin{equation}
M^{\rm tr}_{\bf 00}  = \frac{2}{\pi} 
\frac{1}{\sqrt{ \vert E_0 \vert^2 - 48 t^2 + 16 t \sqrt{ 2 t \vert E_0 \vert + 12 t^2 } }} 
\: {\bf K} \! \left( \sqrt{ \frac{ 32 t \sqrt{ 2 \vert E_0 \vert t + 12 t^2 } }
  { \vert E_0 \vert^2 - 48 t^2 + 16 t \sqrt{ 2 \vert E_0 \vert t + 12 t^2 } } } \right) .  
\label{twopart:eq:fiftyseven}
\end{equation}
The lowest energy of two free particles on the triangular lattice is $E_{11} = -12t$. When $E \rightarrow E_{11} - 0$ from below, the argument of ${\bf K}$ approaches 1 and $M^{\rm tr}_{\bf 00}$ diverges logarithmically. Utilizing Eq.~(\ref{twopart:eq:twentytwo}), one concludes that a bound pair is formed for any attractive $U$. 

Let us derive asymptotic behavior of $E$ at small couplings. Setting $\vert E_0 \vert = 12 t + \Delta$, $\Delta \ll t$ in Eq.~(\ref{twopart:eq:fiftyseven}), one obtains from Eq.~(\ref{twopart:eq:twentytwo}) 
\begin{equation}
\frac{ \vert U \vert }{ 4 \pi \sqrt{3} t } \: {\bf K} \! \left( 1 - \frac{\Delta}{18 t} \right) = 1 \: ,   
\label{twopart:eq:fiftyeight}
\end{equation}
from where the binding energy is 
\begin{equation}
\Delta = 144 t \, \exp{ \left( - \frac{ 8 \pi \! \sqrt{3} \, t }{ \vert U \vert } \right) } .   
\label{twopart:eq:fiftynine}
\end{equation}

\subsection{\label{twopart:sec:threefive}
3D. Simple cubic lattice  
}

The attractive Hubbard model in three dimensions possesses a new qualitative feature: nonzero threshold of pair formation. In 3D, kinetic energy {\em alone} is strong enough to counteract weak attraction. Mathematically, the triple integral in $M_{\bf 00}$ converges when $E \rightarrow E_{11}$ for any ${\bf P}$, rendering the critical potential depth $\vert U_{\rm cr} \vert$ finite. Calculation of $\vert U_{\rm cr} \vert$ requires evaluation of the celebrated Watson integrals~\cite{Watson1939}. Amazingly, on BZ diagonals they are now known analytically for a general lattice point~\cite{Joyce1994,Joyce1998,Glasser2000,Joyce2002,Zucker2011}, see \ref{twopart:sec:appc}. Setting $P_x = P_y = P_z = P$ and $\alpha = 4t \cos{(P/2)}$, one obtains
\begin{equation}
M^{\rm sc}_{\bf 00} =  \frac{1}{(2 \pi)^3}
\int\limits^{\pi}_{-\pi} \!\!\! \int\limits^{\pi}_{-\pi} \!\!\! \int\limits^{\pi}_{-\pi} 
\frac{ {\rm d} q_x \, {\rm d} q_y \, {\rm d} q_z }
{ \vert E \vert - \alpha \, ( \cos{q_x} + \cos{q_y} + \cos{q_z} ) } \: .   
\label{twopart:eq:sixty}
\end{equation}
The minimum energy of two free particles is $E_{11} = -3 \alpha$ at which the classic result of Watson's reads~\cite{Watson1939} 
\begin{equation}
M^{\rm sc}_{\bf 00} (E = - 3 \alpha) 
= \frac{4 ( 18 + 12\sqrt{2} - 10\sqrt{3} - 7\sqrt{6} )}{\alpha \pi^2} 
\: {\bf K}^2 \left[ ( 2 - \sqrt{3} )( \sqrt{3} - \sqrt{2} ) \right] 
= \frac{1}{\alpha} \, 0.505462 \ldots \: .   
\label{twopart:eq:sixtyone}
\end{equation}
The binding threshold then follows from Eq.~(\ref{twopart:eq:twentytwo}):
\begin{equation}
\vert U_{\rm cr} ({\rm diag}) \vert = ( 7.913552 \ldots ) \, t \cos{ \left( P/2 \right) } \: .   
\label{twopart:eq:sixtytwo}
\end{equation}
The $P = 0$ value of $7.914 \ldots$ obtained by numerical integration was reported in Ref.~\cite{Dong1989}. The threshold decreases along the diagonal and becomes zero in the BZ corner. This also implies that for any $\vert U \vert < 7.913552 \ldots$, there is a momentum $P_0$ such that the particles are not bound for $P < P_0$ but bound at $P > P_0$. Thus, pairs become {\em more} stable at large lattice momenta. It is a common property of lattice bound states, which will be encountered many times later in this paper.

\begin{figure}[t]
\begin{center}
\includegraphics[width=0.48\textwidth]{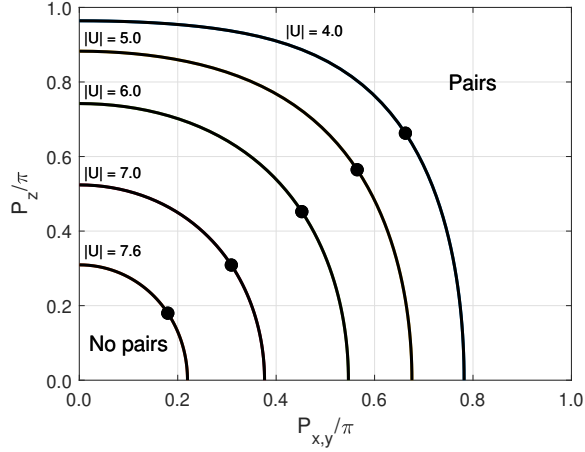}
\end{center}
\caption{Binding threshold surface for the negative Hubbard model in the 3D simple cubic lattice. What is shown is the surface's intersections with plane $P_x = P_y$ for several $\vert U \vert$. Filled circles indicate the threshold momenta on the BZ diagonal, given by $P_z = P_x = P_y = 2 \arccos{ \left( \vert U \vert / 7.913552 \right) }$. }
\label{twopart:fig:three}
\end{figure}

The binding threshold can also be computed on BZ {\em planes} that pass through four BZ corners, for example on the plane $P_x = P_y$. This is possible thanks to the extension of Watson's result to the anisotropic case by Montroll~\cite{Montroll1956,Zucker2011}:  
\begin{align}
M^{\rm tg}_{\bf 00} ( E = - 8t \cos{(P_x/2)} - 4t \cos{(P_z/2)} ) & = 
\frac{1}{\alpha \pi^3} 
\int\limits^{\pi}_0 \!\!\! \int\limits^{\pi}_0 \!\!\! \int\limits^{\pi}_0 
\frac{{\rm d}q_x \, {\rm d}q_y \, {\rm d}q_z}{ 2 - \cos{q_x} - \cos{q_y} + \xi ( 1 - \cos{q_z} ) }
\nonumber \\
& = \frac{4}{\alpha \pi^2} \sqrt{ \frac{ 2 \kappa_1 \kappa_2 }{ \xi } } \, 
{\bf K}(\kappa_1) {\bf K}(\kappa_2) \: ,   
\label{twopart:eq:sixtythree}
\end{align}
where 
\begin{equation}
\kappa_{1,2} = \frac{1}{2\xi} \left( \sqrt{ 4 + 2\xi } \pm 2 \right) \! 
\left( 2 \sqrt{ 1 + \xi } - \sqrt{ 4 + 2\xi } \right) ,   
\label{twopart:eq:sixtyfour}
\end{equation}
\begin{equation}
\alpha \equiv 4t \cos{(P_x/2)} \: , \hspace{1.0cm}
   \xi \equiv \frac{ \cos{(P_z/2)} }{ \cos{(P_x/2)} } \: .
\label{twopart:eq:sixtyfive}
\end{equation}
The threshold value is an inverse of Eq.~(\ref{twopart:eq:sixtythree}) by virtue of Eq.~(\ref{twopart:eq:twentytwo}). Notice how pair's center-of-mass movement induces anisotropy. For fixed $\vert U \vert < 7.913552 \ldots$ and $E = E_{11}$, Eq.~(\ref{twopart:eq:twentytwo}) defines a surface in BZ that separates bound and unbound states. Figure~\ref{twopart:fig:three} shows the intersection of that surface with the $P_x = P_y$ plane for several values of $\vert U \vert$. 

The ground state energy, $E_0 = E({\bf P} = 0)$, is discussed next. A close-form expression for $M^{\rm sc}_{\bf 00}$ in the isotropic simple cubic model was found by Joyce~\cite{Joyce1994,Joyce1998}  
\begin{equation}
M^{\rm sc}_{\bf 00}( E_0 \leq -12 t) = \frac{1}{\vert E_0 \vert} 
\frac{ ( 1 - 9 \zeta^4 ) }{ ( 1 - \zeta )^3 ( 1 + 3\zeta ) } 
\left[  \frac{2}{\pi} {\bf K}(\kappa) \right]^2 ,
\label{twopart:eq:sixtysix}
\end{equation}
\begin{equation}
\kappa^2( \zeta ) = \frac{ 16 \, \zeta^3 }{ ( 1 - \zeta )^3 ( 1 + 3\zeta ) } \: ,
\label{twopart:eq:sixtyseven}
\end{equation}
\begin{equation}
\zeta = \zeta(w) = \left[ 
\frac{ 1 - \sqrt{ 1 - \frac{w^2}{9} } }{ 1 + \sqrt{ 1 - w^2 } } 
\right]^{\frac{1}{2}} , 
\hspace{0.5cm} 
w = \frac{12 \, t}{ \vert E_0 \vert } \: . 
\label{twopart:eq:sixtyeight}
\end{equation}
Using these formulas, Eq.~(\ref{twopart:eq:twentytwo}) defines $E_0$ as a function of $\vert U \vert$. It is plotted in Fig.~\ref{twopart:fig:twozero}(a). 

To derive effective mass, we write $\vert E \vert = \vert E_0 \vert - \frac{3 \hbar^2 P^2}{2 m^{\ast}_p}$ and expand Eq.~(\ref{twopart:eq:sixty}) for small $P$ to get
\begin{equation}
\frac{m_0}{m^{\ast}_p} = \frac{1}{6}
\frac{ \int\limits^{\pi}_{-\pi} \!\!\! \int\limits^{\pi}_{-\pi} \!\!\! \int\limits^{\pi}_{-\pi} 
\frac{ {\rm d}q_x \, {\rm d}q_y \, {\rm d}q_z \, ( \cos{q_x} + \cos{q_y} + \cos{q_z} ) }
{ [ \vert E_0 \vert - 4 t \, ( \cos{q_x} + \cos{q_y} + \cos{q_z} ) ]^2 } }
{ \int\limits^{\pi}_{-\pi} \!\!\! \int\limits^{\pi}_{-\pi} \!\!\! \int\limits^{\pi}_{-\pi} 
\frac{ {\rm d}q_x \, {\rm d}q_y \, {\rm d}q_z}
{ [ \vert E_0 \vert - 4 t \, ( \cos{q_x} + \cos{q_y} + \cos{q_z} ) ]^2 } }  \:  .  
\label{twopart:eq:sixtyeightone}
\end{equation}
Once $E_0$ is found from Eq.~(\ref{twopart:eq:twentytwo}), the effective mass can be obtained by numerical integration.  Alternatively, the denominator in Eq.~(\ref{twopart:eq:sixtyeightone}) is $- \partial M^{\rm sc}_{\bf 00}/\partial \vert E_0 \vert$, while the numerator can be expressed using a similar derivative and a linear transformation to link it also to $M^{\rm sc}_{\bf 00}$. The final formula is 
\begin{equation}
\frac{m_0}{m^{\ast}_p} = \frac{1}{24 t} \left\{ \vert E_0 \vert + 
\frac{ 1 } { \vert U \vert ( \partial M^{\rm sc}_{\bf 00}/\partial \vert E_0 \vert ) }  \right\} ,  
\label{twopart:eq:sixtyeighttwo}
\end{equation}
where we have used $M^{\rm sc}_{\bf 00} = 1/\vert U \vert$. Thus, the effective mass can also be computed by numerical differentiation of $M^{\rm sc}_{\bf 00}$. $m^{\ast}_{p}(U)$ is shown in Fig.~\ref{twopart:fig:twozero}(b).   
 
To compute effective radius, we apply the 3D version of Eq.~(\ref{twopart:eq:twentytwofour}). For ${\bf P} = 0$, it yields 
\begin{equation}
r^{\ast 2}_{pj} = 16 t^2 \: 
\frac{ \int\limits^{\pi}_{-\pi} \!\!\! \int\limits^{\pi}_{-\pi} \!\!\! \int\limits^{\pi}_{-\pi} 
\frac{ \sin^2{q_{j}} \: {\rm d}q_x \, {\rm d}q_y \, {\rm d}q_z }
{ [ \vert E_0 \vert - 4 t \, ( \cos{q_x} + \cos{q_y} + \cos{q_z} ) ]^4 } }
{ \int\limits^{\pi}_{-\pi} \!\!\! \int\limits^{\pi}_{-\pi} \!\!\! \int\limits^{\pi}_{-\pi} 
\frac{ {\rm d}q_x \, {\rm d}q_y \, {\rm d}q_z}
{ [ \vert E_0 \vert - 4 t \, ( \cos{q_x} + \cos{q_y} + \cos{q_z} ) ]^2 } }  \:  .  
\label{twopart:eq:sixtyeightthree}
\end{equation}
$r^{\ast}_{px}(U)$ is shown in Fig.~\ref{twopart:fig:twozero}(c).

\begin{figure*}[t]
\includegraphics[width=0.98\textwidth]{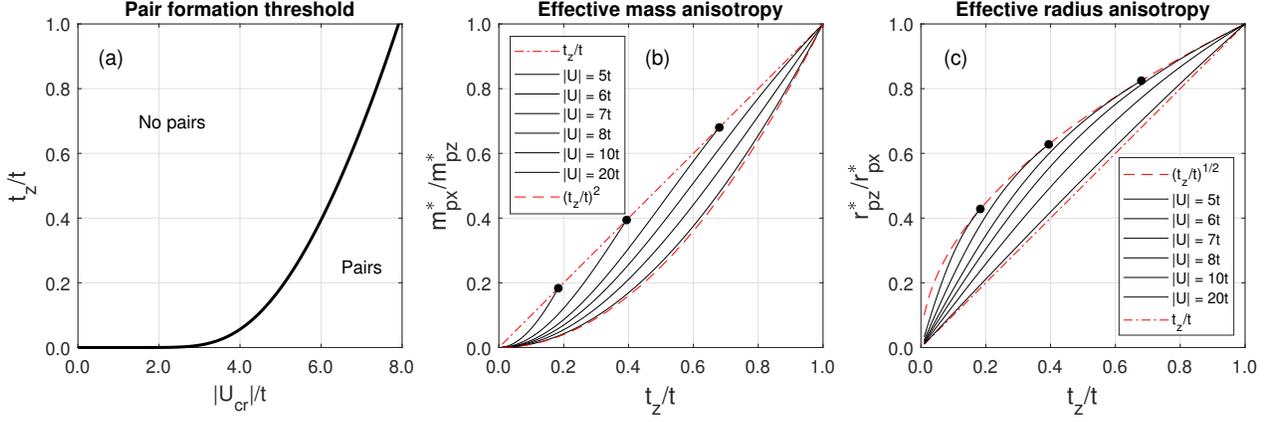}
\caption{Properties of bound pairs in the anisotropic tetragonal attractive Hubbard model. (a) Binding threshold as a function of hopping anisotropy. (b) Anisotropy of effective masses. (c) Anisotropy of effective radii. Circles indicate threshold $t_z$ values for given $\vert U \vert$.}
\label{twopart:fig:four}
\end{figure*}

\subsection{\label{twopart:sec:threesix}
3D. Tetragonal lattice  
}

The tetragonal lattice is of interest because of its relevance to high-$T_c$ superconductors. We explore several effects of hopping anisotropy in this section. Equations~(\ref{twopart:eq:sixtythree}) and (\ref{twopart:eq:sixtyfour}) apply to the tetragonal attractive Hubbard model, where the anisotropy is caused not by pair momentum but by the difference in transfer integrals along the $z$ and $x,y$ directions. Limiting consideration to the ground state, one sets $t_x = t_y = t$, $\alpha = 4 t$ and $\xi = t_z/t$. Figure~\ref{twopart:fig:four}(a) shows the binding threshold as a function of $\xi$. The function is sharp near $\xi = 0$: already at $\xi = 10^{-4}$, $\vert U_{\rm cr} \vert = 1.98 \, t$. Even a tiny interlayer hopping suppresses long-range logarithmic fluctuations and renders the system essentially three-dimensional. An analytical expression for the Green function $M^{\rm tg}_{\bf 00}$ for ${\bf P} = 0$ and arbitrary $E_0$ was derived by Delves, Joyce, and Zucker~\cite{Joyce2001a,Joyce2001b,Joyce2003}. The cumbersome expression is given in \ref{twopart:sec:appctwo}. 

Of physical interest is {\em pair} mass anisotropy $m^{\ast}_{px}/m^{\ast}_{pz}$. Expanding $\vert U \vert M^{\rm tg}_{\bf 00} = 1$ for small ${\bf P}$ one obtains 
\begin{equation}
\frac{m^{\ast}_{px}}{m^{\ast}_{pz}} = \frac{t_z}{t} \: 
\frac{ \int\limits^{\pi}_{-\pi} \!\!\! \int\limits^{\pi}_{-\pi} \!\!\! \int\limits^{\pi}_{-\pi} 
\frac{ \cos{q_{z}} \: {\rm d}q_x \, {\rm d}q_y \, {\rm d}q_z }
{ [ \vert E_0 \vert - 4 t \, ( \cos{q_x} + \cos{q_y} ) - 4 t_z \cos{q_z} ]^2 } }
{ \int\limits^{\pi}_{-\pi} \!\!\! \int\limits^{\pi}_{-\pi} \!\!\! \int\limits^{\pi}_{-\pi} 
\frac{ \cos{q_{x}} \: {\rm d}q_x \, {\rm d}q_y \, {\rm d}q_z}
{ [ \vert E_0 \vert - 4 t \, ( \cos{q_x} + \cos{q_y} ) - 4 t_z \cos{q_z} ]^2 } }  \:  .  
\label{twopart:eq:sixtyeightfour}
\end{equation}
The mass ratio is shown in Fig.~\ref{twopart:fig:four}(b). For a single particle, $m_{x}/m_{z} = t_z/t$, and the graph would be a straight line. Pairing {\em enhances} mass anisotropy which approaches $(t_z/t)^2$ in the strong coupling limit. For intermediate $\vert U \vert$, the mass anisotropy lies between $(t_z/t)$ and $(t_z/t)^2$. 

Another interesting property is the ratio of effective radii. It follows from Eq.~(\ref{twopart:eq:sixtyeightthree}) that
\begin{equation}
\left( \frac{r^{\ast}_{pz}}{r^{\ast}_{px}} \right)^2  = \frac{t^2_z}{t^2} \: 
\frac{ \int\limits^{\pi}_{-\pi} \!\!\! \int\limits^{\pi}_{-\pi} \!\!\! \int\limits^{\pi}_{-\pi} 
\frac{ \sin^2{q_{z}} \: {\rm d}q_x \, {\rm d}q_y \, {\rm d}q_z }
{ [ \vert E_0 \vert - 4 t \, ( \cos{q_x} + \cos{q_y} ) - 4 t_z \cos{q_z} ]^4 } }
{ \int\limits^{\pi}_{-\pi} \!\!\! \int\limits^{\pi}_{-\pi} \!\!\! \int\limits^{\pi}_{-\pi} 
\frac{ \sin^2{q_{x}} \: {\rm d}q_x \, {\rm d}q_y \, {\rm d}q_z}
{ [ \vert E_0 \vert - 4 t \, ( \cos{q_x} + \cos{q_y} ) - 4 t_z \cos{q_z} ]^4 } }  \:  .  
\label{twopart:eq:sixtyeightfive}
\end{equation}
The pair size anisotropy is shown in Fig.~\ref{twopart:fig:four}(c). Near binding threshold, $r^{\ast}_{pz}/r^{\ast}_{px} = \sqrt{t_z/t}$ whereas in general it is confined between $\sqrt{t_z/t}$ and $(t_z/t)$.

\subsection{\label{twopart:sec:threeseven}
3D. Body-centered cubic (BCC) lattice   
}

In a BCC lattice with nearest-neighbor hopping, $\varepsilon_{\bf k} = - 8 t \cos{ \frac{k_x}{2}} \cos{\frac{k_y}{2}} \cos{\frac{k_z}{2}}$. (In this and the following sections, we set the cube length, $a = 1$.) First, we consider the ground state, ${\bf P} = 0$. In calculating $M^{\rm bcc}_{\bf 00}$, integration over the BCC Brillouin zone can be replaced with one fourth of the integral over a cube with side length $4\pi$. Changing momentum variables yields
\begin{equation}
M^{\rm bcc}_{\bf 00} = \frac{1}{(2\pi)^3} 
\int\limits^{\pi}_{-\pi} \!\!\! \int\limits^{\pi}_{-\pi} \!\!\! \int\limits^{\pi}_{-\pi} 
\frac{{\rm d}q_x \, {\rm d}q_y \, {\rm d}q_z}{ \vert E_0 \vert - 16 t \, \cos{q_x} \cos{q_y} \cos{q_z} } \: .   
\label{twopart:eq:sixtynine}
\end{equation}
The integral here is one of the generalized Watson integrals first evaluated by Maradudin~\cite{Maradudin1960,Maradudin1960b} and later studied by other authors~\cite{Katsura1971,Morita1971,Joyce1971}. Application of those results leads to the energy equation  
\begin{equation}
\frac{ \vert U \vert }{ \vert E_0 \vert } \left( \frac{2}{\pi} \right)^2  
{\bf K}^2 \left[ \sqrt{ \frac{1}{2} - \frac{1}{2} 
\sqrt{ 1 - \left( \frac{16 t}{ \vert E_0 \vert } \right)^2 } } \right] = 1 \: .   
\label{twopart:eq:seventy}
\end{equation}
The binding threshold is found by setting $E_0 = - 16 t$, which results in
\begin{equation}
\vert U^{\rm bcc}_{\rm cr} \vert = 
\frac{ 4 \pi^2 \, t }{ \left[ K \!\! \left( \frac{1}{\sqrt{2}} \right) \right]^2  } 
= \frac{ 64 \pi^3 \, t }{ \left[ \Gamma \!\! \left( \frac{1}{4} \right) \right]^4 } 
= ( 11.484320 \ldots ) \, t \: .   
\label{twopart:eq:seventyone}
\end{equation}
Expanding Eq.~(\ref{twopart:eq:seventy}) near $E_0 \approx -16 t$ yields a quadratic dependence of the binding energy near the threshold
\begin{equation}
E_0( |U| \approx |U^{\rm bcc}_{\rm cr}| ) = - 16 \, t - 
\frac{ \left[ \Gamma \!\! \left( \frac{1}{4} \right) \right]^{16} }{ 2^{15} \pi^{10} } 
\frac{ ( |U| - |U^{\rm bcc}_{\rm cr}| )^2 }{ t } \: .   
\label{twopart:eq:seventytwo}
\end{equation}
The numerical coefficient at the quadratic term is $= 0.290501 \ldots$. Formula~(\ref{twopart:eq:seventytwo}) is derived in \ref{twopart:sec:appd}. Finally, expanding Eq.~(\ref{twopart:eq:seventy}) at large $E_0$, one obtains 
\begin{equation}
E_0( \vert U \vert \rightarrow \infty ) = - \vert U \vert - \frac{ 32 \, t^2 }{ \vert U \vert } 
+ o \! \left( \frac{ t^2 }{ \vert U \vert } \right) ,   
\label{twopart:eq:seventythree}
\end{equation}
which is consistent with strong-coupling perturbation theory. 

Nonzero pair momenta are discussed next. Cases with one nonzero component, for example ${\bf P} = ( P_x , 0 , 0 )$, can be reduced to the ground state case. The energy denominator in $M^{\rm bcc}_{\bf 00}$ contains
\begin{equation}
\varepsilon_{\bf q} + \varepsilon_{ {\bf P} - {\bf q} } = 
- 16 \, t \cos{\frac{P_x}{4}} \cos{ \left( \frac{P_x}{4} - q_x \right) } \cos{q_y} \cos{q_z} \: .   
\label{twopart:eq:seventyfour}
\end{equation}
After shifting the integration variable $q_x$, $M^{\rm bcc}_{\bf 00}$ is reduced to the ground state expression, Eq.~(\ref{twopart:eq:sixtynine}), where $t$ is replaced with $t \cos{(P_x/4)}$. Without rederiving all the results given above, let us just mention a generalization of the pair binding condition:
\begin{equation}
\vert U^{\rm bcc}_{\rm cr}( P_x , 0 , 0 ) \vert = 
( 11.484320 \ldots ) \, t \cos{ \left( P_x/4 \right) } \: .   
\label{twopart:eq:seventyfive}
\end{equation}
This expression should be compared with the simple cubic value, Eq.~(\ref{twopart:eq:sixtytwo}).

\begin{table*}[t]
\begin{center}
\renewcommand{\tabcolsep}{0.2cm}
\renewcommand{\arraystretch}{1.5}
\begin{tabular}{|c|c|c|c|}
\hline\hline
 Lattice              &  $z$  &  $|U_{\rm cr}|/t$  & $|U_{\rm cr}|/(zt)$   \\ \hline
\hline 
 Simple cubic         &   6   &  $7.913552 \ldots$  &  $1.318925 \ldots$   \\ \hline
 Body-centered cubic  &   8   & $11.484320 \ldots$  &  $1.435540 \ldots$   \\ \hline
 Face-centered cubic  &  12   & $17.848362 \ldots$  &  $1.487363 \ldots$   \\ \hline
\hline 
\end{tabular}
\end{center}
\caption{
Pair binding thresholds for the attractive Hubbard model in the three cubic lattices with isotropic nearest-neig\-hbor hopping. $z$ is the number of nearest neighbors. ${\bf P} = 0$.   
} 
\label{twopart:tab:one}
\end{table*}

\subsection{\label{twopart:sec:threeeight}
3D. Face-centered cubic (FCC) lattice   
}

Similar to the FCC case, integration over the FCC Brillouin zone can be replaced by integration over the cube with side length $(4\pi)$. In the ground state, $\varepsilon_{\bf q} = \varepsilon_{{\bf P} - {\bf q}}$, and $M^{\rm fcc}_{\bf 00}$ reduces to 
\begin{equation}
M^{\rm fcc}_{\bf 00} = \frac{1}{ 8 t \pi^3 }  
\int\limits^{\pi}_0 \!\! \int\limits^{\pi}_0 \!\! \int\limits^{\pi}_0 
\frac{ {\rm d}q_x \, {\rm d}q_y \, {\rm d}q_z}{ \frac{\vert E_0 \vert}{8 t} -  
\cos{q_x} \cos{q_y} - \cos{q_y} \cos{q_z} - \cos{q_z} \cos{q_x} } \: .   
\label{twopart:eq:seventysix}
\end{equation}
This triple integral was first evaluated by Iwata~\cite{Iwata1969} and later in a different form by Joyce~\cite{Joyce1998}. Joyce's result reads
\begin{equation}
M^{\rm fcc}_{\bf 00}( E_0 < -24 t) = \frac{1}{\vert E_0 \vert} 
\frac{ ( 1 + 3 \zeta^2 )^2 }{ ( 1 - \zeta )^3 ( 1 + 3\zeta ) } 
\left[  \frac{2}{\pi} {\bf K}(\kappa) \right]^2 ,
\label{twopart:eq:seventyseven}
\end{equation}
\begin{equation}
\kappa^2( \zeta ) = \frac{ 16 \, \zeta^3 }{ ( 1 - \zeta )^3 ( 1 + 3\zeta ) } \: ,
\label{twopart:eq:seventyeight}
\end{equation}
\begin{equation}
\zeta = \zeta(w) = \frac{ - 1 + \sqrt{ 1 + \frac{w}{3} } }{ 1 + \sqrt{ 1 - w } } \: , 
\hspace{0.5cm} 
w = \frac{24 \, t}{ \vert E_0 \vert } \: . 
\label{twopart:eq:seventynine}
\end{equation}
Pair formation occurs at $E_0 = -24 \, t$ or $w = 1$. The binding threshold reads   
\begin{equation}
\vert U^{\rm fcc}_{\rm cr} \vert = \frac{ 8 \pi^2 \, t }{ \sqrt{3} 
\left[ {\bf K} \!\! \left( \frac{\sqrt{3}-1}{2\sqrt{2}} \right) \right]^2  } =
( 17.848362 \ldots ) \, t \: .   
\label{twopart:eq:eighty}
\end{equation}
Table~\ref{twopart:tab:one} summarizes threshold values for the three cubic lattices.

\section{\label{twopart:sec:four}
1D. $UV$ model on the one-dimensional chain
}

We now transition to $UV$ models with on-site repulsion $U$ and {\em nearest-neighbor} attraction $V_{\bf b} \equiv - \vert V \vert$. A general feature of these models is the existence of multiple bound states, whose number increases with lattice dimensionality. Because of the complexity of general energy equation, Eq.~(\ref{twopart:eq:thirteen}), it is advantageous to utilize the (anti)symmetrized formalism developed in Section~\ref{twopart:sec:twofour}. We begin with the one-dimensional $UV$ model~\cite{Kornilovitch2004}.

\subsection{\label{twopart:sec:fourone}
Singlet states 
}

The symmetrized set of neighbor vectors consists of two elements: $\{ {\bf b}_{+} \} = \{ (0), (1) \}$. Symmetrized Schr\"odinger equation, Eq.~(\ref{twopart:eq:thirteenseven}), reads:
\begin{equation}
( E_{s} - \varepsilon_{k_1} - \varepsilon_{k_2} ) \, \phi^{+}_{k_1 k_2} =  
U \frac{1}{N} \sum_{ q } \phi^{+}_{ q , \: k_1 + k_2 - q }  
 - \vert V \vert \frac{1}{N} \sum_{ q } \phi^{+}_{ q , \: k_1 + k_2 - q } e^{iq} 
\left( e^{ - i k_1 } + e^{ -i k_2 } \right) .      
\label{twopart:eq:eightyoneone}
\end{equation}
Here $\varepsilon_{k} = -2t \cos{k}$, and subscript in $E_s$ indicates ``spin-singlet''. We introduce two auxiliary functions:
\begin{align}
\Phi^{+}_0(P) & = \frac{1}{N} \sum_{ q } \phi^{+}_{ q , \: k_1 + k_2 - q }           \: ,
\label{twopart:eq:eightyonetwo} \\
\Phi^{+}_1(P) & = \frac{1}{N} \sum_{ q } \phi^{+}_{ q , \: k_1 + k_2 - q } \: e^{iq} \: ,      
\label{twopart:eq:eightyonethree}
\end{align}
so that 
\begin{equation}
\phi^{+}_{k_1 k_2} = 
\frac{ U \Phi^{+}_0(P) - \vert V \vert \Phi^{+}_1(P) \left( e^{ - i k_1 } + e^{ -i k_2 } \right) }
     { E_{s} - \varepsilon_{k_1} - \varepsilon_{k_2} } \: . 
\label{twopart:eq:eightyonefour}
\end{equation}
Substituting Eq.~(\ref{twopart:eq:eightyonefour}) back into the definitions, Eqs.~(\ref{twopart:eq:eightyonetwo}) and (\ref{twopart:eq:eightyonethree}), one obtains:
\begin{align}
\Phi^{+}_0(P) & = U \Phi^{+}_0(P) 
\left( \frac{1}{N} \sum_{ q } \frac{1}{ E_{s} - \varepsilon_{q} - \varepsilon_{P-q} } \right)
- \vert V \vert \Phi^{+}_1(P) 
\left( \frac{1}{N} \sum_{ q } \frac{ e^{ -iq } + e^{ -i (P-q) } }
   { E_{s} - \varepsilon_{q} - \varepsilon_{P-q} } \right)   ,
\label{twopart:eq:eightyonefive} \\
\Phi^{+}_1(P) & = U \Phi^{+}_0(P) 
\left( \frac{1}{N} \sum_{ q } \frac{e^{ iq }}{ E_{s} - \varepsilon_{q} - \varepsilon_{P-q} } \right)
- \vert V \vert \Phi^{+}_1(P) 
\left( \frac{1}{N} \sum_{ q } \frac{ e^{ iq } ( e^{ -iq } + e^{ -i (P-q) } ) }
   { E_{s} - \varepsilon_{q} - \varepsilon_{P-q} } \right)   .      
\label{twopart:eq:eightyonesix}
\end{align}
Next, a change of variables $q' = q - \frac{P}{2}$ under the integrals results in a $( 2 \times 2 )$ matrix equation    
\begin{align}
\Phi^{+}_{0} & = - U M_{0} \cdot \Phi^{+}_{0} +
2 \vert V \vert \, e^{ - i \frac{P}{2} } M_1 \cdot \Phi^{+}_{1}    \: ,  
\label{twopart:eq:eightyone}
\\
\Phi^{+}_{1} & = - U \, e^{ i \frac{P}{2} } M_1 \cdot \Phi^{+}_{0} 
+ \vert V \vert ( M_{0} + M_{2} ) \cdot \Phi^{+}_{1}               \: ,  
\label{twopart:eq:eightytwo}
\end{align}
where 
\begin{equation}
M_n = \frac{1}{N} \sum_{q} \frac{ \cos{nq} }{ \vert E_s \vert - 4t \cos{(P/2)} \cos{q} } 
= \frac{ 1 }{ \sqrt{ \vert E_s \vert^2 - \alpha^2 } } 
    \left[ \frac{ \sqrt{ \vert E_s \vert^2 - \alpha^2 } - \vert E_s \vert }{ - \alpha } \right]^n ,  
\label{twopart:eq:eightythree}
\end{equation}
and $\alpha \equiv 4 t \cos{(P/2)}$. Note that 
\begin{align}
M_1  & = \frac{1}{\alpha} \left( \vert E_s \vert M_{0} - 1 \right) \: , 
\label{twopart:eq:eightythreeone} \\
M_2  & = \frac{ 2 \vert E_s \vert }{\alpha} M_{1} - M_{0}          \: .   
\label{twopart:eq:eightythreetwo}
\end{align}
As a result, everything can be expressed via the basic integral $M_{0}$. The bound state's energy is determined by the consistency condition of Eqs.~(\ref{twopart:eq:eightyone}) and (\ref{twopart:eq:eightytwo}). Expanding the determinant, one obtains 
\begin{equation}
( U M_{0} + 1 ) + \frac{2}{\alpha^2} \, \vert V \vert ( \vert E_{s} \vert + U ) ( 1 - \vert E_s \vert M_{0} ) = 0 \: .   
\label{twopart:eq:eightythreethree}
\end{equation}
This form will be useful later in comparing with similar equations in higher dimensions. Substitution of the explicit form of $M_0$ yields the final expression: 
\begin{equation}
\left[       U       + \sqrt{ \vert E_s \vert^2 - 16 t^2 \cos^2{\frac{P}{2}} } \right] 
\left[ \vert E_s \vert + \sqrt{ \vert E_s \vert^2 - 16 t^2 \cos^2{\frac{P}{2}} } \right] 
- 2 \vert V \vert \left( U + \vert E_s \vert \right) = 0 \: .  
\label{twopart:eq:eightyfour}
\end{equation}
This is a cubic equation for $E_s(P)$, which does not have a simple-form analytical solution. Only at BZ boundary the equation simplifies to $E_s(\pm \pi) = - \vert V \vert$. At strong coupling, $U, \vert V \vert , \vert E_s \vert \gg t$, Eq.~(\ref{twopart:eq:eightyfour}) yields for the ground state
\begin{equation}
E_s(P = 0) = - \vert V \vert - \frac{ 4 t^2 }{ \vert V \vert } 
- \frac{ 8 t^2 }{ U + \vert V \vert } + o(t^2/\vert V \vert)  \: ,   
\label{twopart:eq:eightyfive}
\end{equation}
which is consistent with second-order perturbation theory.

\begin{figure}[t]
\begin{center}
\includegraphics[width=0.48\textwidth]{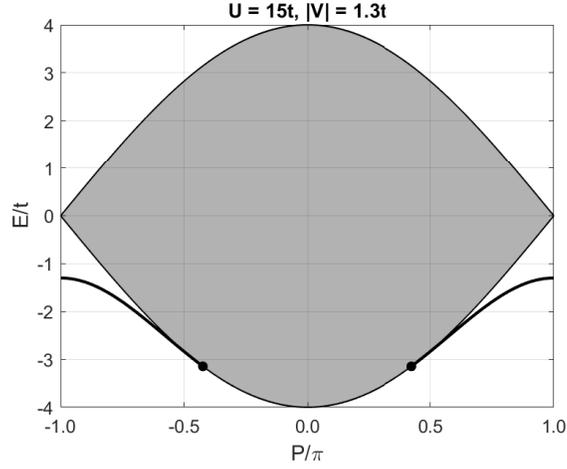}
\end{center}
\caption{Singlet-pair dispersion $E_s$ in the 1D $UV$ model computed from Eq.~(\ref{twopart:eq:eightyfour}) (thick solid line). $U = 15\,t$, $|V| = 1.3\,t$. Note that $|V|$ is below the $P = 0$ binding threshold, $2Ut/(U + 4t)$. The shaded area is the continuum of two-particle scattering states. The circles mark threshold momenta, $P = \pm 2 \arccos{[UV/2/(U-2V)]}$, derived from Eq.~(\ref{twopart:eq:eightysix}). The triplet pair energy $E_t$ given by Eq.~(\ref{twopart:eq:eightynine}) is very close to $E_s$ and therefore not shown in the figure.}
\label{twopart:fig:fiveone}
\end{figure}

In order to obtain binding threshold, set $E_s$ equal to the lowest energy of two free carriers, $E_{11} = - 4 t \cos{(P/2)}$, in Eq.~(\ref{twopart:eq:eightyfour}). It results in 
\begin{equation}
\vert V^{s}_{\rm cr} \vert = \frac{ 2 U t \cos{(P/2)} }{ U + 4 t \cos{(P/2)} }  \: .   
\label{twopart:eq:eightysix}
\end{equation}
This formula possesses several interesting properties. First of all, $\vert V^{s}_{\rm cr} \vert$ is nonzero despite the model being one-dimensional. Here, the attraction competes not only with kinetic energy but also with repulsion $U$ which leads to a nonzero threshold. At weak repulsion, $U < t$, one has $\vert V^{s}_{\rm cr} \vert \approx U/2$ which can be understood from the Born approximation: there are two attractive sites for one repulsive site, hence half as strong $V$ is needed to overcome $U$. In the opposite limit of strong repulsion, $\vert V^{s}_{\rm cr} \vert$ approaches a finite limit. At large $U$, the on-site wave function amplitude, $\Psi( m , m ) \rightarrow 0$, which becomes a boundary condition for the rest of $\Psi$. Once attraction is strong enough to produce a bound state in the presence of this zero, further increase of $U$ has no effect. Finally, $\vert V^{s}_{\rm cr} \vert$ is a strong function of pair momentum. Like in 3D Hubbard models, the pair becomes {\em more} stable at large $P$. At BZ boundary, the pair is always stable for any $U$ however large, and any $\vert V \vert$ however small. 

\begin{figure}[t]
\begin{center}
\includegraphics[width=0.48\textwidth]{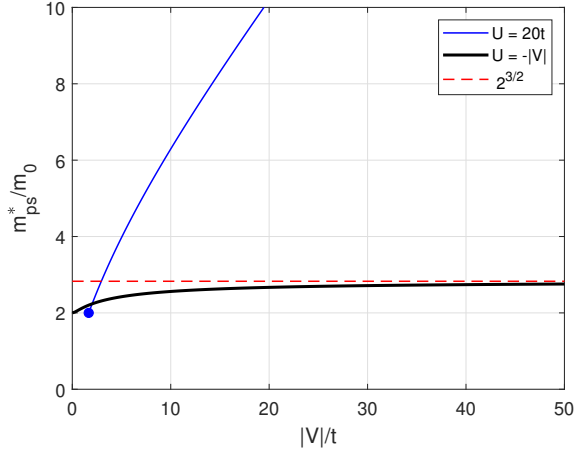}
\end{center}
\caption{Singlet pair mass in the 1D $UV$ model computed from Eqs.~(\ref{twopart:eq:eightyfour}) and (\ref{twopart:eq:eightysixone}). The thin solid line corresponds to a fixed $U = 20 \, t$. The mass grows linearly with $|V|$. The  circle marks threshold value, Eq.~(\ref{twopart:eq:eightysix}), for $P = 0$. The thick solid line corresponds to the {\em light pairs} regime with $U = - \vert V \vert$. The mass stays of order $m_0$ for all $V$. At strong coupling, $\vert V \vert \rightarrow \infty$, the mass approaches the theoretical limit, $m^{\ast}_{ps} = \sqrt{8} \, m_0$ (the dashed line), see Eq.~(\ref{twopart:eq:eightysixnine}). }
\label{twopart:fig:fivezero}
\end{figure}

A typical pair dispersion for $|V| < |V^{s}_{\rm cr}|$ is shown in Fig.~\ref{twopart:fig:fiveone}. Consider weakly bound pairs with $P = 0$. Expanding the exact dispersion relation, Eq.~(\ref{twopart:eq:eightyfour}), for $E_s \approx - 4t$ and utilizing Eq.~(\ref{twopart:eq:eightysix}), one obtains 
\begin{equation}
E_s( P = 0) \approx - 4 t - \frac{ ( \vert V \vert - \vert V^s_{{\rm cr},0} \vert )^2 }{ 2t } \: ,   
\label{twopart:eq:eightyseven}
\end{equation}
where $\vert V^{s}_{{\rm cr},0} \vert = 2Ut/( U + 4t )$. Thus the pair energy varies quadratically near the threshold. Expanding Eq.~(\ref{twopart:eq:eightyfour}) for small $P$, one obtains, after transformations, the singlet's effective mass
\begin{equation}
\frac{m^{\ast}_{ps}}{m_0} \! = \! \frac{( U \! - \! 2E_{s} \! - \! 2 \vert V \vert ) \sqrt{ E^2_{s} \! - \! (4t)^2 } 
\! + \! 2E^2_{s} \! - \! U E_{s} \! - \! 16t^2 }
{ (2t) \left[ 2 \sqrt{E^2_{s} - (4t)^2 } + U - E_{s} \right]} \: .
\label{twopart:eq:eightysixone}
\end{equation}
At threshold, $E_{s} = - 4t$, and the last formula yields $m^{\ast}_{ps}/m_0 = 2$, as expected. At strong attraction, the mass generally grows linearly with coupling, $m^{\ast}_{ps}/m_0 \propto |V|/t$, with the slope depending on $U$ and the $U/V$ ratio. As an example, Fig.~\ref{twopart:fig:fivezero} shows the pair mass for $U = 20 \, t$.

\subsection{\label{twopart:sec:fourthree}
Light bound pairs 
}

In this section, we introduce the topic of {\em light bound pairs} that will be discussed later in several places of this paper. We use this term to describe a situation when pairs are strongly bound, $E \rightarrow -\infty$, but at the same time remain mobile with an effective mass of order $m_0$. It occurs when a pair can move through the lattice without changing its energy, i.e., without breaking the most attractive bond. There are two primary reasons why it can happen. First, because of geometry. In some lattices such as triangular and FCC, one member of a pair can hop to another site while still remaining a nearest neighbor to the second member, which keeps the configuration energy unchanged at $- \vert V \vert$. This is followed by a similar hop by the second member. The two particles hop in turns in a ``crab-like'' fashion, which results in overall movement of the pair through the system. The second origin of light pairs is a flat segment in the attractive part of the inter-particle potential. In this case, the particles can also move in alternating order without changing their energy. 

\begin{figure}[t]
\begin{center}
\includegraphics[width=0.48\textwidth]{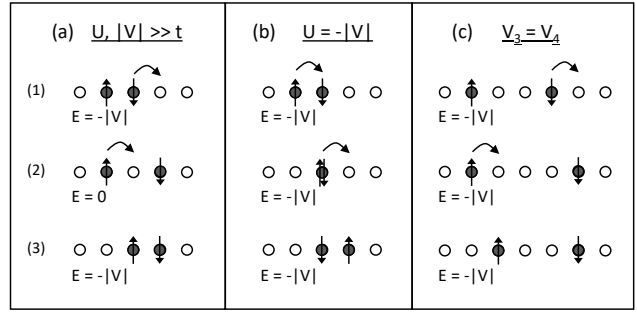}
\end{center}
\caption{Illustration of the light bound pair mechanism. (a) Conventional pair movement. The intermediate configuration has a larger energy, which results in a mass growing linearly with $|V|$. (b) In the resonant case, $U = -|V|$, there are intermediate configurations with the same energy as the initial and starting configurations. (c) An attractive potential with a flat section at a nonzero separation between two particles. In this case, $V_3 = V_4$. }
\label{twopart:fig:sixzero}
\end{figure}

In this section, we use the relative simplicity of the 1D $UV$ model to illustrate the second mechanism. To this end, we set $U = - \vert V \vert$ in the formulas of Section~\ref{twopart:sec:fourone}. Additionally, the strong-coupling limit, $U = V \rightarrow -\infty$, can be treated analytically. Consider Fig.~\ref{twopart:fig:sixzero}(b). It is sufficient to include only two types of spin-singlet configurations:        
\begin{align}
A_{m}  & = \vert \uparrow \downarrow \rangle_{m} \: , 
\label{twopart:eq:eightysixtwo} \\
B_{m}  & = \frac{1}{\sqrt{2}} 
\left( \vert \uparrow   \rangle_{m} \vert \downarrow \rangle_{m+1} + 
       \vert \downarrow \rangle_{m} \vert \uparrow   \rangle_{m+1}  \right)  \: .   
\label{twopart:eq:eightysixthree}
\end{align}
Hamiltonian action within this basis is
\begin{align}
\hat{H} A_{m}  & = - \sqrt{2} t B_{m} - \sqrt{2} t B_{m-1} \: , 
\label{twopart:eq:eightysixfour} \\
\hat{H} B_{m}  & = - \sqrt{2} t A_{m} - \sqrt{2} t A_{m+1} \: .   
\label{twopart:eq:eightysixfive}
\end{align}
The Schr\"odinger equation in momentum space is
\begin{align}
\tilde{E} A_{P}  & =  - \sqrt{2} t \left( 1 + e^{-iPa} \right) B_{P} \: , 
\label{twopart:eq:eightysixsix} \\
\tilde{E} B_{P}  & =  - \sqrt{2} t \left( 1 + e^{iPa} \right) A_{P} \: ,   
\label{twopart:eq:eightysixseven}
\end{align}
where $\tilde{E}$ is pair energy counted from $-|V|$ and $a$ is the lattice constant. Band dispersion is
\begin{equation}
\tilde{E}_{1,2}(P) = \pm 2 \sqrt{2} t \cos{\frac{Pa}{2}} \: ,   
\label{twopart:eq:eightysixeight}
\end{equation}
which corresponds to an effective mass 
\begin{equation}
m^{\ast}_{ps} = \sqrt{2} \, \frac{\hbar^2}{t a^2} = 2 \sqrt{2} \, m_0 \: .    
\label{twopart:eq:eightysixnine}
\end{equation}
Thus, pair mass remains of the order of free-particle mass $m_0$ even in the limit of infinitely strong attraction. Figure~\ref{twopart:fig:fivezero} shows a numerical solution of Eqs.~(\ref{twopart:eq:eightyfour}) and (\ref{twopart:eq:eightysixone}) for the resonant potential $U = -|V|$. Indeed, the pair mass never exceeds the strong-coupling limit $\sqrt{8} \, m_0$. 

One might think of an attractive interaction with $U = -|V|$ as exotic, but it can potentially be realized in cold gases where both $U$ and $V$ can be independently controlled. In crystalline solids, one can envision more realistic potentials comprising a strong repulsive core and a long-range attractive tail. Such a potential will necessarily have a minimum at a finite separation between particles. If the minimum is wide compared with the interatomic distance, then with high probability there will be two separations with equal attractive strengths. Such a situation is illustrated in Fig.~\ref{twopart:fig:sixzero}(c), where attraction on the {\em third} and {\em fourth} nearest neighbors are assumed equal, $V_3 = V_4$. (Importantly, other parts of the potential do not change the argument given below because if the particles are allowed to access configurations outside of the $-|V|$ basis, the pair mass will only decrease!) A proper ground state basis in this example is
\begin{equation}
A_{m} = \vert \bullet \rangle_{m} \vert \bullet \rangle_{m+3}  \: ; 
\hspace{0.5cm} 
B_{m} = \vert \bullet \rangle_{m} \vert \bullet \rangle_{m+4}  \: .   
\label{twopart:eq:eightysixeleven}
\end{equation}
(Since the particles cannot really exchange, there is no need to consider spin degrees of freedom. Singlet and triplet pairs will have the same mass.) The Schr\"odinger equation reads 
\begin{align}
\tilde{E} A_{P}  & =  - t \left( 1 + e^{-iPa} \right) B_{P} \: , 
\label{twopart:eq:eightysixtwelve} \\
\tilde{E} B_{P}  & =  - t \left( 1 + e^{ iPa} \right) A_{P} \: ,   
\label{twopart:eq:eightysixthirteen}
\end{align}
which yields $\tilde{E}_{1,2}(P) = \pm 2t \cos{\frac{Pa}{2}}$ and
\begin{equation}
m^{\ast} = \frac{2 \hbar^2}{t a^2} = 4 \, m_0 \: .    
\label{twopart:eq:eightysixfourteen}
\end{equation}
Thus, the bound pair is {\em no heavier} than just four free particle masses. Note that this conclusion does not depend on the separation distance at which the flat section of the potential occurs.

\subsection{\label{twopart:sec:fourtwo}
Triplet states 
}

\begin{figure}[t]
\begin{center}
\includegraphics[width=0.60\textwidth]{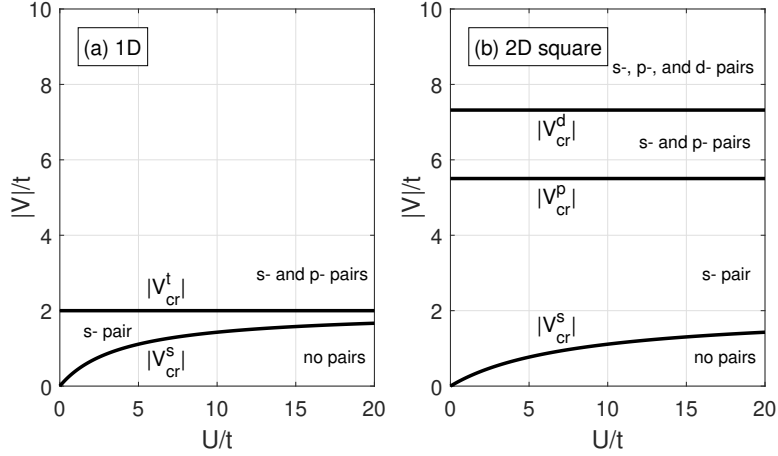}
\end{center}
\caption{Two-fermion phase diagrams in the 1D and 2D square $UV$ models at ${\bf P} = 0$.}
\label{twopart:fig:five}
\end{figure}

The antisymmetrized set of vectors consists of just one element $\{ {\bf b}_{-} \} = \{ (1) \}$ and there is one basis function $\Phi^{-}_{1}(P)$. An antisymmetrized Schr\"odinger equation, Eq.~(\ref{twopart:eq:thirteeneighteen}), reads:
\begin{equation}
( E_{t} - \varepsilon_{k_1} - \varepsilon_{k_2} ) \, \phi^{-}_{k_1 k_2} =    
- \vert V \vert \frac{1}{N} \sum_{ q } \phi^{-}_{ q , \: k_1 + k_2 - q } e^{iq} 
\left( e^{ - i k_1 } - e^{ -i k_2 } \right) .      
\label{twopart:eq:eightysevenone}
\end{equation}
In terms of the auxiliary function
\begin{equation}
\Phi^{-}_1(P) = \frac{1}{N} \sum_{ q } \phi^{-}_{ q , \: k_1 + k_2 - q } \: e^{iq}  \: ,  
\label{twopart:eq:eightyseventwo}
\end{equation}
the pair wave function is expressed as  
\begin{equation}
\phi^{-}_{k_1 k_2} = 
\frac{ - \vert V \vert \Phi^{-}_1(P) \left( e^{ - i k_1 } - e^{ -i k_2 } \right) }
     { E - \varepsilon_{k_1} - \varepsilon_{k_2} } \: . 
\label{twopart:eq:eightyseventhree}
\end{equation}
Substituting Eq.~(\ref{twopart:eq:eightyseventhree}) back in the definition, Eq.~(\ref{twopart:eq:eightyseventwo}), one obtains:
\begin{equation}
\Phi^{-}_1(P) = - \vert V \vert \Phi^{-}_1(P) 
\left( \frac{1}{N} \sum_{ q } \frac{ e^{ iq } ( e^{ -iq } - e^{ -i (P-q) } ) }
   { E - \varepsilon_{q} - \varepsilon_{P-q} } \right) .
\label{twopart:eq:eightysevenfour}
\end{equation}
Changing variables $q' = q - \frac{P}{2}$ in the last equation yields 
\begin{equation}
\left\{ 1 - \vert V \vert \, ( M_{0} - M_{2} ) \right\} \Phi^{-}_{1} = 0    \: .  
\label{twopart:eq:eightyeight}
\end{equation}
Note that it is independent of $U$, as expected for a triplet pair. Direct calculation results in 
\begin{equation}
E_t(P) = - \vert V \vert - \frac{ 4 t^2 \cos^2{\frac{P}{2}} }{ \vert V \vert }    \: ,  
\label{twopart:eq:eightynine}
\end{equation}
for the triplet energy, and in
\begin{equation}
\frac{m^{\ast}_{pt}}{m_0} = \frac{\vert V \vert}{t}  \: ,  
\label{twopart:eq:ninety}
\end{equation}
for the triplet effective mass, where $m_0 = \hbar^2/(2ta^2)$. The triplet pair is stable when 
\begin{equation}
\vert V \vert > \vert V^t_{\rm cr} \vert = 2 t \, \cos{\frac{P}{2}}  \: .   
\label{twopart:eq:ninetyone}
\end{equation}
Notice that in the $U \rightarrow \infty$ limit, $\vert V^s_{\rm cr} \vert = \vert V^t_{\rm cr} \vert$ and $E_s = E_t$. The phase diagram of the 1D $UV$ model at $P = 0$ is shown in Fig.~\ref{twopart:fig:five}(a).

\section{\label{twopart:sec:five}
2D. $UV$ model on the square lattice 
}

Two-dimensional lattice models at low carrier density have been popular in the studies of HTSC because most high-$T_c$ superconductors including the copper oxides are highly anisotropic. According to one point of view, superconductivity in cuprates is essentially two-dimensional. The (repulsive) 2D Hubbard model~\cite{Hubbard1963} and its derivative, the 2D {\it t-J} model~\cite{Chao1978}, were both put forward as capturing the essential physics~\cite{Dagotto1994}. Although this simple picture is being increasingly challenged~\cite{Alexandrov2011,Qin2020,Sherman2021,Jiang2022}, pure 2D models possess rich physics and remain popular in the fields of HTSC~\cite{Romer2020} and cold gases~\cite{Bloch2008}. For the purposes of this review, one should mention that the {\it t-J} model ``in the hole-rich regime''~\cite{Emery1990,Lin1991,Petukhov1992,Kagan1994} bears similarities with the $UV$ model studied here and many of the results derived later in this section apply equally to both models.

\subsection{\label{twopart:sec:fiveone}
Singlet states. $\Gamma$-point 
}

The symmetrized set of neighbor vectors consists of three elements: $\{ {\bf b}_{+} \} = \{ (0,0), (1,0), (0,1) \}$. In writing down the $(3 \times 3)$ system, Eqs.~(\ref{twopart:eq:thirteeneleven}) and (\ref{twopart:eq:thirteentwelve}), it is convenient to shift the inner variables in $M^{+}$: $q'_x = q_x - \frac{P_x}{2}$ and $q'_y = q_y - \frac{P_y}{2}$. That leads to a new set of functions: $\tilde{\Phi}^{+}_{\bf 0} = \Phi^{+}_{00}$, $\tilde{\Phi}^{+}_{\bf x} = e^{ -i (P_x/2) } \Phi^{+}_{10}$, and $\tilde{\Phi}^{+}_{\bf y} = e^{ -i (P_y/2) } \Phi^{+}_{01}$. In terms of the new set, the consistency condition reads
\begin{equation}
\left\vert \begin{array}{ccc}
1 + U M_{00} & - 2 \vert V \vert M_{10}              & - 2 \vert V \vert M_{01} \\
U M_{10}     & 1 - \vert V \vert ( M_{20} + M_{00} ) & - 2 \vert V \vert M_{11} \\
U M_{01}     & - 2 \vert V \vert M_{11}              & 1 - \vert V \vert ( M_{02} + M_{00} )  
\end{array} \right\vert 
= 0 \: ,   
\label{twopart:eq:ninetytwo}
\end{equation}
where
\begin{equation}
M_{nm} = 
\int\limits^{\pi}_{-\pi} \!\! \int\limits^{\pi}_{-\pi} \frac{{\rm d}q_x \, {\rm d}q_y}{(2\pi)^2} 
\frac{\cos{nq_x} \cos{mq_y}}{ \vert E \vert - \alpha \cos{q_x} - \beta \cos{q_y} } \: ,
\label{twopart:eq:ninetythree}
\end{equation}
$\alpha = 4 t \cos{(P_x/2)}$, and $\beta = 4 t \cos{(P_y/2)}$. $M_{00}$ was given in Eq.~(\ref{twopart:eq:thirtyfive}). Other matrix elements in Eq.~(\ref{twopart:eq:ninetytwo}) can also be expressed via complete elliptic integrals, see \ref{twopart:sec:appaone} and \ref{twopart:sec:appafive}. The double integrals can also be computed numerically. 

At the $\Gamma$ point, $P_x = P_y = 0$ and $\alpha = \beta = 4t$. In this case, $M_{10} = M_{01}$, $M_{20} = M_{02}$, and Eq.~(\ref{twopart:eq:ninetytwo}) acquires additional symmetry. Introducing a new basis $\Phi^{+}_{\bf 0}$, $\Phi^{+}_{s} = \frac{1}{2} ( \Phi^{+}_{\bf x} + \Phi^{+}_{\bf y})$, and $\Phi^{+}_{d} = \frac{1}{2} ( \Phi^{+}_{\bf x} - \Phi^{+}_{\bf y} )$, the equation splits into $s$-symmetric and $d$-symmetric sectors. The $s$-sector involves functions $\Phi^{+}_{\bf 0}$ and $\Phi^{+}_{s}$ and its consistency condition reads
\begin{equation}
\left\vert \begin{array}{cc}
1 + U M_{00}  & - 4 \vert V \vert M_{10}                          \\
    U M_{10}  & 1 - \vert V \vert ( M_{20} + M_{00} + 2 M_{11} )
\end{array} \right\vert = 0 \: .   
\label{twopart:eq:onehthree}
\end{equation}
The $d$-sector equation is obtained by subtracting the last two lines of Eq.~(\ref{twopart:eq:ninetytwo}). It involves only one function $\Phi^{+}_{d}$ and does not include $U$:
\begin{equation}
\left\{ 1 - \vert V \vert ( M_{20} + M_{00} - 2 M_{11} ) \right\} \Phi^{+}_{d} = 0 \: .   
\label{twopart:eq:onehsix}
\end{equation}

We begin analysis with the $s$-symmetrical ground state described by Eq.~(\ref{twopart:eq:onehthree}). First, we note that the combination $M_{20} + M_{00} + 2 M_{11} = \frac{2|E|}{\alpha} M_{10}$ can be expressed via $M_{10}$, and the latter can be expressed via $M_{00}$ as $M_{10} = \frac{1}{2\alpha}( |E|M_{00} - 1 )$. Thus, all the matrix elements in Eq.~(\ref{twopart:eq:onehthree}) are expressible via the base integral $M_{00}$. Expanding the determinant, one obtains
\begin{equation}
( U M_{00} + 1 ) + 
\frac{1}{\alpha^2} \vert V \vert ( \vert E_{s} \vert + U ) ( 1 - \vert E_s \vert M_{00} ) = 0 \: ,   
\label{twopart:eq:onehsixone}
\end{equation}
where
\begin{equation}
M_{00} = \frac{2}{\pi |E_s|} \, {\bf K}\left( \frac{2\alpha}{|E_s|} \right) ;   
\hspace{0.5cm} 
\alpha = 4t \: .
\label{twopart:eq:onehsixtwo}
\end{equation}
Equation~(\ref{twopart:eq:onehsixone}) determines the energy of $s$-states in the $\Gamma$ point. Depending on the values of $U$ and $V$, there may be one, two, or no bound states. Equation~(\ref{twopart:eq:onehsixone}) should be compared with its 1D counterpart, Eq.~(\ref{twopart:eq:eightythreethree}). The former has a factor 1 in the second term while the latter has a factor 2. Otherwise, the two energy equations have similar structures.  

Let us determine the pairing threshold for a positive $U$. To this end, set $E_{s} = -8t - 0$ in Eq.~(\ref{twopart:eq:onehsixone}). Then $M_{00}$ logarithmically diverges. This yields a critical coupling strength:
\begin{equation}
\vert V^{s}_{\rm cr} \vert = \frac{2 U t}{ U + 8 t} \: .
\label{twopart:eq:onehsixthree}
\end{equation}
This line separates the regions of ``no pairs'' and ``s-pairs'' in the 2D $UV$ phase diagram, see Fig.~\ref{twopart:fig:five}(b). Using the asymptotic behavior of the elliptic integral 
\begin{equation}
{\bf K}\left( \frac{8t}{ 8t + \Delta } \right) \simeq \frac{1}{2} \log{ \frac{64 \, t}{\Delta} } \: ; 
\hspace{0.5cm} \Delta \ll t \:, 
\label{twopart:eq:onehsixfour}
\end{equation}
one obtains the binding energy near the threshold:
\begin{equation}
\Delta = ( 64 e^{-\pi} ) \, t \, 
\exp{ \left(  - \frac{2 \pi t}{ | V | - | V^{s}_{\rm cr} | } \right) }  \: .  
\label{twopart:eq:onehsixfive}
\end{equation}
Note that the exponent is four times less than in the corresponding expression in the attractive Hubbard model, Eq.~(\ref{twopart:eq:thirtyeight}). This is because the $UV$ model has four attractive sites instead of one. In the $U \rightarrow \infty$ limit, $| V^{s}_{\rm cr} | \rightarrow 2t$, and the general expression simplifies to 
\begin{equation}
\Delta (U = \infty) = 64 \, t \, 
\exp{ \left(  - \frac{\pi |V|}{ | V | - 2t } \right) }  \: .  
\label{twopart:eq:onehsixsix}
\end{equation}
In this form, the binding energy was given in Ref.~\cite{Kagan1994}.

Turning now to the $d$-symmetric state, Eq.~(\ref{twopart:eq:onehsix}), one observes that the combination $M_{20} + M_{00} - 2M_{11}$ converges in the limit $E \rightarrow -8t$. Utilizing explicit expressions in \ref{twopart:sec:appathree}, one derives the $d$-pairing threshold
\begin{equation}
\vert V^{d}_{\rm cr} \vert = \frac{ 2 \pi t }{ 4 - \pi }  
= ( 7.319584 \ldots ) \, t \: .   
\label{twopart:eq:onehsixseven}
\end{equation}
It is independent of $U$, as expected for a $d$-symmetric wave function. General expressions for $M_{00}$, $M_{11}$, and $M_{20}$ at arbitrary $E$ are given in \ref{twopart:sec:appathree}. Using those, Eq.~(\ref{twopart:eq:onehsix}) defines the $d$-state energy as a function of $V$.

\subsection{\label{twopart:sec:fivefour}
Singlet states. Arbitrary momentum ${\bf P}$ 
}

Separation of the singlet dispersion relation, Eq.~(\ref{twopart:eq:ninetytwo}), into $s$ and $d$ sectors is also possible on the BZ diagonal. This is because at $P_x = P_y$, the relations $M_{10} = M_{01}$ and $M_{20} = M_{02}$ continue to be valid. Transformations described in Section~\ref{twopart:sec:fiveone} still apply, leading to the final dispersion relations, Eqs.~(\ref{twopart:eq:onehsixone}) and (\ref{twopart:eq:onehsix}). The only difference is the modified expression for $\alpha = 4t \cos{(P_x/2)}$. This simple dependence on $P$ along BZ diagonals can be used, for example, to extract the effective mass separately for $s$- and $d$-symmetrical pairs. Another consequence is a simple modification of the binding thresholds:
\begin{equation}
\vert V^{s}_{\rm cr}( P_x = P_y ) \vert = 
\frac{ 2 U \, t \cos{\frac{P_x}{2}} }{ U + 8 \, t \cos{\frac{P_x}{2}} } \: .   
\label{twopart:eq:onehfive}
\end{equation}
\begin{equation}
\vert V^{d}_{\rm cr}( P_x = P_y ) \vert = \frac{ 2 \pi \, t}{ 4 - \pi } \, \cos{\frac{P_x}{2}}  \: .   
\label{twopart:eq:oneheight}
\end{equation}
Like in the 1D $UV$ model, see Eq.~(\ref{twopart:eq:eightysix}), these expressions indicate that the pairing thresholds {\em decrease} at large lattice momenta. The energy of bound states still grow with ${\bf P}$ but the lowest energy of two free particles {\em at the same} ${\bf P}$ grows even faster. As a result, a bound pair may form at a finite ${\bf P}$ even if it is unstable at ${\bf P} = 0$. This physics is much richer in 2D than in 1D. Below, we investigate it in some detail~\cite{Kornilovitch2004}.          

\begin{figure}[t]
\begin{center}
\includegraphics[width=0.60\textwidth]{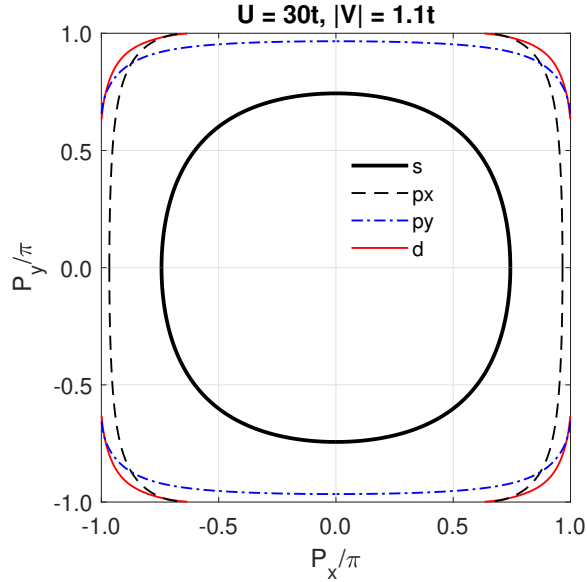}
\end{center}
\caption{The pairing lines in the 2D square $UV$ model for $U = 30\,t$ and $|V| = 1.1\,t$. Bound pairs form {\em outside} the respective lines. The $s$ and $d$ lines are solutions of Eqs.~(\ref{twopart:eq:ninetysix}) and (\ref{twopart:eq:ninetyseven}). The $p_x$ and $p_y$ lines are solutions of Eqs.~(\ref{twopart:eq:onehnine}) and (\ref{twopart:eq:onehten}) at $E = - \alpha - \beta$. At these parameters, both $p$ and $d$ lines terminate at the Brillouin zone boundaries.~\cite{Kornilovitch2004}  } 
\label{twopart:fig:eightone}
\end{figure}

In order to determine the binding threshold at an arbitrary ${\bf P}$, energy $E$ must be set equal to the minimal energy of two free particles $E_{11} = - \alpha - \beta$. Upon substitution $E = E_{11}$, all $M_{nm}$ in Eq.~(\ref{twopart:eq:ninetytwo}) diverge logarithmically. To regularize the determinant, express each $M_{nm}$ as a sum of $M_{00}$ and remaining difference $L_{nm}$:
\begin{equation}
M_{nm} = M_{00} + ( M_{nm} - M_{00} ) \equiv M_{00} + L_{nm} \: . 
\label{twopart:eq:ninetyfour}
\end{equation}
Note that all $L_{nm}$ converge to finite values in the $E = E_{11}$ limit. Explicit analytical expressions are given in \ref{twopart:sec:appaseven}. Insertion of Eq.~(\ref{twopart:eq:ninetyfour}) into Eq.~(\ref{twopart:eq:ninetytwo}) and expansion of the determinant leads to a lengthy expression that is a third-order polynomial in $M_{00}$. However, the $M^3_{00}$ and $M^2_{00}$ terms cancel identically, and the determinant assumes the form
\begin{equation}
A \cdot M_{00} + B = 0 \: , 
\label{twopart:eq:ninetyfive}
\end{equation}
where 
\begin{align}
A = & \; U \vert V \vert^2 
\left[ ( L_{20} - 4 L_{10} )( L_{02} - 4 L_{01} ) - 4 ( L_{10} + L_{01} - L_{11} )^2 \right]   
\nonumber \\
& + U \vert V \vert \left[ ( 4 L_{01} - L_{02} ) + ( 4 L_{10} - L_{20} ) \right] + U  
+ \vert V \vert^2 \left[ 2 ( L_{02} + L_{20} ) - 8 L_{11} \right] - 4 \vert V \vert \: , 
\label{twopart:eq:ninetysix}
\end{align}
and the specific form of $B$ is unimportant. In the limit $E \rightarrow E_{11}$, both $A$ and $B$ remain finite whereas $M_{00}$ diverges. Thus, the binding condition becomes
\begin{equation}
A = 0 \: . 
\label{twopart:eq:ninetyseven}
\end{equation}
Equations~(\ref{twopart:eq:ninetysix}) and (\ref{twopart:eq:ninetyseven}) determine the binding threshold sought. For a given ${\bf P}$, they define a function $\vert V_{\rm cr} \vert (U)$. Alternatively, for some fixed $U$ and $\vert V \vert$, the threshold defines a line inside BZ that separates unbound states at small ${\bf P}$ from bound pairs at large ${\bf P}$. An example of such boundary lines is shown in Fig.~\ref{twopart:fig:eightone}. Properties of these lines were studied in Ref.~\cite{Kornilovitch2004}.

\begin{figure}[t]
\begin{center}
\includegraphics[width=0.60\textwidth]{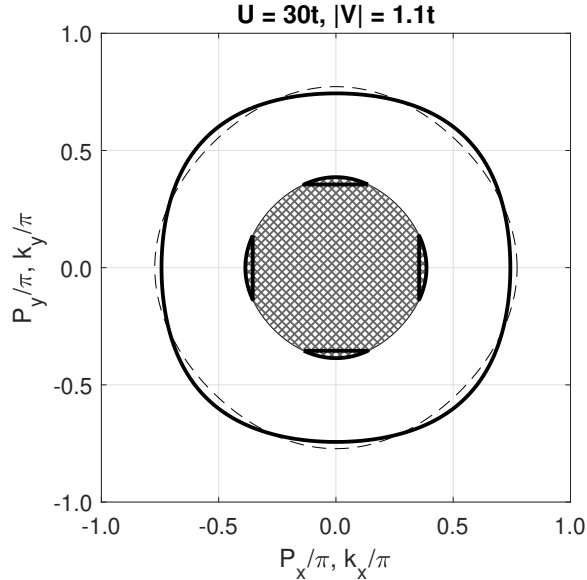}
\end{center}
\caption{Comparison of the pairing and Fermi lines in the 2D square $UV$ model. The thick solid line is the same $s$ pairing line drawn in Fig.~\ref{twopart:fig:eightone}. The hatched area in the center are free-particle states enclosed by a Fermi line with energy $E_F = -2.7\,t$. It corresponds to a filling factor of 0.113 and total particle density of $n = 0.226$. The dashed line is the {\em twice}-Fermi momentum line. The four circular segments outlined by thick lines enclose free particle states that would tend to pair up if they were moving in an empty lattice.} 
\label{twopart:fig:nineone}
\end{figure}

The existence of such lines poses an interesting question about pair formation at a finite but low particle density. Figure~\ref{twopart:fig:nineone} shows the same $s$ pairing line as Fig.~\ref{twopart:fig:eightone}, together with the free-dispersion Fermi line corresponding to a Fermi energy $E_F = -2.7\,t$ and filling factor 0.113. Consider states within one of the four outlined circular segments, for example states with $k_x = 0$ and $k_y = k_F$ (the Fermi momentum). The combined momentum of two such particles would be ${\bf P} = (0,2k_F)$ which would land {\em beyond} the $s$-pairing line. Therefore, the particles would tend to form a bound state. Intriguingly, the states near the Fermi line close to BZ diagonals do not show the same tendency because their combined momentum lies inside the pairing surface, see Fig.~\ref{twopart:fig:nineone}. The entire Fermi line splits into eight disconnected segments: four with pairing and four without. The segments without pairing should produce sharp features in photoemission experiments (``Fermi arcs''), while those with pairing will produce emission lines separated from the Fermi energy by a gap. Such a behavior is consistent with photoemission signatures of some cuprates superconductors~\cite{Damascelli2003}.       

Clearly, the central question is whether this empty-lattice picture retains its qualitative features at small but finite densities. That requires a reliable many-body method that can handle large systems such as Quantum Monte Carlo~\cite{Marsiglio1990,LeBlanc2015} or low-density $T$-matrix approximation~\cite{Micnas1995,Kyung1998}. Such a treatment is beyond the scope of the present work and is not analyzed here.

\begin{figure}[t]
\begin{center}
\includegraphics[width=0.48\textwidth]{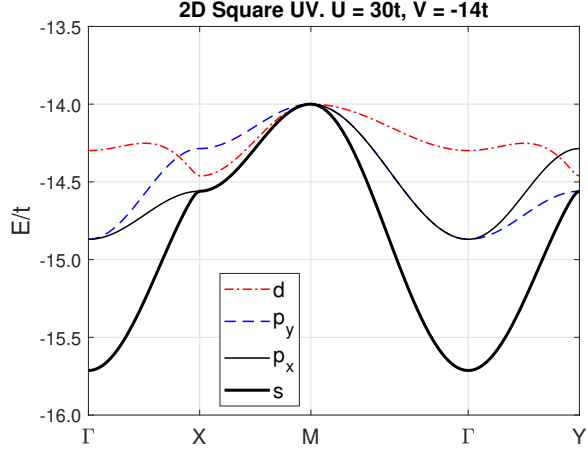}
\end{center}
\caption{Dispersion of bound pairs in the 2D square $UV$ model for $U = 30\,t$ and $V = -14\,t$. The $s$ and $d$ energies are solutions of Eq.~(\ref{twopart:eq:ninetytwo}). The $s$ dispersion along the ${\rm X}-{\rm M}$ line is given by Eq.~(\ref{twopart:eq:ninetysevenone}). $p_x$ and $p_y$ energies are solutions of Eqs.~(\ref{twopart:eq:onehnine}) and (\ref{twopart:eq:onehten}), respectively. Notice $p_x$, $p_y$ degeneracy along the ${\rm M}-\Gamma$ line and $s$, $p_x$ degeneracy along the ${\rm X}-{\rm M}$ line. } 
\label{twopart:fig:ninetwo}
\end{figure}

Next, we discuss full pair dispersion $E({\bf P})$ for all ${\bf P}$. Due to complexity of the main dispersion relation, Eq.~(\ref{twopart:eq:ninetytwo}), very little can be done analytically beyond the separation of $s$ and $d$ energies on BZ diagonals. Some simplification occurs at BZ edges. Let us set for definitiveness, $P_x = \pm \pi$. Then $M_{10} = M_{20} = M_{11} = 0$ by symmetry. The remaining matrix elements, $M_{00}$, $M_{01}$, and $M_{02}$, become one-dimensional integrals given by Eq.~(\ref{twopart:eq:eightythree}). Upon expansion, the determinant splits in two factors. One has an explicit solution
\begin{equation}
E_{1}( \pm \pi, P_y ) = - \sqrt{ |V|^2 + \beta^2 } \: , 
\label{twopart:eq:ninetysevenone}
\end{equation}
where $\beta = 4t \cos{(P_y/2)}$. The second factor can be brought to the following form
\begin{equation}
2 |V| ( U + |E| )( |E| - \sqrt{ |E|^2 - \beta^2 } )          
- \beta^2 ( U + \sqrt{ |E|^2 - \beta^2 } ) = 0 \: . 
\label{twopart:eq:ninetyseventwo}
\end{equation}
This equation does not have a simple-form analytical solution. An example of numerical solution of Eq.~(\ref{twopart:eq:ninetytwo}) is shown in Fig.~\ref{twopart:fig:ninetwo}. A bound pair behaves as a single particle with a fairly complex dispersion. Notice degeneracies along some high-symmetry lines.

\subsection{\label{twopart:sec:fivetwo}
Triplet states 
}

There are two antisymmetrized vectors $\{ {\bf b}_{-} \} = \{ (1,0), $ $(0,1) \}$ and two functions $\Phi^{-}$. Upon constructing the system, Eq.~(\ref{twopart:eq:thirteentwentyone}), the integration variable in $M^{-}_{nm}$ is shifted as ${\bf q}' = {\bf q} - \frac{\bf P}{2}$. Off-diagonal terms vanish by symmetry, and the $( 2 \times 2 )$ system splits into two separate equations:
\begin{align}
\left\{ 1 - \vert V \vert \left( M_{00} - M_{20} \right) \right\} \Phi^{-}_{p_x} & = 0 \: ,
\label{twopart:eq:onehnine} \\
\left\{ 1 - \vert V \vert \left( M_{00} - M_{02} \right) \right\} \Phi^{-}_{p_y} & = 0 \: .
\label{twopart:eq:onehten}
\end{align}
Note that such a decomposition takes place over the entire BZ. Since $M_{20} \neq M_{02}$, $p_x$ and $p_y$ energies are different, as shown Fig.~\ref{twopart:fig:ninetwo}. Along the diagonals, $M_{20} = M_{02}$, and the dispersion becomes double-degenerate. 

Let us derive the binding condition along BZ diagonals. In the limit $E \rightarrow E_{11}$, the difference $M_{00} - M_{20}$ converges. Making use of Eq.~(\ref{twopart:eq:appafortytwo}), one obtains: 
\begin{equation}
\vert V^{p}_{\rm cr} \vert = \frac{ 2 \pi }{ \pi - 2 } \, t \cos{\frac{P_x}{2}} 
= ( 5.503876 \ldots ) \, t \cos{\frac{P_x}{2}} \: .   
\label{twopart:eq:oneheleven}
\end{equation}
The two-particle phase diagram of the square $UV$ model for ${\bf P} = 0$ is shown in Fig.~\ref{twopart:fig:five}(b). The binding condition at arbitrary ${\bf P}$ can be obtained from Eqs.~(\ref{twopart:eq:onehnine}) and (\ref{twopart:eq:onehten}) if analytical expressions for the $M$ differences, Eqs.~(\ref{twopart:eq:appaeightyone}) and (\ref{twopart:eq:appaeightytwo}), are utilized. An example of $p_x$ and $p_y$ pairing lines is given in Fig.~\ref{twopart:fig:eightone}. Analytical properties of the $p$ pairing lines were studied in Ref.~\cite{Kornilovitch2004}. 

Consider now triplet energy at arbitrary ${\bf P}$. Again, certain simplification takes place at BZ edges. At $P_x = \pm \pi$, $M_{20} = 0$, and the energy of $p_x$ pair is given by Eq.~(\ref{twopart:eq:ninetysevenone}). Thus, the dispersion at BZ edges is always double-degenerate. This can also be seen in Fig.~\ref{twopart:fig:ninetwo}. At the $\Gamma$-point, the $p$ energy can be determined from the following equation
\begin{equation}
\pi - 2 {\bf E}\! \left( \frac{ 8 t }{ \vert E_p \vert } \right) = 
\frac{2 \pi t }{ |V| } \frac{8t}{|E_{p}|} \: ,   
\label{twopart:eq:onehtwelve}
\end{equation}
where Eqs.~(\ref{twopart:eq:appathirtyone}) and (\ref{twopart:eq:appathirtyfour}) have been utilized. Equation~(\ref{twopart:eq:onehtwelve}) does not have a simple analytical solution for $E_p$.

\subsection{\label{twopart:sec:fivethree}
Longer-range attractions  
}

The square $UV$ model is the rare case when the two-body problem was solved for interactions beyond the nearest neighbors~\cite{Kornilovitch1995,Bak1999,Adebanjo2022b}. In Refs.~\cite{Bak1999,Adebanjo2022b}, next-nearest interaction and next-nearest hopping were considered. In Ref.~\cite{Kornilovitch1995}, attraction was extended up to the {\em seventh} nearest neighbors (with the potential depth being constant within the radius of attraction $R$) but the hopping was limited to the first neighbors only. Below, we provide the reasoning behind the model analyzed in Ref.~\cite{Kornilovitch1995}.       

\begin{figure}[t]
\begin{center}
\includegraphics[width=0.48\textwidth]{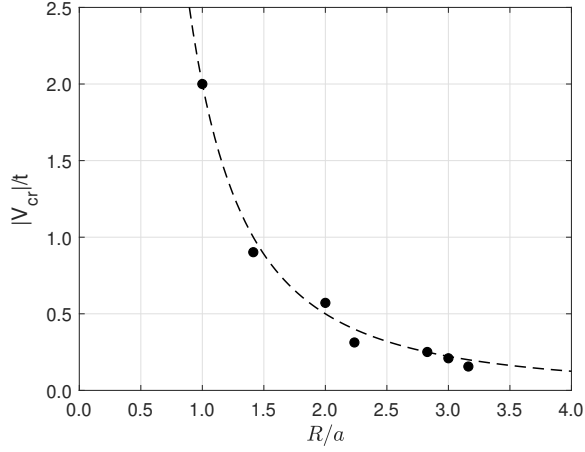}
\end{center}
\caption{Binding threshold of the $UV$ model on the square lattice versus the distance of $n$th nearest neighbors from the origin. $U = \infty$. Solid symbols are exact values obtained from solving two-body Schr\"odinger equations. (The values are 2.0, 0.90212, 0.57139, 0.31280, 0.25075, 0.20967, and 0.15584.) The dashed line is a phenomenological dependence $\vert V_{\rm cr} \vert = 2/R^2$ and is guide to the eye. Adapted from Ref.~\cite{Kornilovitch1995}.} 
\label{twopart:fig:six}
\end{figure}

In order to form a bound pair, attraction must exceed a threshold to overcome the strong repulsive core. For nearest-neighbor attraction, the threshold is about $2t$, see Eq.~(\ref{twopart:eq:onehsixthree}). In physical units, it may be quite a large number. Let us assume the effective hopping of {\em holes} in the cuprates to be $t_{h} \sim 0.1$ eV~\cite{Leung1995,Harrison2023}. Then, the attraction must be of order $V \simeq 0.1 - 0.2$ eV, which is arguably quite large. At the same time, the cuprates are anisotropic polar solids with low carrier density and poorly screened electron-phonon interactions. This favors formation of bipolarons~\cite{Alexandrov2002A,Alexandrov2002B,Zhang1991,Catlow1998,Alexandrov2002C}. It also results in longer-range shallow attractive potentials within the copper--oxygen planes. Spreading attraction to many sites allows lowering the threshold on individual sites. Since it is the ``total power'' of the potential that matters for binding, one expects the threshold to scale approximately inversely to the number of sites participating in attraction, i.e., inversely with the potential {\em area}. (In continuous quantum mechanics the scaling is also $V_{\rm cr} \propto 1/R^2$.) This is what was confirmed by exact calculations~\cite{Kornilovitch1995}. The dependence of $\vert V_{cr} \vert (R)$ is shown in Fig.~\ref{twopart:fig:six}. It approximately follows the $1/R^2$ scaling, as expected for a continuum problem, with fluctuations around the line which reflects the discrete nature of the lattice. One can see that by the 6th or 7th nearest neighbors, the threshold falls by an order of magnitude to about $0.2\,t \simeq 0.02$ eV, which would be easier to attain in real solids.

One can add to this argument the {\rm light pairs} mechanism already discussed in Section~\ref{twopart:sec:fourthree}. In shallow long-range attractive potentials, two or more neighbors will have equal or close attractive strengths with high probability. That will enable resonant movement of pairs without breaking attractive bonds. As a result, pair effective mass will remain of order $m_0$ even in the strong coupling limit. We illustrate this mechanism for an attractive potential extended to the second nearest neighbors with $V_2 = V_1$. We introduce the strong-coupling singlet dimer basis~\cite{Adebanjo2022b}
\begin{equation}
D_{i,{\bf m}} = \frac{1}{\sqrt{2}} \left( 
\vert   \uparrow \rangle_{\bf m} \vert \downarrow \rangle_{ {\bf m} + {\bf b}_i } + 
\vert \downarrow \rangle_{\bf m} \vert   \uparrow \rangle_{ {\bf m} + {\bf b}_i } \right) \: ,   
\label{twopart:eq:onehtwelveone}
\end{equation}
with ${\bf b}_1 = (1,0)$, ${\bf b}_2 = (1,1)$, ${\bf b}_3 = (0,1)$, and ${\bf b}_4 = (-1,1)$. Hamiltonian action within this space is given by 
\begin{align}
\hat{H} D_{1,{\bf m}} & = - t \left( 
D_{2,{\bf m}}              \! + \! D_{2,{\bf m} - {\bf b}_3 } \! + \!  
D_{4,{\bf m} + {\bf b}_1 } \! + \! D_{4,{\bf m} - {\bf b}_4 }  \right)  , \makebox[0.2cm]{}   
\nonumber                         \\
\hat{H} D_{2,{\bf m}} & = - t \left( 
D_{1,{\bf m}}              \! + \! D_{1,{\bf m} + {\bf b}_3 } \! + \!  
D_{3,{\bf m}}              \! + \! D_{3,{\bf m} + {\bf b}_1 }  \right)  ,   
\nonumber                         \\
\hat{H} D_{3,{\bf m}} & = - t \left( 
D_{4,{\bf m}}              \! + \! D_{4,{\bf m} + {\bf b}_1 } \! + \!  
D_{2,{\bf m}}              \! + \! D_{2,{\bf m} - {\bf b}_1 }  \right)  ,   
\nonumber                         \\
\hat{H} D_{4,{\bf m}} & = - t \left( 
D_{3,{\bf m}             } \! + \! D_{3,{\bf m} - {\bf b}_1 } \! + \!  
D_{1,{\bf m} - {\bf b}_1 } \! + \! D_{1,{\bf m} + {\bf b}_4 }  \right)  .   
\label{twopart:eq:onehtwelvefour}
\end{align}
Converting the corresponding Schr\"odinger equation to momentum space and expanding the determinant, one obtains dimer dispersion:
\begin{align}
\tilde{E}_{1,2}({\bf P}) & = \pm (2t) \sqrt{ 2 + \cos{(P_x a)} + \cos{(P_y a)} } \:  ,   
\label{twopart:eq:onehtwelveeight} \\
\tilde{E}_{3,4}({\bf P}) & = 0                                                   \:  ,   
\label{twopart:eq:onehtwelvenine} 
\end{align}
where $\tilde{E}$ is referenced from $-|V|$. At small ${\bf P}$, $\tilde{E}_{1} \approx -4t + (t a^2 P^2)/4$, from where the pair mass is
\begin{equation}
m^{\ast}_{p} = \frac{2 \hbar^2}{t a^2} = 4 \, m_0 \: ,    
\label{twopart:eq:onehtwelveten}
\end{equation}
where $m_0 = \hbar^2/(2 t a^2)$. Thus, in this model the mass enhancement is limited by the same factor of 4 as in the 1D model with long-range attraction, cf. Eq.~(\ref{twopart:eq:eightysixfourteen}). The dimer analysis can also be extended to nonzero second-neighbor hopping~\cite{Adebanjo2022b}. 

Next, we rigorously solve the $UV$ model with attraction extended to second nearest neighbors. In deviation from \cite{Kornilovitch1995}, the second neighbor attraction $-\vert V_2 \vert$ will be different from the first neighbor attraction $- \vert V_1 \vert$. We consider only singlet pairs and begin with ${\bf P} = 0$. According to the general theory, there are {\em five} symmetrized neighbor vectors. They can be chosen, for example, as $\{ {\bf b}_{+} \} = \{ (0,0), (1,0), (0,1) , (1,-1) , (1,1) \}$. Hence, there are five equations, Eqs.~(\ref{twopart:eq:thirteeneleven}) and (\ref{twopart:eq:thirteentwelve}), for five functions $\Phi^{+}$. The equations mix $s$- and $d$-symmetric pair states. To untangle them, add and subtract the equations for $\Phi^{+}_{(1,0)}$ and $\Phi^{+}_{(0,1)}$, and then do the same for the equations for $\Phi^{+}_{(1,-1)}$ and $\Phi^{+}_{(1,1)}$. The two differences are 
\begin{align}
& \left\{ 1 - \vert V_1 \vert ( M_{20} + M_{00} - 2 M_{11} ) \right\} 
\left( \Phi^{+}_{10} - \Phi^{+}_{01} \right)  =  0 \, ,   
\label{twopart:eq:onehfifteen} \\
& \left\{ 1 - \vert V_2 \vert ( M_{22} + M_{00} - 2 M_{20} ) \right\} 
\left( \Phi^{+}_{1-\!1} - \Phi^{+}_{11} \right) = 0 \, .  
\label{twopart:eq:onehsixteen}
\end{align}
The first equation is equivalent to Eq.~(\ref{twopart:eq:onehsix}) and describes a $d$-pair previously discussed. It has lobes along $x$ and $y$ axes, and therefore may be called a $d_{x^2-y^2}$ state. This state does not depend on $V_2$ and has the same pairing threshold as before, Eq.~(\ref{twopart:eq:onehsixseven}). Equation~(\ref{twopart:eq:onehsixteen}) describes another $d$ state with lobes along the square diagonals. It may be called a $d_{xy}$ state. Note that the two $d$ states do not mix. The $d_{xy}$ state does not involve potential $V_1$ because the wave function has nodes on the first nearest neighbors. The linear combination of $M$'s in Eq.~(\ref{twopart:eq:onehsixteen}) converges in the $E \rightarrow E_{11}$ limit. Applying Eqs.~(\ref{twopart:eq:appafortytwo}) and (\ref{twopart:eq:appafortyeight}), one derives the $d_{xy}$ pairing threshold 
\begin{equation}
\vert V^{d_{xy}}_{2,{\rm cr}} \vert = \frac{ 3 \pi }{ 3\pi - 8 } \, t  
= ( 6.614909 \ldots ) \, t \: .   
\label{twopart:eq:onehseventeen}
\end{equation}
It is {\em smaller} than the $d_{x^2-y^2}$ threshold, Eq.~(\ref{twopart:eq:onehsixseven}), by $\approx 10$\%. Because of a node at ${\bf m} = (0,0)$, the pair wave function is larger at the second nearest neighbors than at the first nearest neighbors. As a result, potential $V_2$ has more ``power'' and produces a bound state at a slightly lower value that $V_1$.       

Returning to the general system, Eqs.~(\ref{twopart:eq:thirteeneleven}) and (\ref{twopart:eq:thirteentwelve}), the two equation sums combine with the equation for $\Phi^{+}_{(0,0)}$ and form the $s$-sector. The full system reads 
\begin{equation}
\left[ \begin{array}{ccc}
1 + U M_{00}  & - 4 \vert V_1 \vert M_{10}                         & - 4 \vert V_2 \vert M_{11}              \\
    U M_{10}  & 1 - \vert V_1 \vert ( M_{20} + M_{00} + 2 M_{11} ) & - 2 \vert V_2 \vert ( M_{10} + M_{21} ) \\
    U M_{11}  & - 2 \vert V_1 \vert ( M_{10} + M_{21} )            & 1 - \vert V_2 \vert ( M_{22} + M_{00} + 2 M_{20} ) 
\end{array} \right] 
\left[ \begin{array}{c}
\Phi^{+}_{00}               \\ 
\frac{1}{2} \Phi^{+}_{10}  + \frac{1}{2} \Phi^{+}_{01}  \\
\frac{1}{2} \Phi^{+}_{1-1} + \frac{1}{2} \Phi^{+}_{11}
\end{array} \right]
= \left[ \begin{array}{c}
 0 \\  0  \\  0 
\end{array} \right] .   
\label{twopart:eq:oneheighteen}
\end{equation}
The binding threshold is derived by equating the determinant to zero and setting $E = -8t$. Because all $M_{mn}$ logarithmically diverge, one needs to apply the subtractive procedure, Eq.~(\ref{twopart:eq:ninetyfour}). After the substitution, the determinant is expanded into the form $A \cdot M_{00} + B$, with complex expressions for $A$ and $B$ that are both convergent. Again, terms with $M^2_{00}$ and $M^3_{00}$ cancel to zero. Since only $M_{00}$ remains divergent, the binding condition is simply $A = 0$. Utilizing Eqs.~(\ref{twopart:eq:appafortyone})-(\ref{twopart:eq:appafiftyone}), the final result reads:
\begin{equation}
( 32 - 9 \pi ) U \vert V_1 \vert \vert V_2 \vert - 6 \pi U \vert V_1 \vert t 
+ ( 12 \pi - 64 ) U \vert V_2 \vert t + ( 256 - 72 \pi ) \vert V_1 \vert \vert V_2 \vert t 
+ 12 \pi U t^2 - 48 \pi \vert V_1 \vert t^2 - 48 \pi \vert V_2 \vert t^2 = 0 .
\label{twopart:eq:onehnineteen}
\end{equation}
For $V_2 = 0$, it reduces to Eq.~(\ref{twopart:eq:onehsixthree}). For $V_1 = 0$, Eq.~(\ref{twopart:eq:onehnineteen}) yields 
\begin{equation}
\vert V_{2,{\rm cr}} \vert = \frac{ 3 \pi U t }{ ( 16 - 3 \pi ) U + 12 t } \: . 
\label{twopart:eq:onehnineteenone}
\end{equation}
For $V_1 = V_2$ and $U \rightarrow \infty$, Eq.~(\ref{twopart:eq:onehnineteen}) reduces to 
\begin{equation}
( 32 - 9 \pi ) \vert V \vert^2 + ( 6 \pi - 64 ) \vert V \vert + 12 \pi = 0 \: , 
\label{twopart:eq:onehtwenty}
\end{equation}
that was reported in Ref.~\cite{Kornilovitch1995}. The smallest root of this equation is $V_{\rm cr}(2) = 0.902120$, which is plotted in Fig.~\ref{twopart:fig:six}. $V_{\rm cr}(2)$ should be compared with the $U \rightarrow \infty$ limit of Eq.~(\ref{twopart:eq:onehsixthree}), which is $V_{\rm cr}(1) = 2.0$.

\begin{figure}[t]
\begin{center}
\includegraphics[width=0.48\textwidth]{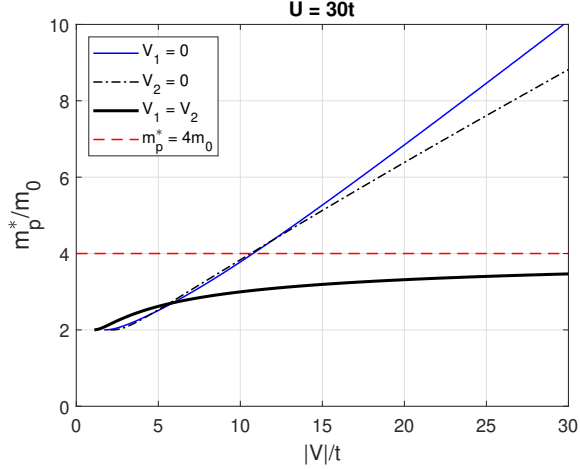}
\end{center}
\caption{Pair effective mass in the 2D square $UV_1V_2$ model with attraction on the first {\em and} second nearest neighbors. The masses are obtained by numerically solving Eq.~(\ref{twopart:eq:oneheighteen}) for $E$ and then applying Eq.~(\ref{twopart:eq:onehtwentyone}). $U = 30\,t$. The dashed line marks the light-pair limit, Eq.~(\ref{twopart:eq:onehtwelveten}). } 
\label{twopart:fig:twelveone}
\end{figure}

Decomposition of the full dispersion relation into $s$ and $d$ sectors also occurs on BZ diagonals, ${\bf P} = (P,P)$. Matrix elements $M_{nm}$ are still given by their isotropic expressions listed in \ref{twopart:sec:appathree} but with $\alpha = 4t \cos{(P/2)}$. This allows for numerical calculation of effective masses using the formula:
\begin{equation}
\frac{m^{\ast}_{p}}{m_0} = 4t \left[ \frac{\partial^2 E(P)}{\partial P^2} \right]^{-1} . 
\label{twopart:eq:onehtwentyone}
\end{equation}
Note that Eq.~(\ref{twopart:eq:oneheighteen}) contains more than one $s$ state~\cite{Bak1999}, but we consider only the lowest one. Figure~\ref{twopart:fig:twelveone} shows the effective mass for three different cases: (i) Attraction on the first neighbors only; (ii) Attraction on the second neighbors only; (iii) Attraction of equal strength on both the first and the second neighbors, $V_{1} = V_{2}$. One can see that in the first two cases the mass grows linearly with $V$ while in the third case the mass is bounded by the light pair limit given by Eq.~(\ref{twopart:eq:onehtwelveten}). We also note that for $|V| < 10\,t$, the mass enhancement is very modest, $m^{\ast}_{p}/m_0 < 4$, in all three cases.

\section{\label{twopart:sec:twenty}
2D. $UV$ model on the rectangular lattice 
}

\begin{figure}[t]
\begin{center}
\includegraphics[width=0.40\textwidth]{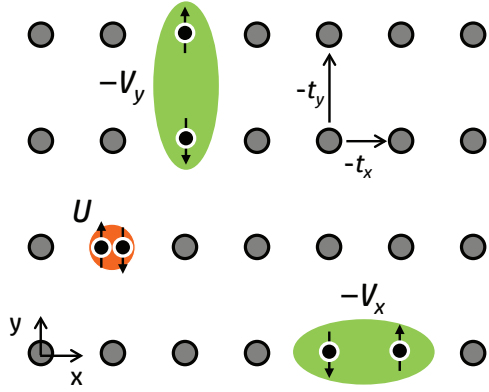}
\end{center}
\caption{The rectangular $UV$ model.} 
\label{twopart:fig:fourteenone}
\end{figure}

In the rectangular $UV$ model, both hopping and interaction along $x$ and $y$ axes are different, as illustrated in Fig.~\ref{twopart:fig:fourteenone}. Such a model can potentially be realized in cold gases. The model is rich in physical content and smoothly interpolates between 1D and 2D square models investigated earlier. In this section, we mostly focus on deriving binding conditions. In difference from preceding sections, we consider the possibility of {\em repulsive} nearest-neighbor interaction $V$. Therefore, the equations are written in terms of $V$ that can be of any sign rather than $-|V|$.

\subsection{\label{twopart:sec:twentyone}
Singlet states. $\Gamma$-point 
}

The dispersion relation is given by Eq.~(\ref{twopart:eq:ninetytwo}) in which $-|V|$ is replaced with $V_x$ and $V_y$ in the second and third columns, respectively: 
\begin{equation}
\left\vert \begin{array}{ccc}
1 + U M_{00} &   2 V_x M_{10}              &   2 V_y M_{01} \\
U M_{10}     & 1 + V_x ( M_{20} + M_{00} ) &   2 V_y M_{11} \\
U M_{01}     &   2 V_x M_{11}              & 1 + V_y ( M_{02} + M_{00} )  
\end{array} \right\vert 
 = 0 \: .   
\label{twopart:eq:onehtwentytwo}
\end{equation}
Matrix elements $M_{nm}$ are still defined by Eq.~(\ref{twopart:eq:ninetythree}) but with $\alpha = 4 t_x \cos{(P_x/2)}$ and $\beta = 4 t_y \cos{(P_y/2)}$. In the following, we only consider the ground state. Hence, $\alpha = 4t_x$ and $\beta = 4 t_y$. At the threshold, $E \rightarrow - \alpha - \beta$, and all $M_{nm}$ logarithmically diverge. To obtain a finite result, we utilize the subtractive procedure defined by Eq.~(\ref{twopart:eq:ninetyfour}). Substitution in Eq.~(\ref{twopart:eq:onehtwentytwo}) and expansion of the determinant results in a third-order polynomial in $M_{00}$. The $M^3_{00}$ and $M^2_{00}$ terms vanish identically, and the determinant assumes the form of Eq.~(\ref{twopart:eq:ninetyfive}). From here, the binding condition is $A = 0$, or, in full form      
\begin{align}
& U V_x V_y \left[ ( L_{20} - 4 L_{10} )( L_{02} - 4 L_{01} ) - 4 ( L_{10} + L_{01} - L_{11} )^2 \right]   
\nonumber \\
& + \: U V_x \left[ L_{20} - 4 L_{10} \right] 
  + U V_y \left[ L_{02} - 4 L_{01} \right]   
  + \: V_x V_y \left[ 2 ( L_{02} + L_{20} ) - 8 L_{11} \right] 
  + \: U + 2 V_x + 2 V_y = 0 \: .  
\label{twopart:eq:onehtwentythree}
\end{align}
Equation~(\ref{twopart:eq:onehtwentythree}) is the general binding condition in the rectangular $UV$ model. It reduces to Eq.~(\ref{twopart:eq:ninetysix}) if $V_x = V_y = - |V|$.

Let us analyze particular cases of Eq.~(\ref{twopart:eq:onehtwentythree}). First, we investigate isotropic hopping, $t_y = t_x = t$. Utilizing Eqs.~(\ref{twopart:eq:appaseventynine})-(\ref{twopart:eq:appaeightythree}), the binding condition becomes 
\begin{equation}
\frac{ 4 - \pi }{ 4 \pi t^2 } \, U V_x V_y + \frac{1}{ \pi t } \,  U ( V_x + V_y )    
      + \frac{ 2 ( 4 - \pi ) }{ \pi t } \,  V_x V_y + U + 2 ( V_x + V_y ) = 0 \: .  
\label{twopart:eq:onehtwentyfour}
\end{equation}
For isotropic attraction, $V_x = V_y = - |V|$, Eq.~(\ref{twopart:eq:onehtwentyfour}) reduces to the product of $s$- and $d$- thresholds, given by Eqs.~(\ref{twopart:eq:onehsixthree}) and (\ref{twopart:eq:onehsixseven}), respectively. To get a sense of the effect of strong nearest neighbor {\em repulsion}, we set $V_x = U$ and then $U \rightarrow \infty$. Obviously, to form a bound state, $V_y$ must be attractive. Equation~(\ref{twopart:eq:onehtwentyfour}) yields the threshold:
\begin{equation}
| V_{y, {\rm cr}} | = \frac{ 4t }{ 4 - \pi } = ( 4.569792 \ldots ) \, t \: .
\label{twopart:eq:onehtwentyfive}
\end{equation}
This should be compared with the $U \to \infty$ limit of the isotropic, four-site attraction, $| V^{s}_{\rm cr}| = 2t$, see Eq.~(\ref{twopart:eq:onehsixthree}). Making two of the attractive sites strongly repulsive increases the necessary strength of the two remaining attractions from $2t$ to $4.57\,t$.  

Next, we consider the case of anisotropic attraction, $V_x = 0$, $V_y < 0$, and arbitrary hopping anisotropy. Equation~(\ref{twopart:eq:onehtwentyfour}) reduces to   
\begin{equation}
| V_{y, {\rm cr}} | = \frac{ U }{ 2 + U ( L_{02} - 4 L_{01} ) } \: .
\label{twopart:eq:onehtwentysix}
\end{equation}
According to \ref{twopart:sec:appaseven}, 
\begin{equation}
L_{02} - 4 L_{01} = \frac{ 1 }{ \pi t_y } \left\{ 
\frac{ t_y - t_x }{ t_y } \arcsin{\sqrt{\frac{t_y}{ t_x + t_y }}} + \sqrt{\frac{t_x}{t_y}}  
\right\} .
\label{twopart:eq:onehtwentyseven}
\end{equation}
Several particular cases are of interest. 

(i) \underline{$t_x = 0$.} In this case, the system splits into individual $y$ chains, each of which hosts a 1D $UV$ model. $L_{02} - 4 L_{01} = (2t_y)^{-1}$ and $|V_{y,{\rm cr}}| = 2Ut_y/(U + 4t_y)$, which coincides with the earlier result, Eq.~(\ref{twopart:eq:eightysix}). 

(ii) \underline{$t_x = t_y = t$.} This is the case of isotropic hopping. $L_{02} - 4 L_{01} = (\pi t)^{-1}$ and 
\begin{equation}
| V_{y, {\rm cr}} | = \frac{ \pi U t }{ U + 2 \pi \, t } \: . 
\label{twopart:eq:onehtwentyeight}
\end{equation}
This expression should be compared with its isotropic attraction counterpart, Eq.~(\ref{twopart:eq:onehsixthree}). The attraction strength must be larger because there are only two attractive sites rather than four. In the $U = \infty$ limit, the threshold value is $\pi t$ rather than $2t$.  
  
(iii) \underline{$t_y = 0$.} In this case, the system splits into individual $x$ chains. The two particles reside on adjacent chains and feel an attraction $V_y$ when their $x$ coordinates coincide. The model is isomorphic to the 1D attractive Hubbard model. In the $t_y \rightarrow 0$ limit, $L_{02} - 4 L_{01}$ diverges as  $\simeq 4 (3\pi)^{-1} ( t_x t_y )^{-1/2}$. Hence, $|V_{y, {\rm cr}} | \rightarrow 0$, as expected for the 1D attractive Hubbard model.

\subsection{\label{twopart:sec:twentytwo}
Triplet states. $\Gamma$-point 
}

Dispersions of $p_x$ and $p_y$ pairs are given by Eqs.~(\ref{twopart:eq:onehnine}) and (\ref{twopart:eq:onehten}) with $V$ replaced by $V_x$ and $V_y$, respectively. The binding thresholds at ${\bf P} = (0,0)$ are
\begin{align}
| V_{x,{\rm cr}} | & = \frac{ \pi t_x }
{ \frac{ t_x + t_y }{ t_x } \arcsin{ \sqrt{\frac{t_x}{ t_x + t_y } } } - \sqrt{ \frac{t_y}{t_x} } } \: ,
\label{twopart:eq:onehtwentynine} \\
| V_{y,{\rm cr}} | & = \frac{ \pi t_y }
{ \frac{ t_x + t_y }{ t_y } \arcsin{ \sqrt{\frac{t_y}{ t_x + t_y } } } - \sqrt{ \frac{t_x}{t_y} } } \: .
\label{twopart:eq:onehthirty}
\end{align}
Consider the $p_y$ pair. In the $t_x = 0$ limit, $| V_{y,{\rm cr}} | = 2t_y$, which is the correct threshold for the 1D $UV$ model, see Eq.~(\ref{twopart:eq:ninetyone}). In the isotropic hopping case, $t_x = t_y = t$, $| V_{y,{\rm cr}} | = 2 \pi t/( \pi - 2 )$, which coincides with Eq.~(\ref{twopart:eq:oneheleven}). Finally, in the $t_y \rightarrow 0$ limit, $| V_{y,{\rm cr}} | = \frac{3\pi}{2} \sqrt{t_x t_y} \rightarrow 0$. The $p_y$ pair forms with a zero threshold.

\section{\label{twopart:sec:six}
2D. $UV$ model on the triangular lattice 
}

\begin{figure}[t]
\begin{center}
\includegraphics[width=0.40\textwidth]{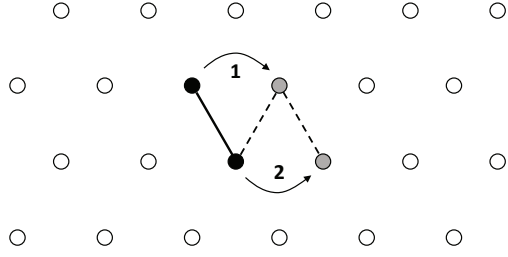}
\end{center}
\caption{In the triangular $UV$ model, the pair moves in first order in hopping even when the attraction is limited to first nearest neighbors. } 
\label{twopart:fig:twelvetwo}
\end{figure}

The triangular $UV$ model possesses another qualitative feature: light pairs are formed due to lattice topology rather than degeneracy of the attractive potential. Consider Fig.~\ref{twopart:fig:twelvetwo}. In a tightly bound pair, the members can move through the lattice while remaining nearest neighbors to each other and without breaking the main attractive bond. As a result, the pair remains light even when attraction is limited to first nearest neighbors. Analysis of dimer motion~\cite{Hague2007,Hague2007C,Hague2008} predicts that the mass of the lowest singlet pair is $m^{\ast}_{p} < 6 \, m_0$.    

This situation is more general than it might seem. There exist other lattices that can support light pairs with first neighbor attraction. One example is the staggered ladder~\cite{Alexandrov2002A,Hague2007}, where the two ladder chains are shifted relative to each other by half of lattice constant. Similarly, two square lattices stuck one on top of each other and shifted by half diagonal of elementary plaquette will support ``crab motion'' as long as the two constituent particles reside on different layers. This system will be considered in Section~\ref{twopart:sec:appfone}. In 3D, the face-centered cubic lattice supports light pairs~\cite{Adebanjo2022}. Furthermore, if the range of primary hopping is extended, then more lattices are added to the list. For example, the square lattice with next nearest neighbor hopping (across the elementary plaquette) produces crab motion and light pairs as a result~\cite{Alexandrov2002A,Alexandrov2002B,Alexandrov2002C}.

The light pair effect has implications for the formation of phonon bipolarons. It is by now well understood~\cite{Alexandrov1999} that long-range electron--ion interactions exponentially reduce the effective mass of {\em polarons}, and consequently of polaron pairs, bipolarons. On triangular-like lattices, the bipolarons acquire additional lightness due to crab motion and become ``superlight''~\cite{Alexandrov2002C,Hague2007,Hague2007B,Hague2010}. Based on that and on the BEC formula, Eq.~(\ref{twopart:eq:one}), it was suggested that triangular-like lattices could provide even higher-$T_c$ superconductivity than square-like lattices~\cite{Hague2007B}.       

The $UV$ model on triangular lattice was solved by Bak~\cite{Bak2007} using a method similar to our unsymmetrized solution, and then in Ref~\cite{Hague2010}. Here we analyze the model with emphasis on the pair mass, dispersion, and binding conditions.

\subsection{\label{twopart:sec:sixone}
General dispersion relations 
}

In the singlet sector, there are four symmetrized vectors: $\{ {\bf b}_{+} \} = \{ (0,0), (1,0), (\frac{1}{2},\frac{\sqrt{3}}{2}) , (-\frac{1}{2},\frac{\sqrt{3}}{2}) \} \equiv \{ {\bf 0} , {\bf 1} , {\bf 2} , {\bf 3} \}$. Changing momentum variable, ${\bf q}' = {\bf q} - \frac{\bf P}{2}$, and redefining amplitudes as $\tilde{\Phi}^{+}_{\bf b} = e^{-i ({\bf P}/2) {\bf b}} \Phi^{+}_{\bf b}$, the master equations, Eqs.~(\ref{twopart:eq:thirteeneleven}) and (\ref{twopart:eq:thirteentwelve}), take the form
\begin{equation}
\left[ \begin{array}{cccc}
1 + U M^{+}_{\bf 00} &   - |V| M^{+}_{\bf 01} &   - |V| M^{+}_{\bf 02} &   - |V| M^{+}_{\bf 03} \\
    U M^{+}_{\bf 10} & 1 - |V| M^{+}_{\bf 11} &   - |V| M^{+}_{\bf 12} &   - |V| M^{+}_{\bf 13} \\
    U M^{+}_{\bf 20} &   - |V| M^{+}_{\bf 21} & 1 - |V| M^{+}_{\bf 22} &   - |V| M^{+}_{\bf 23} \\
    U M^{+}_{\bf 30} &   - |V| M^{+}_{\bf 31} &   - |V| M^{+}_{\bf 32} & 1 - |V| M^{+}_{\bf 33} 
\end{array} \right] 
\left[ \begin{array}{c}
\tilde{\Phi}^{+}_0 \\ \tilde{\Phi}^{+}_1 \\ \tilde{\Phi}^{+}_2 \\ \tilde{\Phi}^{+}_3  
\end{array} \right] = 0  \: .  
\label{twopart:eq:onehthirtytwo}
\end{equation}
Here
\begin{align}
M^{+}_{\bf 00}                & = \frac{1}{N} \sum_{\bf q} \frac{1}{W}                             \: ,
\label{twopart:eq:onehthirtythree} \\
M^{+}_{{\bf 0} {\bf b}'_+}    & = \frac{1}{N} \sum_{\bf q} \frac{2 \cos{({\bf q} {\bf b}'_+)} }{W} \: ,
\label{twopart:eq:onehthirtyfour}  \\
M^{+}_{{\bf b}_+ {\bf 0}}     & = \frac{1}{N} \sum_{\bf q} \frac{ e^{i {\bf q} {\bf b}_+ } }{W}    \: ,
\label{twopart:eq:onehthirtyfive}  \\
M^{+}_{{\bf b}_+ {\bf b}'_+ } & = \frac{1}{N} \sum_{\bf q} 
                         \frac{ 2 \cos{({\bf q} {\bf b}'_+)} \, e^{i {\bf q} {\bf b}_+ } }{W}        \: ,
\label{twopart:eq:onehthirtysix}  
\end{align}
\begin{equation}
W = \vert E \vert - \alpha \cos{q_x} - \beta \cos{\frac{q_x}{2}} \cos{\frac{\sqrt{3} q_y}{2}} 
  - \gamma \sin{\frac{q_x}{2}} \sin{\frac{\sqrt{3} q_y}{2}} \: ,
\label{twopart:eq:onehthirtyseven}  
\end{equation}
and $\alpha$, $\beta$, and $\gamma$ are defined in Eqs.~(\ref{twopart:eq:appbthree})-(\ref{twopart:eq:appbfive}). Integration over BZ can be replaced with integration over the rectangle $0 \leq q_x \leq 2\pi$, $0 \leq q_y \leq (4\pi)/\sqrt{3}$. The expressions for $M^{+}$ are given in \ref{twopart:sec:appbfour}. The following relations hold for any ${\bf P}$:
\begin{align}
M^{+}_{\bf 01} & =  2 M^{+}_{\bf 10} \: , \hspace{0.1cm} 
M^{+}_{\bf 02}   =  2 M^{+}_{\bf 20} \: , \hspace{0.1cm} 
M^{+}_{\bf 03}   =  2 M^{+}_{\bf 30} \: ,   
\label{twopart:eq:onehthirtysevenone} \\
M^{+}_{\bf 12} & =  M^{+}_{\bf 21}   \: , \hspace{0.1cm} 
M^{+}_{\bf 13}   =  M^{+}_{\bf 31}   \: , \hspace{0.1cm}  
M^{+}_{\bf 23}   =  M^{+}_{\bf 32}   \: .
\label{twopart:eq:onehthirtyseventwo}  
\end{align}
Additional relations hold along BZ symmetry lines.

\begin{figure}[t]
\begin{center}
\includegraphics[width=0.48\textwidth]{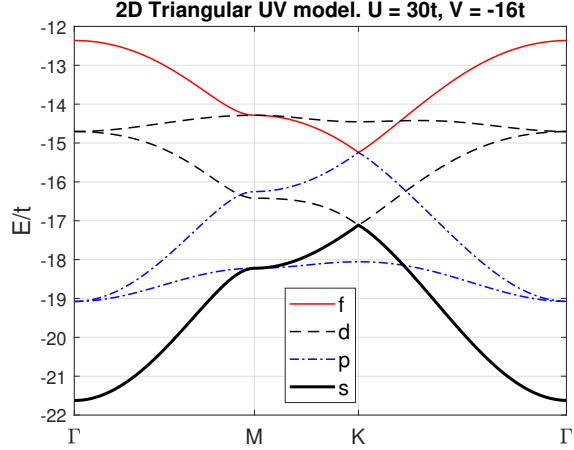}
\end{center}
\caption{Dispersion of bound pairs in the triangular $UV$ model for $U = 30\,t$ and $V = -16\,t$. $E = -12\,t$ is the lowest energy of two free particles at the $\Gamma$-point. } 
\label{twopart:fig:twelvethree}
\end{figure}

In the triplet sector, there are three anti-symmetrized vectors: $\{ {\bf b}_{-} \} = \{ {\bf 1}, {\bf 2} , {\bf 3} \}$. Performing the same variable change, ${\bf q}' = {\bf q} - \frac{\bf P}{2}$, and amplitude redefinition $\tilde{\Phi}^{-}_{{\bf b}_{-}} = e^{-i ({\bf P}/2) {\bf b}_{-}} \Phi^{-}_{{\bf b}_{-}}$, the master equations yield
\begin{equation}
\left[ \begin{array}{ccc}
1 - |V| M^{-}_{\bf 11} &   - |V| M^{-}_{\bf 12} &   - |V| M^{-}_{\bf 13} \\
  - |V| M^{-}_{\bf 21} & 1 - |V| M^{-}_{\bf 22} &   - |V| M^{-}_{\bf 23} \\
  - |V| M^{-}_{\bf 31} &   - |V| M^{-}_{\bf 32} & 1 - |V| M^{-}_{\bf 33} 
\end{array} \right] 
\left[ \begin{array}{c}
\tilde{\Phi}^{-}_1 \\ \tilde{\Phi}^{-}_2 \\ \tilde{\Phi}^{-}_3  
\end{array} \right] = 0   \: , 
\label{twopart:eq:onehthirtyeight}
\end{equation}
where 
\begin{equation}
M^{-}_{{\bf b}_{-} {\bf b}'_{-} } = \frac{1}{N} \sum_{\bf q} 
  \frac{ (-2i) \sin{({\bf q} {\bf b}'_{-})} \, e^{i {\bf q} {\bf b}_{-} } }{W} \: .
\label{twopart:eq:onehthirtynine}  
\end{equation}
The expressions for $M^{-}$ are also given in \ref{twopart:sec:appbfour}, Eqs.~(\ref{twopart:eq:appbthirtyone})-(\ref{twopart:eq:appbthirtynine}). One has:
\begin{equation}
M^{-}_{\bf 12} = M^{-}_{\bf 21}   \: , \hspace{0.3cm} 
M^{-}_{\bf 13} = M^{-}_{\bf 31}   \: , \hspace{0.3cm}  
M^{-}_{\bf 23} = M^{-}_{\bf 32}   \: .
\label{twopart:eq:onehthirtynineone}  
\end{equation}
Thus, the triplet dispersion is described by a symmetric matrix for all ${\bf P}$. A typical pair dispersion is shown in Fig.~\ref{twopart:fig:twelvethree}. Of note is the fact that attraction $V = -16\,t$ is barely above the pairing threshold for the highest, $f$, state given by Eq.~(\ref{twopart:eq:onehsixtytwo}). Thus, all six pair states are well defined in the entire BZ. At weaker attractions, the bound states begin to disappear into the two-particle continuum one-by-one.  

\begin{figure}[t]
\begin{center}
\includegraphics[width=0.47\textwidth]{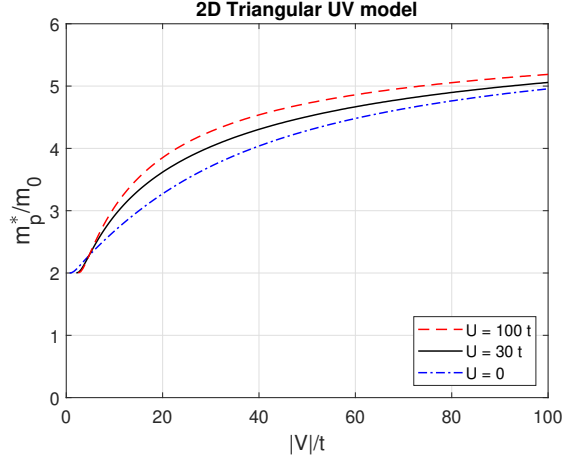}
\end{center}
\caption{Effective mass of the lowest, $s$-symmetric bound pair in the triangular $UV$ model. The strong coupling limit is $m^{\ast}_{p}/m_0 = 6$. } 
\label{twopart:fig:twelvefour}
\end{figure}

The effective mass of the lowest, $s$-symmetric pair is shown in Fig.~\ref{twopart:fig:twelvefour}. One can observe that the mass indeed approaches the strong-coupling limit~\cite{Hague2007,Hague2007C,Hague2008}, $m^{\ast}_{p}/m_0 = 6$, for all $U$. However, the approach is slow. For most realistic attractions, $|V| < 20\,t$, the pair mass is no heavier than just 4 free-particle masses.

\subsection{\label{twopart:sec:sixtwo}
Pairing thresholds at the $\Gamma$-point  
}

Determination of pairing thresholds from the general dispersion relations, Eqs.~(\ref{twopart:eq:onehthirtytwo}) and (\ref{twopart:eq:onehthirtyeight}), at arbitrary ${\bf P}$ can be done only numerically. At the $\Gamma$-point, the systems acquire additional symmetries and the thresholds can be derived analytically. Analysis is greatly aided by the group theory. All the necessary information is given in \ref{twopart:sec:appbthree}.

For the singlet states, we combine $\Phi_0$, which is unchanged, with symmetrized combinations of basis functions given by Eqs.~(\ref{twopart:eq:onehfortynine}) and (\ref{twopart:eq:onehfiftytwo}) into a new basis:
\begin{equation}
\left[ \! \begin{array}{c}
\Phi_{0} \\ \Phi_{s}  \\ \Phi_{d_1} \\  \Phi_{d_2} 
\end{array} \! \right] =  
\left[ \begin{array}{rrrr}
 1 & 0 &  0 &  0 \\ 
 0 & 1 &  1 &  1 \\ 
 0 & 2 & -1 & -1 \\  
 0 & 0 &  1 & -1 
\end{array} \right] \! 
\left[ \! \begin{array}{c}
\Phi_{0} \\ \Phi^{+}_{1}  \\ \Phi^{+}_{2} \\  \Phi^{+}_{3}  
\end{array} \! \right] \equiv 
\hat{A}_{S} \left[ \! \begin{array}{c}
\Phi_{0} \\ \Phi^{+}_{1}  \\ \Phi^{+}_{2} \\  \Phi^{+}_{3}  
\end{array} \! \right] .
\label{twopart:eq:onehfiftyfour}  
\end{equation}
In terms of the new basis, dispersion equation, Eq.~(\ref{twopart:eq:onehthirtytwo}), transforms into a block-diagonal form: 
\begin{equation}
\hat{A}_{S} \cdot [ \ldots ] \cdot \hat{A}^{-1}_{S} = 
\left[ \! \begin{array}{cccc}
1 + U M^{+}_{\bf 00} &   - 2 |V| M^{+}_{\bf 10}                      &          0         &           0          \\
  3 U M^{+}_{\bf 10} & 1 - |V| ( M^{+}_{\bf 11} + 2 M^{+}_{\bf 12} ) &          0         &           0          \\
  0   &   0          & 1 - |V| ( M^{+}_{\bf 11} -   M^{+}_{\bf 12} ) &          0                                \\
  0   &   0          &          0                                    & 1 - |V| ( M^{+}_{\bf 11} - M^{+}_{\bf 12} )   
\end{array} \right] 
\left[ \begin{array}{c}
\Phi_0 \\ \Phi_{s} \\ \Phi_{d_1} \\ \Phi_{d_2}  
\end{array} \right] =  0   \: .
\label{twopart:eq:onehfiftyfive}
\end{equation}
The top-left $2 \times 2$ block describes an $s$-symmetric ground state. To find the binding condition, set $E \rightarrow - 12 \, t$, at which all $M^{+}$ diverge logarithmically. To obtain a finite result, introduce the differences:
\begin{align}
L^{+}_{\bf 10} & =  M^{+}_{\bf 10} -   M^{+}_{\bf 00} \: ,
\label{twopart:eq:onehfiftyfiveone}  \\
L^{+}_{\bf 11} & =  M^{+}_{\bf 11} - 2 M^{+}_{\bf 00} \: ,
\label{twopart:eq:onehfiftyfivetwo}  \\
L^{+}_{\bf 12} & =  M^{+}_{\bf 12} - 2 M^{+}_{\bf 00} \: .
\label{twopart:eq:onehfiftyfivethree}
\end{align}
Next, express $M^{+}$ via $L^{+}$ and $M^{+}_{\bf 00}$ and expand the $2 \times 2$ determinant. The $(M^{+}_{\bf 00})^2$ term vanishes identically while the coefficient at $M^{+}_{\bf 00}$ must be zero. This leads to a binding condition
\begin{equation}
\vert V^{s}_{\rm cr} \vert = \frac{U}{ ( L^{+}_{\bf 11} + 2 L^{+}_{\bf 12} - 12 L^{+}_{\bf 10} ) \, U + 6 } \: .
\label{twopart:eq:onehfiftysix}
\end{equation}
Analytical expressions for $L^{+}$ are given in \ref{twopart:sec:appbtwo}. The final result reads
\begin{equation}
\vert V^{s}_{\rm cr} \vert = \frac{ 2 U t }{ U + 12 \, t } \: .
\label{twopart:eq:onehfiftyseven}
\end{equation}
The other two $1 \times 1$ blocks in Eq.~(\ref{twopart:eq:onehfiftyfive}) describe a $d$-symmetric dublet. Direct calculation yields the pairing threshold
\begin{equation}
\vert V^{d}_{\rm cr} \vert = \frac{ 4 \pi \, t }{ 3 ( 2 \sqrt{3} - \pi ) } = ( 12.998135\ldots) \, t \: .
\label{twopart:eq:onehfiftyeight}
\end{equation}

We now turn to the triplet dispersion, Eq.~(\ref{twopart:eq:onehthirtyeight}), at the $\Gamma$ point. Combining Eqs.~(\ref{twopart:eq:onehfiftythree}) and (\ref{twopart:eq:onehfifty}) into one transformation, one obtains a new basis
\begin{equation}
\left[ \! \begin{array}{c}
\Phi_{p_1}  \\ \Phi_{p_2} \\  \Phi_{f} 
\end{array} \! \right] =  
\left[ \begin{array}{rrr}
 0 & -1 & -1 \\ 
 2 &  1 & -1 \\  
 1 & -1 &  1 
\end{array} \right] \! 
\left[ \! \begin{array}{c}
\Phi^{-}_{1}  \\ \Phi^{-}_{2} \\  \Phi^{-}_{3}  
\end{array} \! \right] \equiv 
\hat{A}_{T} \left[ \! \begin{array}{c}
\Phi^{-}_{1}  \\ \Phi^{-}_{2} \\  \Phi^{-}_{3}  
\end{array} \! \right] .
\label{twopart:eq:onehfiftynine}  
\end{equation}
A transformed dispersion equation reads
\begin{equation}
\hat{A}_{T} \cdot [ \ldots ] \cdot \hat{A}^{-1}_{T} =
\left[ \begin{array}{ccc}
 1 - |V| ( M^{-}_{\bf 11} + M^{-}_{\bf 12} ) &            0         &          0          \\
         0       & 1 - |V| ( M^{-}_{\bf 11} + M^{-}_{\bf 12} )      &          0          \\
         0       &            0              & 1 - |V| ( M^{-}_{\bf 11} - 2 M^{-}_{\bf 12} )   
\end{array} \right] 
\left[ \begin{array}{c}
\Phi_{p_1} \\ \Phi_{p_2} \\ \Phi_{f}  
\end{array} \right] =  0   \: .
\label{twopart:eq:onehsixty}
\end{equation}
The first two blocks describe a $p$-symmetric doublet, whereas the lower-right block describes one $f$-symmetric state. Note that both $M^{-}_{\bf 11}$ and $M^{-}_{\bf 12}$ {\em converge} at threshold, and no subtraction procedure is necessary. Starting with Eqs.~(\ref{twopart:eq:appbthirtyone}) and (\ref{twopart:eq:appbthirtytwo}), setting $\alpha = 4t$, $\beta = 8t$, $\gamma = 0$, $E = -12 t$, elementary integration yields
\begin{align}
M^{-}_{\bf 11}( \Gamma, E = -12\,t ) & =  \frac{ 2 \pi - 3\sqrt{3} }{ 3 \pi t }  \: ,
\label{twopart:eq:onehsixtyoneone}  \\
M^{-}_{\bf 12}( \Gamma, E = -12\,t ) & =  \frac{  2 \sqrt{3} - \pi }{ 4 \pi t }  \: .
\label{twopart:eq:onehsixtyonetwo}
\end{align}
Consequently, the pairing thresholds are 
\begin{align}
\vert V^{p}_{\rm cr} \vert & =  \frac{ 12 \pi t }{ 5 \pi -  6 \sqrt{3} } = (  7.092087\ldots) \, t \: ,
\label{twopart:eq:onehsixtyone}  \\
\vert V^{f}_{\rm cr} \vert & =  \frac{  6 \pi t }{ 7 \pi - 12 \sqrt{3} } = ( 15.622833\ldots) \, t \: .
\label{twopart:eq:onehsixtytwo}
\end{align}

\section{\label{twopart:sec:seven}
3D. $UV$ model on the simple cubic lattice 
}

One qualitatively new feature of 3D models is a larger role of kinetic energy. A finite attraction is needed to form a pair even at $U = 0$. The $UV$ model on the simple cubic lattice was solved in Ref.~\cite{Micnas1990,Davenport2012}. In this section, those results are rederived and extended using the (anti)-symmetrized method developed here.

\subsection{\label{twopart:sec:sevenone}
Singlet states 
}

The symmetrized set of vectors consists of four elements: $\{ {\bf b}_{+} \} = \{ (0,0,0), (1,0,0), (0,1,0), (0,0,1) \}$. Similar to the square lattice, Eqs.~(\ref{twopart:eq:thirteeneleven}) and (\ref{twopart:eq:thirteentwelve}) are transformed by changing integration variables, $q'_j = q_j - \frac{P_j}{2}$, and functions, $\tilde{\Phi}^{+}_{\bf x} = e^{ -i (P_x/2) } \Phi^{+}_{100}$, and so on. The resulting $4 \times 4$ linear system reads  
\begin{equation}
\left[ \begin{array}{cccc}
1 + U M_{000}  & - 2 \vert V \vert M_{100}    & - 2 \vert V \vert M_{010}               &  - 2 \vert V \vert M_{001}  \\
    U M_{100}  & 1 - \vert V \vert ( M_{000} + M_{200} ) & - 2 \vert V \vert M_{110}    &  - 2 \vert V \vert M_{101}  \\
    U M_{010}  & - 2 \vert V \vert M_{110}    & 1 - \vert V \vert ( M_{000} + M_{020} ) &  - 2 \vert V \vert M_{011}  \\
    U M_{001}  & - 2 \vert V \vert M_{101}    & - 2 \vert V \vert M_{011}      &  1 - \vert V \vert ( M_{000} + M_{002} )
\end{array} \right]  
\left[ \begin{array}{c}
\tilde{\Phi}_{0} \\ \tilde{\Phi}^{+}_{\bf x}  \\ \tilde{\Phi}^{+}_{\bf y} \\  \tilde{\Phi}^{+}_{\bf z}  
\end{array} \right] 
= 0 \: ,   
\label{twopart:eq:onesixtythree}
\end{equation}
\begin{equation}
M_{nmk} = 
\int\limits^{\pi}_{-\pi} \!\! \int\limits^{\pi}_{-\pi} \!\! \int\limits^{\pi}_{-\pi} 
\frac{ {\rm d}q_x \, {\rm d}q_y \, {\rm d}q_z }{(2\pi)^3} 
\frac{\cos{nq_x} \cos{mq_y} \cos{kq_z}}{ \vert E \vert - \alpha \cos{q_x} - \beta \cos{q_y} - \gamma \cos{q_z} } \: ,
\label{twopart:eq:onesixtyfour}
\end{equation}
where $\alpha = 4 t \cos{(P_x/2)}$, $\beta = 4 t \cos{(P_y/2)}$, and $\gamma = 4 t \cos{(P_z/2)}$. Pair dispersion is obtained by equating the $ 4 \times 4 $ determinant to zero. The quantities $M_{nmk}$ are generalized Watson integrals.  For arbitrary ${\bf P}$, they can be computed numerically. On BZ diagonals, including the $\Gamma$ point, $\alpha = \beta = \gamma$. In this case, all $M_{nmk}$ in Eq.~(\ref{twopart:eq:onesixtythree}) can be expressed via the complete elliptic integrals in closed form. The expressions are given in \ref{twopart:sec:appcone}.      

At the $\Gamma$-point, $M_{100} = M_{010} = M_{001}$, $M_{200} = M_{020} = M_{002}$, $M_{110} = M_{101} = M_{011}$, and the matrix in Eq.~(\ref{twopart:eq:onesixtythree}) acquires additional symmetries. A point group analysis suggests a new basis
\begin{equation}
\left[ \begin{array}{c}
\Phi_{0} \\ \Phi_{s}  \\ \Phi_{d_1} \\  \Phi_{d_2} 
\end{array} \right] =  
\left[ \begin{array}{rrrr}
 1 & 0 &  0 &  0 \\ 
 0 & 1 &  1 &  1 \\ 
 0 & 2 & -1 & -1 \\  
 0 & 0 &  1 & -1 
\end{array} \right]
\left[ \begin{array}{c}
\Phi_{0} \\ \Phi^{+}_{\bf x}  \\ \Phi^{+}_{\bf y} \\  \Phi^{+}_{\bf z}  
\end{array} \right] ,
\label{twopart:eq:onehsixtyseven}  
\end{equation}
in terms of which the energy equation becomes block-diagonal:
\begin{multline}
\hspace{-0.4cm}
\left[ \begin{array}{cccc}
1 + U M_{000} &   - 2 V M_{100}                        &          0         &           0          \\
  3 U M_{100} & 1 - V ( M_{000} + M_{200} + 4M_{110} ) &          0         &           0          \\
  0   &   0          & 1 - V ( M_{000} + M_{200} - 2M_{110} ) &   0                                \\
  0   &   0          &          0                             & 1 - V ( M_{000} + M_{200} - 2M_{110} )   
\end{array} \right] \times 
\\
\times \left[ \begin{array}{c}
\Phi_0 \\ \Phi_{s} \\ \Phi_{d_1} \\ \Phi_{d_2}  
\end{array} \right] \! =  0  \:  .
\label{twopart:eq:onehsixtyeight}
\end{multline}
The upper-left corner describes an $s$-symmetric ground state. Remarkably, all the matrix elements can be expressed via the basic integral $M_{000}$:
\begin{align}
M_{100} & =  \frac{1}{3\alpha} \left(  \vert E \vert M_{000} - 1 \right) ,
\label{twopart:eq:onesixtynine} \\
M_{000} + M_{200} + 4 M_{110} & =  
\frac{2 \vert E \vert}{3 \alpha^2} \left(  \vert E \vert M_{000} - 1 \right) .
\label{twopart:eq:oneseventy}        
\end{align}
Expanding the $2 \times 2$ determinant, the $s$-pair dispersion equation becomes
\begin{equation}
( U M_{000} + 1 ) + \frac{2}{3\alpha^2} \: \vert V \vert 
\left( \vert E_{s} \vert + U \right)( 1 - \vert E_{s} \vert M_{000} ) = 0 \: ,
\label{twopart:eq:oneseventyone}
\end{equation}
where $M_{000}$ is given in Eq.~(\ref{twopart:eq:sixtysix}) or (\ref{twopart:eq:appfthree}). It is instructive to compare Eq.~(\ref{twopart:eq:oneseventyone}) with its 1D and 2D counterparts, Eqs.~(\ref{twopart:eq:eightythreethree}) and (\ref{twopart:eq:onehsixone}), respectively, which suggests generalizations for $UV$ models on the primitive hyper-cubic lattices in any dimension. This topic is not pursued further here.

\begin{figure}[t]
\begin{center}
\includegraphics[width=0.60\textwidth]{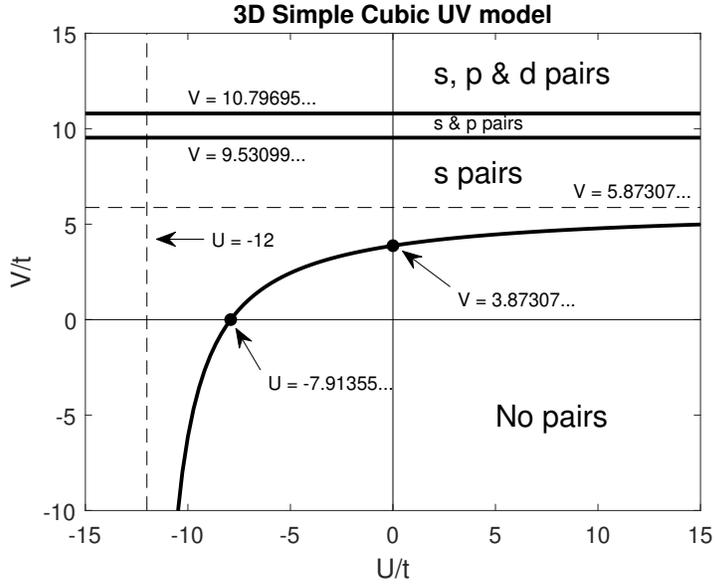}
\end{center}
\caption{Two-fermion phase diagram of the $UV$ model on the simple cubic lattice for ${\bf P} = 0$.} 
\label{twopart:fig:nine}
\end{figure}

From Eq.~(\ref{twopart:eq:oneseventyone}), pair energy can be found numerically for any given $U$, $V$, and $\alpha$. Let us determine the binding threshold along BZ diagonals. To that end, set $\vert E \vert = 3\alpha$. The numerical value of $M_{000}$ at this energy was given in Eq.~(\ref{twopart:eq:sixtyone}). It can be written as 
\begin{equation}
M_{000}(3\alpha) = \frac{ 4 t }{ \alpha U_0 } = \frac{1}{ U_0 \cos{(P_{x}/2)} }  \: ,
\label{twopart:eq:oneseventyoneone}
\end{equation}
where $U_0 \equiv (7.913552\ldots) t$ is the pairing threshold in the attractive Hubbard model at ${\bf P} = 0$, see Eq.~(\ref{twopart:eq:sixtytwo}). Using $\alpha = 4t \cos{(P_{x}/2)}$, one obtains from Eq.~(\ref{twopart:eq:oneseventyone}) 
\begin{equation}
| V^{s}_{{\rm cr}}({\rm diag})| = \frac{24 t^2 \cos{(P_{x}/2)}}{( 12 t - U_0 )} 
\frac{[ U + U_0 \cos{(P_{x}/2)} ]}{ [ U + 12 t \cos{(P_{x}/2)} ] } \: .
\label{twopart:eq:oneseventytwo}
\end{equation}

Next, consider the $d$-symmetric doublet described by the low-right corner of Eq.~(\ref{twopart:eq:onehsixtyeight}). To determine the binding threshold, use Eqs.~(\ref{twopart:eq:appfeleven}) and (\ref{twopart:eq:appfthirteen}) to find 
\begin{equation}
M_{000}(3\alpha) + M_{200}(3\alpha) - 2 M_{110}(3\alpha) = 
7 M_{000}(3\alpha) + \frac{6}{\pi^2 \alpha^2 M_{000}(3\alpha)} - \frac{4}{\alpha} \: ,
\label{twopart:eq:oneseventythree}
\end{equation}
from where
\begin{equation}
\vert V^{d}_{\rm cr}({\rm diag}) \vert = \frac{16 \pi^2 U_0 \, t^2 \cos{(P_{x}/2)}}
{ 56 \pi^2 t^2 + 3 U^2_0 - 8 \pi^2 U_0 t }  
= ( 10.796952 \ldots ) \, t \, \cos{(P_{x}/2)} \: .
\label{twopart:eq:oneseventyfour}
\end{equation}
The full phase diagram of the simple cubic $UV$ model is shown in Fig.~\ref{twopart:fig:nine}.

\subsection{\label{twopart:sec:seventwo}
Triplet states 
}

The anti-symmetrized set of vectors consists of three elements: $\{ {\bf b}_{-} \} = \{ (1,0,0), (0,1,0), (0,0,1) \}$. The dispersion equation splits into three independent equations describing three $p$-symmetric pairs:
\begin{align}
\left[ 1 - \vert V \vert ( M_{000} - M_{200} ) \right] \Phi^{-}_{\bf x} & =  0  \: ,  
\label{twopart:eq:oneseventyfive} \\
\left[ 1 - \vert V \vert ( M_{000} - M_{020} ) \right] \Phi^{-}_{\bf y} & =  0  \: ,
\label{twopart:eq:oneseventysix}  \\
\left[ 1 - \vert V \vert ( M_{000} - M_{002} ) \right] \Phi^{-}_{\bf z} & =  0  \: .
\label{twopart:eq:oneseventyseven}
\end{align}
Note that the decomposition into three independent equations occurs at any pair momentum ${\bf P}$. On BZ diagonals, the spectrum is triple degenerate because $M_{200} = M_{020} = M_{200}$. To obtain pair energies, Eqs.~(\ref{twopart:eq:oneseventyfive})-(\ref{twopart:eq:oneseventyseven}) ought to be solved numerically. On BZ diagonals, analytical expressions for $M_{000}$ and $M_{200}$ are given in \ref{twopart:sec:appcone}. To find the binding threshold, set $\vert E \vert = 3\alpha$ and apply Eqs.~(\ref{twopart:eq:appfeleven}) and (\ref{twopart:eq:oneseventyoneone}). The result is
\begin{equation}
\vert V^{p}_{\rm cr}({\rm diag}) \vert  =  \frac{24 \pi^2 U_0 \, t^2 \cos{(P_{x}/2)}}
{ 8 \pi^2 U_0 t - 56 \pi^2 t^2 - 3 U^2_0 }  
 =  ( 9.530994 \ldots ) \, t \, \cos{(P_{x}/2)} \: .
\label{twopart:eq:oneseventyeight}
\end{equation}
This critical value is plotted in Fig.~\ref{twopart:fig:nine}.

\begin{figure}[t]
\begin{center}
\includegraphics[width=0.60\textwidth]{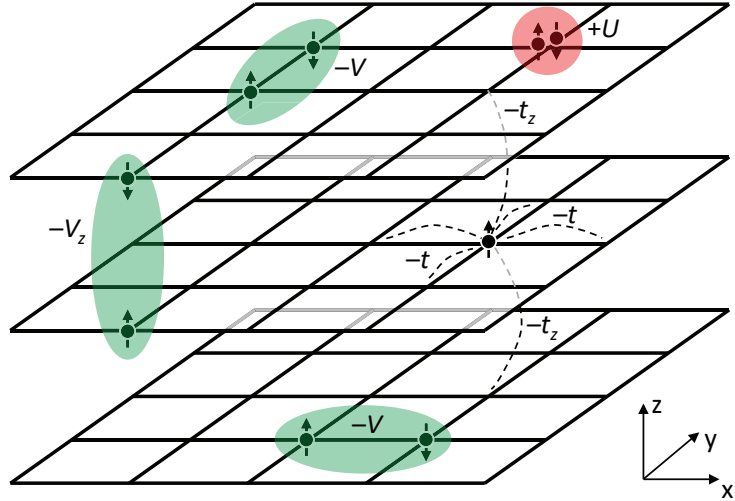}
\end{center}
\caption{Tetragonal $UV$ model.} 
\label{twopart:fig:ten}
\end{figure}

\section{\label{twopart:sec:eight}
3D. Tetragonal $UV$ model 
}

In the {\em tetragonal} $UV$ model, attractive potential $V_z$ and hopping integral $t_z$ along $z$ axis differ from their $xy$ counterparts, see Fig.~\ref{twopart:fig:ten}. The two extra parameters bring considerable richness and complexity. The tetragonal model smoothly interpolates between the quasi-2D limit $t_z \ll t$, $V_z \ll V$ (Section~\ref{twopart:sec:five}) and the quasi-1D limit $t_z \gg t$, $V_z \gg V$ (Section~\ref{twopart:sec:four}) via the isotropic 3D case (Section~\ref{twopart:sec:seven}). The quasi-2D sector is most relevant to the physics of high-temperature superconductors, as discussed in the Introduction. From that standpoint, the special case of $0 \leq t_z \leq t$ and $V_z = 0$ was analyzed in Ref.~\cite{Kornilovitch2015}. It was argued that both $t_z$ and $V$ have non-monotonic effects on preformed-pair superconductivity. A small $V$ cannot form pairs whereas a large $V$ produces pairs that are too heavy. In both cases, superconductivity is suppressed. (This may help understand why the highest $T_c$ occurs at intermediate electron-phonon coupling in HTSC~\cite{Mihailovic2022}.) Likewise, a large $t_z$ destroys the pairs because large kinetic energy overcomes a moderate $V$. In the opposite limit of very small $t_z$, pairs lose 3D coherence, $z$-axis mass becomes very large, and the condensation temperature drops. Thus, it was argued, preformed pair superconductivity is optimal at intermediate values of $V/t$ and $t_z/t$.          

In this section, the general case of nonzero $V_z$ is considered. Due to the model's complexity, very few results can be derived analytically. Bulk of the results presented below are obtained by solving pair dispersion equations numerically.

\subsection{\label{twopart:sec:eightone}
Singlet states. Pairing thresholds at the $\Gamma$-point 
}

Derivation of pair dispersion proceeds along the same lines as the simple cubic case of Section~\ref{twopart:sec:seven}. The result~\cite{Kornilovitch2015} is again Eq.~(\ref{twopart:eq:onesixtythree}) in which the last column contains $V_z$ instead of $V$. A second difference concerns the expression for integrals $M_{nmk}$, Eq.~(\ref{twopart:eq:onesixtyfour}): the parameter $\gamma$ is now given by $\gamma = 4 t_z \cos{(P_z/2)}$. We note in passing that replacing $V$ with another parameter $V_y$ in the {\em third} column of Eq.~(\ref{twopart:eq:onesixtythree}) and setting $\beta = 4t_y \cos{(P_y/2)}$ in Eq.~(\ref{twopart:eq:onesixtyfour}) results in a dispersion relation for the {\em orthorhombic} $UV$ model. The latter is not studied in this paper. 

\begin{figure}[t]
\begin{center}
\includegraphics[width=0.48\textwidth]{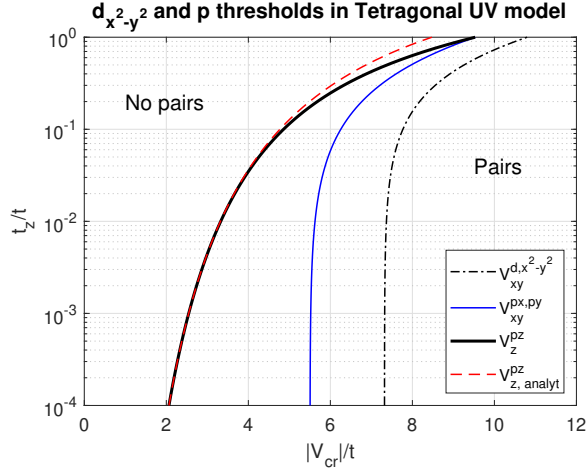}
\end{center}
\caption{Binding thresholds for $d_{x^2-y^2}$ and $p$-symmetrical states at the $\Gamma$-point of the tetragonal $UV$ model as a function of lattice anisotropy. The $d_{x^2-y^2}$ line interpolates between the 2D limit, Eq.~(\ref{twopart:eq:onehsixseven}), and the isotropic 3D limit, Eq.~(\ref{twopart:eq:oneseventyfour}). Similarly, the $p_x,p_y$ line interpolates between Eq.~(\ref{twopart:eq:oneheleven}) and Eq.~(\ref{twopart:eq:oneseventyeight}). The dashed line is Eq.~(\ref{twopart:eq:oneeightytwo}). } 
\label{twopart:fig:eleven}
\end{figure}

We begin with analysis of the binding conditions at the $\Gamma$ point where $M_{100} = M_{010}$, $M_{200} = M_{020}$, and $M_{101} = M_{011}$. Instead of doing a full point symmetry analysis, it is easier to proceed by observing that taking a sum and a difference of the second and third equations in Eq.~(\ref{twopart:eq:onesixtythree}) splits off one $d$-symmetric state. The difference can be written in terms of $\Phi^{+}_{\bf x} - \Phi^{+}_{\bf y}$:
\begin{equation}
\left[ 1 - \vert V_{xy} \vert \left( M_{000} + M_{200} - 2 M_{110} \right) \right] 
\left( \Phi^{+}_{\bf x} - \Phi^{+}_{\bf y} \right) = 0 \: .
\label{twopart:eq:oneseventynine}
\end{equation}
This equation describes a $d_{x^2-y^2}$-symmetric solution with a threshold that smoothly interpolates from the isotropic cubic case, Eq.~(\ref{twopart:eq:oneseventyfour}), to the pure 2D limit, Eq.~(\ref{twopart:eq:oneheight}), shown in Fig.~\ref{twopart:fig:eleven}. The other three equations are 
\begin{equation}
\left[ \begin{array}{cccc}
1 + U M_{000}  &   - 2 \vert V_{xy} \vert M_{100}                 &  - 2 \vert V_z \vert M_{001}            \\
  2 U M_{100}  & 1 -   \vert V_{xy} \vert ( M_{000} + M_{200} + 2 M_{110} ) &  - 4 \vert V_z \vert M_{101}  \\
    U M_{001}  &   - 2 \vert V_{xy} \vert M_{101}                 &  1 - \vert V_z \vert ( M_{000} + M_{002} )
\end{array} \right] 
\left[ \begin{array}{c}
\Phi_0 \\  \Phi^{+}_{\bf x} + \Phi^{+}_{\bf y} \\ \Phi^{+}_{\bf z}  
\end{array} \right] = 0 \: .   
\label{twopart:eq:oneeighty}
\end{equation}
This system describes a mixture of one $s$-symmetric and one $d$-symmetric states. Upon setting $E = E_{11} = - 8t - 4t_z$, the consistency condition of Eq.~(\ref{twopart:eq:oneeighty}) links four model parameters: $U$, $V_{xy}$, $V_z$ and $t_z$. One of them can be expressed via the other three. A new feature of this model relative to the cases considered before is the ability to tune the degree of 3D anisotropy. Therefore, we will be mostly interested in $t_z$-dependence of binding conditions. It is convenient to expand the determinant in Eq.~(\ref{twopart:eq:oneeighty}) in powers of $U$ and $V$. This is done in \ref{twopart:sec:appcfour}. From here, one potential can be expressed via the other two. Out of all possibilities, we consider three special cases: (A) $V_{z} = V_{xy}$, (B) $V_{z} = 0$, and (C) $V_{xy} = 0$, all at a fixed $U$.

\begin{figure}[t]
\begin{center}
\includegraphics[width=0.48\textwidth]{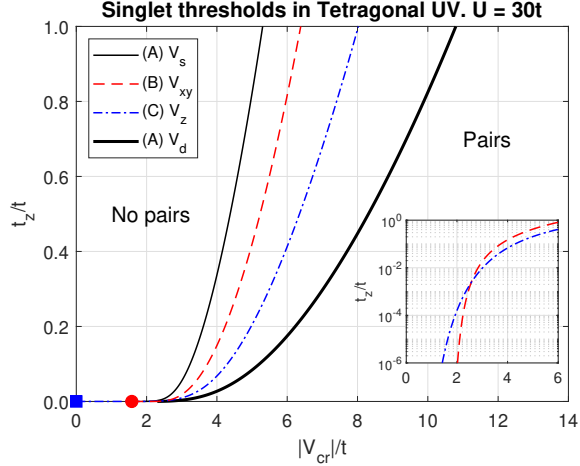}
\end{center}
\caption{Binding thresholds of singlet pairs in the tetragonal $UV$ model at the $\Gamma$-point and $U = 30 \, t$. Pairs are formed to the right of their respective lines. (A) The solid lines describe formation of an $s$- and a $d$-symmetric pair when $V_{xy} = V_{z}$. At $t_{z}/t = 1$, their values are given by Eqs.~(\ref{twopart:eq:oneseventytwo}) and (\ref{twopart:eq:oneseventyfour}), respectively. (B) In-plane attraction only. The dashed line is Eq.~(\ref{twopart:eq:oneeightythree}). In the $t_z \rightarrow 0$ limit, the model reduces to the square $UV$ model. The $t_z =0$ value is given by Eq.~(\ref{twopart:eq:onehsixthree}) and is marked by a circle. (C) Out-of-plane attraction only. The dot-dashed line is Eq.~(\ref{twopart:eq:oneeightyfour}). In the $t_z \rightarrow 0$ limit, the model reduces to the 2D {\em attractive} Hubbard model with zero binding threshold. The square marks the binding threshold for out-of-plane attraction (zero). Inset: comparison of cases B and C at extreme anisotropy. } 
\label{twopart:fig:twelve}
\end{figure}

(A) \underline{$V_{xy} = V_{z} \equiv V$.} In this case, Eq.~(\ref{twopart:eq:appfthirtysix}) becomes a quadratic equation for $|V_{\rm cr}|$. Two real roots correspond to the formation of two bound states: a low-energy one with $s$ orbital symmetry and a high-energy one with $( d_{xz} + d_{yz} )$ symmetry. Both thresholds are shown in Fig.~\ref{twopart:fig:twelve} as functions of $t_z$. At $t_z/t = 1$, this model is equivalent to the simple cubic $UV$ model studied in Section~\ref{twopart:sec:sevenone}.

(B) \underline{$V_z = 0$.} In-plane attraction only. Set $V_z = 0$ and expand the remaining $2 \times 2$ determinant in Eq.~(\ref{twopart:eq:oneeighty}) to express $V_{xy, {\rm cr}}$ vs. $U$:  
\begin{equation}
| V^{s}_{xy, {\rm cr}} | = \frac{ U M_{000} + 1 }
{ U \, [ M_{000} ( M_{000} + M_{200} + 2 M_{110}) - 4 M^2_{100} ] + ( M_{000} + M_{200} + 2 M_{110} ) } \: .
\label{twopart:eq:oneeightythree}
\end{equation}
In the 2D limit, $t_z \rightarrow 0$, all $M_{nmk}$ logarithmically diverge, but utilizing a subtractive procedure $M_{nmk} = M_{000} + L_{nmk}$, one can show that Eq.~(\ref{twopart:eq:oneeightythree}) reduces to Eq.~(\ref{twopart:eq:onehsixthree}). 

(C) \underline{$V_{xy} = 0$.} Out-of-plane attraction only. Set $V_{xy} = 0$ and expand the remaining $2 \times 2$ determinant in Eq.~(\ref{twopart:eq:oneeighty}) to express $V_{z, {\rm cr}}$ vs $U$:  
\begin{equation}
| V^{s}_{z, {\rm cr}} | = \frac{ U M_{000} + 1 }
{ U \, [ M_{000} ( M_{000} + M_{002} ) - 2 M^2_{001} ] + ( M_{000} + M_{002} ) } \: .
\label{twopart:eq:oneeightyfour}
\end{equation}
Both Eqs.~(\ref{twopart:eq:oneeightythree}) and (\ref{twopart:eq:oneeightyfour}) mark the appearance of an $s$-symmetric pair. In case B, the pair is ``disk-like'' extending in the $xy$-plane. In case C, the pair is ``cigar-like'' extending along $z$-axis. Both functions are plotted in Fig.~\ref{twopart:fig:twelve}. It is instructive to compare cases B and C at different degrees of anisotropy. At $t_z/t = 1$, hopping is fully isotropic and all $V$'s contribute equally to binding. There are four attractive bonds in case B but only two in case C. Therefore, one expects the case C threshold to be larger, which can be observed in Fig.~\ref{twopart:fig:twelve}. However, in the opposite 2D limit, $t_z \rightarrow 0$, the situation is reversed. Case B with in-plane attraction reduces to the square $UV$ model studied in Section~\ref{twopart:sec:five}. Accordingly, the threshold line terminates at a $U$-dependent finite value given by Eq.~(\ref{twopart:eq:onehsixthree}). (The limit is marked by a circle in Fig.~\ref{twopart:fig:twelve}.) In contrast, in case C two particles reside on adjacent $z$ planes and attract each other when their $xy$ coordinate coincide. Thus, the model reduces to the 2D {\em attractive} Hubbard model where the threshold is zero, see Section~\ref{twopart:sec:threethree}. The C threshold line extends all the way to $|V| = 0$, which is marked by a square in Fig.~\ref{twopart:fig:twelve}. The zero threshold can also be deduced from Eq.~(\ref{twopart:eq:oneeightyfour}). In the $t_z \rightarrow 0$ limit, both $M_{001}$ and $M_{002}$ tend to zero rather than logarithmically diverge. The threshold reduces to the attractive Hubbard expression $|V_{\rm cr}| = 1/M_{000}$. Since $M_{000}$ diverges, the threshold is zero. The entire line $V_{z}(t_z)$ in Fig.~\ref{twopart:fig:twelve} almost exactly matches the one in the tetragonal attractive Hubbard model, see Section~\ref{twopart:sec:threesix} and Fig.~\ref{twopart:fig:four}(a). The effects of Hubbard repulsion $U$ are barely felt. 

By continuity, the B and C threshold lines must cross, which can be seen in the inset of Fig.~\ref{twopart:fig:twelve}. The crossing happens at high anisotropy, $t_z \simeq 0.002 \, t$. This is a useful anisotropy scale that separates two different regimes. At $t_z < 0.002 \, t$, pair formation is driven by lo\-garithmic divergencies of $M$'s. This is where pairs can be considered purely two-dimensional. At $t_z > 0.002 \, t$, the divergencies are no longer dominant, and pairs become three-dimensional yet strongly anisotropic. This point will be discussed further in Section~\ref{twopart:sec:twelve}.

\subsection{\label{twopart:sec:eighttwo}
Triplet states. Pairing thresholds at the $\Gamma$-point
}

There are three spin-triplet pairs, all with $p$-type orbital symmetries. Their dispersion relations are given by Eqs.~(\ref{twopart:eq:oneseventyfive})-(\ref{twopart:eq:oneseventyseven}). The only difference is that $V_{xy}$ should be used in the $\Phi^{-}_{\bf x}$ and $\Phi^{-}_{\bf y}$ equations but $V_{z}$ in the $\Phi^{-}_{\bf z}$ equation. Like in the simple cubic case, decomposition into three independent dispersion relations takes place over the entire BZ of pair momenta. In this section, we only investigate binding conditions at the $\Gamma$ point. To that end, we set $\alpha = \beta = 4t$, $\gamma = 4t_z$ and $E = -8t - 4t_z$. Analysis of the $p_x$ and $p_y$ thresholds, which are equal, is straightforward. They smoothly interpolate between the pure 2D limit, Eq.~(\ref{twopart:eq:oneheleven}), and the isotropic 3D limit, Eq.~(\ref{twopart:eq:oneseventyeight}). A numerical solution of Eq.~(\ref{twopart:eq:oneseventyfive}) for all $t_z/t$ is plotted in Fig.~\ref{twopart:fig:eleven}.  

The situation with the $p_z$ state is more interesting. Although Eq.~(\ref{twopart:eq:oneseventyseven}) contains a difference of two $M$'s suggesting convergence at $t_z \rightarrow 0$, $M_{002}$ actually tends to zero while $M_{000}$ still diverges, resulting in a zero threshold. A numerical solution of Eq.~(\ref{twopart:eq:oneseventyseven}) plotted in Fig.~\ref{twopart:fig:eleven} confirms the conclusion (thick solid line). Due to a relative simplicity of Eq.~(\ref{twopart:eq:oneseventyseven}), it is possible to derive an asymptotic behavior of $|V^{p_z}_{z,{\rm cr}}|$ in the $t_z \rightarrow 0$ limit. This is done in \ref{twopart:sec:appcfive}. The result is    
\begin{equation}
V^{p_z}_{z,{\rm cr}}( t_z \rightarrow 0 ) \approx  
\frac{ 8 \pi t }{ \ln{ \frac{ 32 \, t }{ \sqrt{e} \, t_z } } } \: .
\label{twopart:eq:oneeightytwo}
\end{equation}
This formula is plotted in Fig.~\ref{twopart:fig:eleven} as the dashed line and is in excellent agreement with the exact result for $t_z/t < 0.1$. Combining this result with the discussion of $s$-pair formation in Section~\ref{twopart:sec:eightone}, one concludes that {\em two} bound states are formed with {\em zero} thresholds in case (C), one cigar-shaped $s$ and one $p_z$. This is an unusual situation in $UV$ models since normally pairs of different symmetries are formed at different thresholds. The tetragonal $UV$ model with out-of-plane attraction is richer than the attractive Hubbard model.

\subsection{\label{twopart:sec:eleven}
Bose-Einstein condensation of real-space pairs   
}

The tetragonal $UV$ model is a convenient system to discuss Bose-Einstein condensation (BEC) of real-space pairs. BEC lies at the core of the preformed-pairs mechanism of high-temperature superconductivity~\cite{Alexandrov1994,Micnas1990,Nozieres1985,Chen2005,Kornilovitch2015,Bogoliubov1970}. One should also mention BEC of mo\-lecules engineered in cold gases~\cite{Greiner2003,Jochim2003,Zwierlein2003}. However, BEC is a many-body effect and as such is out-of-scope of this work. For that reason, the following discussion is only qualitative. 

It was argued in Section~\ref{twopart:sec:eightone} that if $t_z > 0.002 \, t$, then pairs are already {\em three-dimensional} yet highly anisotro\-pic. The above condition is satisfied in most crystalline solids. Hence, the pairs' collective behavior should be described in terms of {\em anisotropic} 3D BEC~\cite{Alexandrov1994,Alexandrov1999b} rather than a pure 2D Berezinskii--Kosterlitz--Thouless transition. A recent neutron scattering study~\cite{Tutueanu2023} supports this viewpoint.

\begin{figure}[t]
\begin{center}
\includegraphics[width=0.60\textwidth]{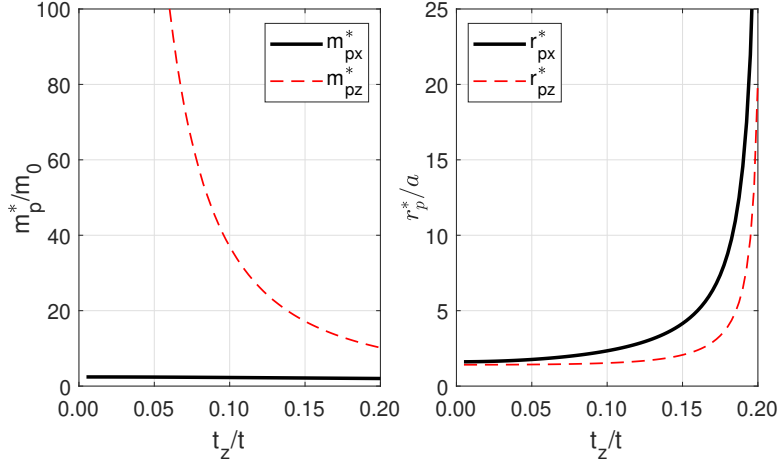}
\end{center}
\caption{Masses (left panel) and radii (right panel) of an $s$-symmetric bound pair in the tetragonal $UV$ model with out-of-plane attraction. $U = 30\,t$, $V_z = 5.0\,t$. The masses are measured in units of the free in-plane mass, $m_0 = \hbar^2/(2 t a^2)$. } 
\label{twopart:fig:twentytwo}
\end{figure}

A convenient starting point is the anisotropic version of the continuous-space BEC formula:
\begin{equation}
k_{B} T_{\rm BEC} = 3.31 \, \frac{\hbar^2 \tilde{n}^{\frac{2}{3}}_{p} }
{ ( \tilde{m}^{\ast 2}_{px} \tilde{m}^{\ast}_{pz} )^{\frac{1}{3}} } \: .
\label{twopart:eq:oneeightysix}
\end{equation}
Here, $\tilde{n}_{p}$ and $\tilde{m}^{\ast}_{p}$ are the density and effective mass of bosons (fermion pairs) in {\em physical} units. Transitioning to relative units one obtains
\begin{equation}
{\cal T}_{\rm BEC} \equiv \frac{1}{t} \: k_{B} T_{\rm BEC} = 6.62 \, \frac{ n^{\frac{2}{3}} }
{ ( m^{\ast 2}_{px} m^{\ast}_{pz} )^{\frac{1}{3}} } \: . 
\label{twopart:eq:oneeightyseven}
\end{equation}
Here, $n$ is the number of pairs per unit cell, and the pair masses are expressed in units of $m_0 = \hbar^2 /(2 t a^2)$. We also assume $m^{\ast}_{px} = m^{\ast}_{py}$. As was argued in the Introduction, a maximum critical temperature is reached at ``close-packing'' when the pair density is approximately equal to the inverse pair volume, $n_{\rm cp} = \Omega^{-1}_{p}$. The close-packing temperature is:
\begin{equation}
{\cal T}^{\ast}_{\rm BEC} = {\cal T}_{\rm BEC}( n_{\rm cp} ) = 
\frac{ 6.62 }{  ( \Omega^{2}_{p} \, m^{\ast 2}_{px} m^{\ast}_{pz} )^{\frac{1}{3}} } \: . 
\label{twopart:eq:oneeightyeight}
\end{equation}
This formula contains only single pair properties supplied by the exact solution. The continuum approximation can be superseded by a more rigorous lattice treatment. The pair density is computed as a full Bose integral and then equated to an inverse pair volume. Instead of Eq.~(\ref{twopart:eq:oneeightyeight}), one has
\begin{equation}
\int_{\rm BZ} \frac{{\rm d}^3 {\bf P}}{(2\pi)^3} 
\frac{1}{ \exp{ \left\{ \frac{ E({\bf P}) - E_{0} }{ {\cal T}^{\ast}_{\rm BEC}} \right\} } - 1 }
 = \frac{1}{\Omega_{p} } \: . 
\label{twopart:eq:oneninetyone}
\end{equation}
$E_{\bf P}$ is the full pair dispersion also provided by the exact solution. This equation~\cite{Kornilovitch2015} does not resort to the effective mass approximation.  

Here, we apply the simplified formula, Eq.~(\ref{twopart:eq:oneeightyeight}), to the tetragonal $UV$ model with out-of-plane attraction only. Only $s$-symmetric pairs are included. The effective masses are calculated numerically from the full singlet dispersion relation [take Eq.~(\ref{twopart:eq:onesixtythree}) and set $V = 0$ in the two middle columns]: 
\begin{equation}
\left[ \! \begin{array}{cc}
1 + U M_{000}  &   - 2 \vert V_{z} \vert M_{001}                 \\
    U M_{001}  &   1 - \vert V_{z} \vert ( M_{000} + M_{002} )
\end{array} \! \right]  \! 
\left[ \! \begin{array}{c}
\Phi_{0} \\  \Phi^{+}_{\bf z}  
\end{array} \! \right] 
= 0 \: .   
\label{twopart:eq:oneninetytwo}
\end{equation}
For the pair volume, we adopt the following formula~\cite{Kornilovitch2015}:
\begin{equation}
\Omega_{p}  =  r^{\ast}_{px} \, r^{\ast}_{py} \, r^{\ast}_{pz} 
 =  \sqrt{ \left[ 1 + \langle ( \triangle x )^2 \rangle \right] 
            \left[ 1 + \langle ( \triangle y )^2 \rangle \right]
            \left[ 1 + \langle ( \triangle z )^2 \rangle \right] } \: ,   
\label{twopart:eq:oneninetythree}
\end{equation}
which also defines the pair sizes $r^{\ast}_{px}$, $r^{\ast}_{py}$, and $r^{\ast}_{pz}$. This form of $\Omega_{p}$ respects the exclusion principle and ensures the pair volume is not less than 1 even in the strong coupling limit. The mean-squared distances are defined via the full coordinate wave function, Eq.~(\ref{twopart:eq:thirteenone}). For example
\begin{equation}
\langle ( \triangle x )^2 \rangle = 
\frac{ \sum_{{\bf m}_1 , {\bf m}_2} ( m_{1x} - m_{2x} )^2 \left\vert \Psi( {\bf m}_1 , {\bf m}_2 ) \right\vert^2 }
     { \sum_{{\bf m}_1 , {\bf m}_2} \left\vert \Psi( {\bf m}_1 , {\bf m}_2 ) \right\vert^2 } \: .   
\label{twopart:eq:oneninetyfour}
\end{equation}
In taking the momentum sum in Eq.~(\ref{twopart:eq:thirteenone}), ${\bf k}_1 = - {\bf k}_2$ should be used, which corresponds to ${\bf P} = 0$. In principle, all the sums can be computed by brute-force numerical calculation utilizing FFT. However, for the tetragonal dispersion, $\langle ( \triangle {\bf r}_i )^2 \rangle$ can be analytically reduced to ratios of two one-dimensional integrals (see Ref.~\cite{Kornilovitch2015}, Appendix C), which makes calculation of $\Omega_{p}$ much more efficient.

\begin{figure}[t]
\begin{center}
\includegraphics[width=0.48\textwidth]{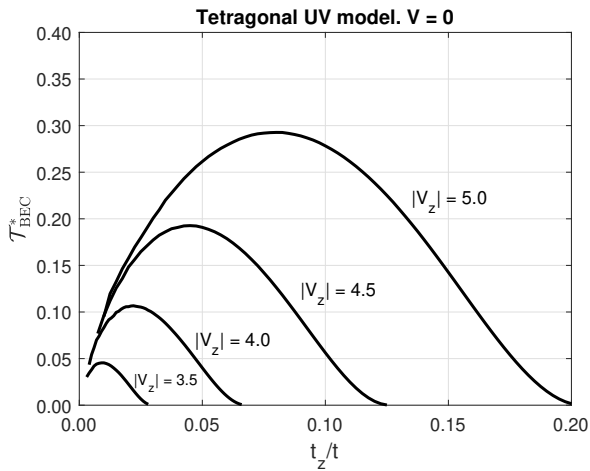}
\end{center}
\caption{The close-packed BEC temperature, Eq.~(\ref{twopart:eq:oneeightyeight}), vs. lattice anisotropy for several out-of-plane attractive strengths $V_z$. Notice how the peak ${\cal T}^{\ast}_{\rm BEC}$ increases with $|V_z|$. $V_z$ values are given in units of in-plane hopping $t$. In-plane attraction $V = 0$.} 
\label{twopart:fig:twentythree}
\end{figure}

Figure~\ref{twopart:fig:twentytwo} shows the effective pair mass and radius as functions of $t_z$ for $U = 30\,t$ and $V_z = 5\,t$. For this attraction, the pair exists in the interval $0 < t_z/t < 0.205$, see Fig.~\ref{twopart:fig:twelve}, case (C). One expects the pair to be strongly bound at $t_z \rightarrow 0$ and loosely bound at $t_z \rightarrow 0.205\,t$. This can be seen in the effective radius plots: both $r^{\ast}_{px}$ and $r^{\ast}_{pz}$ tend to 1 in the 2D limit but diverge near the threshold. The in-plane pair mass, $m^{\ast}_{px}$, is weakly affected by out-plane anisotropy and stays almost constant as a function of $t_z$. In contrast, the out-of-plane mass strongly depends on $t_z$. Near the binding threshold, it approaches the $z$-mass of two free particles, $2t/t_z$. In the opposite limit, $t_z \rightarrow 0$, it diverges as $\propto V_z/t^2_z$.   

Now consider the close-packed BEC temperature given by Eq.~(\ref{twopart:eq:oneeightyeight}). It is bounded by two divergencies: a divergent $m^{\ast}_{pz}$ at $t_z \rightarrow 0$ and a divergent $\Omega_{p}$ near the threshold. Thus, we expect ${\cal T}^{\ast}_{\rm BEC}$ to have a maximum at an intermediate $t_z$. This is shown in Fig.~\ref{twopart:fig:twentythree}. {\em The present model predicts the existence of an optimal degree of anisotropy.} If $t_z$ is too large, the kinetic energy is large, pairs are loosely bound, their volume is large, the packing density is small, and ${\cal T}^{\ast}_{\rm BEC}$ is small as a result. If $t_z$ is too small, the the out-of-plane mass is very large and ${\cal T}^{\ast}_{\rm BEC}$ is small again. A similar conclusion was reached in Ref.~\cite{Kornilovitch2015} for in-plane attraction. Another interesting feature of Fig.~\ref{twopart:fig:twentythree} is that the {\em peak value} of ${\cal T}^{\ast}_{\rm BEC}$ increases with the attraction strength. A stronger $V_z$ creates more compact pairs which boosts the close-packed density. However, large $V_z$'s cause phase separation as discussed in Section~\ref{twopart:sec:ten}.

\section{ \label{twopart:sec:appf}
Light pairs in the strong coupling limit    
}

We have already discussed various aspects of light bound pairs in the 1D chain (Section~\ref{twopart:sec:fourthree}), in the 2D square lattice (Section~\ref{twopart:sec:fivethree}), and in the 2D triangular lattice (Section~\ref{twopart:sec:six}). In this section, we study several other systems that host mobile pairs in the strong-coupling limit.

\subsection{ \label{twopart:sec:appfone}
Staggered square planes   
}

\begin{figure*}[t]
\begin{center}
\includegraphics[width=0.85\textwidth]{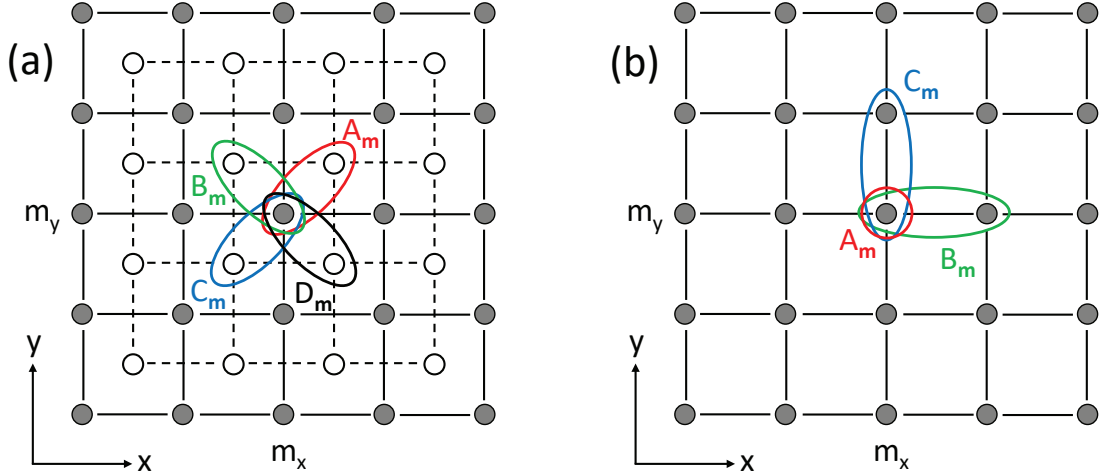}
\end{center}
\caption{ (a) The two-plane staggered square lattice. A square lattice of open circles are shifted out-of-plane (in the $z$ direction) relative to a square lattice of filled circles. Attraction $V$ is {\em between} the planes. In the $U \rightarrow \infty$, $|V| \rightarrow \infty$ limit, the two particles reside on different planes. The ovals mark four basic dimer configurations. (b) The square lattice with resonant attraction, $U < 0$, $U = V$. The circle and ovals mark three basic dimer states that are relevant in the $|U|, |V| \rightarrow \infty$ limit. } 
\label{twopart:fig:appeone}
\end{figure*}

Consider a lattice consisting of two square lattices shifted out-of-plane and staggered in the $(xy)$ plane relative to each other, see Fig.~\ref{twopart:fig:appeone}(a). In the limit of very strong {\em out-of-plane} attraction, one particle will reside on the first plane and another particle on the second plane. The magnitudes of $U$ and inter-plane hopping $t_z$ are irrelevant. At the same time, the bound pair can still move in the first order in in-plane hopping $t$. In the strong-coupling limit, there are four non-equivalent dimer configurations shown in the figure. Because the particles cannot exchange, there is no need to consider symmetrized basis states. Acting on the dimer states by the Hamiltonian yields:  
\begin{align}
\hat{H} A_{\bf m} & =  - t \left( B_{\bf m} + B_{{\bf m} + {\bf x}} \right)   
                  - t \left( D_{\bf m} + D_{{\bf m} + {\bf y}} \right)  , 
\nonumber                         \\
\hat{H} B_{\bf m} & =  - t \left( A_{\bf m} + A_{{\bf m} - {\bf x}} \right)   
                  - t \left( C_{\bf m} + C_{{\bf m} + {\bf y}} \right)  , 
\nonumber                         \\
\hat{H} C_{\bf m} & =  - t \left( B_{\bf m} + B_{{\bf m} - {\bf y}} \right)   
                  - t \left( D_{\bf m} + D_{{\bf m} - {\bf x}} \right)  , 
\nonumber                         \\
\hat{H} D_{\bf m} & =  - t \left( A_{\bf m} + A_{{\bf m} - {\bf y}} \right)   
                  - t \left( C_{\bf m} + C_{{\bf m} + {\bf x}} \right)  . 
\label{twopart:eq:appeone} 
\end{align}
Next, compose a Schr\"odinger equation and transform it to momentum space. That results in a consistency condition
\begin{equation}
\left\vert \begin{array}{cccc}
  \tilde{E}  &   t \left( 1 + e^{i P_x a} \right)   &   0   &   \left( 1 + e^{i P_y a} \right)           \\
t \left( 1 + e^{ - i P_x a} \right)    &   \tilde{E}   &   \left( 1 + e^{i P_y a} \right)   &   0        \\
  0  &   t \left( 1 + e^{ - i P_y a} \right)     &   \tilde{E}   &   t \left( 1 + e^{ - i P_x a} \right) \\
 t \left( 1 + e^{ - i P_y a} \right)   &    0    &   t \left( 1 + e^{i P_x a} \right)   &   \tilde{E}   
\end{array} \right\vert = 0 \: ,
\label{twopart:eq:appetwo} 
\end{equation}
where $\tilde{E}$ is pair energy counted from $-|V|$. Expansion of the determinant yields four dispersion bands
\begin{equation}
\tilde{E}_{1-4} = \pm (2t) \left\vert \cos{\frac{P_x a}{2}} \pm \cos{\frac{P_y a}{2}} \right\vert \: . 
\label{twopart:eq:appethree} 
\end{equation}
Near $P_x = P_y = 0$, the ground state energy is 
\begin{equation}
\tilde{E}_{1} = - (2t) \left( \cos{\frac{P_x a}{2}} + \cos{\frac{P_y a}{2}} \right) \: . 
\label{twopart:eq:appefour} 
\end{equation}
Expanding at small momentum, one finds the effective mass 
\begin{equation}
m^{\ast}_{px} = m^{\ast}_{py} = \frac{ 2 \hbar^2 }{ t \, a^2 } = 4 \, m_0 \: , 
\label{twopart:eq:appefive} 
\end{equation}
where $m_0 = \hbar^2/( 2 t a^2 )$ is the free particle mass.

\subsection{ \label{twopart:sec:appftwo}
Square $UV$ model with resonant attraction, $U = V$    
}

The simple square $UV$ model with only nearest-neighbor attraction does not support light pairs. However, light pairs become possible for resonant attraction $U < 0$, $U = V$. In this case, the energy of on-site dimer and nearest-neighbor dimer are equal, and the pair can move in the first order in $t$. Figure~\ref{twopart:fig:appeone}(b) shows the three relevant basis dimers. Since the particles can exchange, it is essential to consider symmetrized states:
\begin{align}
A_{\bf m}  & =  \vert \uparrow \downarrow \rangle_{\bf m} \: , 
\label{twopart:eq:appesix}    \\
B_{\bf m}  & =  \frac{1}{\sqrt{2}} 
\left( \vert \uparrow   \rangle_{\bf m} \vert \downarrow \rangle_{{\bf m} + {\bf x}} + 
       \vert \downarrow \rangle_{\bf m} \vert \uparrow   \rangle_{{\bf m} + {\bf x}}  \right)  ,   
\label{twopart:eq:appeseven}   \\
C_{\bf m}  & =  \frac{1}{\sqrt{2}} 
\left( \vert \uparrow   \rangle_{\bf m} \vert \downarrow \rangle_{{\bf m} + {\bf y}} + 
       \vert \downarrow \rangle_{\bf m} \vert \uparrow   \rangle_{{\bf m} + {\bf y}}  \right)  .    
\label{twopart:eq:appeeight}       
\end{align}
The Hamiltonian action is
\begin{align}
\hat{H} A_{\bf m} & =  - \sqrt{2} \, t \left( B_{\bf m} + B_{{\bf m} - {\bf x}} \right)   
                  - \sqrt{2} \, t \left( C_{\bf m} + C_{{\bf m} - {\bf y}} \right)  , 
\label{twopart:eq:appethirteen}                          \\
\hat{H} B_{\bf m} & =  - \sqrt{2} \, t \left( A_{\bf m} + A_{{\bf m} + {\bf x}} \right)  , 
\label{twopart:eq:appefourteen}                          \\
\hat{H} C_{\bf m} & =  - \sqrt{2} \, t \left( A_{\bf m} + A_{{\bf m} + {\bf y}} \right)  .  
\label{twopart:eq:appenine} 
\end{align}
Notice the special role of dimer $A$: it is coupled to both $B$ and $C$ while the latter two are not coupled directly but only through $A$. That will allow for generalization to other lattices to be discussed in Section~\ref{twopart:sec:appfthree}. The dispersion relation reads 
\begin{equation}
\left\vert \begin{array}{ccc}
\tilde{E} & \sqrt{2} \, t \left( 1 + e^{ - i P_x a} \right)  &  \sqrt{2} \, t \left( 1 + e^{ - i P_y a} \right) \\
\sqrt{2} \, t \left( 1 + e^{ i P_x a} \right)    &   \tilde{E}    &      0                                      \\
\sqrt{2} \, t \left( 1 + e^{ i P_y a} \right)    &       0        &  \tilde{E}      
\end{array} \right\vert = 0 \: , 
\label{twopart:eq:appeten} 
\end{equation}
which yields three bands: ${\tilde E}_2 = 0$, and 
\begin{equation}
\tilde{E}_{1,3} = \pm (2 \sqrt{2} \, t ) \sqrt{ \cos^2{\frac{P_x a}{2}} + \cos^2{\frac{P_y a}{2}} } \: . 
\label{twopart:eq:appeeleven} 
\end{equation}
Expanding the lowest branch at small ${\bf P}$ yields the mass
\begin{equation}
m^{\ast}_{px} = m^{\ast}_{py} = \frac{ 2 \hbar^2 }{ t \, a^2 } = 4 \, m_0 \: , 
\label{twopart:eq:appetwelve} 
\end{equation}
where $m_0 = \hbar^2/( 2 t a^2 )$ is the free particle mass.

\subsection{ \label{twopart:sec:appfthree}
Generalization to other lattices with resonant attraction, $U = V$    
}

The preceding calculation is readily generalized to other lattices with resonant contact and nearest-neighbor attraction. Consider a set of nearest-neighbor vectors ${\bf b}_{+}$ such that the dimers $D$ built on sites ${\bf m}$ and ${\bf m} + {\bf b}_{+}$ are coupled only to $A_{\bf m}$ but not to each other. Then, instead of Eqs.~(\ref{twopart:eq:appethirteen})-(\ref{twopart:eq:appenine}) we have:
\begin{align}
\hat{H} A_{\bf m} & =  - \sqrt{2} \, t \sum_{{\bf b}_{+}} \left( B_{\bf m} + B_{{\bf m} - {\bf b}_{+}} \right) , 
\label{twopart:eq:appefifteen}                          \\
\hat{H} B_{\bf m} & =  - \sqrt{2} \, t \left( A_{\bf m} + A_{{\bf m} + {\bf b}_{+}} \right)  .  
\label{twopart:eq:appesixteen}                           
\end{align}
This simple form enables analytical expressions for several other lattices. For example, in the 3D simple cubic lattice, ${\bf b}_{+} = {\bf x}$, ${\bf y}$, and ${\bf z}$. The lowest energy band is 
\begin{equation}
\tilde{E}_{1} = - ( 2 \sqrt{2} \, t ) 
\sqrt{ \cos^2{\frac{P_x a}{2}} + \cos^2{\frac{P_y a}{2}} + \cos^2{\frac{P_z a}{2}} } \: . 
\label{twopart:eq:appeseventeen} 
\end{equation}
The effective mass is
\begin{equation}
m^{\ast}_{px} = m^{\ast}_{py} = m^{\ast}_{pz} = 
\sqrt{6} \, \frac{ \hbar^2 }{ t \, a^2 } = 2 \sqrt{6} \, m_0 \: .  
\label{twopart:eq:appeeighteen} 
\end{equation}
Comparison between Eqs.~(\ref{twopart:eq:eightysixeight}), (\ref{twopart:eq:appeeleven}), and (\ref{twopart:eq:appeseventeen}) suggests an obvious generalization to hyper-cubic lattices in higher dimensions $D > 3$. 

In the body-centered cubic lattice, there are four vectors: ${\bf b}_{+} = \frac{1}{2} ( {\bf x} \pm {\bf y} \pm {\bf z} )$. Dimer dispersion has five energy bands with ${\tilde E}_{2,3,4} = 0$ and 
\begin{equation}
\tilde{E}_{1,5} = \pm ( 4 \, t ) 
\sqrt{ 1 + \cos{\frac{P_x a}{2}} \cos{\frac{P_y a}{2}} \cos{\frac{P_z a}{2}} } \: . 
\label{twopart:eq:appenineteen} 
\end{equation}
The effective mass of the lowest band is
\begin{equation}
m^{\ast}_{px} = m^{\ast}_{py} = m^{\ast}_{pz} = 
2 \sqrt{2} \, \frac{ \hbar^2 }{ t \, a^2 } = 4 \sqrt{2} \, m_0 \: .  
\label{twopart:eq:appetwenty} 
\end{equation}

It is noteworthy that Eqs.~(\ref{twopart:eq:eightysixnine}), (\ref{twopart:eq:appetwelve}), (\ref{twopart:eq:appeeighteen}), and (\ref{twopart:eq:appetwenty}) can be unified as
\begin{equation}
m^{\ast}_{p} = 2 \sqrt{z} \, m_0 \: ,  
\label{twopart:eq:appetwentyone} 
\end{equation}
where $z$ is the number of nearest neighbors in respective lattices.

\section{\label{twopart:sec:ten}
More than two particles  
}

The properties of one bound pair, which this review summarizes, make physical sense only if the pairs keep their identity in a macroscopic system with finite particle density. For most of the results to remain valid, the pairs should not aggregate in trions, quads, and larger clusters. In other words, the effective interaction between pairs should be either repulsive or weakly attractive when such a weak attraction is unable to bind pairs into larger complexes (in 3D). The situation is clear in attractive Hubbard models. Two fermions forming an on-site real-space pair have two opposite spin projections and prevent other fermions from occupying the same site at the same time. Thus, clusters of three or more fermions are prohibited by the exclusion principle. The attractive Fermi--Hubbard model is stable against phase separation~\cite{Su1996}, and all the results derived in Section~\ref{twopart:sec:three} remain valid. Bosons, on the other hand, can pile up on the same site without limits, sending the total energy to negative infinity. The attractive Bose--Hubbard model is unstable beyond the pair-forming threshold, such as the ones listed in Table~\ref{twopart:tab:one}.    

The situation is more subtle in $UV$ models, {\it t-J} models, and other models where the attractive interaction is of finite range. The exclusion principle does not directly prohibit pairs from sticking to each other side-by-side and forming a macroscopic cluster. Thus, any such system will phase-separate in the large $V$ limit. However, the exclusion principle still plays an important role at intermediate $V$. In the context of large bipolarons, Emin argued~\cite{Emin1994} that when forming a quadpolaron, additional carriers must occupy excited states of the self-trapping well. That leads to an additional short-range repulsion between pairs (large bipolarons in this case). A liquid of real-space pairs must be stable against clustering in a finite attraction interval just above the pairing threshold. Thus, the question becomes quantitative: when exactly does the system phase-separate, and how is the phase separation threshold $V_{\infty}$ related to the pair-forming threshold $V_{2}$ studied in this paper? Since the discovery of high-temperature superconductivity, these issues have been discussed mostly in the context of phase-separation in the {\it t-J} model~\cite{Emery1990,Dagotto1993} and in bipolaronic superconductors~\cite{Alexandrov2002A,Alexandrov2002B}. Those investigations were based on comparing the energy of a collection of pairs with the energy of an infinite cluster. Because both energies were determined either by applying approximations or on small lattice segments, the phase boundaries were approximate.

\begin{figure}[t]
\begin{center}
\includegraphics[width=0.60\textwidth]{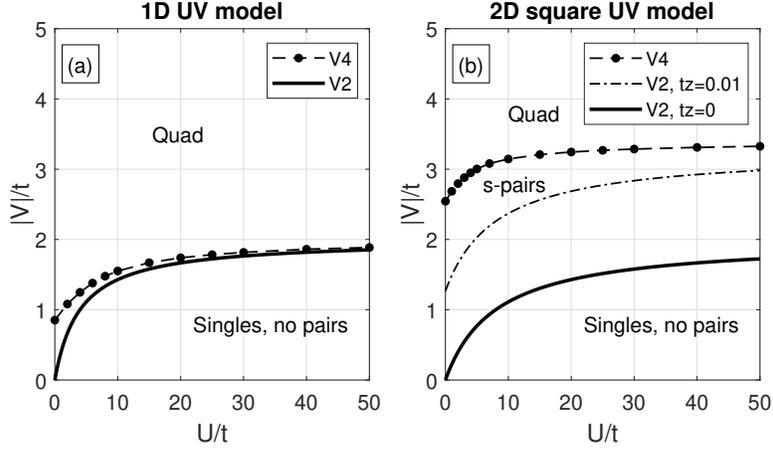}
\end{center}
\caption{Phase diagrams~\cite{Kornilovitch2022} of four fermions in the 1D and 2D $UV$ models at zero total momentum. Singlet pairs are stable against clustering between the solid and dashed lines. The solid lines are Eqs.~(\ref{twopart:eq:eightysix}) and (\ref{twopart:eq:onehsixthree}), respectively. Quad thresholds $V_4$ were obtained by extrapolating to infinite lattices. In 2D, the $U \rightarrow \infty$ limit of $V_4$ is $\approx 3.35\,t$, which is consistent with the $4 \times 4$ values reported in \cite{Emery1990,Dagotto1993}. The dot-dashed line in panel (b) is the pair threshold in the 3D {\em tetragonal} $UV$ model with $V_z = 0$ and $t_z = 0.01\,t$, see Eq.~(\ref{twopart:eq:oneeightythree}). } 
\label{twopart:fig:thirteen}
\end{figure}

Rigorous determination of clustering thresholds and phase boundaries requires solving a few-body or a full many-body problem on an infinite lattice with sufficient accuracy, and is therefore difficult. Berciu reported~\cite{Berciu2011} a zero region of pair stability in a system of 3 to 5 {\em spinless} fermions with nearest-neighbor attraction in an infinite 1D lattice. However, with second-nearest-neighbor {\em repulsion} turned on, the pairs stabilized in a wide interval of parameters. In this model, pairs need to be stabilized dynamically because Emin's argument does not apply to spinless fermions. Also in 1D, Chakraborty, Tezuka and Min investigated four ``extended-Holstein bipolarons'' on chains up to 24 sites long~\cite{Chakraborty2014}. At high phonon frequencies, this model~\cite{Alexandrov1999} maps to a $UV$ model. The authors reported a narrow but finite interval of bipolaron stability in qualitative agreement with the 1D $UV$ phase diagram shown on Fig.~\ref{twopart:fig:thirteen}(a) and to be discussed below.  

In 2D, Emery, Kivelson, and Lin reported~\cite{Emery1990} pair stability in the interval $2.0\,t < |V| < 3.53\,t$ for a $4 \times 4$ lattice and $U = \infty$. A similar interval, $2.0\,t < |V| < 3.8\,t$ on a $4 \times 4$ lattice, was reported by Dagotto and Riera for the {\it t-J} model~\cite{Dagotto1993}.       

Multi-magnon bound states in a 2D square frustrated magnetic were analyzed in \cite{Jiang2022b}. 

We are unaware of any numerical investigation of clustering in 3D lattices except our own work~\cite{Kornilovitch2020} to be discussed next. 

\begin{figure}[t]
\begin{center}
\includegraphics[width=0.48\textwidth]{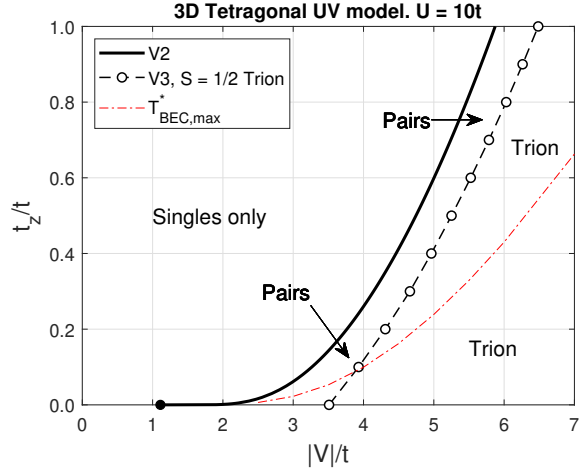}
\end{center}
\caption{Phase diagram of three fermions in the 3D tetragonal $UV$ model for $U = 10\,t$~\cite{Kornilovitch2020}. The pairs are stable against formation of trions between the $V_2$ and $V_3$ lines. The trion threshold $V_3$ was computed on a $12 \times 12 \times 12$ lattice. The dot-dashed line marks the coordinates of {\em peak} close-packed BEC temperature, ${\cal T}^{\ast}_{\rm BEC}$, determined by locating the maxima of plots like those shown in Fig.~\ref{twopart:fig:twentythree}. } 
\label{twopart:fig:fourteen}
\end{figure}

We now summarize our own results on pair liquid stability in $UV$ models~\cite{Kornilovitch2022,Kornilovitch2013,Kornilovitch2014,Kornilovitch2020}. First off, {\em bosonic} $UV$ models are unstable: as soon as pairs form, trions~\cite{Kornilovitch2013} and quads~\cite{Kornilovitch2022} form as well, triggering phase separation. The same conclusion applies to {\em spinless fermions}: those pairs are unstable, too. Phase diagrams of four {\em spinful fermions} in 1D and 2D are shown in Fig.~\ref{twopart:fig:thirteen}(a) and Fig.~\ref{twopart:fig:thirteen}(b), respectively~\cite{Kornilovitch2022}. In 1D, the pair stability region is narrow; it shrinks to zero in the $U \rightarrow \infty$ limit. In 2D, the pair stability region is $\approx 2t$ wide for all $U$. In 3D, exact solutions are computationally more demanding. Only a three-fermion problem has been solved in a 3D $UV$ model so far~\cite{Kornilovitch2020}. Figure~\ref{twopart:fig:fourteen} shows the phase diagram of three fermions in the tetragonal $UV$ model for $U = 10 \, t$. Pairs are stable against formation of trions between the solid and dashed lines. One important feature of Fig.~\ref{twopart:fig:fourteen} is the sharp increase of $V_{2}$ for small but nonzero $t_z$. Going back to Fig.~\ref{twopart:fig:thirteen}(b), one should be careful when interpreting the pure 2D case as an approximation to a highly anisotropic 3D case. Even a small $t_z$ elevates the $V_{2}$ line and significantly reduces the domain of pair stability, which is illustrated by the $t_z = 0.01 \, t$ line. Although the domain of pair stability shrinks significantly upon turning on interlayer hopping, {\em the former remains finite for all $U$.}

A high-level conclusion from this analysis is the existence of finite domains of pair stability in various forms of the $UV$ model in all lattice dimensionalities. {\em It is within such domains is where the results derived in the bulk of this paper are valid.} The fundamental reason for pair stability is the Fermi statistics and associated nodes in the many-fermion wave functions~\cite{Emin1994,Kornilovitch2022}. Since these reasons are general and applicable to any system, we expect them to hold for other types of potentials including, for example, large-radius attractions considered in Section~\ref{twopart:sec:fivethree}.

\section{\label{twopart:sec:twelve}
Summary and discussion   
}

\subsection{\label{twopart:sec:twelveone}
Common properties of lattice bound pairs   
}

In this paper, $UV$ models (extended attractive Hubbard models) of increasing complexity have been analyzed in detail. Taken together, the results point to several common properties of lattice bound states. They are summarized in this section.

\subsubsection{\label{twopart:sec:twelveoneone}
Mathematically exact solutions
}

The most important property of two-particle lattice problems is their exact solvability. This is a big advantage as all the results can be easily verified and cross-checked, both analytically and numerically. When it comes to physical applications, typical arguments are centered around justifications of a starting Hamiltonian or relevancy to a particular real system. But once the model is agreed upon, the conclusions that follow are not usually questioned.     

At the same time, the complexity of a solution increases rapidly with interaction range and lattice dimensionality. As shown in Section~\ref{twopart:sec:twotwo}, the size of a matrix that defines pair dispersion, Eq.~(\ref{twopart:eq:thirteen}), is equal to the number of interacting lattice sites. Separation into singlet and triplet states described in Sections~\ref{twopart:sec:twofour} and \ref{twopart:sec:twofive} cuts the matrix size by about half, which is helpful. Still, the ability to derive analytical results is largely limited to nearest-neighbor interactions. The longer-range square model considered in Section~\ref{twopart:sec:fivethree} and Ref.~\cite{Kornilovitch1995} is a rare exception. Additionally, the matrix elements $M$ that populate Eq.~(\ref{twopart:eq:thirteen}) are two-body lattice Green's functions whose complexity increases with lattice dimensionality. In general, $M$'s can be analytically evaluated in 1D, see Eqs.~(\ref{twopart:eq:twentyeighttwo}) and (\ref{twopart:eq:appbfourteenone})-(\ref{twopart:eq:appbfourteenfive}), and in 2D, see \ref{twopart:sec:appa}. In 3D, analytical expressions are only known for the isotropic simple cubic lattice, \ref{twopart:sec:appcone}. In other cases, numerical integration is necessary. The ability to perform two integrations analytically, as shown in \ref{twopart:sec:appcthree}, reduces $M$'s to one-dimensional integrals which makes subsequent numerics efficient.   

The exact dispersion equation defines pair energy over the entire BZ of pair momenta. It provides singlet--triplet splitting, energy gaps between various sub-bands of the same parity, and a total bandwidth. The dispersion possesses degeneracies at high-symmetry points, which can be cross-checked with group theory classifications. By setting the pair energy equal to the lowest energy of two free particles {\em with the same total momentum} a binding threshold is determined. At this special energy, matrix elements $M$ simplify considerably. In many cases, and especially at the $\Gamma$ point and along BZ diagonals, the thresholds can be calculated analytically. Many explicit examples have been derived in this paper.    

The exact solution also provides pair effective mass and radius. The mass is rigorously defined via a second derivative of the total energy near the band minimum and as such it can be extracted from the dispersion relation. In some cases, one can utilize the fact that the dispersion relation simplifies along BZ diagonals and derive analytical expressions. See, for example, Es.~(\ref{twopart:eq:fortysix}). Calculation of effective radius is more involved as it requires knowledge or the entire pair wave function and its subsequent integration. Still, analytic expressions have been obtained in simple cases, see Eqs.~(\ref{twopart:eq:thirtyfour}) and (\ref{twopart:eq:fiftyoneone}). In more complicated models, analytical manipulations enable reductions of multidimensional integrals to one-dimensional integrals, as was done, for example, in Ref.~\cite{Kornilovitch2015}. The effective radius yields the pair volume and the {\em close-packed density of pairs} which is an important parameter in the theory of Bose-Einstein condensation of pairs. Together with pair mass and full dispersion, it provides estimates of the close-packed BEC temperatures and enables conclusions about optimal model parameters.

\subsubsection{\label{twopart:sec:twelveonetwo}
Pair binding energy depends on its momentum
}

This is purely a lattice effect that is absent in continuous space~\cite{Mattis1986}. Since the lattice provides a reference frame, movement through it is not Galilean-invariant. Binding energies and masses of bound complexes become dependent on their momenta ${\bf P}$. The pair energy and the minimum energy of two free particles both rise with ${\bf P}$ but the latter rises faster. It means that the binding energy {\em increases} with ${\bf P}$. In systems with a finite threshold, it leads to situations when there are no bound pairs at ${\bf P} = 0$ but one or more bound pair at finite ${\bf P}$. A similar effect has long been known in the theory of quantum spin waves~\cite{Wortis1963}. Increasing stability of pairs can also be seen in threshold formulas with explicit dependence on ${\bf P}$, see for example Eqs.~(\ref{twopart:eq:eightysix}), (\ref{twopart:eq:onehfive}), and (\ref{twopart:eq:oneseventytwo}), where thresholds decrease to zero in BZ corners. That implies there is {\em always} at least one bound pair in BZ corners for {\em any} $U$ as long as $|V| > 0$. 

An important consequence is the existence of a {\em binding surface} in momentum space which separates unbound and bound pairs. As shown in Section~\ref{twopart:sec:fivefour}, the binding surface does not in general match the Fermi surface of {\em free} electrons. The mismatch leads to segmentation of the Fermi surface into areas that are more or less prone to pairing and potential appearance of Fermi arcs. We note that such a pair would have formed with a nonzero total momentum. This is opposite to conventional Cooper pairs that are formed with zero total momentum. The described effect may also have links to the FFLO phase. A key question is how the binding surface is modified by a finite fermion density. This is a many-body problem that goes beyond the scope of this work.

\subsubsection{\label{twopart:sec:twelveonethree}
Light pairs
}

Effective mass is an important characteristic of a bound pair. Broadly, it defines pair mobility and response to external perturbations. In the context of superfluidity, the mass determines a BEC temperature in a system of many pairs, as argued in the Introduction. It is important to know what makes pairs either light or heavy. 

Near threshold, the mass is always close to two free-particle masses. In the opposite, strong coupling limit, the mass generally scales as $m^{\ast}_{p}/m_0 \propto |V|/t$, i.e., increases linearly with the binding energy. This is because the pair moves in the second order in hopping $t$ and breaks an attractive bond in the intermediate state, see Fig.~\ref{twopart:fig:sixzero}(a). However, there are two circumstances when a pair can move in the {\em first} order in $t$ and the above scaling is not followed. Instead, pair mass remains of order $m_0$ at all energies including the strong coupling limit. We refer to such pairs as {\em light}. 

The first situation arises for purely geometric reasons. One pair member can hop to a neighbor site without breaking an attractive bond. In other words, both the starting site and the ending site are nearest neighbors to the second pair member. Then the second member moves in the same way and the process repeats. The constituent particles hop in turns and the entire pair moves through the lattice in a ``crab-like'' fashion without ever breaking a bond. A good example is provided by the triangular lattice investigated in Section~\ref{twopart:sec:sixone} (see also Refs.~\cite{Hague2008,Hague2007B}). Pair movement is illustrated in Fig.~\ref{twopart:fig:twelvetwo} and the mass is plotted in Fig.~\ref{twopart:fig:twelvefour}. Observe that even at extreme attractive strengths, $|V| > 60\,t$, the pair mass remains below $5m_0$. For more realistic attractions, $|V| < 10\,t$, the mass is below $3m_0$. That is, mass increase due to binding is less than 50\%. It was suggested in Ref.~\cite{Hague2007B} in the context of bipolaron superconductivity that for that very reason triangular and hexagonal lattices may host high-temperature superconductivity. Other lattices that support crab motion and light pairs include staggered ladders~\cite{Alexandrov2002A,Hague2007C,Hague2007} ($m^{\ast}_{p} < 4 m_0$), staggered square planes (Section~\ref{twopart:sec:appfone}, $m^{\ast}_{p} < 4 m_0$), and face-centered cubic lattice~\cite{Adebanjo2022} ($m^{\ast}_{p} < 6 m_0$). One should add to this list the {\em body-centered} tetragonal lattice that supports light pairs in $xy$ plane but not in $z$ direction. This lattice is of interest because of potential connection to cuprate superconductivity. Light pairs in this lattice are yet to be studied. Inclusion of second-neighbor hopping induces light pairs in other lattices~\cite{Alexandrov2002B,Alexandrov2002C,Hague2007}.   

The second situation leading to light pairs arises when two or more lattice sites have equal attractive potentials. One such model was considered in Section~\ref{twopart:sec:fourthree}. If the on-site interaction is attractive rather than repulsive and $U = V$, then a first-order resonant motion is possible as illustrated in Fig.~\ref{twopart:fig:sixzero}(b). The analysis is readily generalized to other lattices that do not support light pairs for geometric reasons, see Sections~\ref{twopart:sec:appftwo} and \ref{twopart:sec:appfthree}. The pair effective mass is $m^{\ast}_{p} < 2 \sqrt{z} \, m_0$, where $z$ is the number of nearest neighbors in a lattice. Such a potential is unlikely to arise in a solid crystal but can be engineered in cold gases where $U$ and $V$ are controlled independently~\cite{Joerdens2008,Strohmaier2007,Hague2012,Lahaye2009}. Another class of potentials involves equal-strength attractions at finite separation between particles. We considered a 1D example in Section~\ref{twopart:sec:fourthree}, see Fig.~\ref{twopart:fig:sixzero}(c), and a 2D example in Section~\ref{twopart:sec:fivethree}. In both cases, $m^{\ast}_{p} < 4 \, m_0$ was found. One might think that such conditions require fine-tuning and is therefore unlikely in real systems. However, as was argued in Section~\ref{twopart:sec:fivethree}, realistic potentials are combinations of decaying repulsions and long-range attractions. They necessarily have broad and shallow attractive minima. In those situations, the likelihood of two sites near the minimum to have equal attractive strengths is quite high. More instances of this mechanism are investigated in Ref.~\cite{Adebanjo2022b}.     

An overall conclusion of this analysis is that in many circumstances, binding does not lead to significant increase of pair mass. Pair mass is not an impediment for a high mobility or a high BEC temperature.

\subsubsection{\label{twopart:sec:twelveonefour}
Anisotropic 3D pairs
}

The role played by lattice anisotropy is important in the context of cuprate superconductivity. All cuprate superconductors have a layered structure which has led to a popular point of view that the superconductivity is essentially two-dimensional. A number of pure 2D theoretical models have been proposed and extensively studied. In the preformed pair mechanism, however, superconductivity is still considered three-dimensional albeit highly anisotropic. Investigations conducted in the present work support this picture. We analyzed two 3D anisotropic models: tetragonal attractive Hubbard model in Section~\ref{twopart:sec:threesix} and tetragonal $UV$ model in Section~\ref{twopart:sec:eight}. We found in that in both cases the pair formation line splits in two regions, see Figs.~\ref{twopart:fig:four}(a) and \ref{twopart:fig:twelve}. At extremely high anisotropy, $t_z < 0.002\,t$, pair formation is indeed dominated by logarithmic divergencies of 2D integrals. By that measure, the pairs may indeed be regarded as pure two-dimensional. However, when $t_z > 0.002\,t$ the divergencies are no longer dominant and the threshold is on the order of its value in an isotropic 3D lattice. For that reason, real-space pairs in this regime are already three-dimensional. Thus, two-body solutions may help to clarify the nature of superconductivity depending on the level of anisotropy observed in a system.

\subsubsection{\label{twopart:sec:twelveonefive}
Stability of fermion pairs against phase separation
}

Phase separation is a ubiquitous feature of all models with finite-range attraction such as $UV$ or {\it t-J} mo\-dels. Physical reasoning suggests that any system with a strong enough attraction will form clusters and eventually phase-separate. It raises the question about stability of real-space pairs in many-body systems. A positive answer is provided by the Fermi statistics~\cite{Kornilovitch2022,Emin1994,Kornilovitch2013,Kornilovitch2014,Kornilovitch2020}. The ground state of a fermion pair is usually a spin singlet with a nodeless wave function. When two pairs attempt to coalesce into a quad, the full wave function must develop nodes. It is equivalent to a short-range {\em repulsion} between pairs. This additional repulsion keeps the pairs separate as long as attraction is not too strong. (Since this reasoning does not apply to bosons, Bose-$UV$ models are {\em not} stable against phase separation~\cite{Kornilovitch2022}.)     

The derived conclusion is very general and applicable to a wide class of inter-particle potentials including three-dimensional and long-range ones. It has been confirmed by exact numerical calculations for {\it t-J} models~\cite{Emery1990,Dagotto1993}, $UV$ models~\cite{Kornilovitch2022,Kornilovitch2013,Kornilovitch2014,Kornilovitch2020}, and electron-phonon models~\cite{Chakraborty2014}. Real-space pairs retain their identity in a {\em finite} interval of attraction above the binding threshold.

\subsection{\label{twopart:sec:twelvetwo}
Relevance to cuprate superconductivity    
}

Real-space pairs are known to adequately describe many aspects of high-$T_c$ phenomenology. This topic is not discussed here and the reader is referred to available reviews~\cite{Alexandrov1994,Alexandrov1999b,Alexandrov2013,Micnas1990,Alexandrov1993b}. We also note the recent observation of pairs in copper oxides~\cite{Zhou2019} and iron-based superconductors~\cite{Seo2019,Kang2020}. 

In this section, we discuss insights into cuprate superconductivity provided by two-body problems. The first one concerns the role of anisotropy. It was shown in Sec.~\ref{twopart:sec:eleven} and Ref.~\cite{Kornilovitch2015} that the preformed pair mechanism naturally leads to the notion of {\em optimal anisotropy}. If $t_z$ is large, then kinetic energy is large, attraction is below the threshold, and pairs do not form. Even if attraction is above the threshold, the binding is weak, pairs are large, close-packed density is small and ${\cal T}^{\ast}_{\rm BEC}$ is small as a result. In the opposite limit of very small $t_z$, pairs are well formed and are compact but the out-of-plane mass is very large, which also reduces ${\cal T}^{\ast}_{\rm BEC}$. Thus, there {\em must} be an optimal $t_z$. Importantly, this argument does not exclude the existence of fully isotropic 3D superconductors like Ba$_{1-x}$K$_{x}$BiO$_3$. Rather, it means that any isotropic superconductivity with low carrier density can be {\em improved upon} by reducing coupling between layers. That will result in more compact pairs and a higher ${\cal T}^{\ast}_{\rm BEC}$. We also note, in passing, that the level of anisotropy can be dramatically altered by electron-phonon interaction~\cite{Kornilovitch1999} or similar mediators.     

The above argument applies equally to quasi-1D superconductivity where the crystal is composed of weakly bound 1D chains. In this respect, the recent claim of a $T_c = 90$ K superconductivity in a quasi-1D copper oxide is particularly noteworthy~\cite{Rajak2023}. 

Similar to anisotropy, there is an optimal strength of attraction. A weak attraction cannot form pairs but a strong attraction leads to clustering and phase separation. Aggregating available results on the tetragonal $UV$ model~\cite{Kornilovitch2015,Kornilovitch2020}, one finds that a ``working'' interval of $|V|$ is about $0.5\,t$. The interval is actually $U$-dependent, see Fig.~\ref{twopart:fig:thirteen}(b). It grows larger at small $U$ but shrinks at very large $U > 50\,t$. Then an optimal level of anisotropy~\cite{Kornilovitch2020} is $t_z < 0.1\,t$. Let us assume $t = 0.1$ eV as a ballpark estimate of the in-plane hopping in physical units. (A  corresponding total bandwidth is $\approx 1$ eV.) Then the working interval of attraction is $\triangle V \approx 0.05$ eV and the optimal interlayer hopping is $t_z < 0.01$ eV. Thus, the levels of attraction and anisotropy must both fall within narrow intervals to be close to optimal. {\em This may help explain the experimental fact that high-$T_c$ superconductivity is a rare phenomenon only found in a few materials families.}

\begin{figure}[t]
\begin{center}
\includegraphics[width=0.48\textwidth]{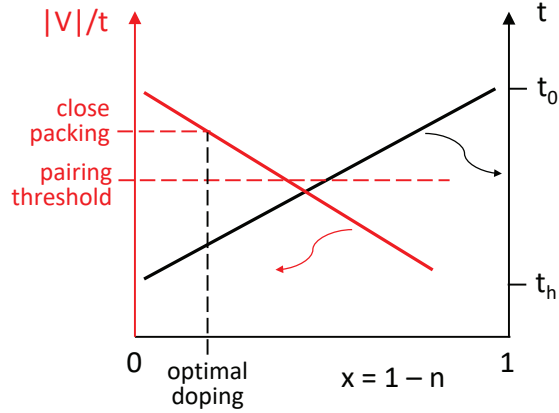}
\end{center}
\caption{Evolution of a $UV$ model with hole doping. The carrier hopping (right axis) increases from $t_{h}$ near half-filling ($x = 0$) to $t_0$ in the empty lattice ($x = 1$). Accordingly, the attraction strength {\em expressed in units of\/} $t$ (left axis) decreases from above threshold to below threshold.} 
\label{twopart:fig:twentysix}
\end{figure}

Further insights can be gained by examining Fig.~\ref{twopart:fig:twentythree}. It clearly shows that the {\em peak} ${\cal T}^{\ast}_{\rm BEC}$ increases with $|V|$. It is beneficial for superconductivity to increase $|V|$ all the way until phase separation. One concludes that {\em highest critical temperatures are expected in systems close to phase separation}. Since the system is close to phase separation globally, it may go over the threshold {\em locally} due to disorder, lattice defects, Jahn--Teller distortions~\cite{Kabanov2002,Mertelj2005}, and other factors. This may help explain proliferation of charge density waves, nematic orders, and other charge instabilities experimentally observed in the cuprates~\cite{Tranquada1995,Chang2012,Ghiringhelli2012,Comin2015,Miao2021}.  

Finally, we discuss the role played by strong electron correlations represented by Hubbard repulsion $U$. Consider the square lattice with bare electron hopping $t_0$. Near half-filling, the effective hopping of {\em holes} is reduced by strong correlations by a factor of $\sim 4 - 10$~\cite{Dagotto1994,Leung1995}. As a consequence, a relatively weak attraction $|V|$ that was unable to bind fast electrons moving with hopping $t_0$ is now able to bind slow holes moving with a smaller effective hopping $t_{h}$. {\em The role of strong in-plane correlation is to further slow down the carriers to enable pair formation by a weak attraction.} In a recent analysis of thermodynamic measurements of several cuprate families, Harrison and Chan argued~\cite{Harrison2023} that the in-plane hopping near half-filling is about $t_{h} = 0.11$ eV while it should be 0.36 eV according to electronic structure calculations. Such a band narrowing is due to electron correlations. Since the band structure value should be accurate in an empty lattice, one concludes that the effective $t$ increases from 0.11 to 0.36 eV with doping. At the same time, quantum chemistry calculations~\cite{Zhang1991,Catlow1998} suggest an attractive potential of about 0.12 eV. Thus, $V/t \approx 1$ near half-filling. This may be sufficient for binding in some highly anisotropic $UV$ models. 

Consider evolution of the system with hole doping starting at half-filling illustrated in Fig.~\ref{twopart:fig:twentysix}. By continuity, the effective carrier hopping must smoothly increase from $t_{h}$ at $x = 0$ to $t_0$ at $x = 1$. Since exact shape of this curve is unknown, we show it schematically as a straight line. We then assume that the attraction strength {\em in physical units} stays approximately constant with doping. Then, $|V|$ will be systematically {\em decreasing} when expressed in units of $t$. If $|V|$ was above a pairing threshold at $x = 0$, it will end up below the threshold at a finite $x$. At this doping, the pairs disappear. At some intermediate doping, the pairs balloon in size to reach close-packing. At this {\em optimal doping}, the critical temperature is maximal. {\em The preformed-pair mechanism naturally explains the existence of optimal doping and eventual disappearance of superconductivity with increasing hole density.} A recent analysis of the magnetic susceptibility and Knight shift in cuprates~\cite{Noat2022} suggests that the pseudogap decreases linearly with doping and coexists with superconductivity until both disappear in the overdoped regime. This conclusion supports our qualitative scenario. Returning to the specific numbers reported in Ref.~\cite{Harrison2023}, let us assume the threshold to be $|V_{\rm cr}| = 0.6 \, t$. When $t$ increases to $\approx 0.20$ eV with doping, the pairs would evaporate. Thus, superconductivity should disappear by $x = 0.36$.

\subsection{\label{twopart:sec:twelvethree}
Future outlook    
}

Several additional two-body problems seem to be of interest. One of them is the body-centered tetragonal (BCT) $UV$ model with out-of-plane attraction. The BCT lattice supports in-plane light pairs (but not out-of-plane ones), which should boost $T^{\ast}_{\rm BEC}$. The negative effects of $U$ on pairing can be minimized by a small $t_z$. Compared with the simple tetragonal lattice, BCT has four times as many attractive bonds, which should reduce the threshold values by about a factor of four. Finally, BCT is close to the crystal structure of some cuprate superconductors.     

It would also be interesting to analyze two-body problems with effective single-particle dispersions arising from strong correlations. For example, holes in nearly half-filled {\it t-J} models acquire the tendency to hop diagonally to second nearest neighbors~\cite{Dagotto1994,Leung1995}. According to Bak and Micnas~\cite{Bak1999}, such a dispersion may result in a $d$-symmetric ground state. Thus, such models may shed light on the symmetry of superconducting order parameter in cuprates.     

Also of interest are multi-orbital models describing CuO$_2$ planes and even full CuO$_6$ octahedra. Only one such investigation has been published so far~\cite{Alexandrov1993}. Extension of this work to 3D and accommodation of the available quantum-mechanical calculations of $t$, $U$, and $V$~\cite{Pickett1989,Zhang1991,Catlow1998,Ramadhan2022}, would make the theory less phenomenological.    

Finally, we mention the need to investigate a four-fermion problem in the tetragonal $UV$ model to complete the picture of phase separation in 3D.

\section*{Declaration of competing interest}

The author declares to not have known competing financial interests or personal relationships that could have appeared to influence the work reported in this paper.

\section*{Data availability}

No data was used for the research described in the article.

\section*{Acknowledgements}

The author wishes to thank James Hague for long-term collaboration and numerous discussions on the subject of this paper. The author also wishes to thank Alexander Chernyshev, Frank Marsiglio, and the anonymous referee for useful comments. This research did not receive any specific grant from funding agencies in the public, commercial, or not-for-profit sectors.

%\end{twocolumn}

%\begin{onecolumn}

\appendix

\section{ \label{twopart:sec:appj}
Omissions and extensions  
}

\subsection{\label{twopart:sec:nine}
Miscellaneous two particle problems  
}

In this section, we briefly describe other two-particle problems that have been left out of the present review. 

Early results on bound pairs in 1D, 2D square, and 3D simple cubic $UV$ models were derived by Micnas and reported in Sec.~III.C of Ref.~\cite{Micnas1990}.  

Two-particle bound states in the square {\it t-J} model and {\it t-J-U} models, which map to the $UV$ model, were studied by Lin~\cite{Lin1991}, Petukhov, Gal\'an, and Verg\'es~\cite{Petukhov1992}, and Kagan and Rice~\cite{Kagan1994}. All those works are consistent with the material of Section~\ref{twopart:sec:five}. 

Bak and Micnas~\cite{Bak1999,Bak2012} considered a square $UV$ model with second nearest-neighbor {\em hopping}. They found that the latter could change the symmetry of the ground state. The $d$-symmetric pair had a lower energy than the $s$-symmetric one when the first-neighbor and second-neighbor hoppings were of opposite signs. Since effective second-neighbor hopping may be induced by strong correlations on an antiferromagnetic background near half-filling, this result may have implications for understanding symmetries of the superconducting order parameter in cuprates.    

Marsiglio and Hirsch~\cite{Marsiglio1990} developed and solved a two-body problem with {\em correlated hopping}, in which hopping of one particle depended on nearby presence of another particle.  

In Ref.~\cite{Hague2020}, a square $UV$ model with {\em anisotropic} next-nearest-neighbor attraction was studied in the context of cold atom quantum simulator of extended Hubbard--Holstein models.  

A two-magnon problem with complicated second-neighbor hopping and interaction was solved in Ref.~\cite{Jiang2022b}.  

A 3D $UV$ model on the BCC lattice was considered in Ref.~\cite{Adebanjo2021}. Pair mass, radius, and binding conditions were computed. Based on those results, a critical temperature of an atomic condensate in a BCC optical lattice was estimated to be around 10 nK. 
 
In Ref.~\cite{Adebanjo2022}, a $UV$ model on the FCC lattice was analyzed in the context of superconducting fullerides. The pairs were found to be of small radius, strongly-bound, but relatively light due to the light-pair effect. It was estimated that such pairs could Bose-condense at high temperatures even if the lattice constant was large, as in the fullerides.     

The following works studied {\em multi-orbital} models with on-site or intersite attractions. 

A ``mixed'' repulsive-attractive one-dimensional Hubbard model was studied in Ref.~\cite{Alexandrov1992}. The unit cell consisted of two non-equivalent sites. The first site hosted a Hubbard repulsion $U_{r}$ whereas the second site hosted a Hubbard attraction $- \vert U_{a} \vert$. The singlet pairs were found to form when 
\begin{equation}
\vert U_{a} \vert > \frac{2 U_{r} t \sqrt{\cos{P}} }{ U_{r} + 2t \sqrt{\cos{P}} } \: .   
\label{twopart:eq:oneninetyfive}
\end{equation}
Note an unusual, square-root, dependence on the pair momentum $P$. The latter is confined to the interval $- \frac{\pi}{2} \leq P \leq \frac{\pi}{2}$ in this case.   

In Ref.~\cite{Ivanov1994}, a $UV$ model on a checker-board square lattice was studied. The symmetry between two sublattices was assumed to be broken by antiferromagnetic correlations of the underlying many-body system. The correlations created a staggered magnetic field $\Delta$ that elevated energy of a spin-up hole on the first sublattice and lowered it on the second sublattice. For a spin-down hole, sublattice energies were reversed. It was found that a nonzero $\Delta$ favored pairing because of two effects. First, the resulting band splitting reduced kinetic energies of the holes. Second, a nonzero $\Delta$ suppressed double occupancy and reduced the influence of $U$. Both factors effectively increased the attraction $V$ relative to $t$ and $U$, which leads to easier pairing. A more detailed analysis of this two-particle lattice problem can be found in Ref.~\cite{Kornilovitch1997}.   

Two-body bound states in multi-orbital models with {\em flat bands} were studied in Ref.~\cite{Torma2018}.

Finally, we mention the mixed Hubbard model on a multi-orbital ${\rm Cu O}_{2}$ plane~\cite{Alexandrov1993}. The model included a repulsion on copper sites, $U_{\rm C}$, and an attraction on oxygen sites, $- \vert U_{\rm O} \vert$. The atomic energies of Cu and O sites were assumed equal, $\varepsilon_{\rm C} = \varepsilon_{\rm O}$, for both spin orientations. The pairing condition was found to be $\vert U_{\rm O} \vert > 8 \sqrt{2} U_{\rm C} t / ( 3 U_{\rm C} + 4\sqrt{2} t )$, where $t$ is the absolute value of the hoping integral between the copper's $d_{x^2-y^2}$ orbital and oxygen's $p_{x,y}$ orbitals. One should add that multiple orbitals introduced significant technical complications in exact two-particle solutions, as explained in \ref{twopart:sec:twothree}. A similar analysis for the case of {\em oxygen--oxygen} attraction~\cite{Kornilovitch1993} resulted in a threshold condition, $|V_{OO}| > U_{C} t/( 0.595 U_{C} + 2t )$. The latter model in its {\em many-body} version was recently studied by constrained-path Monte Carlo in Ref.~\cite{Huang2021}.

\subsection{\label{twopart:sec:twothree}
Extension to non-Bravais lattices and multi-orbital models 
}

When two fermions interact within a complex lattice with $S > 1$ orbitals per unit cell, an exact solution becomes considerably more complicated. Systematic investigation of such models are beyond the scope of this work. In this section, we briefly outline the procedure and point out how it differs from the basic $S = 1$ case.    

Assuming the index ${\bf m}$ continues to number {\em unit cells}, the two-body wave function $\Psi_{\alpha \beta}({\bf m}_1,{\bf m}_2)$ comprises $S^2$ components arranged in an $( S^2 \times 1 )$ array. The Schr\"odinger equation, Eqs.~(\ref{twopart:eq:four}) and (\ref{twopart:eq:six}), comprises $S^2$ coupled equations. The right-hand-side of Eq.~(\ref{twopart:eq:six}) still contains a {\em finite} number of integrals $\Phi({\bf P})$ which,  similarly to Eq.~(\ref{twopart:eq:ten}), allows expressing $\psi_{{\bf k}_1 \alpha , {\bf k}_2 \beta}$ as a linear combination of $\Phi({\bf P})$. However, the respective energy-dependent coefficients are now components of an inverted $( S^2 \times S^2 )$ matrix. Each coefficient is a ratio of an $( S^2 - 1 )$-degree polynomial of $E$ and an $S^2$-degree polynomial of $E$. The denominator can be factorized into a product of $S^2$ factors $( E - \varepsilon_{\alpha {\bf k}_1} - \varepsilon_{\beta {\bf k}_2 } )$, where $\varepsilon_{\alpha {\bf k}}$ are the $S$ bands of the single-particle dispersion. When $\varepsilon_{\alpha {\bf k}}$ is known analytically, factorization can also be performed analytically, at least in principle. Then each coefficient can be expanded into a sum of $S^2$ simple fractions. Finally, substitution of $\psi_{{\bf k}_1 \alpha , {\bf k}_2 \beta}$ into the definitions of $\Phi({\bf P})$ produces a finite set of algebraic equations for $\Phi({\bf P})$ but with matrix elements $M$ being sums of $S^2$ terms where each term has the general form of Eq.~(\ref{twopart:eq:twelve}) with denominators $( - E + \varepsilon_{\alpha {\bf q}} + \varepsilon_{\beta, {\bf P} - {\bf q} } )$ and additional functions of ${\bf q}$ in numerators.    

In practice, this program can be completed for the simplest multiorbital models only~\cite{Ivanov1994,Alexandrov1992,Alexandrov1993,Kornilovitch1997}. Often enough, $\varepsilon_{\alpha {\bf k}}$ cannot be calculated analytically, and the above procedure pauses after the matrix inversion but before factorization. From that stage, everything must be computed numerically.

\section{ \label{twopart:sec:appa}
Green's functions of the 2D square and rectangular lattices  
}

\subsection{ \label{twopart:sec:appaone}
Definitions  
}

In this section, we will be concerned with analytical evaluation of the two dimensional integrals
\begin{equation}
M^{\rm rt}_{nm} ( E ; \alpha , \beta ) = 
\frac{1}{N} \sum_{\bf q} \frac{ \cos{nq_x} \cos{mq_y} }{ \vert E \vert - \alpha \cos{q_x} - \beta \cos{q_y} } =
\int\limits^{\pi}_{-\pi} \!\! \int\limits^{\pi}_{-\pi} \frac{ {\rm d}q_x \, {\rm d}q_y }{(2\pi)^2} 
\frac{ \cos{nq_x} \cos{mq_y} }{ \vert E \vert - \alpha \cos{q_x} - \beta \cos{q_y} } \: .
\label{twopart:eq:appaone}
\end{equation}
Here $\alpha \equiv 4 t_{x} \cos{\frac{P_x}{2}} \geq 0$ and $\beta \equiv 4 t_{y} \cos{\frac{P_y}{2}} \geq 0$. $t_{x,y} > 0$ are absolute values of hopping integrals along $x$ and $y$ axes, respectively. $P_x$ and $P_y$ are two components of a total pair quasi-momentum. Since $-\pi \leq P_x , P_y \leq +\pi$, both $\alpha$ and $\beta$ are non-negative. The quantities $M^{\rm rt}_{nm} ( E ; \alpha , \beta )$ are {\em Green's functions} of the rectangular lattice. However, we prefer not to use the term ``Green's function'' in the main text of this study because our definition, Eq.~(\ref{twopart:eq:appaone}), differs from the standard one by a constant negative factor. The superscript ``rt'' indicates that these $M_{nm}$ belong to the rectangular lattice. When $\alpha = \beta$, a superscript ``sq'' will be used, indicating Green's functions of the square lattice. The rectangular functions, Eq.~(\ref{twopart:eq:appaone}), appear in the {\em square} two-body problem because the center-of-mass motion breaks the {\it x-y} symmetry of the underlying lattice. We will be interested only in the discrete energy spectrum with $E \leq - ( \alpha + \beta )$. For those energies, all $M^{\rm rt}_{nm}$ are real, positive, and logarithmically diverge as $E \rightarrow - ( \alpha + \beta )$ from below. Since $M^{\rm rt}_{-n,m} = M^{\rm rt}_{nm}$ and $M^{\rm rt}_{n,-m} = M^{\rm rt}_{nm}$, only $M^{\rm rt}_{nm}$ with non-negative $n$ and $m$ need to be evaluated.      

All $M^{\rm rt}_{nm}$ are eventually expressible via the three complete elliptic integrals ${\bf K}$, ${\bf E}$, and ${\bf \Pi}$. In this paper, the following definitions are used
\begin{equation}
{\bf K}(\kappa) = \int\limits^{\pi/2}_{0}  
\frac{{\rm d} \phi}{\sqrt{ 1 - \kappa^2 \sin^2{\! \phi} }} \: ;
\hspace{0.3cm}
{\bf E}(\kappa) = \int\limits^{\pi/2}_{0} \sqrt{ 1 - \kappa^2 \sin^2{\! \phi}} \: {\rm d} \phi \: ;
\hspace{0.3cm}
{\bf \Pi}(n , \kappa) = \int\limits^{\pi/2}_{0} 
\frac{{\rm d} \phi}{( 1 - n \sin^2{\! \phi} ) \sqrt{ 1 - \kappa^2 \sin^2{\! \phi} }} \: .
\label{twopart:eq:appatwo}
\end{equation}
Expressions of $M^{\rm rt}_{nm}$ via the generalized hypergeometric function can be found in Ref.~\cite{Katsura1971b}. Rectangular Green's functions with second-neighbor hopping were evaluated in Ref.~\cite{Bak2012}.

\subsection{ \label{twopart:sec:appatwo}
Basic integrals for the square case, $\alpha = \beta$
}

We begin with calculating the basic integral $M^{\rm sq}_{00}$ for the isotropic case, $\alpha = \beta$. First integration utilizes the residue theorem with the result 
\begin{equation}
M^{\rm sq}_{00}( E ; \alpha , \alpha ) = \int^{\pi}_{0} \frac{{\rm d}q_x}{\pi} 
\frac{1}{ \sqrt{ ( \vert E \vert + \alpha - \alpha \cos{q_x} ) 
                 ( \vert E \vert - \alpha - \alpha \cos{q_x} ) } } 
= \frac{1}{\pi\alpha}  \int^{1}_{-1} 
\frac{{\rm d}z}{ \sqrt{ ( 1 - z^2 )( z_1 - z )( z_2 - z ) } } \: ,     
\label{twopart:eq:appafive}
\end{equation}
where $z_{1} = ( \vert E \vert + \alpha )/\alpha$ and $z_{2} = ( \vert E \vert - \alpha )/\alpha$. Then two successive transformations of the integration variable 
\begin{align}
z & =  -1 + \frac{1}{u} \: ,   
\label{twopart:eq:appasix} \\
u & =  \frac{1}{ z_1 + 1 } + \frac{ \frac{1}{2} - \frac{1}{z_1 + 1} }{\sin^2{\phi}} \: ,   
\label{twopart:eq:appaseven}
\end{align}
result in 
\begin{equation}
M^{\rm sq}_{00}( E \leq - 2 \alpha ; \alpha , \alpha ) = \frac{2}{\pi \vert E \vert} \: 
{\bf K} \! \left( \frac{ 2 \alpha }{ \vert E \vert } \right) .
\label{twopart:eq:appaeight}
\end{equation}

To calculate $M^{\rm sq}_{10}$, we first compose the sum rule
\begin{equation}
M^{\rm sq}_{n+1,m}( E ; \alpha , \alpha ) + M^{\rm sq}_{n-1,m}( E ; \alpha , \alpha ) + 
M^{\rm sq}_{n,m-1}( E ; \alpha , \alpha ) + M^{\rm sq}_{n,m+1}( E ; \alpha , \alpha ) = 
- \frac{2}{\alpha} \: \delta_{n0} \delta_{m0} + 
\frac{2 \vert E \vert}{\alpha} \: M^{\rm sq}_{nm}( E ; \alpha , \alpha ) \: , 
\label{twopart:eq:appanine}
\end{equation}
which can be verified by direct substitution. For $n = m = 0$ and $\alpha = \beta$, all four terms in the left-hand-side are equal by symmetry, which leads to 
\begin{equation}
M^{\rm sq}_{10}( E \leq - 2 \alpha ; \alpha , \alpha ) = 
\frac{\vert E \vert}{2 \alpha} \: M^{\rm sq}_{00} - \frac{1}{2 \alpha} =  
\frac{1}{\pi \alpha} \: {\bf K} \! \left( \frac{ 2 \alpha }{ \vert E \vert } \right) -
\frac{1}{2 \alpha} \: .
\label{twopart:eq:appaten}
\end{equation}

Calculation of the next basic integral, $M^{\rm sq}_{20}$, is more involved. Since the following derivation will also be needed in the anisotropic case, $\alpha \neq \beta$, we present here its general version. After integration over $q_y$, which is elementary, 
\begin{equation}
M^{\rm sq}_{20}( E ; \alpha , \beta ) = \int^{\pi}_{0} \frac{{\rm d}q_x}{\pi} 
\frac{ 2 \cos^{2}{\!q_x} - 1 }
{ \sqrt{ ( \vert E \vert + \beta - \alpha \cos{q_x} ) 
         ( \vert E \vert - \beta - \alpha \cos{q_x} ) } } \: .     
\label{twopart:eq:appaeleven}
\end{equation}
The second term here is $M^{\rm sq}_{00}$. It will be moved, temporarily, to the left side of the equation. After the substitution $\cos{q_x} = z$,
\begin{equation}
M^{\rm sq}_{20}(E;\alpha,\beta) + M^{\rm sq}_{00}(E;\alpha,\beta) = 
\frac{2}{\pi\alpha}  \int^{1}_{-1} 
\frac{z^2 \: {\rm d}z}{ \sqrt{ ( 1 - z^2 )( z_1 - z )( z_2 - z ) } } \: ,     
\label{twopart:eq:appatwelve}
\end{equation}
where $z_{1} = ( \vert E \vert + \beta )/\alpha$ and $z_{2} = ( \vert E \vert - \beta )/\alpha$. Next we use substitution $z = z_1 + \frac{1}{u}$ to convert the quartic polynomial under the square root into a cubic polynomial. The result is
\begin{equation}
M^{\rm sq}_{20}(E;\alpha,\beta) + M^{\rm sq}_{00}(E;\alpha,\beta) = 
\frac{2}{\pi\alpha} \sqrt{ a b c } \: 
\left\{ \left( \frac{\vert E \vert + \beta}{\alpha} \right)^2 \!\! J_{0} 
- 2 \frac{\vert E \vert + \beta}{\alpha} \: J_{-1} + J_{-2} \right\} ,     
\label{twopart:eq:appathirteen}
\end{equation}
where
\begin{equation}
a \equiv \frac{\alpha}{2 \beta}                        \: ,  \hspace{1.0cm} 
b \equiv \frac{\alpha}{\vert E \vert + \beta - \alpha} \: ,  \hspace{1.0cm}    
c \equiv \frac{\alpha}{\vert E \vert + \beta + \alpha} \: ,  \hspace{1.0cm}    
a > b > c \: ,
\label{twopart:eq:appafourteen}
\end{equation}
\begin{equation}
J_{m} = \int^{b}_{c} 
\frac{u^m \: {\rm d}u}{ \sqrt{ ( a - u )( b - u )( u - c ) } } \: .     
\label{twopart:eq:appafifteen}
\end{equation}
According to the general theory of elliptic integrals~\cite{Hancock1910}, $J_{-2}$ can expressed as a linear combination of three fundamental integrals $J_{-1}$, $J_{0}$ and $J_{1}$. To this end, call the polynomial under the square root $Q(u)$, calculate the derivative $\frac{d}{du}[ \frac{1}{u} \sqrt{Q(u)} ]$ and integrate the result between $u = c$ and $u = b$. This leads to
\begin{equation}
J_{-2} = \frac{1}{ 2 a b c } \: 
\left\{ \left( ab + ac + bc \right) \! J_{-1} - J_{1} \right\} .     
\label{twopart:eq:appasixteen}
\end{equation}
Substitution of Eq.~(\ref{twopart:eq:appasixteen}) in Eq.~(\ref{twopart:eq:appathirteen}) yields 
\begin{equation}
M^{\rm sq}_{20}(E;\alpha,\beta) + M^{\rm sq}_{00}(E;\alpha,\beta) = 
\frac{2}{\pi\alpha} \sqrt{ a b c } \: 
\left\{ \left( \frac{\vert E \vert + \beta}{\alpha} \right)^2 \!\! J_{0} 
- \frac{\vert E \vert}{\alpha} \: J_{-1} - \frac{1}{2abc} \:J_{1} \right\} .     
\label{twopart:eq:appaseventeen}
\end{equation}
As the next step, we use the substitution 
\begin{equation}
u = c + ( b - c ) \sin^2{\!\phi} \: ,     
\label{twopart:eq:appaeighteen}
\end{equation}
to transform $J_{m}$'s into Legendre normal forms, Eq.~(\ref{twopart:eq:appatwo}). The results are
\begin{align}
J_{0}(\alpha,\beta) & =  \frac{2}{ \sqrt{ a - c } } \: 
{\bf K} \! \left( \sqrt{ \frac{ b - c }{ a - c }} \right) ,   
\label{twopart:eq:appanineteen} \\
J_{1}(\alpha,\beta) & =  \frac{2}{ \sqrt{ a - c } } 
\left\{ a \: {\bf K} \! \left( \sqrt{ \frac{ b - c }{ a - c }} \right) - 
( a - c ) \: {\bf E} \! \left( \sqrt{ \frac{ b - c }{ a - c }} \right) \right\} ,   
\label{twopart:eq:appatwenty}   \\
J_{-1}(\alpha,\beta) & =  \frac{2}{ c \sqrt{ a - c } } \: 
{\bf \Pi} \! \left( - \frac{b-c}{c} , \sqrt{ \frac{ b - c }{ a - c }} \right) .   
\label{twopart:eq:appatwentyone} 
\end{align}
Note that the first argument in ${\bf \Pi}$ is negative. At this point, we return to the isotropic case, $\alpha = \beta$. Using $a$, $b$, $c$ from Eq.~(\ref{twopart:eq:appafourteen}), one obtains
\begin{align}
J_{0}(\alpha,\alpha) & =  2 \sqrt{2}  
\sqrt{\frac{ \vert E \vert + 2\alpha }{ \vert E \vert }} \: 
{\bf K} \! \left( \frac{ 2 \alpha }{ \vert E \vert } \right) ,   
\label{twopart:eq:appatwentytwo} \\
J_{1}(\alpha,\alpha) & =  \sqrt{2}  
\sqrt{\frac{ \vert E \vert + 2\alpha }{ \vert E \vert }} \:  
\left\{ {\bf K} \! \left( \frac{ 2 \alpha }{ \vert E \vert } \right) - 
\frac{ \vert E \vert }{ \vert E \vert + 2 \alpha } \: 
{\bf E} \! \left( \frac{ 2 \alpha }{ \vert E \vert } \right) \right\} ,   
\label{twopart:eq:appatwentythree}   \\
J_{-1}(\alpha,\alpha) & =  
\frac{2 \sqrt{2} \: ( \vert E \vert + 2 \alpha )^{3/2}}{ \alpha \sqrt{ \vert E \vert } } \: 
{\bf \Pi} \! \left( - \frac{ 2 \alpha }{ \vert E \vert } , \frac{ 2 \alpha }{ \vert E \vert } \right) .   
\label{twopart:eq:appatwentyfour} 
\end{align}
Notice that the expression for $J_{-1}(\alpha,\alpha)$ involves ${\bf \Pi}$ with their two arguments being equal by absolute value. In this special case, ${\bf \Pi}$ can be reduced to ${\bf K}$. According to \cite{Abramowitz1972}, \#17.7.22,   
\begin{equation}
2 {\bf \Pi}(-\kappa,\kappa) - {\bf K}(\kappa) = \frac{\pi}{2} \frac{1}{ 1 + \kappa } \: ,     
\label{twopart:eq:appatwentyfive}
\end{equation}
which is valid for any $0 \leq \kappa < 1$. We can prove Eq.~(\ref{twopart:eq:appatwentyfive}) by the following procedure. Go back to $M^{\rm sq}_{10}(E;\alpha,\beta)$ and transform it in the same manner as $M^{\rm sq}_{20}$ starting with Eq.~(\ref{twopart:eq:appaeleven}). Instead of Eq.~(\ref{twopart:eq:appaseventeen}), we have 
\begin{equation}
M^{\rm sq}_{10}(E;\alpha,\beta) = 
\frac{1}{\pi\alpha} \sqrt{ a b c } \: 
\left\{ \frac{ \vert E \vert + \beta }{\alpha} \: J_{0} - J_{-1} \right\} .     
\label{twopart:eq:appatwentysix}
\end{equation}
The isotropic version reads
\begin{equation}
M^{\rm sq}_{10}(E;\alpha,\alpha) = 
\frac{2}{ \pi \alpha \vert E \vert} \: \left\{ \left( \vert E \vert + \alpha \right) 
{\bf K} \! \left( \frac{ 2 \alpha }{ \vert E \vert } \right) - 
\left( \vert E \vert + 2 \alpha \right)
{\bf \Pi} \! \left( - \frac{2\alpha}{\vert E \vert} , \frac{2\alpha}{\vert E \vert} \right) \right\} .     
\label{twopart:eq:appatwentyseven}
\end{equation}
Thus, $M^{\rm sq}_{10}$ also contains a special case of ${\bf \Pi}(-\kappa,\kappa)$. However, we previously derived an expression for $M^{\rm sq}_{10}$ that includes ${\bf K}(\kappa)$ only. By comparing Eq.~(\ref{twopart:eq:appaten}) and Eq.~(\ref{twopart:eq:appatwentyseven}), one obtains 
\begin{equation}
{\bf \Pi} \! \left( - \frac{2\alpha}{\vert E \vert} , \frac{2\alpha}{\vert E \vert} \right) = 
\frac{1}{2} \: {\bf K} \! \left( \frac{2\alpha}{\vert E \vert} \right) + 
\frac{\pi}{4} \frac{\vert E \vert}{ \vert E \vert + 2 \alpha} \: ,     
\label{twopart:eq:appatwentyeight}
\end{equation}
which is Eq.~(\ref{twopart:eq:appatwentyfive}) for $\kappa = \frac{2\alpha}{\vert E \vert}$. Finally, by substituting Eq.~(\ref{twopart:eq:appatwentyeight}) in Eq.~(\ref{twopart:eq:appatwentyfour}), then expressions for $J_{m}(\alpha,\alpha)$, $a$, $b$ and $c$ in Eq.~(\ref{twopart:eq:appaseventeen}), and utilizing the previously derived formula for $M^{\rm sq}_{00}(E,\alpha,\alpha)$, one arrives at the deceptively simple result:
\begin{equation}
M^{\rm sq}_{20}( E \leq - 2 \alpha ; \alpha , \alpha ) = 
\frac{2}{\pi \vert E \vert} \: {\bf K} \! \left( \frac{ 2 \alpha }{ \vert E \vert } \right) +
\frac{\vert E \vert}{ \alpha^2 } \left\{ 
\frac{2}{\pi} \: {\bf E} \! \left( \frac{ 2 \alpha }{ \vert E \vert } \right) - 1 \right\} .
\label{twopart:eq:appatwentynine}
\end{equation}

\subsection{ \label{twopart:sec:appathree}
Recurrence relations for all $M_{nm}$ in the square case, $\alpha = \beta$
}

The sum rule, Eq.~(\ref{twopart:eq:appanine}), and basic integrals $M^{\rm sq}_{00}$, $M^{\rm sq}_{10}$ and $M^{\rm sq}_{20}$, derived in the previous section enable calculation of several more $M^{\rm sq}_{mn}$. Specifically, by setting $(n,m) = (1,0)$, one can obtain $M^{\rm sq}_{11}$. Then, by setting $(n,m) = (1,1)$, one can obtain $M^{\rm sq}_{21} = M^{\rm sq}_{12}$. Finally, by setting $(n,m) = (2,0)$, one can derive $M^{\rm sq}_{30}$. However, after that progress stops. Further extension of this method is impossible. Fortunately, there is a second, remarkable identity due to Morita~\cite{Morita1975}, that involves only $M_{n0}$'s and for our case reads
\begin{equation}
2n \! \left( 2 \vert E \vert^2 - \alpha^2 \right) \! M^{\rm sq}_{n0} - 
2 \alpha \vert E \vert ( 2n + 1 ) M^{\rm sq}_{n+1,0} - 
2 \alpha \vert E \vert ( 2n - 1 ) M^{\rm sq}_{n-1,0} +
\alpha^2 ( n + 1 ) M^{\rm sq}_{n+2,0} +
\alpha^2 ( n - 1 ) M^{\rm sq}_{n-2,0} = 0 \: .
\label{twopart:eq:appathirty}
\end{equation}
This expression is valid for $n \neq 0$. Using expressions for $M^{\rm sq}_{00}$, $M^{\rm sq}_{10}$, $M^{\rm sq}_{20}$ and $M^{\rm sq}_{30}$ derived above, one can validate Eq.~(\ref{twopart:eq:appathirty}) for $n = 1$. Then, by setting $n = 2, 3, \ldots$, one can calculate $M^{\rm sq}_{40}$, $M^{\rm sq}_{50}$, and so on, for any $n$. After that, application of the sum rule, Eq.~(\ref{twopart:eq:appanine}), enables calculation of {\em any} $M^{\rm sq}_{nm}$, at least in principle. Below, we list all $M^{\rm sq}_{nm}$ for $ n + m \leq 4$.   
\begin{align}
M^{\rm sq}_{00}( E \leq - 2\alpha ; \alpha,\alpha)   = & \frac{2}{\pi \vert E \vert} \: 
{\bf K} \! \left( \frac{ 2 \alpha }{ \vert E \vert } \right) ,
\label{twopart:eq:appathirtyone}     \\
M^{\rm sq}_{10}( E \leq - 2\alpha ; \alpha,\alpha)   = 
M^{\rm sq}_{01}( E \leq - 2\alpha ; \alpha,\alpha)   = & 
\frac{1}{\pi \alpha} \: {\bf K} \! \left( \frac{ 2 \alpha }{ \vert E \vert } \right) -
\frac{1}{2 \alpha} \: ,
\label{twopart:eq:appathirtytwo}     \\
M^{\rm sq}_{11}( E \leq - 2\alpha ; \alpha,\alpha)   = & \frac{ \vert E \vert }{ \pi \alpha^2 }
\left\{ \left( 1 - \frac{2\alpha^2}{ \vert E \vert^2 } \right) 
{\bf K} \! \left( \frac{ 2 \alpha }{ \vert E \vert } \right) - 
{\bf E} \! \left( \frac{ 2 \alpha }{ \vert E \vert } \right) \right\} , 
\label{twopart:eq:appathirtythree}   \\
M^{\rm sq}_{20}( E \leq - 2\alpha ; \alpha,\alpha)   = 
M^{\rm sq}_{02}( E \leq - 2\alpha ; \alpha,\alpha)   = & 
\frac{ 2 }{ \pi \vert E \vert } \: 
{\bf K} \! \left( \frac{ 2 \alpha }{ \vert E \vert } \right) + 
\frac{ \vert E \vert }{ \alpha^2 }
\left\{ \frac{2}{\pi} \: {\bf E} \! \left( \frac{ 2 \alpha }{ \vert E \vert } \right) - 1 \right\} , 
\label{twopart:eq:appathirtyfour}    \\
M^{\rm sq}_{21}( E \leq - 2\alpha ; \alpha,\alpha) = 
M^{\rm sq}_{12}( E \leq - 2\alpha ; \alpha,\alpha)   = &
\left\{ \frac{\vert E \vert^2}{\pi \alpha^3} - \frac{3}{\pi \alpha} \right\} 
{\bf K} \! \left( \frac{ 2 \alpha }{ \vert E \vert } \right) - 
\frac{\vert E \vert^2}{\pi \alpha^3} \: 
{\bf E} \! \left( \frac{ 2 \alpha }{ \vert E \vert } \right) +
\frac{1}{2 \alpha} \: ,  
\label{twopart:eq:appathirtyfive}    \\
M^{\rm sq}_{30}( E \leq - 2\alpha ; \alpha,\alpha) = 
M^{\rm sq}_{03}( E \leq - 2\alpha ; \alpha,\alpha)   = &
\left\{ \frac{9}{\pi \alpha} - \frac{2 \vert E \vert^2}{\pi \alpha^3} \right\} 
{\bf K} \! \left( \frac{ 2 \alpha }{ \vert E \vert } \right) + 
\frac{ 6 \vert E \vert^2 }{\pi \alpha^3} \: 
{\bf E} \! \left( \frac{ 2 \alpha }{ \vert E \vert } \right) - 
\frac{ 2 \vert E \vert^2 }{ \alpha^3 } - \frac{1}{2 \alpha} \: ,
\label{twopart:eq:appathirtysix}     \\
M^{\rm sq}_{40}( E \leq - 2\alpha ; \alpha,\alpha)   = 
M^{\rm sq}_{04}( E \leq - 2\alpha ; \alpha,\alpha)   = & 
\left\{ \frac{2}{\pi \vert E \vert } + \frac{80 \vert E \vert}{3 \pi \alpha^2} - 
        \frac{20 \vert E \vert^3 }{3 \pi \alpha^4} \right\} 
{\bf K} \! \left( \frac{ 2 \alpha }{ \vert E \vert } \right) 
\nonumber \\
&  + \left\{ \frac{ 8 \vert E \vert }{ 3 \pi \alpha^2} + 
              \frac{ 44 \vert E \vert^3 }{ 3 \pi \alpha^4 }  \right\}  
{\bf E} \! \left( \frac{ 2 \alpha }{ \vert E \vert } \right) - 
\frac{ 4 \vert E \vert }{ \alpha^2 } - \frac{ 4 \vert E \vert^3 }{ \alpha^4 } \: ,
\label{twopart:eq:appathirtyseven}  \\
M^{\rm sq}_{31}( E \leq - 2\alpha ; \alpha,\alpha)   = 
M^{\rm sq}_{13}( E \leq - 2\alpha ; \alpha,\alpha)   = &
\left\{ - \frac{2}{\pi \vert E \vert } - \frac{13 \vert E \vert}{3 \pi \alpha^2} + 
        \frac{4 \vert E \vert^3 }{3 \pi \alpha^4} \right\} 
{\bf K} \! \left( \frac{ 2 \alpha }{ \vert E \vert } \right) 
\nonumber \\
&  + \left\{ - \frac{ 7 \vert E \vert }{ 3 \pi \alpha^2} - 
                \frac{ 4 \vert E \vert^3 }{ 3 \pi \alpha^4 }  \right\}  
{\bf E} \! \left( \frac{ 2 \alpha }{ \vert E \vert } \right) 
+ \frac{ 2 \vert E \vert }{ \alpha^2 } \: ,
\label{twopart:eq:appathirtyeight}  \\
M^{\rm sq}_{22}( E \leq - 2\alpha ; \alpha,\alpha)   = & 
\left\{ \frac{2}{\pi \vert E \vert } - \frac{8 \vert E \vert}{3 \pi \alpha^2} + 
        \frac{2 \vert E \vert^3 }{3 \pi \alpha^4} \right\} 
{\bf K} \! \left( \frac{ 2 \alpha }{ \vert E \vert } \right) 
\nonumber \\
&  + \left\{ \frac{ 4 \vert E \vert }{ 3 \pi \alpha^2} - 
              \frac{ 2 \vert E \vert^3 }{ 3 \pi \alpha^4 }  \right\}  
{\bf E} \! \left( \frac{ 2 \alpha }{ \vert E \vert } \right) . 
\label{twopart:eq:appathirtynine}
\end{align}

Several comments are in order. (i) $M^{\rm sq}_{nm}(E;\alpha,\alpha)$ describe not only bound $UV$ pairs in the $\Gamma$ point but on the Brillouin zone diagonal as well. This follows from the definitions of $\alpha$ and $\beta$. (ii) All integrals have the analytical structure of $\alpha M^{\rm sq}_{nm} = Q_1(\kappa^{-1}) {\bf K}(\kappa) + Q_2(\kappa^{-1}){\bf E}(\kappa) + Q_3(\kappa^{-1})$, where $Q_{1,2,3}$ are polynomials and $\kappa = \frac{2\alpha}{\vert E \vert}$. The smallest power in all $Q_{i}$ is $\geq -1$ while the highest power grows with $n$ and $m$. The overall complexity of the analytical expressions increases with $n$ and $m$. All $M^{\rm sq}_{nm}$ include a nonzero term $\propto {\bf K}(\kappa)$ that logarithmically diverges when $E \rightarrow - 2\alpha - 0$, as expected from the definition, Eq.~(\ref{twopart:eq:appaone}). However, coefficients by ${\bf K}(\kappa)$ approach a universal limit $\frac{1}{\pi \alpha}$ that is the same for {\em all} $n$ and $m$. As a result, all differences $M^{\rm sq}_{nm} - M^{\rm sq}_{n'm'}$ are finite at $E = -2\alpha$. They are discussed in the next section.

\subsection{ \label{twopart:sec:appafour}
$M^{\rm sq}_{nm} - M^{\rm sq}_{00}$ at threshold. Square case, $\alpha = \beta$
}

As discussed in the main body of this paper, to determine a pairing threshold, energy $E$ must be sent to the lowest energy of two free particles from below. For the isotropic square $UV$ model, it means $E \rightarrow - 2\alpha - 0$. As shown in \ref{twopart:sec:appafour}, all $M^{\rm sq}_{nm}$ logarithmically diverge in that limit and direct determination of the threshold is impossible. The procedure can be regularized by subtracting from all $M^{\rm sq}_{nm}$ a common term that has the same logarithmic divergence. The subtrahend can be any one of $M^{\rm sq}_{nm}$ or any other function that has the same logarithmic divergence. The most obvious choice is $M^{\rm sq}_{00}$. Accordingly, we introduce the following new functions
\begin{equation}
L^{\rm sq}_{nm}(\alpha,\alpha) \equiv 
M^{\rm sq}_{nm}( E = - 2\alpha ; \alpha , \alpha ) - 
M^{\rm sq}_{00}( E = - 2\alpha ; \alpha , \alpha ) \: ,
\label{twopart:eq:appaforty}
\end{equation}
that are finite for all $n$, $m$ and $\alpha$. It can be deduced from the definitions, Eq.~(\ref{twopart:eq:appaone}) and Eq.~(\ref{twopart:eq:appaforty}), that all $L^{\rm sq}_{nm} < 0$.  

There are two principle ways of calculating $L^{\rm sq}_{nm}$. For those $\{ nm \}$ for which full expression $M^{\rm sq}_{nm}$ is known, $L^{\rm sq}_{nm}$ is given by its non-singular part, utilizing ${\bf E}(1) = 1$. But in general, $L^{\rm sq}_{nm}$ can be calculated directly from the definition, Eq.~(\ref{twopart:eq:appaone}), because in this special case the second integration is also elementary. Below, we list $L^{\rm sq}_{nm}$ for $\vert n \vert + \vert m \vert \leq 5$. 
\begin{align}
L^{\rm sq}_{10}(\alpha,\alpha) = L^{\rm sq}_{01}(\alpha,\alpha) & =  
- \frac{1}{2\alpha}                        = - \frac{1}{\alpha} \, 0.5000000000 \, \ldots  \: ,
\label{twopart:eq:appafortyone}    \\
L^{\rm sq}_{20}(\alpha,\alpha) = L^{\rm sq}_{02}(\alpha,\alpha) & =  
- \frac{ 2 \, ( \pi - 2 ) }{ \pi \alpha }  = - \frac{1}{\alpha} \, 0.7267604552 \, \ldots  \: ,
\label{twopart:eq:appafortytwo}    \\
L^{\rm sq}_{11}(\alpha,\alpha) & =  
- \frac{ 2 }{ \pi \alpha }                 = - \frac{1}{\alpha} \, 0.6366197723 \, \ldots  \: ,
\label{twopart:eq:appafortythree}  \\
L^{\rm sq}_{30}(\alpha,\alpha) = L^{\rm sq}_{03}(\alpha,\alpha) & =  
- \frac{ 17 \pi - 48 }{ 2 \pi \alpha }     = - \frac{1}{\alpha} \, 0.8605627315 \, \ldots  \: ,
\label{twopart:eq:appafortyfour}   \\
L^{\rm sq}_{21}(\alpha,\alpha) = L^{\rm sq}_{12}(\alpha,\alpha) & =  
- \frac{ 8 - \pi }{ 2 \pi \alpha }         = - \frac{1}{\alpha} \, 0.7732395447 \, \ldots  \: ,
\label{twopart:eq:appafortyfive}   \\
L^{\rm sq}_{40}(\alpha,\alpha) = L^{\rm sq}_{04}(\alpha,\alpha) & =  
- \frac{ 120 \pi - 368 }{ 3 \pi \alpha }   = - \frac{1}{\alpha} \, 0.9539872947 \, \ldots  \: ,
\label{twopart:eq:appafortysix}    \\
L^{\rm sq}_{31}(\alpha,\alpha) = L^{\rm sq}_{13}(\alpha,\alpha) & =  
- \frac{ 46 - 12 \pi }{ 3 \pi \alpha }     = - \frac{1}{\alpha} \, 0.8807515881 \, \ldots  \: ,
\label{twopart:eq:appafortyseven}  \\
L^{\rm sq}_{22}(\alpha,\alpha) & =  
- \frac{ 8 }{ 3 \pi \alpha  }              = - \frac{1}{\alpha} \, 0.8488263631 \, \ldots  \: ,
\label{twopart:eq:appafortyeight}  \\
L^{\rm sq}_{50}(\alpha,\alpha) = L^{\rm sq}_{05}(\alpha,\alpha) & =  
- \frac{ 1203 \pi - 3760 }{ 6 \pi \alpha } = - \frac{1}{\alpha} \, 1.0258046581 \, \ldots  \: ,
\label{twopart:eq:appafortynine}   \\
L^{\rm sq}_{41}(\alpha,\alpha) = L^{\rm sq}_{14}(\alpha,\alpha) & =  
- \frac{ 160 - 49 \pi }{ 2 \pi \alpha }    = - \frac{1}{\alpha} \, 0.9647908947 \, \ldots  \: ,
\label{twopart:eq:appafifty}       \\
L^{\rm sq}_{32}(\alpha,\alpha) = L^{\rm sq}_{23}(\alpha,\alpha) & =  
- \frac{ 8 + 3 \pi }{ 6 \pi \alpha }       = - \frac{1}{\alpha} \, 0.9244131815 \, \ldots  \: .
\label{twopart:eq:appafiftyone}
\end{align}

\subsection{ \label{twopart:sec:appafive}
Basic integrals in the rectangular case, $\alpha \neq \beta$    
}

Evaluation of basic integrals in the general case, $\alpha \neq \beta$, parallels that of the isotropic case considered in \ref{twopart:sec:appatwo}. First, integration over $q_y$, and two substitutions defined by Eqs.~(\ref{twopart:eq:appasix}) and (\ref{twopart:eq:appaseven}) yield    
\begin{equation}
M^{\rm rt}_{00}( E ; \alpha, \beta ) =   
\frac{2}{\pi \sqrt{ \vert E \vert^2 - ( \alpha - \beta )^2 } } \: {\bf K}(\kappa_0) \: ,
\label{twopart:eq:appafiftytwo}
\end{equation}
where 
\begin{equation}
\kappa_0 \equiv \sqrt{ \frac{ 4 \, \alpha \beta }{ \vert E \vert^2 - ( \alpha - \beta )^2 } } \: .
\label{twopart:eq:appafiftythree}
\end{equation}
In order to derive $M^{\rm rt}_{10}$ and $M^{\rm rt}_{20}$, we apply the transformation sequence, Eqs.~(\ref{twopart:eq:appaeleven})-(\ref{twopart:eq:appatwentyone}). The only additional information needed is a relation between ${\bf \Pi}(-n,\kappa)$ and ${\bf \Pi}(+n,\kappa)$ (\cite{Abramowitz1972}, \# 17.7.17) which for our case reads  
\begin{equation}
{\bf \Pi} \left( - \frac{2\alpha}{ \vert E \vert + \beta - \alpha } , \kappa \right) = 
\frac{ 2 \beta }{ \vert E \vert + \alpha + \beta } \: {\bf K}( \kappa ) + 
\frac{ \vert E \vert - \alpha - \beta }{ \vert E \vert + \alpha + \beta } \: 
{\bf \Pi} \left( \frac{2\alpha}{ \vert E \vert + \alpha - \beta } , \kappa \right) .
\label{twopart:eq:appafiftyfour}
\end{equation}
With that, the results are
\begin{align}
M^{\rm rt}_{10}( E \leq - ( \alpha + \beta ) ; \alpha , \beta ) & =   
\frac{2}{\pi \alpha \sqrt{ \vert E \vert^2 - ( \alpha - \beta )^2 } } 
\left\{  \left( \vert E \vert - \beta \right) {\bf K}(\kappa_0) 
- \left( \vert E \vert - \alpha - \beta \right) {\bf \Pi} \left( n_{10} , \kappa_0 \right) \right\} ,
\label{twopart:eq:appafiftyfive}     \\
M^{\rm rt}_{20}( E \leq - ( \alpha + \beta ) ; \alpha , \beta ) & =   
\frac{2}{\pi \alpha^2 \sqrt{ \vert E \vert^2 - ( \alpha - \beta )^2 } } 
\left\{  \left( \vert E \vert - \beta \right)^2 {\bf K}( \kappa_0 )
+ \left[ \vert E \vert^2 - ( \alpha - \beta )^2 \right] {\bf E}( \kappa_0 ) \right.
\nonumber                            \\
&  \makebox[4.0cm]{} \left. - 2 \vert E \vert \left( \vert E \vert - \alpha - \beta \right) 
{\bf \Pi}\left( n_{10} , \kappa_0 \right) \right\} ,
\label{twopart:eq:appafiftysix}      \\
n_{10} & =  \frac{ 2 \alpha }{ \vert E \vert + \alpha - \beta } \: .
\label{twopart:eq:appafiftyseven}
\end{align}
$M^{\rm rt}_{01}$ and $M^{\rm rt}_{02}$ are obtained by permuting $\alpha \leftrightarrow \beta$:
\begin{align}
M^{\rm rt}_{01}( E \leq - ( \alpha + \beta ) ; \alpha , \beta ) & =   
\frac{2}{\pi \beta \sqrt{ \vert E \vert^2 - ( \alpha - \beta )^2 } } 
\left\{  \left( \vert E \vert - \alpha \right) {\bf K}(\kappa_0) 
- \left( \vert E \vert - \alpha - \beta \right) {\bf \Pi} \left( n_{01} , \kappa_0 \right) \right\} ,
\label{twopart:eq:appafiftyeight}     \\
M^{\rm rt}_{02}( E \leq - ( \alpha + \beta ) ; \alpha , \beta ) & =   
\frac{2}{\pi \beta^2 \sqrt{ \vert E \vert^2 - ( \alpha - \beta )^2 } } 
\left\{  \left( \vert E \vert - \alpha \right)^2 {\bf K}( \kappa_0 )
+ \left[ \vert E \vert^2 - ( \alpha - \beta )^2 \right] {\bf E}( \kappa_0 ) \right.
\nonumber                            \\
&  \makebox[4.0cm]{} \left. - 2 \vert E \vert \left( \vert E \vert - \alpha - \beta \right) 
{\bf \Pi}\left( n_{01} , \kappa_0 \right) \right\} ,
\label{twopart:eq:appafiftynine}      \\
n_{01} & =  \frac{ 2 \beta }{ \vert E \vert + \beta - \alpha } \: .
\label{twopart:eq:appasixty}
\end{align}
Note that $n_{10} n_{01} = \kappa^{2}_{0}$. Next, we compose a sum rule that is an anisotropic version of Eq.~(\ref{twopart:eq:appanine}):
\begin{equation}
\alpha \left\{ M^{\rm rt}_{n+1,m}(\alpha,\beta) + M^{\rm rt}_{n-1,m}(\alpha,\beta) \right\} + 
\beta  \left\{ M^{\rm rt}_{n,m-1}(\alpha,\beta) + M^{\rm rt}_{n,m+1}(\alpha,\beta) \right\} = 
- 2 \, \delta_{n0} \delta_{m0} + 2 \vert E \vert \, M^{\rm rt}_{nm}(\alpha,\beta) \: . 
\label{twopart:eq:appasixtynine}
\end{equation}
It can be verified by direct substitution of the definitions, Eq.~(\ref{twopart:eq:appaone}). The $n = m = 0$ version of the sum rule reads 
\begin{equation}
\vert E \vert \, M^{\rm rt}_{00}(\alpha,\beta)
- \alpha M^{\rm rt}_{10}(\alpha,\beta) - \beta M^{\rm rt}_{01}(\alpha,\beta) = 1 \: , 
\label{twopart:eq:appaseventy}
\end{equation}
which is satisfied by the expressions given above. By setting $(n,m) = (1,0)$ or $(0,1)$ in Eq.~(\ref{twopart:eq:appasixtynine}), one can derive the first diagonal integral 
\begin{equation}
M^{\rm rt}_{11}( E \leq - ( \alpha + \beta ) ; \alpha, \beta ) = 
\frac{1}{\pi \alpha \beta \sqrt{ \vert E \vert^2 - ( \alpha - \beta )^2 }} 
\left\{ \left( \vert E \vert^2 - \alpha^2 - \beta^2 \right) {\bf K}(\kappa_0) - 
        \left[ \vert E \vert^2 - ( \alpha - \beta )^2 \right] {\bf E}(\kappa_0) \right\} . 
\label{twopart:eq:appaseventyone}
\end{equation}
Equations (\ref{twopart:eq:appafiftytwo}), (\ref{twopart:eq:appafiftyfive}), (\ref{twopart:eq:appafiftysix}), (\ref{twopart:eq:appafiftyeight}), (\ref{twopart:eq:appafiftynine}) and (\ref{twopart:eq:appaseventyone}) are sufficient to analyze the square $UV$ model at arbitrary pair momenta away from the threshold. The same equations are instrumental in developing an efficient numerical method of computing similar integrals in 3D $UV$ models, see \ref{twopart:sec:appcthree}.  

By using the identity $\frac{1}{c} = \int^{\infty}_{0} \exp{(-cz)} \, dz$, Eq.~(\ref{twopart:eq:appaone}) can be transformed into an integral of a product of two Bessel functions: 
\begin{equation}
M^{\rm rt}_{nm}( E ; \alpha, \beta ) = 
\int^{\infty}_{0} e^{ - \vert E \vert z } \, I_{n}(\alpha z) I_{m}(\beta z) \, {\rm d}z \: . 
\label{twopart:eq:appaseventytwo}
\end{equation}
Using this representation, expressions for $M^{\rm rt}_{00}(E;\alpha,\beta)$ and $M^{\rm rt}_{11}(E;\alpha,\beta)$ can also be found in integrals handbooks (see, e.g., \cite{Prudnikov1986}, \# 2.15.20.1), albeit without derivation.

\subsection{ \label{twopart:sec:appasix}
Recurrence relations for the rectangular case, $\alpha \neq \beta$    
}

To proceed further, we need anisotropic analogs of the Morita recurrence relation~\cite{Morita1975}. Repeating the derivation twice, first time for $M^{\rm rt}_{n0}$ and second time for $M^{\rm rt}_{0m}$, one obtains
\begin{align}
&   2n( 2 \vert E \vert^2 + \alpha^2 - 2 \beta^2 ) M^{\rm rt}_{n0}  
   - 2 \alpha \vert E \vert ( 2n + 1 ) M^{\rm rt}_{n+1,0} 
   - 2 \alpha \vert E \vert ( 2n - 1 ) M^{\rm rt}_{n-1,0}
\nonumber \\   
&    \makebox[4.0cm]{}    + \alpha^2 ( n + 1 ) M^{\rm rt}_{n+2,0}
                          + \alpha^2 ( n - 1 ) M^{\rm rt}_{n-2,0} = 0 \: ,
\hspace{1.0cm} n \neq 0 \: ,                          
\label{twopart:eq:appaseventythree}     \\
&   2m( 2 \vert E \vert^2 + \beta^2 - 2 \alpha^2 ) M^{\rm rt}_{0m}  
   - 2 \beta \vert E \vert ( 2m + 1 ) M^{\rm rt}_{0,m+1} 
   - 2 \beta \vert E \vert ( 2m - 1 ) M^{\rm rt}_{0,m-1}
\nonumber \\   
&    \makebox[4.0cm]{}    + \beta^2 ( m + 1 ) M^{\rm rt}_{0,m+2}
                          + \beta^2 ( m - 1 ) M^{\rm rt}_{0,m-2} = 0 \: ,
\hspace{1.0cm} m \neq 0 \: .                                    
\label{twopart:eq:appaseventyfour}     
\end{align}
Using Eqs.~(\ref{twopart:eq:appaseventythree})-(\ref{twopart:eq:appaseventyfour}), one can derive all $M^{\rm rt}_{n0}$ and $M^{\rm rt}_{0m}$, at least in principle. For example, 
\begin{align}
M^{\rm rt}_{30}( E \leq - ( \alpha + \beta ) ; \alpha, \beta ) = &  
\frac{2}{\pi \alpha^3 \sqrt{ \vert E \vert^2 - ( \alpha - \beta )^2 }} 
\left\{ \left[ ( \vert E \vert - \beta ) 
( \vert E \vert^2 - 3 \vert E \vert \beta - \alpha^2 + 2 \beta^2 ) 
+ \vert E \vert \alpha^2 \right] {\bf K}(\kappa_0) \right. 
\nonumber \\ 
 & \left.
 + \: 3 \, \vert E \vert \left[ \vert E \vert^2 - ( \alpha - \beta )^2 \right] {\bf E}(\kappa_0) 
 - ( 4 \vert E \vert^2 - \alpha^2 + 2 \beta^2 )( \vert E \vert - \alpha - \beta ) 
 {\bf \Pi}( n_{10} , \kappa_0 ) \right\} . \makebox[0.5cm]{} 
\label{twopart:eq:appaseventyfive}
\end{align}
$M^{\rm rt}_{03}$ is obtained from Eq.~(\ref{twopart:eq:appaseventyfive}) by permuting $\alpha \leftrightarrow \beta$ and $n_{10} \leftrightarrow n_{01}$. 

Then, utilizing the sum rule, Eq.~(\ref{twopart:eq:appasixtynine}), one can derive all other $M^{\rm rt}_{nm}$. For example, 
\begin{align}
M^{\rm rt}_{21}( E \leq - ( \alpha + \beta ) ; \alpha, \beta ) & =   
\frac{ \vert E \vert }{ \beta } \: M^{\rm rt}_{20} 
- \frac{ \alpha }{ 2 \beta } \left( M^{\rm rt}_{10} + M^{\rm rt}_{30} \right) , 
\label{twopart:eq:appaseventysix} \\ 
M^{\rm rt}_{12}( E \leq - ( \alpha + \beta ) ; \alpha, \beta ) & =   
\frac{ \vert E \vert }{ \alpha } \: M^{\rm rt}_{02} 
- \frac{ \beta }{ 2 \alpha } \left( M^{\rm rt}_{01} + M^{\rm rt}_{03} \right) , 
\label{twopart:eq:appaseventyseven}
\end{align}
and so on. These analytical expressions are useful in investigations of square, simple cubic, and tetragonal $UV$ models with attraction beyond nearest neighbors.

\subsection{ \label{twopart:sec:appaseven}
Table of $M_{nm} - M_{00}$ at threshold in the rectangular case, $\alpha \neq \beta$    
}

All $M^{\rm rt}_{nm}$ derived in \ref{twopart:sec:appafive} and \ref{twopart:sec:appasix} diverge at pair binding threshold when $E = - \alpha - \beta$. To obtain meaningful results, a subtractive procedure must be used. We define    
\begin{equation}
L^{\rm rt}_{nm}(\alpha,\beta) \equiv 
M^{\rm rt}_{nm}( E = - \alpha - \beta ; \alpha , \beta ) - 
M^{\rm rt}_{00}( E = - \alpha - \beta ; \alpha , \beta ) \: .
\label{twopart:eq:appaseventyeight}
\end{equation}
$L^{\rm rt}_{nm}(\alpha,\beta)$ can be derived either by taking the $E \rightarrow - \alpha - \beta$ limit in the general expressions for $M^{\rm rt}_{nm}(\alpha,\beta)$, or by direct evaluation of the double integrals in definitions, Eq.~(\ref{twopart:eq:appaseventyeight}). In the latter case, the second integration is elementary but can be laborious. Below, we list several $L^{\rm rt}_{nm}(\alpha,\beta)$ for the lowest $(n,m)$:
\begin{align}
L^{\rm rt}_{10}(\alpha,\beta)  = & 
- \frac{ 2 }{ \pi \alpha } \arcsin{ \sqrt{ \frac{ \alpha }{ \alpha + \beta } } } \: ,   
\label{twopart:eq:appaseventynine}   \\
L^{\rm rt}_{01}(\alpha,\beta)  = & 
- \frac{ 2 }{ \pi \beta  } \arcsin{ \sqrt{ \frac{  \beta }{ \alpha + \beta } } } \: ,     
\label{twopart:eq:appaeighty}        \\
L^{\rm rt}_{20}(\alpha,\beta)  = & 
- \frac{ 4 }{ \pi \alpha } \left[ \frac{ \alpha + \beta }{ \alpha } 
\arcsin{ \sqrt{ \frac{ \alpha }{ \alpha + \beta } } } - \sqrt{ \frac{ \beta }{ \alpha } } \right] ,
\label{twopart:eq:appaeightyone}     \\ 
L^{\rm rt}_{02}(\alpha,\beta)  = & 
- \frac{ 4 }{ \pi \beta  } \left[ \frac{ \alpha + \beta }{ \beta  } 
\arcsin{ \sqrt{ \frac{ \beta  }{ \alpha + \beta } } } - \sqrt{ \frac{ \alpha }{ \beta } } \right] ,
\label{twopart:eq:appaeightytwo}     \\ 
L^{\rm rt}_{11}(\alpha,\beta)  = & 
- \frac{ 2 }{ \pi \sqrt{ \alpha \beta } } \: ,
\label{twopart:eq:appaeightythree}   \\
L^{\rm rt}_{21}(\alpha,\beta)  = & 
- \frac{ 2 }{ \pi \alpha } \left[ \frac{ \alpha + \beta }{ \sqrt{\alpha\beta} } 
- \frac{\beta}{\alpha} \arcsin{ \sqrt{ \frac{ \alpha  }{ \alpha + \beta } } } \right] ,
\label{twopart:eq:appaeightyfour}    \\
L^{\rm rt}_{12}(\alpha,\beta)  = & 
- \frac{ 2 }{ \pi \beta } \left[ \frac{ \alpha + \beta }{ \sqrt{\alpha\beta} } 
- \frac{\alpha}{\beta} \arcsin{ \sqrt{ \frac{ \beta  }{ \alpha + \beta } } } \right] ,
\label{twopart:eq:appaeightyfive}    \\
L^{\rm rt}_{22}(\alpha,\beta)  = & 
- \frac{8}{3\pi \sqrt{\alpha\beta}} \: .
\label{twopart:eq:appaeightysix}
\end{align}
These expressions are needed in deriving the pairing threshold in the square $UV$ model at nonzero pair momenta~\cite{Kornilovitch2004}.

\section{ \label{twopart:sec:appb}
$UV$ model on the 2D triangular lattice  
}

\subsection{ \label{twopart:sec:appbone}
Evaluation of basic integral $M^{\rm tr}_{00}$, Eq.~(\ref{twopart:eq:twelve}), for the triangular lattice   
}

The single-particle dispersion in the nearest-neighbor hopping approximation is 
\begin{equation}
\varepsilon_{\bf k} = - 2 t \cos{k_x} 
- 4 t \cos{ \left( \frac{k_x}{2} \right) \cos{ \left( \frac{\sqrt{3}k_y}{2} \right) }} .   
\label{twopart:eq:appbone}
\end{equation}
Substituting this into Eq.~(\ref{twopart:eq:twelve}), shifting the integration variables ${\bf q} = {\bf q}' + \frac{\bf P}{2}$, and rescaling $q^{\prime}_y = \frac{2}{\sqrt{3}} \, q^{\prime\prime}_{y}$, the double integral $M^{\rm tr}_{00}$ is transformed into an integral over the $-\pi \leq q_x , q_y \leq \pi$ square domain: 
\begin{align}
M^{\rm tr}_{00}(E,{\bf P}) & =   
\int\limits^{\pi}_{-\pi} \!\! \int\limits^{\pi}_{-\pi} \frac{{\rm d}q_x \, {\rm d}q_y}{(2\pi)^2} 
\frac{1}{ \vert E \vert - \alpha \cos{q_x} 
- \beta \cos{\frac{q_x}{2}} \cos{q_y} - \gamma \sin{\frac{q_x}{2}} \sin{q_y} } = 
\nonumber \\
& = \int^{\pi}_{-\pi} 
\frac{{\rm d}q_x}{2 \pi} \frac{1}{ \sqrt{ ( \vert E \vert - \alpha \cos{q_x} )^2 
- ( \beta \cos{\frac{q_x}{2}} )^2 - ( \gamma \sin{\frac{q_x}{2}} )^2 }} 
\nonumber \\
& = \frac{1}{\pi} \int^{\pi}_{0} 
\frac{{\rm d}q_x}{ \sqrt{ \tilde{\alpha} \cos^2{q_x} + \tilde{\beta} \cos{q_x} + \tilde{\gamma} } } \: ,
\label{twopart:eq:appbtwo}
\end{align}
where
\begin{align}
\alpha  & =  4 t \cos{\frac{P_x}{2}} \: ,
\label{twopart:eq:appbthree}    \\
\beta   & =  8 t \cos{\frac{P_x}{4}} \cos{ \frac{\sqrt{3} P_y}{4}} \: ,
\label{twopart:eq:appbfour}     \\
\gamma  & =  8 t \sin{\frac{P_x}{4}} \sin{ \frac{\sqrt{3} P_y}{4}} \: .
\label{twopart:eq:appbfive}     \\
\tilde{\alpha} & =  \alpha^2                                                          \: ,
\label{twopart:eq:appbsix}      \\
\tilde{\beta}  & =  - 2 \alpha \vert E \vert - \frac{\beta^2}{2} + \frac{\gamma^2}{2} \: , 
\label{twopart:eq:appbseven}    \\
\tilde{\gamma} & =  \vert E \vert^2 - \frac{\beta^2}{2} - \frac{\gamma^2}{2}          \: .
\label{twopart:eq:appbeight}
\end{align}
Integration over $q_y$ in Eq.~(\ref{twopart:eq:appbtwo}) was done by utilizing the residue theorem. The next step depends on whether the expression under the square root can be factorized into a product of two real factors. If ${\tilde{\beta}}^2 > 4 \tilde{\alpha} \tilde{\gamma}$, it is possible, and 
\begin{align}
\sqrt{ \tilde{\alpha} \cos^2{q_x} + \tilde{\beta} \cos{q_x} + \tilde{\gamma} } & =  
\alpha \sqrt{ ( z_1 - \cos{\phi} )( z_2 - \cos{\phi} ) } \: , 
\label{twopart:eq:twoheight} \\
z_{1,2} & =  \frac{1}{ 2 \tilde{\alpha} } 
\left( - \tilde{\beta} \pm \sqrt{ {\tilde{\beta}}^2 - 4 \tilde{\alpha} \tilde{\gamma} } \right) .
\label{twopart:eq:twohnine}
\end{align}
Then the problem is reduced to Eq.~(\ref{twopart:eq:appafive}), and the same transformations, Eqs.~(\ref{twopart:eq:appasix}) and (\ref{twopart:eq:appaseven}), result in 
\begin{equation}
M^{\rm tr}_{00} = \frac{2}{\pi \alpha} \frac{1}{\sqrt{ ( z_1 - 1 )( z_2 + 1 ) }} \: 
{\bf K} \! \left( \sqrt{ \frac{ 2 ( z_1 - z_2 ) }{ ( z_1 - 1 )( z_2 + 1 ) } } \right) .
\label{twopart:eq:twohten}
\end{equation}
In the ground state, $P_{x} = P_{y} = 0$, $\alpha = 4t$, $\beta = 8t$, $\gamma = 0$, and 
\begin{equation}
z_{1,2} = \frac{1}{\alpha} \left\{ ( \vert E \vert + \alpha ) 
\pm \sqrt{ 2 \alpha \vert E \vert + 3 \alpha^2 }  \right\} .
\label{twopart:eq:appbtwelve}
\end{equation}
Substitution in Eq.~(\ref{twopart:eq:twohten}) yields 
\begin{equation}
M^{\rm tr}_{00} = \frac{2}{\pi} 
\frac{1}{\sqrt{ \vert E_0 \vert^2 - 48 t^2 + 16 t \sqrt{ 2 t \vert E_0 \vert + 12 t^2 } }} \: 
{\bf K} \! \left( \sqrt{ \frac{ 32 t \sqrt{ 2 \vert E_0 \vert t + 12 t^2 } }
  { \vert E_0 \vert^2 - 48 t^2 + 16 t \sqrt{ 2 \vert E_0 \vert t + 12 t^2 } } } \right) .
\label{twopart:eq:appbthirteen}
\end{equation}

If ${\tilde{\beta}}^2 < 4 \tilde{\alpha} \tilde{\gamma}$, factorization is not possible. Reduction to Legendre standard form is achieved by consecutive application of Eq.~(\ref{twopart:eq:appasix}) and 
\begin{align}
u & =  \frac{1}{2} + \sqrt{ \frac{1}{4} + \frac{p}{2} + q } \cdot \tan^2{ \frac{\phi}{2} } \: ,
\label{twopart:eq:twoheleven}  \\
p & =  \frac{ \tilde{\beta} - 2 \tilde{\alpha} }
             { \tilde{\alpha} - \tilde{\beta} + \tilde{\gamma} }                   \: , 
\label{twopart:eq:twohtwelve}      \\
q & =  \frac{ \tilde{\alpha} }{ \tilde{\alpha} - \tilde{\beta} + \tilde{\gamma} } \: .
\label{twopart:eq:twohthirteen}
\end{align}
The final result is
\begin{equation}
M^{\rm tr}_{00} = \frac{2}{\pi} 
\frac{1}{ \left[ ( \tilde{\alpha} + \tilde{\gamma} )^2 - {\tilde{\beta}}^2 \right]^{1/4}} \: 
{\bf K} \! \left( \sqrt{ \frac{1}{2} 
\left[ 1 - \frac{ \tilde{\gamma} - \tilde{\alpha} }
{\sqrt{ ( \tilde{\alpha} + \tilde{\gamma} )^2 - {\tilde{\beta}}^2 } } \right] } \right) .
\label{twopart:eq:twohfourteen}
\end{equation}

\subsection{ \label{twopart:sec:appbfour}
Numerical evaluation of $M^{\pm}_{\bf nm}$ for the triangular lattice   
}

In this section, the following integrals are utilized:
\begin{align}
\int^{\pi}_{-\pi} \frac{{\rm d}x}{2\pi} \frac{1}{ a - b \cos{x} - c \sin{x} }          & =  
\frac{1}{\sqrt{ a^2 - b^2 - c^2 }} \: ,
\label{twopart:eq:appbfourteenone}      \\
\int^{\pi}_{-\pi} \frac{{\rm d}x}{2\pi} \frac{ \cos{x} }{ a - b \cos{x} - c \sin{x} }  & =  
\frac{ b }{ b^2 + c^2 } \left( \frac{a}{\sqrt{ a^2 - b^2 - c^2 }} - 1 \right) ,
\label{twopart:eq:appbfourteentwo}      \\
\int^{\pi}_{-\pi} \frac{{\rm d}x}{2\pi} \frac{ \sin{x} }{ a - b \cos{x} - c \sin{x} }  & =  
\frac{ c }{ b^2 + c^2 } \left( \frac{a}{\sqrt{ a^2 - b^2 - c^2 }} - 1 \right) ,
\label{twopart:eq:appbfourteenthree}    \\
\int^{\pi}_{-\pi} \frac{{\rm d}x}{2\pi} \frac{ \cos{(2x)} }{ a - b \cos{x} - c \sin{x} }  & =  
\frac{ b^2 - c^2 }{ \sqrt{ a^2 - b^2 - c^2 } } \left( 
\frac{ a - \sqrt{ a^2 - b^2 - c^2 } }{ b^2 + c^2 } \right)^2 ,
\label{twopart:eq:appbfourteenfour}    \\  
\int^{\pi}_{-\pi} \frac{{\rm d}x}{2\pi} \frac{ \sin{(2x)} }{ a - b \cos{x} - c \sin{x} }  & =  
\frac{ 2 b c }{ \sqrt{ a^2 - b^2 - c^2 } } \left( 
\frac{ a - \sqrt{ a^2 - b^2 - c^2 } }{ b^2 + c^2 } \right)^2 ,
\label{twopart:eq:appbfourteenfive}    \\  
\int^{\pi}_{0} \frac{{\rm d}x}{\pi} \frac{ 1 }{ \sqrt{ a \cos^2{x} + b \cos{x} + c } } & =  
\frac{2}{\pi} \frac{1}{[ ( a + c )^2 - b^2  ]^{1/4}} \:
{\bf K} \! \left[ \sqrt{ \frac{1}{2} 
\left( 1 - \frac{ c - a }{ \sqrt{ ( a + c )^2 - b^2 } } \right) } \right] ,
\label{twopart:eq:appbfourteensix}
\end{align}
where $a, b, c > 0$, and $a^2 > b^2 + c^2$ is assumed. We also define
\begin{equation}
S \equiv \left( \vert E \vert - \alpha \cos{q_x} \right)^2 
- \beta^2 \cos^2{\frac{q_x}{2}} - \gamma^2 \sin^2{\frac{q_x}{2}}   \: , 
\label{twopart:eq:appbh}
\end{equation}
to simplify notation. The parameters $\alpha$, $\beta$, and $\gamma$ are defined in Eqs.~(\ref{twopart:eq:appbthree})-(\ref{twopart:eq:appbfive}).

The spin-singlet matrix elements are defined in Eqs.~(\ref{twopart:eq:onehthirtythree})-(\ref{twopart:eq:onehthirtyseven}). The nearest-neighbor vectors are defined as: ${\bf 0} = (0,0)$; ${\bf b}_{+1} \equiv {\bf 1} = (1,0)$; ${\bf b}_{+2} \equiv {\bf 2} = \left( \frac{1}{2} , \frac{\sqrt{3}}{2} \right)$; ${\bf b}_{+3} \equiv {\bf 3} = \left( - \frac{1}{2} , \frac{\sqrt{3}}{2} \right)$. Integration over $q_y$ yields
\begin{align}
% M^{+}_{00}
M^{+}_{\bf 00}  = &  
\int\limits^{\pi}_{-\pi} \!\! \int\limits^{\pi}_{-\pi} \frac{{\rm d}q_x \, {\rm d}q_y}{(2\pi)^2} 
\frac{1}{ \vert E \vert - \alpha \cos{q_x} 
- \beta \cos{\frac{q_x}{2}} \cos{q_y} - \gamma \sin{\frac{q_x}{2}} \sin{q_y} } = 
\int^{\pi}_{0} \frac{{\rm d}q_x}{\pi} \frac{1}{ \sqrt{ S } } \: ,
\label{twopart:eq:appbfifteen}    \\
% M^{+}_{01} = 2*M^{+}_{10}
M^{+}_{\bf 01} = 2 M^{+}_{\bf 10}  = &  
\int\limits^{\pi}_{-\pi} \!\! \int\limits^{\pi}_{-\pi} \frac{{\rm d}q_x \, {\rm d}q_y}{(2\pi)^2} 
\frac{2 \cos{q_x}}{ \vert E \vert - \alpha \cos{q_x} 
- \beta \cos{\frac{q_x}{2}} \cos{q_y} - \gamma \sin{\frac{q_x}{2}} \sin{q_y} } = 
2 \int^{\pi}_{0} \frac{{\rm d}q_x}{\pi} \frac{ \cos{q_x} }{ \sqrt{ S }} \: ,
\label{twopart:eq:appbsixteen}    \\
% M^{+}_{02} = 2*M^{+}_{20}
M^{+}_{\bf 02} = 2 M^{+}_{\bf 20}  = &   
\int\limits^{\pi}_{-\pi} \!\! \int\limits^{\pi}_{-\pi} \frac{{\rm d}q_x \, {\rm d}q_y}{(2\pi)^2} 
\frac{2 \cos{ \left( \frac{q_x}{2} + q_y \right) } }{ \vert E \vert - \alpha \cos{q_x} 
- \beta \cos{\frac{q_x}{2}} \cos{q_y} - \gamma \sin{\frac{q_x}{2}} \sin{q_y} } = 
\nonumber                         \\
  & = 2 \int^{\pi}_{0} \frac{{\rm d}q_x}{\pi} 
\frac{ \beta   \cos^2{\frac{q_x}{2}} - \gamma   \sin^2{\frac{q_x}{2}} }
     { \beta^2 \cos^2{\frac{q_x}{2}} + \gamma^2 \sin^2{\frac{q_x}{2}} }
\left( \frac{ |E| - \alpha \cos{q_x} }{ \sqrt{ S } } - 1 \right) ,
\label{twopart:eq:appbseventeen}  \\
% M^{+}_{03} = 2*M^{+}_{30}
M^{+}_{\bf 03} = M^{+}_{\bf 30}  = &   
\int\limits^{\pi}_{-\pi} \!\! \int\limits^{\pi}_{-\pi} \frac{{\rm d}q_x \, {\rm d}q_y}{(2\pi)^2} 
\frac{2 \cos{ \left( - \frac{q_x}{2} + q_y \right) } }{ \vert E \vert - \alpha \cos{q_x} 
- \beta \cos{\frac{q_x}{2}} \cos{q_y} - \gamma \sin{\frac{q_x}{2}} \sin{q_y} } = 
\nonumber                         \\
 & = 2 \int^{\pi}_{0} \frac{{\rm d}q_x}{\pi} 
\frac{ \beta   \cos^2{\frac{q_x}{2}} + \gamma   \sin^2{\frac{q_x}{2}} }
     { \beta^2 \cos^2{\frac{q_x}{2}} + \gamma^2 \sin^2{\frac{q_x}{2}} }
\left( \frac{ |E| - \alpha \cos{q_x} }{ \sqrt{ S } } - 1 \right) ,
\label{twopart:eq:appbeighteen}
\end{align}
\begin{align}
% M^{+}_{11}
M^{+}_{\bf 11}  = &  
\int\limits^{\pi}_{-\pi} \!\! \int\limits^{\pi}_{-\pi} \frac{{\rm d}q_x \, {\rm d}q_y}{(2\pi)^2} 
\frac{2 e^{iq_x} \cos{q_x} }{ \vert E \vert - \alpha \cos{q_x} 
- \beta \cos{\frac{q_x}{2}} \cos{q_y} - \gamma \sin{\frac{q_x}{2}} \sin{q_y} } = 
2 \int^{\pi}_{0} \frac{{\rm d}q_x}{\pi} \frac{ \cos^2{q_x} }{ \sqrt{ S } } \: ,
\label{twopart:eq:appbnineteen}   \\
% M^{+}_{21}
M^{+}_{\bf 21}  = &  
\int\limits^{\pi}_{-\pi} \!\! \int\limits^{\pi}_{-\pi} \frac{{\rm d}q_x \, {\rm d}q_y}{(2\pi)^2} 
\frac{2 e^{i \left( \frac{q_x}{2} + q_y \right) } \cos{q_x}}{ \vert E \vert - \alpha \cos{q_x} 
- \beta \cos{\frac{q_x}{2}} \cos{q_y} - \gamma \sin{\frac{q_x}{2}} \sin{q_y} } = 
\nonumber                         \\
 & = 2 \int^{\pi}_{0} \frac{{\rm d}q_x}{\pi} 
\frac{ \cos{q_x} ( \beta   \cos^2{\frac{q_x}{2}} - \gamma   \sin^2{\frac{q_x}{2}} ) }
     {             \beta^2 \cos^2{\frac{q_x}{2}} + \gamma^2 \sin^2{\frac{q_x}{2}}   }
\left( \frac{ |E| - \alpha \cos{q_x} }{ \sqrt{ S }} - 1 \right) ,
\label{twopart:eq:appbtwenty}     \\
% M^{+}_{12}
M^{+}_{\bf 12}  = &  
\int\limits^{\pi}_{-\pi} \!\! \int\limits^{\pi}_{-\pi} \frac{{\rm d}q_x \, {\rm d}q_y}{(2\pi)^2} 
\frac{2 e^{iq_x} \cos{ \left( \frac{q_x}{2} + q_y \right)} }{ \vert E \vert - \alpha \cos{q_x} 
- \beta \cos{\frac{q_x}{2}} \cos{q_y} - \gamma \sin{\frac{q_x}{2}} \sin{q_y} } = M^{+}_{\bf 21}  \: ,
\label{twopart:eq:appbtwentyone}  \\
% M^{+}_{31}
M^{+}_{\bf 31}  = &  
\int\limits^{\pi}_{-\pi} \!\! \int\limits^{\pi}_{-\pi} \frac{{\rm d}q_x \, {\rm d}q_y}{(2\pi)^2} 
\frac{2 e^{i \left( - \frac{q_x}{2} + q_y \right) } \cos{q_x}}{ \vert E \vert - \alpha \cos{q_x} 
- \beta \cos{\frac{q_x}{2}} \cos{q_y} - \gamma \sin{\frac{q_x}{2}} \sin{q_y} } = 
\nonumber                         \\
 & = 2 \int^{\pi}_{0} \frac{{\rm d}q_x}{\pi} 
\frac{ \cos{q_x} ( \beta   \cos^2{\frac{q_x}{2}} + \gamma   \sin^2{\frac{q_x}{2}} ) }
     {             \beta^2 \cos^2{\frac{q_x}{2}} + \gamma^2 \sin^2{\frac{q_x}{2}}   }
\left( \frac{ |E| - \alpha \cos{q_x} }{ \sqrt{ S }} - 1 \right) ,
\label{twopart:eq:appbtwentytwo}     \\
% M^{+}_{13}
M^{+}_{\bf 13}  = &  
\int\limits^{\pi}_{-\pi} \!\! \int\limits^{\pi}_{-\pi} \frac{{\rm d}q_x \, {\rm d}q_y}{(2\pi)^2} 
\frac{2 e^{iq_x} \cos{ \left( - \frac{q_x}{2} + q_y \right)} }{ \vert E \vert - \alpha \cos{q_x} 
- \beta \cos{\frac{q_x}{2}} \cos{q_y} - \gamma \sin{\frac{q_x}{2}} \sin{q_y} } = M^{+}_{\bf 31}  \: ,
\label{twopart:eq:appbtwentythree}
\end{align}
\begin{align}
% M^{+}_{22}
M^{+}_{\bf 22}  = &  
\int\limits^{\pi}_{-\pi} \!\! \int\limits^{\pi}_{-\pi} \frac{{\rm d}q_x \, {\rm d}q_y}{(2\pi)^2} 
\frac{2 e^{i \left( \frac{q_x}{2} + q_y \right) } 
       \cos{ \left( \frac{q_x}{2} + q_y \right) } }
{ \vert E \vert - \alpha \cos{q_x} - \beta \cos{\frac{q_x}{2}} \cos{q_y} - \gamma \sin{\frac{q_x}{2}} \sin{q_y} } = 
\nonumber                           \\
 & = \int^{\pi}_{0} \frac{{\rm d}q_x}{\pi} \left\{
\frac{1}{\sqrt{S}} + 
\frac{ ( \beta + \gamma )^2 \cos^2{q_x} + ( \beta^2 - \gamma^2 ) \cos{q_x} - 2 \beta \gamma }{ 2 \sqrt{S}}
\left( \frac{ |E| - \alpha \cos{q_x} - \sqrt{S} }{ \beta^2 \cos^2{\frac{q_x}{2}} + \gamma^2 \sin^2{\frac{q_x}{2}}}
\right)^2 \right\} ,
\label{twopart:eq:appbtwentyfour}  
\end{align}
\begin{align}
% M^{+}_{23}
M^{+}_{\bf 23}  = &  
\int\limits^{\pi}_{-\pi} \!\! \int\limits^{\pi}_{-\pi} \frac{{\rm d}q_x \, {\rm d}q_y}{(2\pi)^2} 
\frac{2 e^{i \left(   \frac{q_x}{2} + q_y \right) } 
       \cos{ \left( - \frac{q_x}{2} + q_y \right) } }
{ \vert E \vert - \alpha \cos{q_x} - \beta \cos{\frac{q_x}{2}} \cos{q_y} - \gamma \sin{\frac{q_x}{2}} \sin{q_y} } = 
\nonumber                           \\
 & = \int^{\pi}_{0} \frac{{\rm d}q_x}{\pi} \left\{
\frac{ \cos{q_x} }{ \sqrt{S} } + 
\frac{ ( \beta^2 - \gamma^2 ) + ( \beta^2 + \gamma^2 ) \cos{q_x} }{ 2 \sqrt{S} }
\left( \frac{ |E| - \alpha \cos{q_x} - \sqrt{S} }{ \beta^2 \cos^2{\frac{q_x}{2}} + \gamma^2 \sin^2{\frac{q_x}{2}}}
\right)^2 \right\} ,
\label{twopart:eq:appbtwentyfive}   \\
% M^{+}_{32}
M^{+}_{\bf 32}  = &  
\int\limits^{\pi}_{-\pi} \!\! \int\limits^{\pi}_{-\pi} \frac{{\rm d}q_x \, {\rm d}q_y}{(2\pi)^2} 
\frac{2 e^{i \left( - \frac{q_x}{2} + q_y \right) } 
       \cos{ \left(   \frac{q_x}{2} + q_y \right) } }
{ \vert E \vert - \alpha \cos{q_x} - \beta \cos{\frac{q_x}{2}} \cos{q_y} - \gamma \sin{\frac{q_x}{2}} \sin{q_y} }  
= M^{+}_{\bf 23} \: ,              
\label{twopart:eq:appbtwentysix}
\end{align}
\begin{align}
% M^{+}_{33}
M^{+}_{\bf 33} = &  
\int\limits^{\pi}_{-\pi} \!\! \int\limits^{\pi}_{-\pi} \frac{{\rm d}q_x \, {\rm d}q_y}{(2\pi)^2} 
\frac{2 e^{i \left( - \frac{q_x}{2} + q_y \right) } 
       \cos{ \left( - \frac{q_x}{2} + q_y \right) } }
{ \vert E \vert - \alpha \cos{q_x} - \beta \cos{\frac{q_x}{2}} \cos{q_y} - \gamma \sin{\frac{q_x}{2}} \sin{q_y} } = 
\nonumber                           \\
 & = \int^{\pi}_{0} \frac{{\rm d}q_x}{\pi} \left\{
\frac{1}{\sqrt{S}} + 
\frac{ ( \beta - \gamma )^2 \cos^2{q_x} + ( \beta^2 - \gamma^2 ) \cos{q_x} + 2 \beta \gamma }{ 2 \sqrt{S}}
\left( \frac{ |E| - \alpha \cos{q_x} - \sqrt{S} }{ \beta^2 \cos^2{\frac{q_x}{2}} + \gamma^2 \sin^2{\frac{q_x}{2}}}
\right)^2 \right\} .
\label{twopart:eq:appbtwentyseven}  
\end{align}
These expressions can be used for efficient numerical evaluation of $M^{+}_{\bf nm}$.  

The spin-triplet matrix elements $M^{-}_{\bf nm}$ are defined in Eq.~(\ref{twopart:eq:onehthirtynine}). The nearest-neighbor vectors are defined as: ${\bf b}_{-1} \equiv {\bf 1} = (1,0)$; ${\bf b}_{-2} \equiv {\bf 2} = \left( \frac{1}{2} , \frac{\sqrt{3}}{2} \right)$; ${\bf b}_{-3} \equiv {\bf 3} = \left( - \frac{1}{2} , \frac{\sqrt{3}}{2} \right)$. Integration over $q_y$ yields
\begin{align}
% M{-}_{11}
M^{-}_{\bf 11}  = & \; ( - 2 \, i ) 
\int\limits^{\pi}_{-\pi} \!\! \int\limits^{\pi}_{-\pi} \frac{{\rm d}q_x \, {\rm d}q_y}{(2\pi)^2} 
\frac{ e^{iq_x} \sin{q_x} }{ \vert E \vert - \alpha \cos{q_x} 
- \beta \cos{\frac{q_x}{2}} \cos{q_y} - \gamma \sin{\frac{q_x}{2}} \sin{q_y} } = 
2 \int^{\pi}_{0} \frac{{\rm d}q_x}{\pi} \frac{ \sin^2{q_x} }{ \sqrt{ S } } \: ,
\label{twopart:eq:appbthirtyone}     \\
% M^{-}_{12}
M^{-}_{\bf 12}  = & \; ( - 2 \, i ) 
\int\limits^{\pi}_{-\pi} \!\! \int\limits^{\pi}_{-\pi} \frac{{\rm d}q_x \, {\rm d}q_y}{(2\pi)^2} 
\frac{ e^{iq_x} \sin{ \left( \frac{q_x}{2} + q_y \right)} }{ \vert E \vert - \alpha \cos{q_x} 
- \beta \cos{\frac{q_x}{2}} \cos{q_y} - \gamma \sin{\frac{q_x}{2}} \sin{q_y} }
\nonumber                            \\
 & = \int^{\pi}_{0} \frac{{\rm d}q_x}{\pi} 
\frac{ ( \beta + \gamma )  \sin^2{q_x} }
     { \beta^2 \cos^2{\frac{q_x}{2}} + \gamma^2 \sin^2{\frac{q_x}{2}}   }
\left( \frac{ |E| - \alpha \cos{q_x} }{ \sqrt{ S }} - 1 \right) ,
\label{twopart:eq:appbthirtytwo}     \\
% M^{-}_{21}
M^{-}_{\bf 21}  = & \; ( - 2 \, i ) 
\int\limits^{\pi}_{-\pi} \!\! \int\limits^{\pi}_{-\pi} \frac{{\rm d}q_x \, {\rm d}q_y}{(2\pi)^2} 
\frac{ e^{ i \left( \frac{q_x}{2} + q_y \right) } \sin{ q_x } }{ \vert E \vert - \alpha \cos{q_x} 
- \beta \cos{\frac{q_x}{2}} \cos{q_y} - \gamma \sin{\frac{q_x}{2}} \sin{q_y} } = M^{-}_{\bf 12} \: ,
\label{twopart:eq:appbthirtythree}   \\
% M^{-}_{13}
M^{-}_{\bf 13}  = & \; ( - 2 \, i ) 
\int\limits^{\pi}_{-\pi} \!\! \int\limits^{\pi}_{-\pi} \frac{{\rm d}q_x \, {\rm d}q_y}{(2\pi)^2} 
\frac{ e^{iq_x} \sin{ \left( - \frac{q_x}{2} + q_y \right)} }{ \vert E \vert - \alpha \cos{q_x} 
- \beta \cos{\frac{q_x}{2}} \cos{q_y} - \gamma \sin{\frac{q_x}{2}} \sin{q_y} }
\nonumber                            \\
 & = \int^{\pi}_{0} \frac{{\rm d}q_x}{\pi} 
\frac{ ( \gamma - \beta )  \sin^2{q_x} }
     { \beta^2 \cos^2{\frac{q_x}{2}} + \gamma^2 \sin^2{\frac{q_x}{2}}   }
\left( \frac{ |E| - \alpha \cos{q_x} }{ \sqrt{ S }} - 1 \right) ,
\label{twopart:eq:appbthirtyfour}    \\
% M^{-}_{31}
M^{-}_{\bf 31}  = & \; ( - 2 \, i ) 
\int\limits^{\pi}_{-\pi} \!\! \int\limits^{\pi}_{-\pi} \frac{{\rm d}q_x \, {\rm d}q_y}{(2\pi)^2} 
\frac{ e^{ i \left( - \frac{q_x}{2} + q_y \right) } \sin{ q_x } }{ \vert E \vert - \alpha \cos{q_x} 
- \beta \cos{\frac{q_x}{2}} \cos{q_y} - \gamma \sin{\frac{q_x}{2}} \sin{q_y} } = M^{-}_{\bf 13} \: ,
\label{twopart:eq:appbthirtyfive}    \\
% M^{-}_{22}
M^{-}_{\bf 22}  = & \; ( - 2 \, i ) 
\int\limits^{\pi}_{-\pi} \!\! \int\limits^{\pi}_{-\pi} \frac{{\rm d}q_x \, {\rm d}q_y}{(2\pi)^2} 
\frac{ e^{ i \left( \frac{q_x}{2} + q_y \right) } 
       \sin{ \left( \frac{q_x}{2} + q_y \right) } }
{ \vert E \vert - \alpha \cos{q_x} - \beta \cos{\frac{q_x}{2}} \cos{q_y} - \gamma \sin{\frac{q_x}{2}} \sin{q_y} } = 
\nonumber                           \\
 & = \int^{\pi}_{0} \frac{{\rm d}q_x}{\pi} \left\{
\frac{1}{\sqrt{S}} - 
\frac{ ( \beta + \gamma )^2 \cos^2{q_x} + ( \beta^2 - \gamma^2 ) \cos{q_x} - 2 \beta \gamma }{ 2 \sqrt{S}}
\left( \frac{ |E| - \alpha \cos{q_x} - \sqrt{S} }{ \beta^2 \cos^2{\frac{q_x}{2}} + \gamma^2 \sin^2{\frac{q_x}{2}}}
\right)^2 \right\} ,
\label{twopart:eq:appbthirtysix}    
\end{align}
\begin{align}
% M^{-}_{23}
M^{-}_{\bf 23}  = & \; ( - 2 \, i ) 
\int\limits^{\pi}_{-\pi} \!\! \int\limits^{\pi}_{-\pi} \frac{{\rm d}q_x \, {\rm d}q_y}{(2\pi)^2} 
\frac{ e^{i \left(   \frac{q_x}{2} + q_y \right) } 
       \sin{ \left( - \frac{q_x}{2} + q_y \right) } }
{ \vert E \vert - \alpha \cos{q_x} - \beta \cos{\frac{q_x}{2}} \cos{q_y} - \gamma \sin{\frac{q_x}{2}} \sin{q_y} } = 
\nonumber                           \\
 & = \int^{\pi}_{0} \frac{{\rm d}q_x}{\pi} \left\{
\frac{ \cos{q_x} }{ \sqrt{S} } - 
\frac{ ( \beta^2 - \gamma^2 ) + ( \beta^2 + \gamma^2 ) \cos{q_x} }{ 2 \sqrt{S} }
\left( \frac{ |E| - \alpha \cos{q_x} - \sqrt{S} }{ \beta^2 \cos^2{\frac{q_x}{2}} + \gamma^2 \sin^2{\frac{q_x}{2}}}
\right)^2 \right\} ,
\label{twopart:eq:appbthirtyseven}   \\
% M^{-}_{32}
M^{-}_{\bf 32}  = & \; ( - 2 \, i )  
\int\limits^{\pi}_{-\pi} \!\! \int\limits^{\pi}_{-\pi} \frac{{\rm d}q_x \, {\rm d}q_y}{(2\pi)^2} 
\frac{ e^{i \left( - \frac{q_x}{2} + q_y \right) } 
      \sin{ \left(   \frac{q_x}{2} + q_y \right) } }
{ \vert E \vert - \alpha \cos{q_x} - \beta \cos{\frac{q_x}{2}} \cos{q_y} - \gamma \sin{\frac{q_x}{2}} \sin{q_y} }  
= M^{-}_{\bf 23} \: ,              
\label{twopart:eq:appbthirtyeight}  \\
% M^{-}_{33}
M^{-}_{\bf 33}  = & \; ( - 2 \, i ) 
\int\limits^{\pi}_{-\pi} \!\! \int\limits^{\pi}_{-\pi} \frac{{\rm d}q_x \, {\rm d}q_y}{(2\pi)^2} 
\frac{ e^{ i \left( - \frac{q_x}{2} + q_y \right) } 
       \sin{ \left( - \frac{q_x}{2} + q_y \right) } }
{ \vert E \vert - \alpha \cos{q_x} - \beta \cos{\frac{q_x}{2}} \cos{q_y} - \gamma \sin{\frac{q_x}{2}} \sin{q_y} } = 
\nonumber                           \\
 & = \int^{\pi}_{0} \frac{{\rm d}q_x}{\pi} \left\{
\frac{1}{\sqrt{S}} - 
\frac{ ( \beta - \gamma )^2 \cos^2{q_x} + ( \beta^2 - \gamma^2 ) \cos{q_x} + 2 \beta \gamma }{ 2 \sqrt{S}}
\left( \frac{ |E| - \alpha \cos{q_x} - \sqrt{S} }{ \beta^2 \cos^2{\frac{q_x}{2}} + \gamma^2 \sin^2{\frac{q_x}{2}}}
\right)^2 \right\} .
\label{twopart:eq:appbthirtynine}    
\end{align}

\subsection{ \label{twopart:sec:appbtwo}
Table of $M^{\rm tr}_{\bf nm} - M^{\rm tr}_{\bf 00}$ at threshold at the $\Gamma$-point   
}

We set $P_x = P_y = 0$ and $E = -12 \, t$. Then $\alpha = 4t$, $\beta = 8t$ and $\gamma = 0$. $L^{+}_{\bf nm}$ are defined in Eqs.~(\ref{twopart:eq:onehfiftyfiveone})-(\ref{twopart:eq:onehfiftyfivethree}). Utilizing Eqs.~(\ref{twopart:eq:appbh})-(\ref{twopart:eq:appbtwentytwo}), elementary integration yields:
\begin{align}
L^{+}_{\bf 10}  = & \int^{\pi}_{0} \frac{{\rm d}q_x}{\pi} \frac{ \cos{q_x} - 1 }
{ \sqrt{ ( 12 t - 4t \cos{q_x} )^2 - 64 t^2 \cos^2{\frac{q_x}{2}}}} = 
- \frac{1}{12 \, t} \: ,
\label{twopart:eq:appbforty}         \\
L^{+}_{\bf 11}  = & \int^{\pi}_{0} \frac{{\rm d}q_x}{\pi} \frac{ \cos^2{q_x} - 1 }
{ \sqrt{ ( 12 t - 4t \cos{q_x} )^2 - 64 t^2 \cos^2{\frac{q_x}{2}}}} = 
\frac{ 3 \sqrt{3} - 2\pi }{ 3 \pi t } \: ,
\label{twopart:eq:appbfortyone}      \\
L^{+}_{\bf 12}  = & \int^{\pi}_{0} \frac{{\rm d}q_x}{\pi} \frac{ \cos{q_x} ( 3 - \cos{q_x} ) - 2 }
{ \sqrt{ ( 12 t - 4t \cos{q_x} )^2 - 64 t^2 \cos^2{\frac{q_x}{2}}}} = 
\frac{ \pi - 6 \sqrt{3} }{ 12 \pi t }  \: .
\label{twopart:eq:appbfortyonetwo}    
\end{align}

\subsection{ \label{twopart:sec:appbthree}
Application of group theory   
}

The dispersion relations, Eqs.~(\ref{twopart:eq:onehthirtytwo}) and (\ref{twopart:eq:onehthirtyeight}), are quite complex and in general can be solved only numerically. However, at high symmetry points of the Brillouin zone there is enough simplification that some results can be derived analytically. The analysis is greatly aided by the theory of point groups. This section is devoted to the exposition of this subject. The triangular lattice is used as an example; $UV$ models on other lattices can be analyzed similarly.   

At the $\Gamma$-point (${\bf P} = 0$), the following relations among singlet matrix elements hold: 
\begin{align}
M^{+}_{\bf 10}(\Gamma) & =  M^{+}_{\bf 20}(\Gamma) = M^{+}_{\bf 30}(\Gamma) \: ,
\label{twopart:eq:onehforty}   \\
M^{+}_{\bf 01}(\Gamma) & =  M^{+}_{\bf 02}(\Gamma) = M^{+}_{\bf 03}(\Gamma) = 2 \, M^{+}_{\bf 10}(\Gamma) \: ,
\label{twopart:eq:onehfortyone}    \\
M^{+}_{\bf 11}(\Gamma) & =  M^{+}_{\bf 22}(\Gamma) = M^{+}_{\bf 33}(\Gamma) \: ,
\label{twopart:eq:onehfortytwo}         \\
M^{+}_{\bf 12}(\Gamma) & =  M^{+}_{\bf 21}(\Gamma) = M^{+}_{\bf 13}(\Gamma) =  
M^{+}_{\bf 31}(\Gamma) = M^{+}_{\bf 23}(\Gamma) = M^{+}_{\bf 32}(\Gamma)  \: .
\label{twopart:eq:onehfortythree}         
\end{align}
Thus, there remains only four independent integrals $M^{+}_{\bf 00}(\Gamma)$, $M^{+}_{\bf 10}(\Gamma)$, $M^{+}_{\bf 11}(\Gamma)$, and $M^{+}_{\bf 12}(\Gamma)$. Similarly, the following relations hold between triplet matrix elements:
\begin{align}
M^{-}_{\bf 11}(\Gamma) & =  M^{-}_{\bf 22}(\Gamma) =   M^{-}_{\bf 33}(\Gamma) \: ,
\label{twopart:eq:onehfortyfour}   \\
M^{-}_{\bf 12}(\Gamma) & =  M^{-}_{\bf 21}(\Gamma) =   M^{-}_{\bf 23}(\Gamma) = M^{-}_{\bf 32}(\Gamma) \: ,
\label{twopart:eq:onehfortyfive}    \\
M^{-}_{\bf 13}(\Gamma) & =  M^{-}_{\bf 31}(\Gamma) = - M^{-}_{\bf 12}(\Gamma) \: ,
\label{twopart:eq:onehfortysix}         
\end{align}
leaving only two independent integrals $M^{-}_{\bf 11}(\Gamma)$ and $M^{-}_{\bf 12}(\Gamma)$. As a result, the general Eqs.~(\ref{twopart:eq:onehthirtytwo}) and (\ref{twopart:eq:onehthirtyeight}) acquire additional symmetries. It is clear that a bound pair at ${\bf P} = 0$ should possess a hexagonal symmetry, which implies the solutions can be classified according to the irreducible representations (irreps) of the point group $C_{6v}$. We now execute a textbook algorithm~\cite{Petrashen1969} of constructing linear combinations of functions $\Phi$ that form bases of the irreps of $C_{6v}$.

\begin{figure}[t]
\begin{center}
\includegraphics[width=0.60\textwidth]{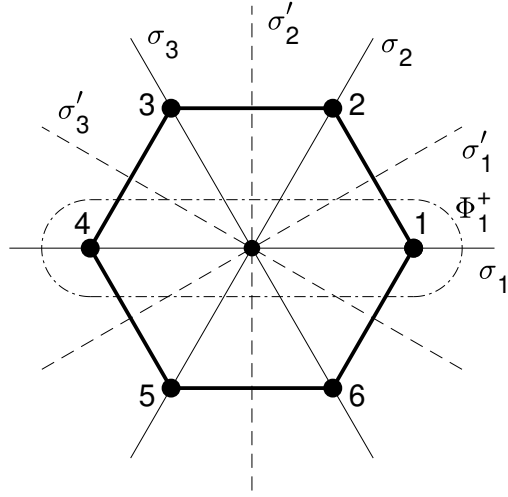}
\end{center}
\caption{Symmetry operations of a bound pair in the triangular $UV$ model (point group $C_{6v}$). Rotations are defined in the anti-clockwise direction. $\sigma$ denote mirror reflections with respect to indicated lines. The dot-dashed oval illustrates the symmetric basis function $\Phi^{+}_{1}$.} 
\label{twopart:fig:eight}
\end{figure}
\begin{table*}[t]
\renewcommand{\tabcolsep}{0.2cm}
\renewcommand{\arraystretch}{1.5}
\begin{center}
\begin{tabular}{|c|rrrrrrrrrrrr|}
\hline
 $C_{6v}$ & $E$ & $C^1_6$ & $C^1_3$ & $C_2$ & $C^2_3$ & $C^5_6$ & $\sigma_1$ & $\sigma'_1$ & $\sigma_2$ & $\sigma'_2$ & $\sigma_3$ & $\sigma'_3$ \\ \hline 
 $A_1$    &  1  &    1    &    1    &    1  &    1    &    1    &      1     &     1       &      1     &      1      &      1     &      1      \\ 
 $A_2$    &  1  &    1    &    1    &    1  &    1    &    1    &    $-1$    &   $-1$      &    $-1$    &    $-1$     &    $-1$    &    $-1$     \\ 
 $B_2$    &  1  &  $-1$   &    1    &  $-1$ &    1    &  $-1$   &      1     &   $-1$      &      1     &    $-1$     &      1     &    $-1$     \\
 $B_1$    &  1  &  $-1$   &    1    &  $-1$ &    1    &  $-1$   &    $-1$    &     1       &    $-1$    &      1      &    $-1$    &      1      \\ 
 $E_2$    &  2  &  $-1$   &  $-1$   &    2  &  $-1$   &  $-1$   &      0     &     0       &      0     &      0      &      0     &      0      \\
 $E_1$    &  2  &    1    &  $-1$   &  $-2$ &  $-1$   &    1    &      0     &     0       &      0     &      0      &      0     &      0      \\  
\hline 
 $D^{+}$  &  3  &    0    &    0    &    3  &    0    &    0    &      1     &     1       &      1     &      1      &      1     &      1      \\
 $D^{-}$  &  3  &    0    &    0    &  $-3$ &    0    &    0    &      1     &   $-1$      &      1     &    $-1$     &      1     &    $-1$     \\  
\hline
\end{tabular}
\end{center}
\caption{
Character table for representations $D^{+}$, $D^{-}$ and for all six irreps of $C_{6v}$.     
} 
\label{twopart:tab:two}
\end{table*}

The starting point is the three basis functions $\Phi^{+}_{1-3}$ in the singlet case ($\Phi_0$ will be added later) and $\Phi^{-}_{1-3}$ in the triplet case. By acting on both bases by the symmetry operations of $C_{6v}$, see Fig.~\ref{twopart:fig:eight}, two {\em reducible} representations $D^{+}$ and $D^{-}$ are constructed. For example, 
\begin{equation}
D^{+}(C^1_6) = \left( \begin{array}{ccc}
 0  &  0  &  1  \\
 1  &  0  &  0  \\
 0  &  1  &  0 
\end{array} \right) , \hspace{0.3cm}
D^{-}(\sigma^{\prime}_2) = \left( \begin{array}{ccc}
 -1  &  0  &  0  \\
  0  &  0  &  1  \\
  0  &  1  &  0 
\end{array} \right) ,
\label{twopart:eq:onehbone}  
\end{equation}
and so on. The characters of $D^{\pm}$ are given in Table~\ref{twopart:tab:two}. Applying the orthogonality theorem, one obtains the decompositions
\begin{align}
D^{+} & =  A_1 \oplus E_2 \: ,
\label{twopart:eq:onehfortyseven}  \\
D^{-} & =  B_2 \oplus E_1 \: .
\label{twopart:eq:onehfortyeight}
\end{align}
Next, we construct symmetrized linear combinations for the one-dimensional irreps $A_1$ and $B_2$ according to the formula~\cite{Petrashen1969} (the overall normalization factors are omitted)
\begin{align}
\phi^{A_1} & =  \sum_{g \in C_{6v}} \chi^{A_1}(g) \, \hat{g} \Phi^{+}_1 =  \Phi^{+}_1 + \Phi^{+}_2 + \Phi^{+}_3 \: ,
\label{twopart:eq:onehfortynine} \\
\phi^{B_2} & =  \sum_{g \in C_{6v}} \chi^{B_2}(g) \, \hat{g} \Phi^{-}_1 =  \Phi^{-}_1 - \Phi^{-}_2 + \Phi^{-}_3 \: ,
\label{twopart:eq:onehfifty}   
\end{align}
where $g$ are all the symmetry operations of $C_{6v}$ listed in the top row of Table~\ref{twopart:tab:two}. For the two-dimensional irreps $E_1$ and $E_2$, the characters $\chi$ need to be replaced with full $2 \times 2$ matrices $D^{E}(g)$ representing group elements $g$   
\begin{equation}
\phi^{E_{2,1}}_{ik} = \sum_{g \in C_{6v}} D^{E_{2,1}}_{ik}(g) \, \hat{g} \Phi^{\pm}_1  \: .
\label{twopart:eq:onehfiftyone}  
\end{equation}
The matrices $D$ for both $E_2$ and $E_1$ are given below. For a fixed index $k$, application of Eq.~(\ref{twopart:eq:onehfiftyone}) yields two linear combinations (one for $i = 1$ and another for $i = 2$) that together form a basis sought. Thus, each irrep gets two bases (one for $k = 1$ and one for $k = 2$), one of which may be trivial. In the case of $E_2$, the $k = 2$ basis turns out to be trivial (identical zero) whereas $k = 1$ produces a nontrivial basis
\begin{equation}
\phi^{E_2} = \left\{ \begin{array}{r}
2 \Phi^{+}_1 - \Phi^{+}_2 - \Phi^{+}_3  \\
               \Phi^{+}_2 - \Phi^{+}_3
\end{array} \right.  .
\label{twopart:eq:onehfiftytwo}  
\end{equation}
Equations~(\ref{twopart:eq:onehfortynine}) and (\ref{twopart:eq:onehfiftytwo}) are combined to form a new basis, Eq.~(\ref{twopart:eq:onehfiftyfour}). A similar treatment of irrep $E_1$ yields a trivial basis for $k = 1$ and a nontrivial one for $k = 2$:
\begin{equation}
\phi^{E_1} = \left\{ \begin{array}{r}
             - \Phi^{-}_2 - \Phi^{-}_3  \\
2 \Phi^{-}_1 + \Phi^{-}_2 - \Phi^{-}_3
\end{array} \right.  .
\label{twopart:eq:onehfiftythree}  
\end{equation}
This completes the analysis of symmetry at the $\Gamma$ point. Equations~(\ref{twopart:eq:onehfifty}) and (\ref{twopart:eq:onehfiftythree}) are combined to form a new basis, Eq.~(\ref{twopart:eq:onehfiftynine}).

The matrices of $E_2$ are:
\begin{equation}
D^{E_2}(E) = \left[ \begin{array}{rr} 
1 & 0 \\
0 & 1 
\end{array} \right] \: , 
\hspace{0.5cm} 
D^{E_2}(C^1_6) = \left[ \begin{array}{rr} 
- \frac{1}{2}        & - \frac{\sqrt{3}}{2}  \\
  \frac{\sqrt{3}}{2} & - \frac{1}{2}  
\end{array} \right] \: , 
\hspace{0.5cm} 
D^{E_2}(C^1_3) = \left[ \begin{array}{rr} 
- \frac{1}{2}        &   \frac{\sqrt{3}}{2}  \\
- \frac{\sqrt{3}}{2} & - \frac{1}{2}  
\end{array} \right] \: , 
\label{twopart:eq:fivehone}
\end{equation}
\begin{equation}
D^{E_2}(C_2) = \left[ \begin{array}{rr} 
1 & 0 \\
0 & 1 
\end{array} \right] \: , 
\hspace{0.5cm} 
D^{E_2}(C^2_3) = \left[ \begin{array}{rr} 
- \frac{1}{2}        & - \frac{\sqrt{3}}{2}  \\
  \frac{\sqrt{3}}{2} & - \frac{1}{2}  
\end{array} \right] \: , 
\hspace{0.5cm} 
D^{E_2}(C^5_6) = \left[ \begin{array}{rr} 
- \frac{1}{2}        &   \frac{\sqrt{3}}{2}  \\
- \frac{\sqrt{3}}{2} & - \frac{1}{2}  
\end{array} \right] \: , 
\label{twopart:eq:fivehtwo}
\end{equation}
\begin{equation}
D^{E_2}(\sigma_1) = \left[ \begin{array}{rr} 
1 &  0 \\
0 & -1 
\end{array} \right] \: , 
\hspace{0.5cm} 
D^{E_2}(\sigma'_1) = \left[ \begin{array}{rr} 
- \frac{1}{2}        &   \frac{\sqrt{3}}{2}  \\
  \frac{\sqrt{3}}{2} &   \frac{1}{2}  
\end{array} \right] \: , 
\hspace{0.5cm} 
D^{E_2}(\sigma_2) = \left[ \begin{array}{rr} 
- \frac{1}{2}        & - \frac{\sqrt{3}}{2}  \\
- \frac{\sqrt{3}}{2} &   \frac{1}{2}  
\end{array} \right] \: , 
\label{twopart:eq:fivehthree}
\end{equation}
\begin{equation}
D^{E_2}(\sigma'_2) = \left[ \begin{array}{rr} 
1 &  0 \\
0 & -1 
\end{array} \right] \: , 
\hspace{0.5cm} 
D^{E_2}(\sigma_3) = \left[ \begin{array}{rr} 
- \frac{1}{2}        &   \frac{\sqrt{3}}{2}  \\
  \frac{\sqrt{3}}{2} &   \frac{1}{2}  
\end{array} \right] \: , 
\hspace{0.5cm} 
D^{E_2}(\sigma'_3) = \left[ \begin{array}{rr} 
- \frac{1}{2}        & - \frac{\sqrt{3}}{2}  \\
- \frac{\sqrt{3}}{2} &   \frac{1}{2}  
\end{array} \right] \: . 
\label{twopart:eq:fivehfour}
\end{equation}

The matrices of $E_1$ are:
\begin{equation}
D^{E_1}(E) = \left[ \begin{array}{rr} 
1 & 0 \\
0 & 1 
\end{array} \right] \: , 
\hspace{0.5cm} 
D^{E_1}(C^1_6) = \left[ \begin{array}{rr} 
  \frac{1}{2}        & - \frac{\sqrt{3}}{2}  \\
  \frac{\sqrt{3}}{2} &   \frac{1}{2}  
\end{array} \right] \: , 
\hspace{0.5cm} 
D^{E_1}(C^1_3) = \left[ \begin{array}{rr} 
- \frac{1}{2}        & - \frac{\sqrt{3}}{2}  \\
  \frac{\sqrt{3}}{2} & - \frac{1}{2}  
\end{array} \right] \: , 
\label{twopart:eq:fivehfive}
\end{equation}
\begin{equation}
D^{E_1}(C_2) = \left[ \begin{array}{rr} 
-1 &  0 \\
 0 & -1 
\end{array} \right] \: , 
\hspace{0.5cm} 
D^{E_1}(C^2_3) = \left[ \begin{array}{rr} 
- \frac{1}{2}        &   \frac{\sqrt{3}}{2}  \\
- \frac{\sqrt{3}}{2} & - \frac{1}{2}  
\end{array} \right] \: , 
\hspace{0.5cm} 
D^{E_1}(C^5_6) = \left[ \begin{array}{rr} 
  \frac{1}{2}        &   \frac{\sqrt{3}}{2}  \\
- \frac{\sqrt{3}}{2} &   \frac{1}{2}  
\end{array} \right] \: , 
\label{twopart:eq:fivehsix}
\end{equation}
\begin{equation}
D^{E_1}(\sigma_1) = \left[ \begin{array}{rr} 
-1 &  0 \\
 0 &  1 
\end{array} \right] \: , 
\hspace{0.5cm} 
D^{E_2}(\sigma'_1) = \left[ \begin{array}{rr} 
- \frac{1}{2}        & - \frac{\sqrt{3}}{2}  \\
- \frac{\sqrt{3}}{2} &   \frac{1}{2}  
\end{array} \right] \: , 
\hspace{0.5cm} 
D^{E_1}(\sigma_2) = \left[ \begin{array}{rr} 
  \frac{1}{2}        & - \frac{\sqrt{3}}{2}  \\
- \frac{\sqrt{3}}{2} & - \frac{1}{2}  
\end{array} \right] \: , 
\label{twopart:eq:fivehseven}
\end{equation}
\begin{equation}
D^{E_1}(\sigma'_2) = \left[ \begin{array}{rr} 
1 &  0 \\
0 & -1 
\end{array} \right] \: , 
\hspace{0.5cm} 
D^{E_1}(\sigma_3) = \left[ \begin{array}{rr} 
  \frac{1}{2}        &   \frac{\sqrt{3}}{2}  \\
  \frac{\sqrt{3}}{2} & - \frac{1}{2}  
\end{array} \right] \: , 
\hspace{0.5cm} 
D^{E_1}(\sigma'_3) = \left[ \begin{array}{rr} 
- \frac{1}{2}        &   \frac{\sqrt{3}}{2}  \\
  \frac{\sqrt{3}}{2} &   \frac{1}{2}  
\end{array} \right] \: . 
\label{twopart:eq:fiveheight}
\end{equation}

\section{ \label{twopart:sec:appc}
$UV$ model on the simple cubic and tetragonal lattices    
}

\subsection{ \label{twopart:sec:appcone}
Green's functions, Eq.~(\ref{twopart:eq:onesixtyfour}), in the simple cubic case, $\alpha = \beta = \gamma$  
}

For $\alpha = \beta = \gamma$, the Watson integrals 
\begin{equation}
M^{\rm sc}_{nmk} = M^{\rm sc}_{nmk}( E ; \alpha , \alpha , \alpha ) =  
\int\limits^{\pi}_{-\pi} \!\! \int\limits^{\pi}_{-\pi} \!\! \int\limits^{\pi}_{-\pi} 
\frac{{\rm d}q_x \, {\rm d}q_y \, {\rm d}q_z}{(2\pi)^3} 
\frac{\cos{nq_x} \cos{mq_y} \cos{kq_z}}{ \vert E \vert - \alpha ( \cos{q_x} + \cos{q_y} + \cos{q_z} ) } \: ,
\label{twopart:eq:appfone}
\end{equation}
were evaluated by Joyce~\cite{Joyce1994,Joyce1998,Joyce2002}. First, one defines three auxiliary variables
\begin{equation}
w = \frac{3 \, \alpha}{ \vert E \vert }                                       \: , \hspace{0.5cm} 
\zeta = \zeta(w) = \left[ 
\frac{ 1 - \sqrt{ 1 - \frac{w^2}{9} } }{ 1 + \sqrt{ 1 - w^2 } } 
\right]^{\frac{1}{2}}                                                         \: , \hspace{0.5cm}
\kappa^2( \zeta ) = \frac{ 16 \, \zeta^3 }{ ( 1 - \zeta )^3 ( 1 + 3\zeta ) }  \: .
\label{twopart:eq:appftwo}
\end{equation}
Then the basic integral $M^{\rm sc}_{000}$ is~\cite{Joyce1994,Joyce1998} 
\begin{equation}
M^{\rm sc}_{000} = \frac{1}{\vert E \vert} 
\frac{ ( 1 - 9 \zeta^4 ) }{ ( 1 - \zeta )^3 ( 1 + 3\zeta ) } 
\left[  \frac{2}{\pi} {\bf K}(\kappa) \right]^2 .
\label{twopart:eq:appfthree}
\end{equation}
The nearest neighbor integrals can be expressed via $M^{\rm sc}_{000}$ after a simple linear transformation of the numerator
\begin{equation}
M^{\rm sc}_{100} = M^{\rm sc}_{010} = M^{\rm sc}_{001} = \frac{1}{3 \, \alpha} 
\left( \vert E \vert M_{000} - 1 \right) \: .
\label{twopart:eq:appffour}
\end{equation}
The second nearest neighbor integrals are
\begin{align}
M^{\rm sc}_{200} = M^{\rm sc}_{020} = M^{\rm sc}_{002} & = \frac{\vert E \vert}{9 \, \alpha^2} \left\{ 
- 6 + \frac{ 3( 5 - 21\zeta^2 - 21\zeta^4 - 27\zeta^6) }{( 1 - \zeta )^3 ( 1 + 3\zeta ) ( 1 + 3\zeta^2 )} 
\left[  \frac{2}{\pi} {\bf K}(\kappa) \right]^2  \right. 
\nonumber \\
& \left. - \frac{ 36( 1 - 5\zeta^2 ) }{( 1 - 9\zeta^4 )} 
\left[  \frac{2}{\pi} {\bf K}(\kappa) \right] \left[  \frac{2}{\pi} {\bf E}(\kappa) \right] 
+ \frac{ 27( 1 - \zeta )^3 ( 1 + 3\zeta ) }{( 1 - 9\zeta^4 )} 
\left[  \frac{2}{\pi} {\bf E}(\kappa) \right]^2 \right\} ,
\label{twopart:eq:appffive}   \\
M^{\rm sc}_{110} = M^{\rm sc}_{101} = M^{\rm sc}_{011} & = \frac{\vert E \vert}{\alpha^2} \left\{ 
- \frac{ ( 1 + \zeta )^2 ( 1 - 3\zeta ) }{ 4 ( 1 - \zeta ) ( 1 - 3\zeta^2 )} 
\left[  \frac{2}{\pi} {\bf K}(\kappa) \right]^2  \right. 
\nonumber \\
& \left. + \frac{ ( 1 - 5\zeta^2 ) }{( 1 - 9\zeta^4 )} 
\left[  \frac{2}{\pi} {\bf K}(\kappa) \right] \left[  \frac{2}{\pi} {\bf E}(\kappa) \right] 
- \frac{ 3( 1 - \zeta )^3 ( 1 + 3\zeta ) }{ 4 ( 1 - 9\zeta^4 ) } 
\left[  \frac{2}{\pi} {\bf E}(\kappa) \right]^2 \right\} .
\label{twopart:eq:appfsix}
\end{align}
The strong-coupling limits of the integrals are 
\begin{align}
M^{\rm sc}_{000}( \vert E \vert \gg \alpha ) & =  
\frac{1}{\vert E \vert}            + o \left( \frac{1}{\vert E \vert} \right)          ,
\label{twopart:eq:appfseven} \\
M^{\rm sc}_{100}( \vert E \vert \gg \alpha ) & =  
\frac{\alpha}{2 \vert E \vert^2}   + o \left( \frac{\alpha}{\vert E \vert^2} \right)   ,
\label{twopart:eq:appfeight} \\
M^{\rm sc}_{200}( \vert E \vert \gg \alpha ) & =  
\frac{\alpha^2}{4 \vert E \vert^3} + o \left( \frac{\alpha^2}{\vert E \vert^3} \right) ,
\label{twopart:eq:appfnine}  \\
M^{\rm sc}_{110}( \vert E \vert \gg \alpha ) & =  
\frac{\alpha^2}{2 \vert E \vert^3} + o \left( \frac{\alpha^2}{\vert E \vert^3} \right) . 
\label{twopart:eq:appften}   
\end{align}

Joyce~\cite{Joyce2002} also derived an explicit expression for $M^{\rm sc}_{300}$ and provided an algorithm how to systematically derive $M^{\rm sc}_{nmk}$ from $M^{\rm sc}_{000}$, $M^{\rm sc}_{200}$ and $M^{\rm sc}_{300}$ using several recurrence relations. Thus, {\em any} $M^{\rm sc}_{nmk}$ can be analytically expressed with products of the complete elliptic integrals ${\bf K}$ and ${\bf E}$, at least in principle.  

At threshold, $\vert E \vert = 3 \alpha$, {\em all} $M^{\rm sc}_{nmk}$ can be expressed via $M^{\rm sc}_{000}$ using the remarkable identities derived by Glasser and Boersma~\cite{Glasser2000} and Joyce~\cite{Joyce2002}
\begin{equation}
M^{\rm sc}_{200}(3\alpha) = M^{\rm sc}_{020}(3\alpha) = M^{\rm sc}_{002}(3\alpha) = 
- \frac{2}{\alpha} + \frac{10}{3} \, M^{\rm sc}_{000}(3\alpha) + \frac{2}{\pi^2 \alpha^2 M^{\rm sc}_{000}(3\alpha)} \: ,
\label{twopart:eq:appfeleven}
\end{equation}
\begin{equation}
M^{\rm sc}_{300}(3\alpha) = M^{\rm sc}_{030}(3\alpha) = M^{\rm sc}_{003}(3\alpha) = 
- \frac{13}{\alpha} + \frac{35}{2} \, M^{\rm sc}_{000}(3\alpha) + \frac{21}{\pi^2 \alpha^2 M^{\rm sc}_{000}(3\alpha)} \: .
\label{twopart:eq:appftwelve}
\end{equation}
All other $M^{\rm sc}_{nmk}(3\alpha)$ can be obtained from the recurrence relations~\cite{Glasser2000,Joyce2002}. For example, 
\begin{equation}
M^{\rm sc}_{110}(3\alpha) = M^{\rm sc}_{101}(3\alpha) = M^{\rm sc}_{011}(3\alpha) = 
\frac{1}{4} \left\{ 5 M^{\rm sc}_{000}(3\alpha) - M^{\rm sc}_{200}(3\alpha) - \frac{2}{\alpha} \right\} .
\label{twopart:eq:appfthirteen}
\end{equation}

\subsection{ \label{twopart:sec:appctwo}
Green's functions, Eq.~(\ref{twopart:eq:onesixtyfour}), in the tetragonal case, $\alpha = \beta \neq \gamma$  
}

The tetragonal Green's functions are defined as 
\begin{equation}
M^{\rm tg}_{nmk} = 
\int\limits^{\pi}_{-\pi} \!\! \int\limits^{\pi}_{-\pi} \!\! \int\limits^{\pi}_{-\pi} 
\frac{{\rm d}q_x \, {\rm d}q_y \, {\rm d}q_z}{(2\pi)^3} 
\frac{\cos{nq_x} \cos{mq_y} \cos{kq_z}}{ \vert E \vert - \alpha ( \cos{q_x} + \cos{q_y} ) - \gamma \cos{q_z} } \: .
\label{twopart:eq:appffourteen}
\end{equation}
As far as we are aware, only the basic integral $M^{\rm tg}_{000}$ has been evaluated analytically for arbitrary $|E|$, $\alpha$, and $\gamma$ by Delves, Joyce and Zucker~\cite{Joyce2001a,Joyce2001b,Joyce2003}:  
\begin{equation}
M^{\rm tg}_{000} = \frac{1}{|E|} 
\frac{2}{ \left[ \sqrt{ 1 - \frac{\alpha^2}{|E|^2} \left( 2 - \frac{\gamma}{\alpha} \right)^2 } + 
                 \sqrt{ 1 - \frac{\alpha^2}{|E|^2} \left( 2 + \frac{\gamma}{\alpha} \right)^2 } \right] 
                 \sqrt{ 1 - \kappa^2_{+} } \sqrt{ 1 - \kappa^2_{-} } } 
\!   \left[ \frac{2}{\pi} {\bf K} \left( \sqrt{ \frac{ - \kappa^2_{+} }{ 1 - \kappa^2_{+}} } \right) \! \right] 
\!\! \left[ \frac{2}{\pi} {\bf K} \left( \sqrt{ \frac{ - \kappa^2_{-} }{ 1 - \kappa^2_{-}} } \right) \! \right] 
\label{twopart:eq:appffifteen}
\end{equation}
where
\begin{align}
\kappa^2_{\pm} & =  \frac{1}{2} - \frac{1}{2} 
\frac{ \left[ \sqrt{ 1 + \frac{\alpha}{|E|} \left( 2 - \frac{\gamma}{\alpha} \right) }  
              \sqrt{ 1 - \frac{\alpha}{|E|} \left( 2 + \frac{\gamma}{\alpha} \right) } +
              \sqrt{ 1 - \frac{\alpha}{|E|} \left( 2 - \frac{\gamma}{\alpha} \right) }  
              \sqrt{ 1 + \frac{\alpha}{|E|} \left( 2 + \frac{\gamma}{\alpha} \right) } \right] }
{ \left[ \sqrt{ 1 - \frac{\alpha^2}{|E|^2} \left( 2 - \frac{\gamma}{\alpha} \right)^2 } + 
         \sqrt{ 1 - \frac{\alpha^2}{|E|^2} \left( 2 + \frac{\gamma}{\alpha} \right)^2 } \right]^3 } \times 
\label{twopart:eq:appfsixteen} \\
               &   \times  \left\{ \pm \frac{ 16 \alpha^2 }{ |E|^2 } + \sqrt{ 1 - \frac{\gamma^2}{|E|^2}} 
\left[ \sqrt{ 1 + \frac{\alpha}{|E|} \left( 2 - \frac{\gamma}{\alpha} \right) }  
       \sqrt{ 1 + \frac{\alpha}{|E|} \left( 2 + \frac{\gamma}{\alpha} \right) } +
       \sqrt{ 1 - \frac{\alpha}{|E|} \left( 2 - \frac{\gamma}{\alpha} \right) }  
       \sqrt{ 1 - \frac{\alpha}{|E|} \left( 2 + \frac{\gamma}{\alpha} \right) } \right]^2  \right\} .     
\nonumber
\end{align}
Compared with the original papers, we have applied the identity
\begin{equation}
{\bf K}(i \kappa) = \frac{ 1 }{ \sqrt{ 1 - \kappa^2 } } \: 
{\bf K} \! \left( \sqrt{ \frac{ - \kappa^2 }{ 1 - \kappa^2 } } \right) ,
\label{twopart:eq:appfseventeen}
\end{equation}
since $\kappa^2_{\pm} < 0$. Other integrals $M^{\rm tr}_{nmk}$ can be calculated numerically using the recipes of ~\ref{twopart:sec:appcthree}.

\subsection{ \label{twopart:sec:appcthree}
Green's functions, Eq.~(\ref{twopart:eq:onesixtyfour}), in the orthorhombic case, $\alpha \neq \beta \neq \gamma$  
}

The general integrals, 
\begin{equation}
M_{nmk} = 
\int\limits^{\pi}_{-\pi} \!\! \int\limits^{\pi}_{-\pi} \!\! \int\limits^{\pi}_{-\pi} 
\frac{{\rm d}q_x \, {\rm d}q_y \, {\rm d}q_z}{(2\pi)^3} 
\frac{\cos{nq_x} \cos{mq_y} \cos{kq_z}}{ \vert E \vert - \alpha \cos{q_x} - \beta \cos{q_y} - \gamma \cos{q_z} } \: ,
\label{twopart:eq:appftwenty}
\end{equation}
apply to the simple cubic $UV$ model at arbitrary ${\bf P}$, tetragonal $UV$ model ($t_x = t_y \neq t_z$) at arbitrary ${\bf P}$, and {\em orthorhombic} $UV$ model ($t_x \neq t_y \neq t_z$, not studied in this paper) also at arbitrary ${\bf P}$. They cannot be evaluated analytically. However, two out of three integrations can be carried out analytically. Thus, only one integration remains to be done numerically which leads to an efficient numerical scheme. The first integration in Eq.~(\ref{twopart:eq:appftwenty}) is elementary while the second produces complete elliptic integrals. First, we introduce three elliptic moduli:
\begin{align}
\kappa_x & \equiv  \sqrt{ \frac{ 4 \beta  \gamma }{ ( |E| - \alpha \cos{q_x} )^2 - ( \beta  - \gamma )^2 } } \: ,
\label{twopart:eq:appftwentyone}  \\
\kappa_y & \equiv  \sqrt{ \frac{ 4 \alpha \gamma }{ ( |E| - \beta  \cos{q_y} )^2 - ( \alpha - \gamma )^2 } } \: ,
\label{twopart:eq:appftwentytwo}  \\
\kappa_z & \equiv  \sqrt{ \frac{ 4 \alpha \beta  }{ ( |E| - \gamma \cos{q_z} )^2 - ( \alpha - \beta  )^2 } } \: .
\label{twopart:eq:appftwentythree}
\end{align}
Next, we apply Eqs.~(\ref{twopart:eq:appafiftytwo}) and (\ref{twopart:eq:appaseventyone}) to perform the double integration. For $M_{000}$, three equivalent representations are possible, depending on which two variables are integrated over:
\begin{align}
M_{000} & =  \int^{\pi}_{0} \frac{{\rm d} q_x}{ \pi }
\frac{2 \, {\bf K}(\kappa_x) }{\pi \sqrt{ ( |E| - \alpha \cos{q_x} )^2 - ( \beta  - \gamma )^2 } } 
\makebox[0.9cm]{}
\label{twopart:eq:appftwentyfour}  \\
        & =  \int^{\pi}_{0} \frac{{\rm d} q_y}{ \pi }
\frac{2 \, {\bf K}(\kappa_y) }{\pi \sqrt{ ( |E| - \beta  \cos{q_y} )^2 - ( \alpha - \gamma )^2 } } 
\label{twopart:eq:appftwentyfive}  \\
        & =  \int^{\pi}_{0} \frac{{\rm d} q_z}{ \pi }
\frac{2 \, {\bf K}(\kappa_z) }{\pi \sqrt{ ( |E| - \gamma \cos{q_z} )^2 - ( \alpha - \beta )^2 } }  \: .
\label{twopart:eq:appftwentysix}
\end{align}
This equivalence can be used to validate the numerical method. Other integrals are  
\begin{align}
M_{100} & =  \int^{\pi}_{0} \frac{{\rm d} q_x}{ \pi }
\frac{2 \cos{q_x} \, {\bf K}(\kappa_x) }{\pi \sqrt{ ( |E| - \alpha \cos{q_x} )^2 - ( \beta  - \gamma )^2 } } \: ,
\makebox[0.8cm]{}
\label{twopart:eq:appftwentyseven}  \\
M_{010} & =  \int^{\pi}_{0} \frac{{\rm d} q_y}{ \pi }
\frac{2 \cos{q_y} \, {\bf K}(\kappa_y) }{\pi \sqrt{ ( |E| - \beta  \cos{q_y} )^2 - ( \alpha - \gamma )^2 } } \: ,
\label{twopart:eq:appftwentyeight}  \\
M_{001} & =  \int^{\pi}_{0} \frac{{\rm d} q_z}{ \pi }
\frac{2 \cos{q_z} \, {\bf K}(\kappa_z) }{\pi \sqrt{ ( |E| - \gamma \cos{q_z} )^2 - ( \alpha - \beta )^2 } }  \: .
\label{twopart:eq:appftwentynine}   \\
M_{200} & =  \int^{\pi}_{0} \frac{{\rm d} q_x}{ \pi }
\frac{2 \cos{(2q_x)} \, {\bf K}(\kappa_x) }{\pi \sqrt{ ( |E| - \alpha \cos{q_x} )^2 - ( \beta  - \gamma )^2 } } \: ,
\makebox[0.8cm]{}
\label{twopart:eq:appfthirty}      \\
M_{020} & =  \int^{\pi}_{0} \frac{{\rm d} q_y}{ \pi }
\frac{2 \cos{(2q_y)} \, {\bf K}(\kappa_y) }{\pi \sqrt{ ( |E| - \beta  \cos{q_y} )^2 - ( \alpha - \gamma )^2 } } \: ,
\label{twopart:eq:appftthirtyone}  \\
M_{002} & =  \int^{\pi}_{0} \frac{{\rm d} q_z}{ \pi }
\frac{2 \cos{(2q_z)} \, {\bf K}(\kappa_z) }{\pi \sqrt{ ( |E| - \gamma \cos{q_z} )^2 - ( \alpha - \beta )^2 } }  \: .
\label{twopart:eq:appfthirtytwo}    \\
M_{110} & =  \int^{\pi}_{0} \frac{{\rm d} q_z}{ \pi }
\frac{ ( 2 - \kappa^2_z ) {\bf K}(\kappa_z) - 2 \, {\bf E}(\kappa_z) }{\pi \kappa_z \sqrt{ \alpha \beta  } } \: ,
\makebox[1.0cm]{}
\label{twopart:eq:appfthirtythree}  \\
M_{101} & =  \int^{\pi}_{0} \frac{{\rm d} q_y}{ \pi }
\frac{ ( 2 - \kappa^2_y ) {\bf K}(\kappa_y) - 2 \, {\bf E}(\kappa_y) }{\pi \kappa_y \sqrt{ \alpha \gamma } } \: ,
\label{twopart:eq:appftthirtyfour}  \\
M_{011} & =  \int^{\pi}_{0} \frac{{\rm d} q_x}{ \pi }
\frac{ ( 2 - \kappa^2_x ) {\bf K}(\kappa_x) - 2 \, {\bf E}(\kappa_x) }{\pi \kappa_x \sqrt{ \beta  \gamma } } \: .
\label{twopart:eq:appfthirtyfive}
\end{align}

\subsection{ \label{twopart:sec:appcfour}
Determinant of Eq.~(\ref{twopart:eq:oneeighty})   
}

We expand the determinant of Eq.~(\ref{twopart:eq:oneeighty}) in powers of $U$, $V_{xy}$, and $V_{z}$:
\begin{eqnarray}
D & = & U \cdot |V_{xy}| \cdot |V_z| \cdot \left\vert
\begin{array}{ccc}
  M_{000}  & - 2M_{100}                          & - 2M_{001}               \\
 2M_{100}  & - ( M_{000} + M_{200} + 2M_{110} )  & - 4M_{101}               \\
  M_{001}  & - 2M_{101}                          & - ( M_{000} + M_{002} )   
\end{array} \right\vert
\nonumber \\
  &   & + \, U \cdot |V_{xy}| \cdot \left\vert 
\begin{array}{cc}  
  M_{000}  & - 2M_{100}                                                     \\
 2M_{100}  & - ( M_{000} + M_{200} + 2M_{110} )                   
\end{array} \right\vert
        + \, U \cdot |V_{z}| \cdot \left\vert 
\begin{array}{cc}  
  M_{000}  & - 2M_{001}                                                     \\
  M_{001}  & - ( M_{000} + M_{002} )                   
\end{array} \right\vert
\nonumber \\
  &   & + \, |V_{xy}| \cdot |V_{z}| \cdot \left\vert 
\begin{array}{cc}  
 - ( M_{000} + M_{200} + 2M_{110} )  & - 4M_{101}                           \\
 - 2M_{101}                          & - ( M_{000} + M_{002} ) 
\end{array} \right\vert
\nonumber \\
  &   & +    U     \cdot \left[     M_{000}                        \right]
        + |V_{xy}| \cdot \left[ - ( M_{000} + M_{200} + 2M_{110} ) \right]
        +  |V_{z}| \cdot \left[ - ( M_{000} + M_{002} )            \right] + 1 \: .
\label{twopart:eq:appfthirtysix}
\end{eqnarray}
From here, one potential can be expressed via the other two.

\subsection{ \label{twopart:sec:appcfive}
Derivation of Eq.~(\ref{twopart:eq:oneeightytwo})   
}

For convenience, we introduce an anisotropy parameter, $\sigma \equiv t_z/t$. Starting with Eq.~(\ref{twopart:eq:oneseventyseven}), we wish to evaluate 
\begin{equation}
M_{000} - M_{002} = 
\int\limits^{\pi}_{-\pi} \!\! \int\limits^{\pi}_{-\pi} \!\! \int\limits^{\pi}_{-\pi} 
\frac{{\rm d}q_x \, {\rm d}q_y \, {\rm d}q_z}{(2\pi)^3} 
\frac{ 1 - \cos{2q_z}}{ 8t + 4t_z - 4t \cos{q_x} - 4t \cos{q_y} - 4t_z \cos{q_z} } \: , 
\label{twopart:eq:appfthirtyseven}
\end{equation}
in the limit $t_z \rightarrow 0$. Application of Eqs.~(\ref{twopart:eq:appftwentysix}) and (\ref{twopart:eq:appfthirtytwo}) yields
\begin{equation}
M_{000} - M_{002} = \frac{1}{ 2 \pi t }
\int^{\pi}_{0} \frac{{\rm d} q_z}{\pi} \frac{ 1 - \cos{2q_z}}{ 2 - \sigma ( 1 - \cos{q_z} ) } \:
{\bf K} \left[ \frac{2}{ 2 + \sigma ( 1 - \cos{q_z} ) } \right] .
\label{twopart:eq:appfthirtyeight}
\end{equation}
At $\sigma \rightarrow 0$, ${\bf K}$ logarithmically diverges. In the main nonvanishing order, one can neglect the $\sigma$ term in the denominator outside of ${\bf K}[\ldots]$. Using the asymptote, ${\bf K}(z \rightarrow 1) = \frac{1}{2} \ln{ \frac{16}{1-z^2} }$, and changing variables, Eq.~(\ref{twopart:eq:appfthirtyeight}) is brought to the form
\begin{equation}
M_{000} - M_{002} \approx \frac{1}{ 8 \pi t }
\left\{ \ln{ \frac{16}{\sigma} - \frac{2}{\pi} }
\int^{2}_{0} {\rm d}u \, \sqrt{u (2 - u)} \: \ln{(2-u)} \right\} \equiv 
\frac{1}{ 8 \pi t } \left\{ \ln{ \frac{16}{\sigma} } - J  \right\} .
\label{twopart:eq:appfthirtynine}
\end{equation}
The remaining integral is expressible via $\Gamma$-function (\cite{Prudnikov1998}, \#2.6.10.25)
\begin{equation}
J = \frac{2}{ \pi } \: 4 \frac{ \Gamma(\frac{3}{2}) \Gamma(\frac{3}{2}) }{ \Gamma(3) } 
\left[ \ln{2} + \frac{ \Gamma^{\prime}(\frac{3}{2}) }{ \Gamma(\frac{3}{2}) } - 
                \frac{ \Gamma^{\prime}(3) }{ \Gamma(3) }                       \right] =
\frac{1}{2} - \ln{2} \: .
\label{twopart:eq:appfforty}
\end{equation}
Finally,
\begin{equation}
M_{000} - M_{002} \approx \frac{1}{ 8 \pi t } \ln{ \frac{32}{\sigma \sqrt{e} } } =
\frac{1}{ 8 \pi t } \ln{ \frac{32 \, t}{\sqrt{e} \: t_z } } \: ,
\label{twopart:eq:appffortyone}
\end{equation}
from where Eq.~(\ref{twopart:eq:oneeightytwo}) follows by inversion.

\section{ \label{twopart:sec:appd}
$UV$ model on the BCC lattice  
}

\subsection{ \label{twopart:sec:appdone}
Derivation of Eq.~(\ref{twopart:eq:seventytwo}) 
}

Introducing potential increment: $\vert U \vert = \vert U^{\rm bcc}_{\rm cr} \vert + u$, where $u \ll \vert U^{\rm bcc}_{\rm cr} \vert$, and binding energy $E_0 = -16 t ( 1 + \Delta )$, where $\Delta \ll 1$, and expanding Eq.~(\ref{twopart:eq:seventy}) for small $u$ and $\Delta$, one obtains
\begin{equation}
\sqrt{\Delta} = \frac{u}{\vert U^{\rm bcc}_{\rm cr} \vert} 
\frac{{\bf K} \!\! \left( \frac{1}{\sqrt{2}} \right) }
{{\bf K}^{\prime} \!\! \left( \frac{1}{\sqrt{2}} \right) } \: ,
\label{twopart:eq:threehone}
\end{equation}
where the prime denotes derivative with respect to ${\bf K}$'s argument. Applying Eq.~(\ref{twopart:eq:fortyfive}) of the main text, the last expression is rewritten as 
\begin{equation}
\sqrt{\Delta} = \frac{u}{\sqrt{2} \vert U^{\rm bcc}_{\rm cr} \vert} \cdot 
\frac{{\bf K} \!\! \left( \frac{1}{\sqrt{2}} \right) }
{2 {\bf E} \!\! \left( \frac{1}{\sqrt{2}} \right) - {\bf K} \!\! \left( \frac{1}{\sqrt{2}} \right)} \: .
\label{twopart:eq:threehtwo}
\end{equation}
It is known that both ${\bf K}(2^{-1/2})$ and ${\bf E}(2^{-1/2})$ can be expressed via gamma function, $\Gamma(1/4)$. Whittaker and Watson~\cite{Whittaker1920} derive the following identities:
\begin{equation}
{\bf K} \!\! \left( \frac{1}{\sqrt{2}} \right) = \frac{1}{4 \sqrt{\pi}} 
\left[ \Gamma \left( \frac{1}{4} \right) \right]^2 ,
\label{twopart:eq:threehthree}
\end{equation}
\begin{equation}
2 {\bf E} \!\! \left( \frac{1}{\sqrt{2}} \right) - {\bf K} \!\! \left( \frac{1}{\sqrt{2}} \right) 
= 4 \pi^{\frac{3}{2}} \left[ \Gamma \left( \frac{1}{4} \right) \right]^{-2} .
\label{twopart:eq:threehfour}
\end{equation}
Substituting them into Eq.~(\ref{twopart:eq:threehtwo}) and using $\vert U^{\rm bcc}_{\rm cr} \vert$ from Eq.~(\ref{twopart:eq:seventyone}), Eq.~(\ref{twopart:eq:threehone}) becomes 
\begin{equation}
\sqrt{\Delta} = \frac{u}{t} \cdot \frac{1}{2^9 \sqrt{2} \pi^5}
\left[ \Gamma \left( \frac{1}{4} \right) \right]^{8} .
\label{twopart:eq:threehfive}
\end{equation}
Finally, applying definitions of $\Delta$ and $u$, total pair energy $E_0$ can be expressed as Eq.~(\ref{twopart:eq:seventytwo}) with the $u^2$ coefficient given by
\begin{equation}
\frac{ \left[ \Gamma \!\! \left( \frac{1}{4} \right) \right]^{16} }{ 2^{15} \pi^{10} } = 
\frac{ 2 }{ \pi^{6} } \left[ {\bf K} \!\! \left( \frac{1}{\sqrt{2}} \right) \right]^8 = 
0.29050160675... \: .
\label{twopart:eq:threehsix}
\end{equation}
%

%\end{onecolumn}

\end{document}